%% file: muon-g2-review-xiv.tex
\documentclass[12pt]{iopart}
\usepackage{graphics}
\usepackage{subfigure}
\usepackage{epsfig}
\usepackage[noxcolor]{pstricks}
\usepackage{color,pst-plot}
\newcommand{\lbl}[1]{\label{eq:#1}}
\newcommand{ \rf}[1]{(\ref{eq:#1})}
\newcommand{\setl}{\setlength\arraycolsep{2pt}} 
\newcommand{\noi}{\noindent}
\newcommand{\nn}{\nonumber}
\newcommand{\ra}{\rightarrow}
\newcommand{\Ra}{\Rightarrow}
\newcommand{\lesssim}{ {\
\lower-1.2pt\vbox{\hbox{\rlap{$<$}\lower5pt\vbox{\hbox{$\sim$}}}}\ } 
}
\newcommand{\gtrsim}{ {\
\lower-1.2pt\vbox{\hbox{\rlap{$>$}\lower5pt\vbox{\hbox{$\sim$}}}}\ } 
} 
\newcommand{\cA}{{\cal A}}

\newcommand{\cF}{{\cal F}}

\newcommand{\cH}{{\cal H}}

\newcommand{\cO}{{\cal O}}

\newcommand{\cW}{{\cal W}}

\newcommand{\Imm}{\mbox{\rm Im}}

\newcommand{\MeV}{\mbox{\rm MeV}}
\newcommand{\GeV}{\mbox{\rm GeV}}

\newcommand{\annd}{\mbox{\rm and}}

\newcommand{\als}{\alpha_{\mbox{\rm {\scriptsize s}}}}

\newcommand{\GF}{G_{\mbox{\rm {\tiny F}}}}

\newcommand{\ksls}{\not \! k}
\newcommand{\psls}{\not \! p}

\newcommand{\pslsin}{\not \! p_{1}}
\newcommand{\pslsout}{\not \! p_{2}}
\def\ms{{$\mu$s} }
\def\g2{{$(g-2)$} }
\def\be{\begin{equation}}
\def\ee{\end{equation}}
\def\ben{\begin{enumerate}}
\def\een{\end{enumerate}}
\def\bi{\begin{itemize}}
\def\ei{\end{itemize}}
\def\bea{\begin{eqnarray}}
\def\eea{\end{eqnarray}}
\def\bc{\begin{center} }
\def\ec{\end{center} }

\begin{document}
\hfill Preprint CPT-P07-2007

\title[Muon g-2: Review of Theory and Experiment]
{Muon $(g-2)$: Experiment and Theory}

\author{James P. Miller\ddag, Eduardo de Rafael\dag, and 
B. Lee Roberts\ddag  
\footnote[3]{To
whom correspondence should be addressed (roberts@bu.edu)}
}

\address{\dag\ Centre de Physique Th\'eorique, CNRS-Luminy, Case 907, 
F-13288 Marseille Cedex 9, France}

\address{\ddag\ Department of Physics, Boston University, Boston, MA 02215, USA}

\begin{abstract}

A review of the experimental and theoretical
determinations of the anomalous magnetic moment of the muon is given.
The anomaly is defined by $a=(g-2)/2$, where the Land\'e $g$-factor is  the
proportionality constant that relates the spin to the magnetic moment. 
For the muon, as well as for  the electron
and tauon, the anomaly $a$ differs slightly from zero 
(of order $10^{-3}$) because of
radiative corrections.
In the Standard Model, contributions to the anomaly
come from virtual `loops' containing 
photons and the known massive particles.
The relative contribution from heavy  particles
scales as the square of the lepton mass over the heavy  mass, 
leading to small differences
in the anomaly for $e$, $\mu$, and $\tau$.  If 
there are heavy new particles outside the
 Standard Model which couple to photons and/or leptons, the relative effect on
the  muon anomaly will be 
$ \sim (m_\mu/ m_e)^2 \approx 43\times 10^3$
larger compared with the electron anomaly. 
Because both the theoretical and experimental values 
of the muon anomaly are determined
to high precision, it is an excellent place to search for the effects of new
physics, or to constrain speculative extensions to the Standard Model.
Details of the current theoretical evaluation, and of the series of
experiments that culminates with E821
at the Brookhaven National Laboratory are given. 
 At present the theoretical  and the experimental values are known with a similar relative precision of 0.5~ppm. There is, however, 
a 3.4 standard deviation difference between the two, 
strongly suggesting the need for continued experimental and theoretical study.

\end{abstract}



\maketitle

% \tableofcontents
%\newpage

\null
This page is left blank intentionally.

\newpage

\input intro026feb07a.tex

\input E821-exp20feb07.tex

\input det-jim-lee-20feb07.tex

\input beam-dyn20feb07.tex

\input omega-p-anal20feb07.tex

\input omega-a-20feb07.tex

\input EDMforreview26feb07a.tex

\input E821summary20feb07.tex

\input theory27feb07.tex

\input future-exp20feb07.tex

\input summary20fev07.tex
\newpage

\section{References}

\end{document}

%% file: intro026feb07a.tex
\section{Introduction and History of $g$-Factors}
%
%%%%%%%%%%%%%%%%%%%%%%%%%%%%%%%%%%%%%%%%%%%%%
%  26 FEB 2007
%
% 
%
%%%%%%%%%%%%%%%%%%%%%%%%%%%%%%%%%%
%
A charged elementary particle with half-integer intrinsic spin has
a magnetic dipole moment
$\vec \mu$ aligned with its spin $\vec s$:
\be
\vec \mu = g_s \left( {q \over 2m}\right) \vec s,
\label{eq:gfact}
\ee 
where $q = \pm e$ is the charge of the particle in terms of the
magnitude of the electron charge $e$, and the proportionality constant
$g_s$ is the  Land\'e $g$-factor.
 For the charged leptons, $e$, $\mu$, or $\tau$,  $g_s$ is 
slightly greater than 2. From a quantum-mechanical view,
the magnetic moment must be directed along the spin, since in the
absence of an external electric or magnetic field, the spin provides
the only preferred direction in space.

While the behavior of the spin of an individual particle must
be treated quantum mechanically, the polarization, 
the average behavior of the spins of a large
ensemble of particles,
can be treated to a large extent as a classical collection of
spinning bar magnets.\footnote{Following the custom in the literature,
in this article
we will often refer to the motion
of the 'spin' when more correctly we are speaking of the
motion of the polarization of an ensemble of a large number of
particles which have spin.}
An externally applied magnetic field, $\vec B$,  will exert a torque
on the magnetic moment
which tends to align the polarization with the direction of the field.
However, because the magnetic moments are associated with the
intrinsic angular momentum,
the polarization precesses in this case like
a classical collection of gyroscopes in the plane perpendicular to the field.
In both the classical and quantum-mechanical
cases,  the torque on the particle is given by
$\vec N =\vec \mu \times \vec B$,
the energy depends on the orientation of the magnetic dipole,
${\mathcal H} = - \vec \mu \cdot \vec B$,
and there is a net force on the particle if the field is non-uniform. 

The study of atomic and sub-atomic magnetic moments, which
continues to this day, began in 1921 with the famous "Stern-Gerlach" 
experiments\cite{stern,sg}.
A beam of silver atoms was passed through a gradient
magnetic field to separate the individual quantum states. Two 
spatially separated bands of
atoms were observed, signifying two quantum states. From this
separation, the magnetic
moment of the silver atom was determined to be one Bohr magneton\cite{sgap},
${e\hbar / 2 m_e }$, to within 10\%.
 Phipps and Taylor\cite{phipps27} repeated the
experiment with a hydrogen beam in 1927, and they also
 observed two bands.  From the splitting of the bands
they concluded that, like silver, the magnetic moment of the hydrogen
atom was one Bohr magneton. In terms of spin, which was proposed
independently by Compton\cite{compton21}, and by Uhlenbeck
and Goudsmit\cite{UG}, the two-band structure 
indicated that the $z$-component of the
angular momentum had two values,  $s_z = \pm \hbar/2$.
Later, the magnetic moments of both atoms would be traced to the un-paired
 spin of an atomic electron, implying (from Equation~\ref{eq:gfact}) 
that the $g$-factor of the
electron is 2. 

In 1928 Dirac published his relativistic
wave equation for the 
electron. It employed four-vectors and the famous 
$4\times 4$ matrix formulation, with the spin degree of freedom
emerging in a natural way.   Dirac pointed out that his wave equation for
an electron in external electric and magnetic fields has 
``the two extra terms\footnote{This expression 
uses Dirac's original notation.} 
\be
{e h \over c}\left({\mathbf \sigma} , {\mathbf H} \right) 
+ i{e h \over c}\rho_1 \left( {\mathbf \sigma} , {\mathbf E} \right) ,
\label{eq:dirac-dpm}
\ee
\dots when divided by the factor $2m$ can be regarded as the
additional potential energy of the electron due to its new
degree of freedom\cite{Dirac28}.''
 These terms represent the magnetic
dipole (Dirac) moment and electric dipole moment interactions with
the external magnetic and electric fields. 
Dirac theory predicts that the electron
magnetic moment is one Bohr-magneton ($g_s = 2$), consistent
with the value measured by the Stern-Gerlach and Phipps-Taylor experiments.
Dirac later commented: 
``It gave just the properties that one needed for an electron.
That was an unexpected bonus for me, completely unexpected\cite{Pais98}.''
The non-relativistic reduction of the Dirac equation for the
electron in a weak magnetic field\cite{BD},
\be
i \hbar {\partial \phi \over \partial t} =
\left[ {(\vec p)^2 \over 2 m} - {e \over 2 m}
\left( \vec L + 2 \vec S \right) \cdot \vec B \right] \phi
\ee
shows clearly that $g_s = 2$, and the $g$-factor for orbital angular
momentum, $g_\ell=1$.

With the success of the Dirac equation, it was believed that the proton
should also have a $g$-factor of 2.  However, in
1933 Stern and his collaborators\cite{sternp}
showed that the $g$-factor of the proton was $\sim5.5$.
 In 1940 Alvarez and Bloch\cite{nmdm} 
found that the neutron likewise 
had a large  magnetic moment, which was a complete
surprise,  since $q=0$ for the neutron. These two results
remained quite mysterious for many years, 
but we now understand that the baryon magnetic moments are
related to their internal quark-gluon structure.

Theoretically it is useful to break the  magnetic moment into two pieces:
\be
\mu = \left(1 + a\right){q \hbar \over 2m}\quad {\rm where} \quad
a = {{g - 2} \over 2}.
\ee
The first piece, predicted by the Dirac equation
and called the Dirac moment, is $1$ in units of
the appropriate magneton, $ {q \hbar / 2m}$. The second piece is
the anomalous (Pauli) moment\cite{Bethe57}, where the dimensionless quantity
$a$ is referred to as  {\it the anomaly}.

\begin{figure}[h!]
\begin{center}
  \includegraphics[width=0.75\textwidth,angle=0]{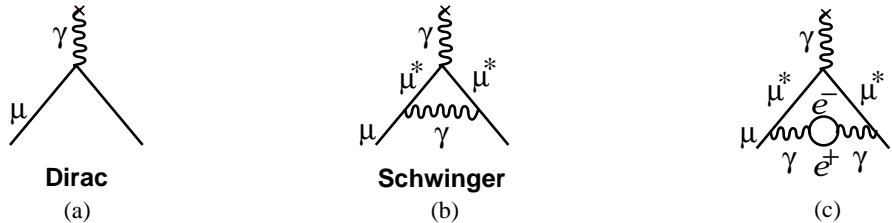}
  \caption{The Feynman graphs for: (a) $g=2$; (b) the lowest-order radiative
correction first calculated by Schwinger; and (c) the vacuum polarization
contribution, which is an example of the next-order term.
The * emphasizes that in the loop the muon is off-shell.
  \label{fg:sch}}
\end{center}
\end{figure}

In 1947, motivated by measurements of the hyperfine structure
in hydrogen that obtained splittings larger than expected from the Dirac 
theory\cite{nafe,nagel,kf1},
Schwinger\cite{Sch48}
 showed that from a theoretical viewpoint 
these ``discrepancies can be accounted for by a small additional
electron spin magnetic moment'' that arises from the lowest-order
radiative correction to the Dirac
 moment.\footnote{In response to Nafe, et al.\cite{nafe},
Breit\cite{Breit47} conjectured that
this discrepancy could be explained by the presence of
a small Pauli moment.  It's not 
clear whether this paper influenced Schwinger's work, but in a footnote
Schwinger states: ``However, Breit has not correctly drawn the consequences
of his empirical hypothesis.''}
This radiative correction, which we
 now call the one-loop correction to $g=2$, is shown diagrammatically in
Figure~\ref{fg:sch}(b).
Schwinger obtained the value
 $a_e = \alpha/(2 \pi)\simeq 0.00116\cdots$ 
 (which is also true for $a_\mu$
and $a_\tau$). In the same year, Kusch and Foley\cite{kf} measured
$a_e$ with 4\% precision, and found that the measured electron anomaly 
agreed well with Schwinger's prediction.
In the intervening time since the Kusch and Foley paper,
many improvements have been made in the precision of the
electron anomaly\cite{Rich72}.  Most recently $a_e$ 
has been measured to a relative precision
of 0.65~ppb  (parts per billion)\cite{gabeg2}, a factor of 6
improvement over the
famous experiments at the University of Washington\cite{eg2}.

The ability to calculate loop diagrams such as those shown
in Figure~\ref{fg:sch} (b) and (c)
is intimately tied to the
renormalizability of the theory, which provides a prescription to
deal with the infinities encountered in calculating radiative
corrections, and was important to the
development of quantum electrodynamics.  In his original paper
Schwinger\cite{Sch48} 
described a new procedure that transformed the Dirac Hamiltonian
to include the electron self-energy which arises from the emission and
absorption of virtual photons.  By doing so, he eliminated the
divergences encountered in calculating the lowest-order radiative correction.
He  pointed to three important features
of his new Hamiltonian: ``it involves the experimental electron mass''
(known today as the `dressed or physical
mass'); ``an electron now interacts with the
radiation field only in the presence of an external field;''
(i.e. the virtual photons from the self-energy are absent)  and
 ``the interaction energy of an electron
with an external field is now subject to a {\it finite} 
radiative correction''\cite{Sch48}.
This concept of renormalization also played an important
role in the development of the Standard Model, and the 
lowest-order contribution 
from virtual $W$ and $Z$ gauge bosons to $a_\mu$
was calculated very soon after
the electroweak theory was shown to be renormalizable\cite{EW's}.

The  diagram in Figure~\ref{fg:sch}(a) corresponds to $g=2$, 
and the first-order (Schwinger) correction which dominates the
anomaly is given in diagram~\ref{fg:sch}(b).  More generally, 
the Standard-Model value of the electron, muon or tauon
anomaly, $a({\rm SM})$, arises from  loops (radiative corrections)
containing virtual leptons, hadrons and gauge bosons.
By convention, these contributions are divided into three classes:
the dominant QED terms, like Schwinger's correction, which contain 
only leptons and photons;
 terms which involve hadrons, particularly
the hadronic vacuum polarization correction to the Schwinger term;
and electroweak terms, which contain the Higgs, $W$ and $Z$. 
Some of the terms
are identical for all three leptons, but as noted below, there are 
mass-dependent terms which are significant for the muon and tauon but
not for the electron. As a result, the muon anomaly is slightly larger
than that of the electron.

Both the QED and electroweak contributions can be calculated to high
precision. For the muon,
the former  has been calculated to 8th order (4th order in 
$\alpha/\pi$)
and the 10th order terms are currently being evaluated 
(see \S\ref{qed}). The
electroweak contributions have been calculated to two loops, as discussed
in \S\ref{EW}.
 
In contrast, the hadronic contribution to $a_\mu$ cannot
be accurately calculated from low-energy quantum chromodynamics
(QCD), and contributes the dominant theoretical
uncertainty on the Standard-Model prediction.
The lowest-order hadronic contribution, determined using
experimental data via a dispersion relation as discussed 
in \S\ref{HVPlowest}, accounts for over half of 
this uncertainty. The Hadronic light by light scattering 
contribution (see \S\ref{HLLS}) contributes the rest.
This latter contribution cannot be related to 
existing data, but rather must  be calculated 
using models that incorporate the features of QCD.

Since the value of $a_\mu$ arises from all particles 
that couple, directly or indirectly,
 to the muon,  a precision measurement of the muon anomaly 
serves as an excellent probe for new physics. Depending on their mass
and coupling strength, as yet unknown particles could make a significant
contribution to $a_\mu$. Lepton or $W$-boson substructure
might also have a measurable effect. Conversely, the comparison between the
experimental and Standard-Model values of the anomaly can be used to
constrain the parameters of speculative 
theories\cite{Czar01,Martin03,Stockinger_07}.

While the current experimental uncertainty of $\pm0.5$~ppm 
(parts per million) on the muon anomaly  is 770 times
larger than that on the electron anomaly, the former is far more sensitive
to the effects of high mass scales. In the lowest-order diagram where
mass effects appear, the contribution of heavy virtual particles
scales as 
$(m_{\rm lepton}/m_{\rm HV})^2$,  giving the muon a factor 
of $(m_\mu/m_e)^2\simeq 43000$  increase in sensitivity over the electron. 
Thus, at a precision of 0.5 ppm, the muon anomaly
is sensitive to physics at a few hundred GeV scale, while the 
reach of the electron anomaly at the current experimental uncertainty of
0.65 ppb is limited to the 
few hundred MeV range.

To observe effects of new physics in $a_e$, 
it is necessary to have a sufficiently precise  standard-model value
 to compare with the experimental value. 
Since the QED contribution  depends directly on
a power series expansion in the fine-structure constant
 $\alpha$,  a meaningful
comparison requires that $\alpha$ be known
 {\it from an independent experiment} to the same relative precision 
as $a_e$.
At present, independent measurements of $\alpha$  have a precision
of $\sim 6.7$~ppb\cite{alpha1},  about ten times larger than the error
on $a_e$. 
On the other hand, if one assumes that there are no
new physics contributions to $a_e$,
i.e. the  Standard-Model 
radiative corrections are the {\it only} source of $a_e$, 
one can use the measured value of $a_e$ and the standard-model calculation
to obtain the most precise 
determination of $\alpha$\cite{gabalpha}.

\subsection{Muon Decay \label{sct:mudecay}}

Since the kinematics of muon decay are central to
the measurements of $a_\mu$,  we discuss the general features
 in this section.  Specific issues that relate to the design of
E821 at Brookhaven are discussed in the detector 
section, \S~\ref{sct:calodesign}.

After the discovery of parity violation in $\beta$-decay\cite{wu}
and in muon decay\cite{garwin1,fried}, 
it was realized that one could
make beams of polarized muons
 in  the pion decay reactions 
$$
\pi^{-} \rightarrow \mu^{-} + \bar \nu_{\mu}
\quad {\rm or} \quad 
\pi^{+} \rightarrow \mu^{+} +  \nu_{\mu}.
$$
The pion has spin zero, the neutrino  (antineutrino) has helicity of
-1 (+1), and the forces in the decay process are very
short-range, so the orbital angular momentum in the 
final state is zero. Thus conservation of angular momentum requires that
the $\mu^-$  ($\mu^+$) helicity be +1 (-1)
 in the pion rest frame.  The muons from pion decay at rest are
always polarized.

From a beam of pions traversing a straight beam-channel consisting of
focusing and defocusing elements (FODO), a beam of polarized,
high energy muons can be produced by selecting the "forward"
or "backward" decays.
The forward muons are
those produced, in the pion rest frame, nearly parallel to the
pion laboratory momentum and are the decay muons with the highest laboratory
momenta. The backward muons are those produced
nearly anti-parallel to the pion momentum and have the lowest laboratory
momenta. The forward $\mu^-$ ($\mu^+$) are polarized along
(opposite)  their lab momenta respectively; the polarization
reverses for backward muons.  The
E821 experiment uses forward muons,
produced by a pion beam with an average momentum of
 $p_{\pi}\approx 3.15$ GeV/c. Under a Lorentz transformation from the
 pion rest frame to the laboratory frame, 
the decay muons have momenta in the range $0< p_{\mu}<3.15 $ GeV/c.
After momentum selection, the average momentum of muons stored in
the ring is 3.094 GeV/c,
and the average polarization is in excess of 95\%.

The pure $(V-A)$ three-body weak decay of the muon, 
$\mu^{-} \rightarrow e^{-}+ \nu_{\mu}+ \bar \nu_e$ or
$\mu^{+} \rightarrow e^{+}+ \bar \nu_{\mu}+  \nu_e$,
is ``self-analyzing'',
that is, the parity-violating correlation between the
directions in the muon rest frame (MRF) of the decay electron
and the muon spin can provide information on the muon spin orientation at the
time of the decay.  (In the following text, we use `electron' generically for 
either $e^-$ and $e^+$ from the decay of the $\mu^\mp$.) 
Consider the case when the decay electron
has the maximum allowed energy in the MRF,
$E'_{\rm max} \approx (m_\mu c^2)/2= 53$~MeV.
The neutrino and anti-neutrino
are directed parallel to each other and
at $180^\circ$ relative to the electron direction. The $\nu\bar\nu$
pair carry
zero total angular momentum, since the neutrino is left-handed and the 
anti-neutrino is right-handed;
 the electron  carries the muon's angular momentum of
${1/ 2}$. The electron, being a lepton,
is preferentially emitted left-handed in a weak
decay, and thus has a larger probability to be emitted with
its momentum {\it anti-parallel}
rather than parallel to the
$\mu^-$ spin. By the same line of reasoning, in $\mu^+$ decay,
the highest-energy 
positrons are emitted {\it parallel} to the muon spin in the MRF.

In the other extreme, when the electron kinetic energy 
is zero in the MRF, the neutrino and anti-neutrino are emitted back-to-back
 and carry a total
angular momentum of one. 
In this case, the electron spin is directed opposite to the muon spin
in order to
conserve angular momentum.
 Again, the electron is preferentially emitted with
helicity -1, however in this case its momentum
will be preferentially directed {\it parallel} to the $\mu^-$ spin. 
The positron, in $\mu^+$ decay, is preferentially emitted with
helicity +1, and therefore its momentum
will be preferentially directed {\it anti-parallel} to the $\mu^+$ spin. 

\begin{figure}[h]
\begin{center}
\subfigure[Muon Rest Frame]
{\includegraphics[width=0.45\textwidth,angle=0]{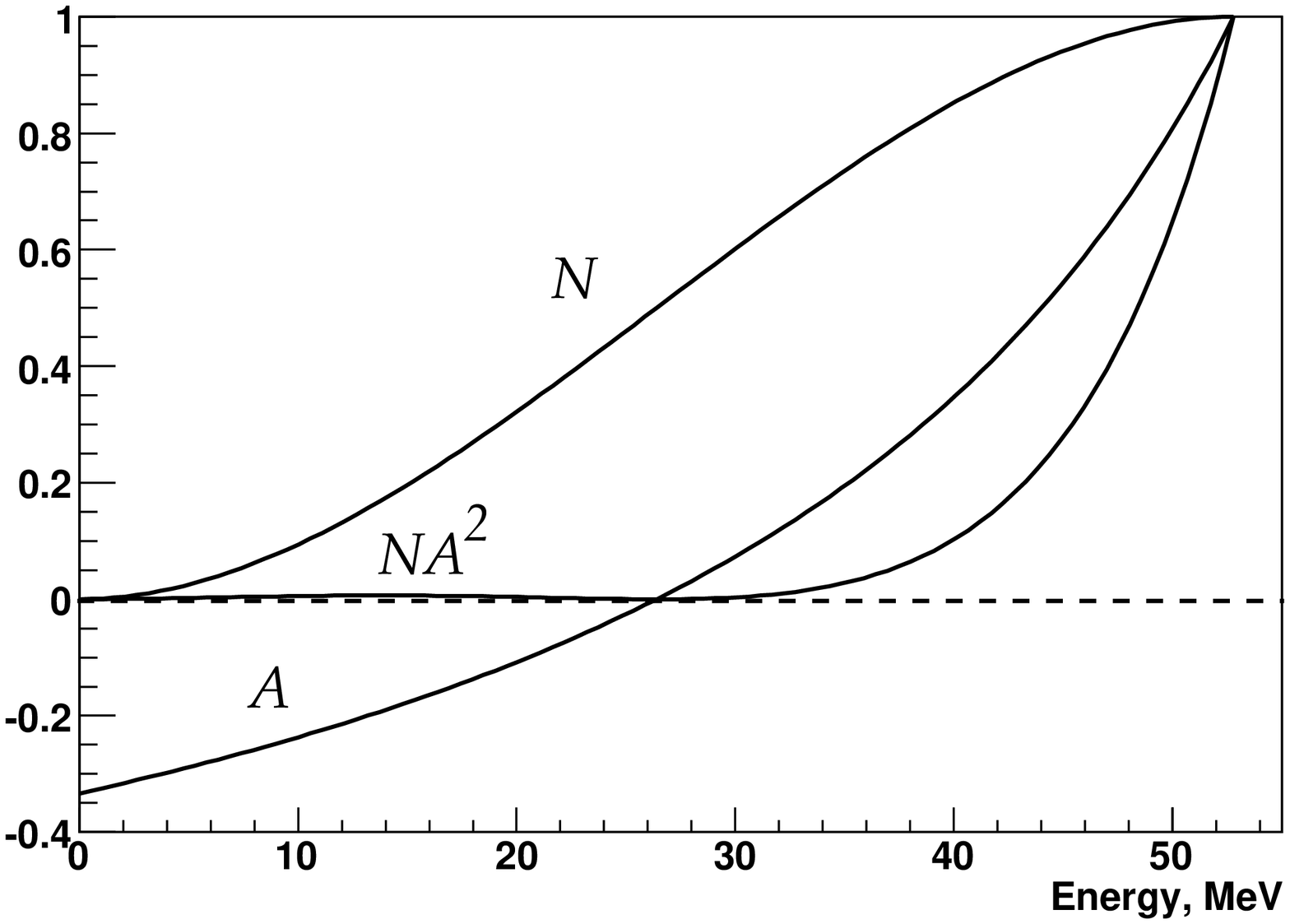}}
\subfigure[Laboratory Frame]
{\includegraphics[width=0.45\textwidth,angle=0]{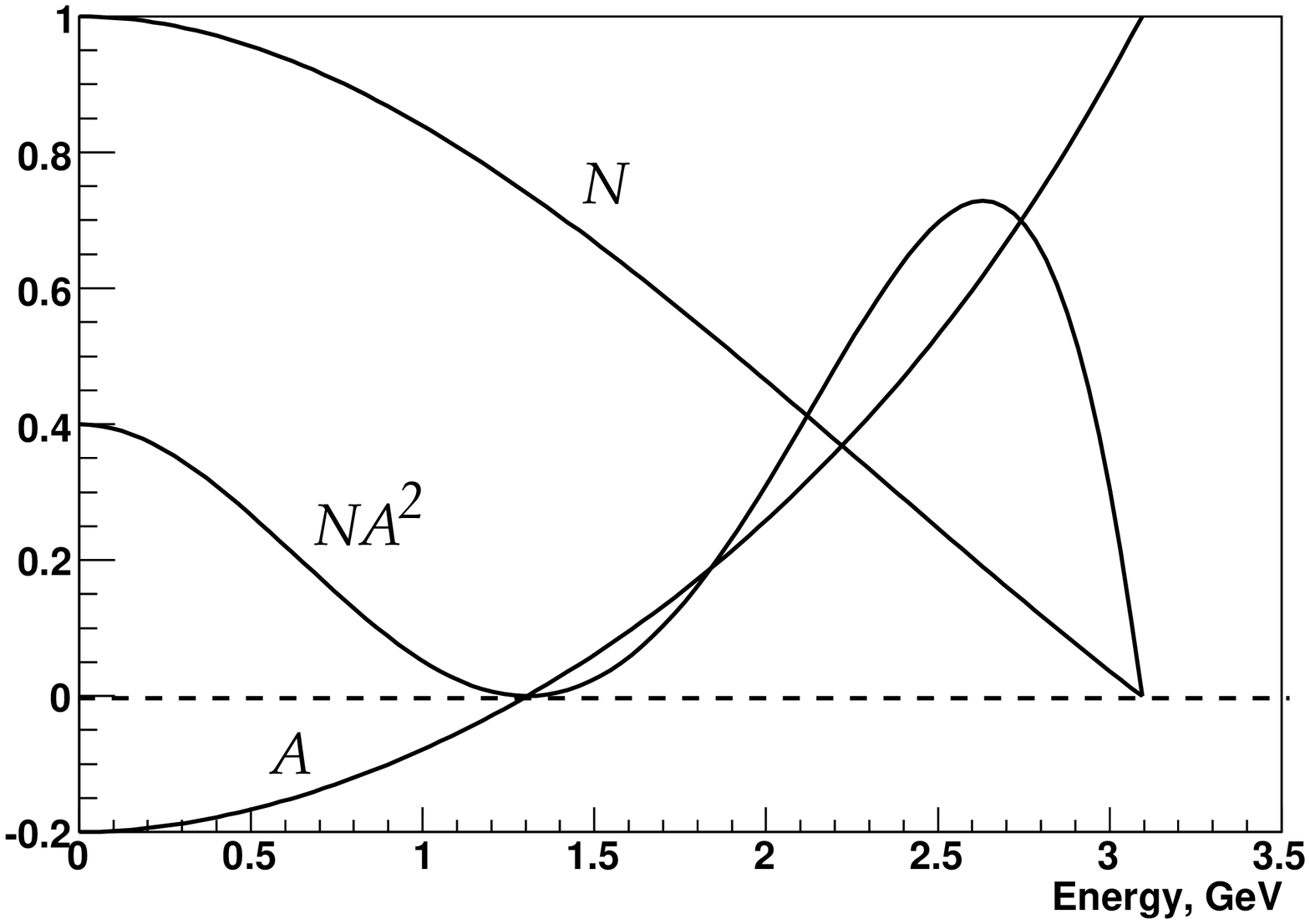}}
\end{center}
\caption{Number of decay electrons per unit energy, 
N (arbitrary units),
 value of the asymmetry $A$, and relative figure of merit
$NA^2$ (arbitrary units) as a function of electron energy.
Detector
acceptance has not been incorporated, and the polarization is unity.
For the third CERN experiment and E821, $E_{max}\approx 3.1$~GeV
($p_\mu = 3.094$~GeV/c) in the laboratory frame.
}
\label{fg:differential_na}
\end{figure}

With the approximation that the energy of the 
decay electron
$E'>> m_ec^2$,
the differential decay distribution in the muon rest frame
is given 
by\cite{kono},
\be
dP(y',\theta') \propto n'(y')
 \left[ 1 \pm { \mathcal A} (y')\cos \theta'\right] dy' d\Omega' 
\label{eq:cmdecaydist}
\ee
where $y'$ is the momentum fraction of the electron, 
$y' = p'_e / p'_{e\ \rm  max}$, $d \Omega'$ is the solid angle,
 $\theta' = \cos^{-1}{(\hat p_e'\cdot \hat s)}$ is
the angle between the muon spin and $\vec p\ '_e$,
$ p'_{e\ \rm  max}c \approx E_{max}'$,
and the $(-)$ sign is
for negative muon decay.  The number distribution
$n(y')$ and the decay asymmetry ${\mathcal A}(y')$ are given by
\be
n(y') = 2y'^2(3-2y') \quad {\rm and}
 \quad { \mathcal A }(y') = { 2y' - 1 \over 3 - 2 y'}.
\label{eq:decayna}
\ee
Note that both the number and
asymmetry reach their maxima at $y'=1$, and the asymmetry changes sign at
$y'={1\over 2}$, as shown in Figure~\ref{fg:differential_na}(a).

The CERN and Brookhaven based muon $(g-2)$ experiments stored relativistic
muons in a uniform magnetic field, which resulted in the muon spin precessing 
with constant frequency
$\vec \omega_a$, while the muons traveled in circular orbits.
If {\it all} decay electrons were counted, the number detected as a function
of time would be a pure exponential; 
therefore we seek cuts on the laboratory observables to 
select subsets of decay electrons whose numbers oscillate at the precession
frequency. Recalling that the number of decay electrons in the MRF varies
with the angle between the electron and spin directions, the electrons in the
subset should have a preferred direction in the MRF when weighted according
to their asymmetry as given in Equation~\ref{eq:cmdecaydist}.
At $p_\mu \approx 3.094$~GeV/c the directions
of the electrons resulting from muon decay
in the laboratory frame are
very nearly parallel to the muon momentum regardless of
their energy or direction in the MRF.
Therefore the
only practical remaining cut is on the electron's laboratory energy.
Typically, selecting an energy subset will have the desired effect:
there will be a net component of
electron MRF
momentum either parallel or antiparallel to the laboratory muon direction.
For example, suppose that we only count electrons with the highest laboratory
energy, around 3.1 GeV.
Let $\hat z$ indicate the direction of the muon laboratory momentum. 
The highest-energy electrons in the laboratory are those
near the maximum MRF energy of 53 MeV, and with MRF directions
nearly parallel to $\hat z$. 
There are more of these high-energy electrons when the $\mu^-$ spins
are in the direction opposite to $\hat z$ than when the spins are parallel to
$\hat z$. 
Thus the number of
decay electrons reaches a maximum when the muon spin direction is
opposite to $\hat z$, and a minimum when they are parallel.
As the spin precesses the number of high-energy electrons will oscillate
with frequency $\omega_a$.
More generally, at laboratory energies above
$\sim 1.2$~GeV, the electrons have a preferred average MRF
direction
parallel to $\hat z$ (see Figure~\ref{fg:differential_na}). 
 In this discussion, it is assumed that
the spin precession vector, $\vec\omega_a$,
is independent of time, and therefore the angle between the spin component in
the orbit plane and the muon momentum direction is given by
$\omega_{a}t+\phi$, where $\phi$ is a constant.

Equations~\ref{eq:cmdecaydist} and ~\ref{eq:decayna}
can be transformed to the laboratory
frame to give the electron number oscillation with time
as a function of electron energy,
\be
N_d(t,E) = 
N_{d0}(E)e^{-t/\gamma\tau}[1+A_d(E)\cos(\omega_{a}t+\phi_d(E))],
\label{eq:fivepd} 
\ee
or, taking all electrons above threshold energy $E_{th}$,
\be
N(t,E_{th}) = 
N_{0}(E_{th})e^{-t/\gamma\tau}[1+A(E_{th})\cos(\omega_{a}t+\phi(E_{th}))].
\label{eq:fivep} 
\ee

In Equation~\ref{eq:fivepd} the differential quantities are,
\be
A_d(E)={\mathcal P}  {-8y^2+y+1 \over 4y^2-5y-5},\quad
N_{d0}(E)\propto (y-1)(4y^2-5y-5),
\ee
and in Equation~\ref{eq:fivep},

\be
N(E_{th})\propto (y_{th}-1)^2(-y_{th}^2+y_{th}+3),
\qquad
A(E_{th})={ \mathcal P} {y_{th}(2y_{th}+1) \over -y_{th}^2+y_{th}+3}.
\ee

In the above equations, $ y={E / E_{max}}$, $ y_{th}={E_{th} / E_{max}}$,
 ${\mathcal P}$ is the
polarization of the muon beam,
and $E$, $E_{th}$, and $E_{max}=3.1$ GeV
are the electron laboratory energy, threshold energy, and 
maximum energy, respectively.

\begin{figure}[h]
\begin{center}
\subfigure[No detector acceptance or energy resolution included]
{\includegraphics[width=0.45\textwidth,angle=0]{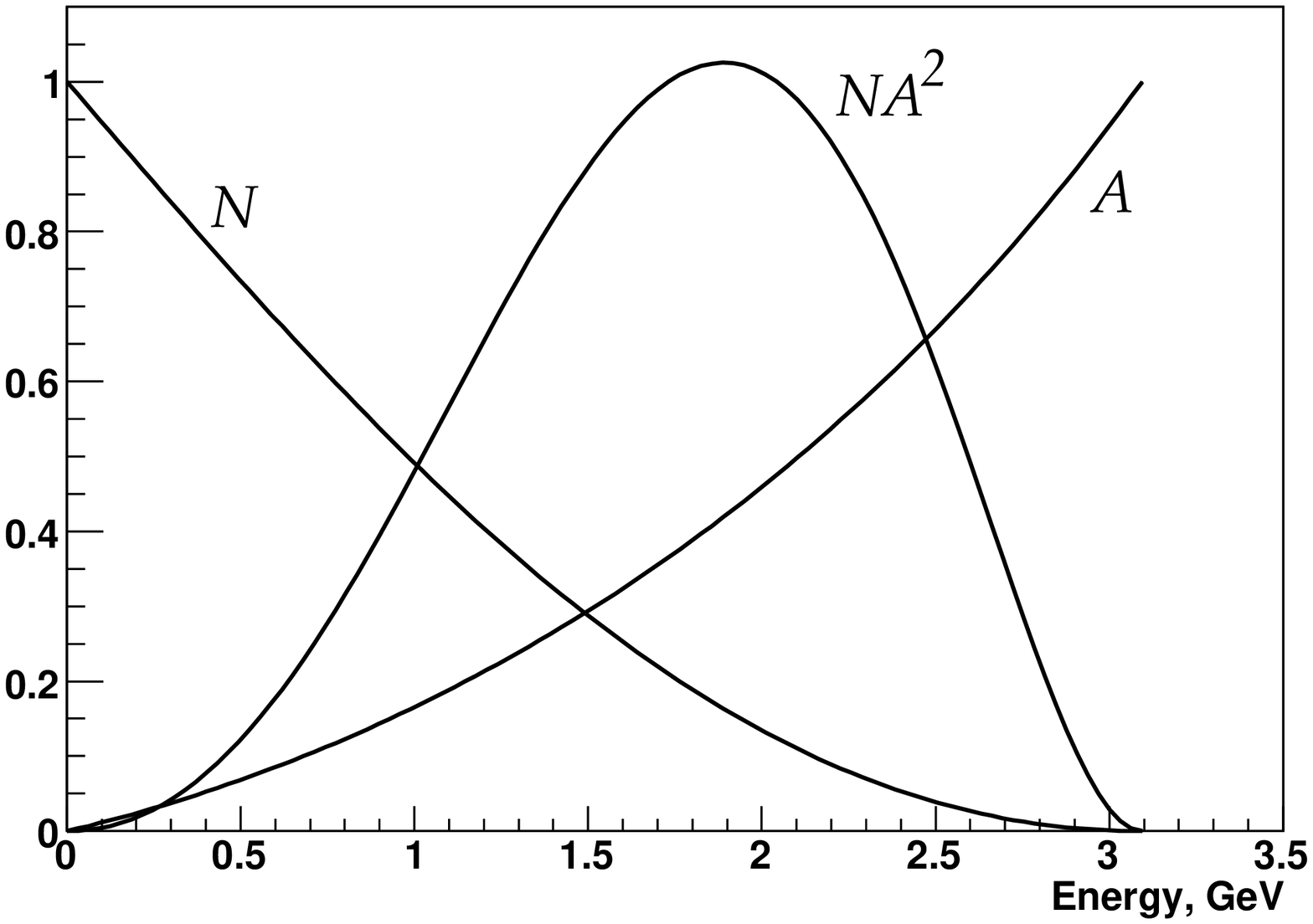}}
\subfigure[Detector acceptance and energy resolution included]
{\includegraphics[width=0.45\textwidth,angle=0]{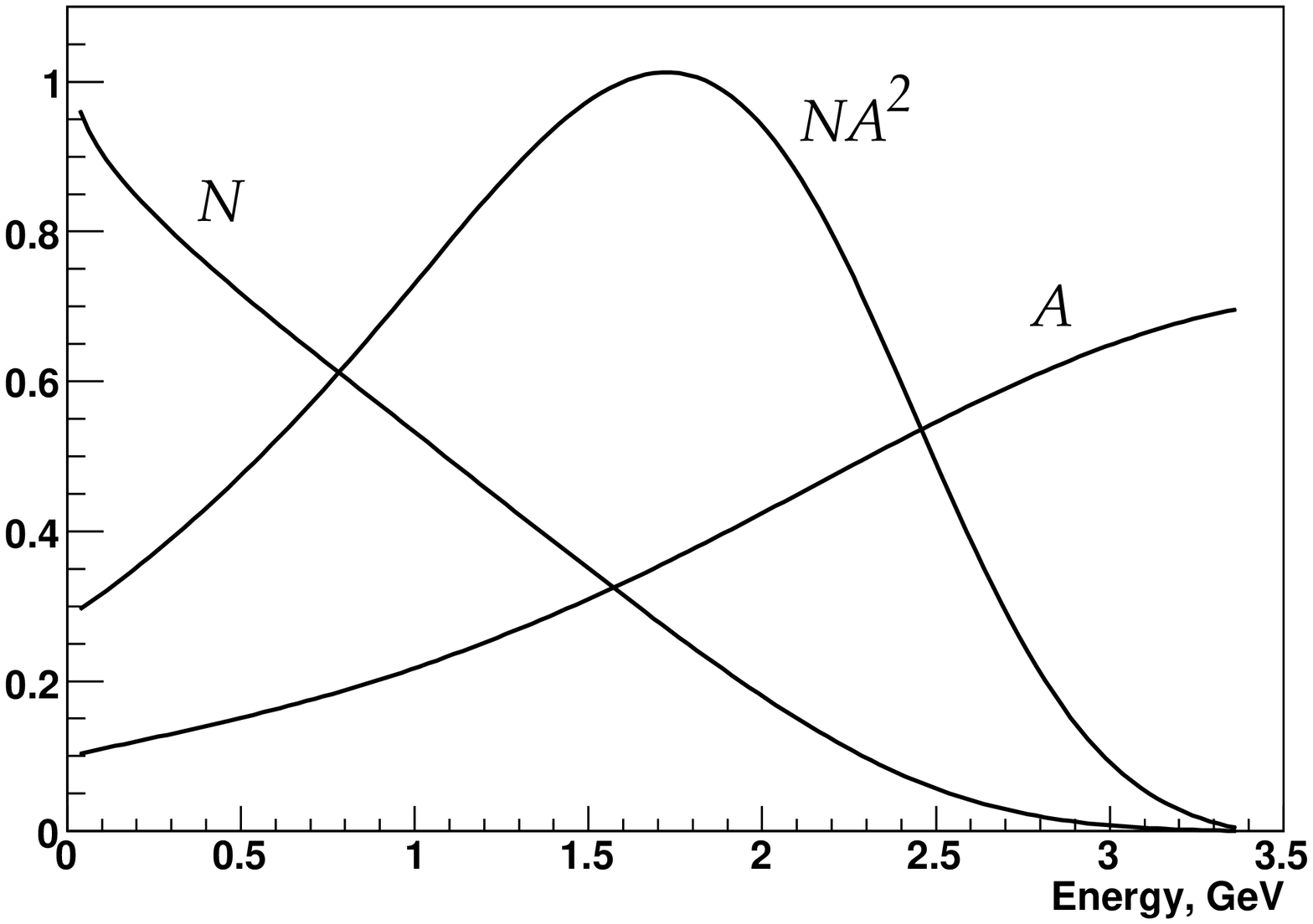}}
\caption{The integral $N$, $A$, and
$NA^2$ (arbitrary units)
for a single energy-threshold as a function of the threshold energy;
(a) in the laboratory frame, not including  and 
(b) including 
the effects of detector acceptance and
energy resolution for the E821 calorimeters discussed below. 
For the third CERN experiment and E821, $E_{max}\approx 3.1$~GeV
($p_\mu = 3.094$~GeV/c) in the laboratory frame.
\label{fg:integral_na}}
\end{center}
\end{figure}

The 
fractional statistical error on the precession frequency, 
when fitting data collected over many muon lifetimes
to the five-parameter function (Equation~\ref{eq:fivep}), is given by
\be
\delta \epsilon = { \delta \omega_a \over \omega_a} =
{\sqrt{2} \over 2 \pi f_a \tau_{\mu} N^{1\over 2}A}.
\label{eq:fracterr}
\ee
where $N$ is the total number of electrons,
and $A$ is the asymmetry, in the given data sample. 
For a fixed magnetic field and muon momentum,
 the statistical figure of merit is $NA^2$, 
the quantity to be maximized in order to minimize the statistical
 uncertainty.

The energy dependences of the numbers and asymmetries used in
Equations~\ref{eq:fivepd} and ~\ref{eq:fivep}, along with
the figures of merit $NA^2$,
are plotted in Figures ~\ref{fg:differential_na} and ~\ref{fg:integral_na}
for the case of E821.
The statistical power is greatest for electrons
at 2.6 GeV (Figure~\ref{fg:differential_na}). 
When a fit is made to all electrons above some energy threshold, 
the optimal threshold energy 
is about 1.7-1.8~GeV (Figure~\ref{fg:integral_na}).

\subsection{History of the Muon $(g-2)$ Experiments}

In 1957, at the Columbia-Nevis cyclotron, the spin rotation of a muon
in a magnetic field was observed for the first time. The torque exerted
by the magnetic field on the
muon's magnetic moment produces a spin precession frequency
\be
\vec \omega_S= - {qg \vec B \over 2m}  
- {q \vec B \over \gamma m }(1-\gamma),
\label{eq:freq-spin}
\ee
where $\gamma = (1 - \beta^2)^{-{1\over 2}}$, with $\beta = v/c$.
 Garwin et al.\cite{garwin1} found that the observed rate of spin rotation was
consistent with $g=2$.

In a subsequent paper, 
Garwin et al.\cite{garwin2} reported the results of a second experiment,
 a measurement of the
muon anomaly to a relative precision of 6.6\%,
$g_\mu \geq 2(1.00122 \pm 0.00008)$, with the inequality coming from the poor
knowledge of the muon mass.  
 In a note added in proof,
the authors reported that a new measurement of the muon mass 
permitted them to conclude
that $g_\mu = 2(1.00113_{-0.00012}^{+0.00016})$.
 Within the experimental
uncertainty, the muon's anomaly was equal to that of the electron.  
More generally, the muon was shown to behave like a heavy electron, a spin 1/2 
fermion obeying QED.

In 1961 the first of three experiments to be
carried out at CERN reported a more precise result obtained
at the CERN synchrocyclotron\cite{cern1}.  In this experiment, highly
polarized muons of momentum
90 MeV/c were injected into a 6-meter long magnet with a graded magnetic
field. As the muons moved in almost circular orbits which drifted transverse
to
the gradient, their spin vectors precessed with respect to their momenta. 
The rate of spin precession is readily calculated.
Assuming that $\vec \beta \cdot \vec B =0$, 
the momentum vector of a muon undergoing 
cyclotron motion rotates with frequency
\be
\vec  \omega_C = - {q \vec B \over m \gamma}.
\label{eq:freq-cyc}
\ee
The spin precession {\it relative to the momentum} 
occurs at the difference frequency, $\omega_a$, between the spin frequency
in Equation~\ref{eq:freq-spin} and the cyclotron frequency,
\be
\vec \omega_a = \vec \omega_S - \vec \omega_C 
= - \left( {g-2 \over 2} \right) {q \vec B \over m}
= - a_\mu { q \vec B \over m}.
\label{eq:diffreq}
\ee
The precession frequency $\omega_a$ has the important property that it is 
{\it independent} of the muon momentum.
When the muons reached the end of the magnet, they were
extracted and their polarizations measured.
The polarization measurement exploited the
self-analyzing property of the muon:
more electrons are emitted opposite than along the muon spin.
For an ensemble of muons, $\omega_a$ is the average observed frequency,
and  $B$ is the average 
magnetic field obtained by folding the muon distribution
with the magnetic field map.

The result from the first CERN experiment was\cite{cern1} 
$ a_{\mu^+} = 0.001\,145(22)$ (1.9\%), which can be compared with
${\alpha/ 2\pi} = 0.001\,161\,410 \cdots$.  With additional data
this technique resulted in the first observation of the effects
of the $(\alpha/\pi)^2$ 
term in the QED expansion\cite{cern1a}.

The second CERN experiment used a 
muon storage ring operating at 1.28~GeV/$c$.   Vertical focusing 
was achieved with magnetic gradients in the storage-ring field. 
While the use of magnetic gradients to focus a 
charged particle beam is quite common, it makes a precision determination of
the (average) magnetic field which enters into 
Equation~\ref{eq:diffreq} rather 
difficult for two reasons.  
Since the field is not uniform, 
information on where the muons are in the storage ring is needed
to correct the average field for the gradients encountered.  Also,
the presence of gradient magnetic fields broadens the NMR line-shape,
which reduces the precision on the NMR measurement of the magnetic field.

A temporally narrow bunch of  $10^{12}$ protons at 10.5~GeV/$c$
from the CERN proton synchrotron (PS) struck a  target inside the
storage ring, producing pions, a few of which decay in such
a way that their 
daughter muons are stored in the ring. A huge flux of 
other hadrons was also produced, which presented a 
challenge to the decay electron detection system.
The electron detectors could only be placed in positions around the ring
well-removed from the production target, which
limited their geometric coverage.
Of the pions which circulated in the ring for several turns and then decayed, 
only one in a thousand produced a stored muon, resulting in
about 100 stored muons per injected proton bunch.
The polarization of the stored muons was 26\%\cite{cern2}.

In all of the experiments discussed in this review,
the magnetic field was measured by observing the Larmor frequency
of stationary protons, $\omega_p$, in nuclear magnetic resonance (NMR) probes. 
Ignoring the signs of the muon and proton charges, 
we have the two equations:
\be
 \omega_a =   {e \over m} a_{\mu}  B \quad {\rm and} \quad
\omega_p = g_p \left( {e B \over 2m_p}\right),
\ee
where $\omega_p$ is the Larmor frequency for a {\it free} proton.
Dividing and solving for $a_{\mu}$ we find
\be
a_{\mu} 
= { {\tilde \omega_a / \omega_p} \over \lambda 
-  {\tilde \omega_a / \omega_p}}=
{ {\mathcal R} \over \lambda -  {\mathcal R}}  ,
\label{eq:lambda}
\ee
where the fundamental constant $\lambda = \mu_{\mu} / \mu_p$, and
${\mathcal R}= \tilde \omega_a / \omega_p$.
 The tilde on $\tilde \omega_a$ indicates that
the measured frequency has been adjusted for any necessary
(small) corrections, such as the pitch and radial electric field corrections
discussed  in \S\ref{sct:beam-dyn}.
Equation~\ref{eq:lambda} was used in the second CERN experiment, and in all of
the subsequent muon \g2 experiments, to determine the value of the anomaly. 
The most accurate value of
the ratio $\lambda$ is derived from measurements of the
muonium (the $\mu^+e^-$ atom) hyperfine structure\cite{liu,PDG}, i.e., 
from measurements made with the $\mu^+$, so that it is more properly  written
$\lambda_+ = \mu_{\mu^+} / \mu_p$. Application of Equation~\ref{eq:lambda}
to determine $a_{\mu^-}$ requires the assumption of CPT invariance.

The second CERN experiment  obtained
$a_\mu= (11661.6 \pm 3.1)\times 10^{-7}$ (0.27\%)\cite{cern2}, 
testing quantum electrodynamics for the muon to the three-loop level.
Initially this result was 1.8 standard deviations above the QED
value, which stimulated a new calculation of the QED
light-by-light scattering contribution\cite{Kinetal69}
that brought theory and experiment into agreement.

In the third experiment at CERN, several of the undesirable features
of the second experiment were eliminated.  A secondary pion beam 
produced by the PS was brought into the storage ring, which was 
placed a suitable distance away from the production target.
The $\pi \rightarrow \mu$ decay was used to kick muons onto a 
stable orbit, producing a stored muon beam with 
polarization of 95\%.  
The ``decay kick'' had an injection efficiency of 125~ppm, with
many of the remaining pions striking objects in the storage ring and
causing background in the detectors after injection. While still significant,
the background in the electron calorimeters
associated with beam injection was
much reduced compared to the second CERN experiment.

 Another very important improvement in the third CERN experiment,
adopted by the Brookhaven based
experiment as well, was the use of 
electrostatic focusing\cite{cern3a}, which represented a major 
conceptual breakthrough. The magnetic gradients
used by the second CERN experiment made  the precise determination of
the average magnetic field difficult, as discussed above.
The more uniform field greatly improves the performance of NMR techniques,
to provide a measure of the field at the
$10^{-7}$ to  $10^{-8}$ level.  With electrostatic focusing
Equations~\ref{eq:freq-spin} and \ref{eq:freq-cyc}
must be modified.
The cyclotron  frequency becomes 
 \be
\vec  \omega_C =- {q  \over m}\left[ { \vec B \over \gamma} 
- {\gamma \over \gamma^2 -1} 
\left({ \vec \beta \times \vec E \over c}\right) \right],
\label{eq:cyc-E}
\ee
 and the spin precession  frequency becomes\cite{TBMT}
\be
\hspace*{-2.5cm} \vec \omega_S = -{q \over  m } \left[
\left({g \over 2} -1 + {1\over \gamma}\right) \vec B
- \left( {g \over 2} -1 \right){\gamma \over \gamma + 1}(\vec \beta \cdot
\vec B)\vec \beta -
\left( {g \over 2} - {\gamma \over \gamma + 1}\right) 
\left( { \vec \beta \times \vec E \over c}\right )
\right]\,.
\label{eq:fullfreq}
\ee
Substituting for  $a_{\mu} = (g_{\mu} -2)/2$, we find that
the spin difference frequency is
\be
\vec \omega_a 
=  - {q \over m}
\left[ a_{\mu} \vec B -  a_{\mu}\left( {\gamma \over \gamma + 1}\right)
(\vec \beta \cdot \vec B)\vec \beta 
- \left( a_{\mu}- {1 \over \gamma^2 - 1} \right) 
{ {\vec \beta \times \vec E }\over c }\right]\,.
\label{eq:Ediffreq}
\ee
If $\vec \beta \cdot \vec B =0 $, this reduces to
\be
\vec \omega_a = 
 - \ {q \over m} 
\left[ a_{\mu} \vec B -
\left( a_{\mu}- {1 \over \gamma^2 - 1} \right)
{ {\vec \beta \times \vec E }\over c }
\right]\,.
\label{eq:omega}
\ee
For $\gamma_{\rm magic} = 29.3$ ($p_{\mu} = 3.09$~GeV/$c$), the 
second term vanishes; one is left with the simpler result of
Equation~\ref{eq:diffreq}, and the electric
field does not contribute to the spin precession relative to the 
momentum. The great
experimental advantage of using the magic $\gamma$ is clear.
By using electrostatic focusing, a gradient B field is not needed for
focusing and B can be made as uniform as
possible. The spin precession is determined almost completely by
Equation~\ref{eq:diffreq}, which is independent of muon momentum; {\it all}
muons precess at the same rate. Because of the high uniformity of
the B-field, a precision knowledge of the stored beam trajectories
in the storage region is not required.
  
Since the spin precession period of 4.4~$\mu$s  is much 
longer than the cyclotron period of 149~ns,
during a single precession period a muon samples the magnetic
field over the entire azimuth 29 times.
Thus the important quantity to be made uniform
is the magnetic field averaged over 
azimuth.  The CERN magnet was shimmed to an average azimuthal
uniformity of $\pm 10$~ppm, with the
 total systematic error from all issues related to the magnetic field
of $\pm 1.5$~ppm\cite{cern3b}.

Two small corrections to $\omega_a$ are required to form
$\tilde\omega_a$, which is used in turn to determine $a_{\mu}$
in  Equation~\ref{eq:lambda}.
1) The vertical pitching motion (betatron oscillations)
about the mid-plane of the storage region 
means that $\vec \beta \cdot \vec B$ differs slightly from zero
(see Equation~ref{eq:Ediffreq}).
2) Only muons with
the central radius of 7112~mm are at the  magic value of $\gamma$, so
that a radial electric field will effect the spin precession of muons
with $\gamma \neq \gamma_{magic}$. The
small corrections are described in \S\ref{sct:p-E}.

The design and shimming techniques  of the CERN III storage ring
were chosen to produce a  very stable magnet mechanically.
 After cycling the magnet power, the field was reproducible
to a few ppm\cite{farleypc}.
The 14~m-diameter storage ring ( $B_0 = 1.5$~T) 
was constructed using 40 identical C-magnets, 
bolted together in a regular polygon to make a ring,
and the field  was excited by four
concentric coils connected in series.  
The coils that excited the field in the CERN storage ring were
conventional warm coils, and it took about four days  after powering
for thermal and mechanical equilibrium to be reached\cite{farleypc,cernmag}.
The shimming procedure involved  grinding steel from the pole pieces, which
introduced a periodic shape to the contour map averaged over 
azimuth (see Figure 5 of Reference~\cite{cern3b}). 
The magnetic field was mapped once or twice per running period, by
 removing the vacuum chamber and stepping NMR probes through the storage 
region. 

The electrostatic quadrupoles had a two-fold symmetry, covered approximately 
80\% of the azimuth, and defined a rectangular storage region 
120~mm (radial) by  80~mm (vertical).  The ring behaved as 
a weak focusing storage ring  (betatron)\cite{wied,edwards,cp}, 
with a field index of $\sim 0.135$ (see \S\ref{sct:beam-dyn}).

The pion beam was brought into the storage ring through a pulsed co-axial 
``inflector'', which produced a magnetic field that canceled the 1.5~T
storage-ring
field and permitted the beam to arrive undeflected, tangent to the storage
circle but displaced 76 mm
radially outward from the central
orbit.  The inflector 
current pulse rose to a peak current of 300~kA in 12~$\mu$s.
The transient effects from this pulsed
device on the magnetic field seen by the stored muons during the precession
measurement were estimated to 
be small\cite{cern3b}.

The decay electrons were detected by 24 lead-scintillator sandwiches each
viewed through an air light-guide by a single 5-inch photomultiplier tube.
 The photomultiplier signal from each detector was discriminated at 
several analog thresholds,
thus providing a coarse pulse height measurement as well as the time
of the pulse.
Each of the discriminated
signals was assigned to a time bin, in a manner independent
of both the ambient rate and the electron arrival time relative to the clock 
time-boundary.
 The final precision of 7.3~ppm (statistics dominated) 
convincingly confirmed 
the effect of hadronic vacuum polarization, which was predicted to 
contribute to $a_\mu$ at the level of 60~ppm. The factor of 35 increase in
sensitivity between the second and third experiments
can be traced to the improved statistical power provided by the
injected pion beam, both in the number and
asymmetry of the muon sample, and in reduced systematic errors resulting
from the use of electrostatic focusing with
the central momentum set to $\gamma_{magic}$.

\begin{table*}[h!]
\caption[Results]{Measurements of the muon anomalous magnetic moment.
When the uncertainty on the measurement is the size of the next term
in the QED expansion, or the hadronic or weak contributions, the 
term is
listed under ``sensitivity''.  The ``?'' indicates a result that
differs by greater than two standard deviations with the Standard Model.
For completeness, we include the experiment of Henry, et al.,\cite{Henry69},
which is not discussed in the text.
}
\begin{center}

{\small
\begin{tabular}{|c|l|c| c|l|} \hline \hline 
{\bf $\pm$} & {\bf Measurement} & $\sigma_{a_{\mu}} / a_{\mu}$ & {\bf Sensitivity} &
{\bf Reference}
\\ \hline \hline
$\mu^+$ & $g = 2.00 \pm 0.10$ & &$g=2$ &Garwin {\it et al}\cite{garwin1}, Nevis (1957) \\
$\mu^+$ &$0.001\,13_{-0.00012}^{+0.00016}$ & { 12.4\%} & ${\alpha \over  \pi}$ 
     &Garwin {\it et al}\cite{garwin2}, Nevis (1959)\\
\hline
$\mu^+$ &$0.001\,145(22)$ & { 1.9\%} &  ${\alpha \over  \pi}$  
     &Charpak {\it et al}\cite{cern1} CERN 1 (SC) (1961) \\
$\mu^+$ & $0.001\,162(5)$ & {\ 0.43\%} 
     &  $\left({\alpha \over  \pi}\right)^2$ 
     &Charpak {\it et al}\cite{cern1a} CERN 1 (SC) (1962) \\
\hline
$\mu^\pm$ &$0.001\,166\,16(31)$ & 265~ppm 
     & $\left({\alpha \over \pi}\right)^3$
     &Bailey {\it et al}\cite{cern2} CERN 2 (PS) (1968) \\
\hline
$\mu^+$ & $0.001\,060(67)$ & 5.8\% &  ${\alpha \over  \pi}$ &
     Henry{\it et al}\cite{Henry69} solenoid (1969) \\
\hline
$\mu^{\pm}$ & 
  $0.001\,165\,895(27)$ & 23~ppm 
     & $\left( {\alpha \over  \pi} \right)^3$ + Hadronic 
     &Bailey {\it et al}\cite{cern3a} CERN 3 (PS) (1975) \\
$\mu^{\pm} $ & $0.001\,165\,911(11)$  &  7.3~ppm  
      & $\left( {\alpha \over  \pi}\right)^3$ + Hadronic  
      &Bailey {\it et al}\cite{cern3b} CERN 3 (PS) (1979)\\
\hline
$\mu^+$ & $0.001\,165\,919\,1(59)$  & { 5~ppm}  
      & $\left( {\alpha \over  \pi}\right)^3$ + Hadronic  
      &Brown {\it et al}\cite{brown1} BNL (2000)\\
$\mu^+$ & $0.001\,165\,920\,2(16)$  & { 1.3~ppm}   
      & $\left( {\alpha \over  \pi}\right)^4$ + Weak
      &Brown {\it et al}\cite{brown2} BNL (2001)\\
$\mu^+$ & $0.001\,165\,920\,3(8)$  & { 0.7~ppm}  
      &$\left( {\alpha \over  \pi}\right)^4$ + Weak + ?   
      &Bennett {\it et al}\cite{bennett1} BNL (2002)\\
$\mu^-$ & $0.001\,165\,921\,4(8)(3)$  & {0.7~ppm}   
      & $\left( {\alpha \over  \pi} \right)^4$ + Weak + ?  
      &Bennett {\it et al}\cite{bennett2} BNL (2004)\\
\hline\hline
$\mu^{\pm}$  & $0.001\,165\,920\,80(63)$  & { 0.54~ppm} 
      &$\left( {\alpha \over  \pi} \right)^4$ + Weak + ?  
      &Bennett {\it et al}\cite{bennett2,bennett3} BNL WA (2004)\\
\hline\hline
\end{tabular}
}
\end{center}

\label{tb:measurements}
\end{table*} 
%%%%%%%%%%%%

Around 1984 a new collaboration (E821) was formed with the
aim of  improving on the CERN result to a relative precision of
 0.35~ppm, a goal chosen
because it was 20\% of the predicted
first-order electroweak contribution to $a_\mu$. 
During the subsequent twenty-year period, over
100 scientists and engineers contributed to E821, which was
built and performed at the Brookhaven National
Laboratory (BNL) Alternating Gradient Synchrotron (AGS).
A measurement at this level of precision would
test the electroweak renormalization
procedure for the Standard Model and serve as 
a very sensitive constraint on new 
physics\cite{Czar01,Martin03,Stockinger_07}.

The E821 experiment at 
Brookhaven\cite{carey,brown1,brown2,bennett1,bennett2,bennett3}
has advanced the
relative experimental precision of $a_\mu$ to 0.54 ppm,
with a several standard-deviation difference from the prediction of the
Standard 
Model. 
Assuming {\sl CPT} symmetry, viz. $a_{\mu^+} = a_{\mu^-} $, the new
world average
is
\begin{equation}
  a_\mu(\mathrm{Expt}) = 11\,659\,208.0(6.3) \times
  10^{-10}~~\mbox{(0.54\,ppm)}.\label{eq:E821wa}
\end{equation}
The total uncertainty includes a 0.46~ppm statistical uncertainty
and a 0.28~ppm systematic uncertainty, combined in quadrature.

A summary showing the physics reach of each of the successive muon (g-2)
experiments  is given in 
Table~\ref{tb:measurements}. 
In the subsequent sections we review the most recent experiment, E821 at 
Brookhaven, and the Standard-Model theory with which it is compared.

%% file: E821-exp20feb07.tex
\section{The Brookhaven experiment E821: Experimental Technique }
%
% This is the recovered version, after the loss of two days editing.
%

In general terms, such as the use of electrostatic focusing and the magic 
$\gamma$, E821 was modeled closely on the third CERN experiment. 
However, the  statistical error goal and the ability of
the Brookhaven AGS to provide the (necessary) tremendous increase in beam flux 
posed enormous challenges to the injection system as well as to the detectors,
electronics and data acquisition. At the same time, systematic errors,
in the measurements of both  the field and anomalous precession frequencies,
had to to be reduced significantly compared the the third CERN 
experiment. The challenges
included:

\bi
\item How to allow multiple beam bunches from the AGS  to pass undeflected
through the back leg of the storage ring magnet. The use of multiple 
bunches reduces the instantaneous rates in the detectors, but it was judged
technically impractical to rapidly pulse an
inflector like that used at CERN.
\item How to provide a homogeneous storage ring field, uniform to 1~ppm when
averaged over azimuth.
\item How to store an injected {\it muon} beam, which would greatly reduce
the hadronic flash seen by the detectors.
\item How to maximize the statistical power of the detected electrons
\item How to make good measurements on the decay 
electrons - good energy resolution,
and, in the face of high data rates with a wide dynamic 
range, how to maintain the
stability of signal gain and timing pickoff

\ei

Some of the solutions were the natural product of  technical advances 
over the previous twenty years. Others, such as the superconducting
inflector, were truly novel, and required considerable development.
 A comparison of E821 and the third CERN
experiment is given in Table~\ref{tb:compar}.

\begin{table*}[h]
\caption[Results]{A comparison of the features of the E821 and the 
third CERN muon $(g-2)$ experiment\cite{cern3b}. Both 
experiments operated at the
``magic'' $\gamma = 29.3$, and used electrostatic quadrupoles for
vertical focusing. Bailey, et al. \cite{cern3b}, do not quote a systematic
error on the muon frequency $\omega_a$}.
\begin{center}

{\small
\begin{tabular}{|c|l|l|} \hline \hline 
{\bf Quantity} & {\bf E821 } & CERN \\ \hline \hline
Magnet & Superconducting & Room Temperature \\
Yoke Construction   & Monolithic Yoke & 40 Separate Magnets \\
Magnetic Field & 1.45~T & 1.47~T  \\
Magnet Gap &  180~mm & 140~mm \\
Stored Energy & 6~MJ &  \\ 
Field mapped in situ? &yes  & no\\
Central Orbit Radius & 7112~mm &7000~mm \\
Averaged Field Uniformity & $\pm1$~ppm & $\pm 10$~ppm \\
Muon Storage Region& 90~mm Diameter Circle &$ 120\times 80$~mm$^2$ Rectangle \\
\hline
Injected Beam & Muon  & Pion \\
Inflector &Static Superconducting & Pulsed Coaxial Line \\
\hline
Kicker & Pulsed Magnetic & $\pi \rightarrow \mu \  \nu_\mu $ decay\\
Kicker Efficiency & $~\sim 4$\% & 125 ppm \\
Muons stored/fill & $10^4$  & 350 \\
\hline
Ring Symmetry &Four-fold & Two-fold \\
$\sqrt {\beta_{\rm max} / \beta_{\rm min}}$&  1.03 & 1.15\\
\hline
Detectors & Pb-Scintillating Fiber & Pb-Scintillator ``Sandwich''\\
Electronics & Waveform Digitizers & Discriminators \\
\hline
Systematic Error on B-field& 0.17~ppm & 1.5~ppm\\
Systematic Error on $\omega_a$ &0.21~ppm & Not given  \\
Total Systematic Error & 0.28~ppm &  1.5~ppm \\
Statistical Error on $\omega_a$ & 0.46~ppm  & 7.0~ppm \\
\hline
Final Total Error on $a_{\mu}$ & 0.54~ppm & 7.3~ppm \\
\hline\hline
\end{tabular}
}
\end{center}

\label{tb:compar}
\end{table*} 
%%%%%%%%%%%%

An engineering run with pion injection occurred in 1997, a brief 
$\mu$-injection run where 
 the new muon kicker was commissioned happened in 1998, and 
major data-collection periods of 3 to 4 months duration
took place in 1999, 2000 and 
2001. In each period,
protons were accelerated by a linear accelerator, accelerated further by a
booster synchrotron, and then injected  into the BNL 
alternating gradient synchrotron (AGS).
Radio-frequency cavities in the AGS ring provide acceleration
to a momentum of 24~GeV/$c$, and 
maintain the protons in a number of
discrete, equally spaced bunches. 
The number of bunches (harmonic number) 
in the AGS during a 2.7~s acceleration
cycle was different for each of these periods:
eight in 1999, six in 2000 and  twelve in 2001. 
The AGS has the ability to deliver up to 
 $70 \times 10^{12}$ protons (70 Tp)
in one AGS cycle,  providing a proton intensity per hour
180 times greater than that available at CERN
in the 1970s.  

\begin{figure}[h!]
  \includegraphics[width=0.75\textwidth,angle=0]{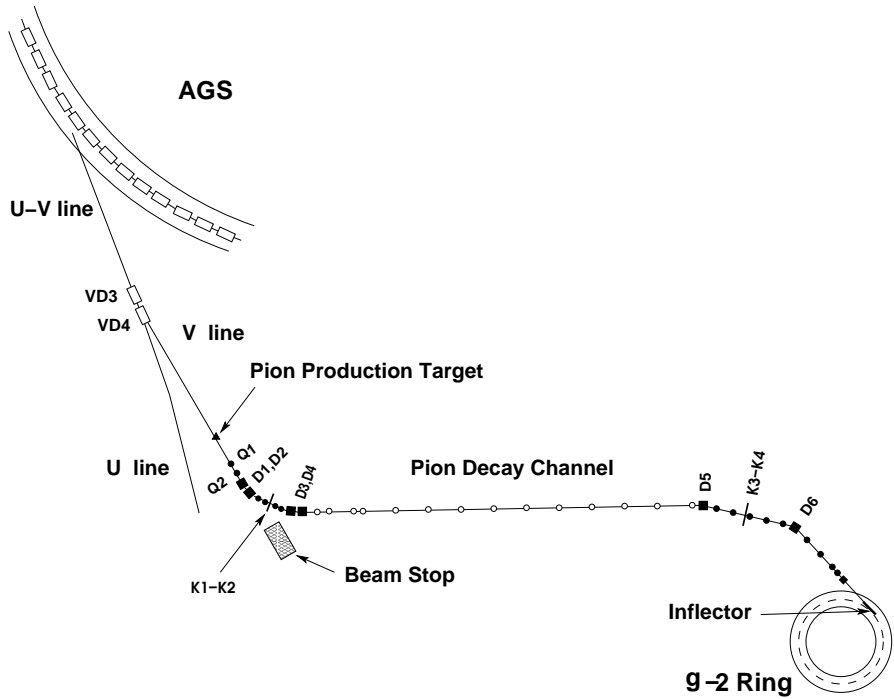}
  \caption{The E821 beamline and storage ring.  Pions produced at $0^{\circ}$
are collected by the quadrupoles Q1-Q2 and the momentum is 
selected by the collimators
K1-K2.  The pion decay channel is 72 m in length.  Forward muons at
the magic momentum are selected by the collimators K3-K4.
  \label{fg:bmline}}
\end{figure}

The proton  beam is
extracted from the AGS one bunch at a time at 33 ms intervals.
Each proton bunch results in a narrow time bunch of muons which 
is injected into the storage ring, and then the electrons from 
muon decays are measured for about 10 muon  lifetimes, or about
 640~$\mu$s. A plan view of the Brookhaven Alternating Gradient Synchrotron, 
injection line and storage ring are shown in Figure~\ref{fg:bmline}.
Because the maximum
total intensity available from the AGS is $\leq 70$~Tp, the bunch intensity
and the resulting pile-up (accidental coincidences between two electrons)
in the detectors is minimized by maximizing the
number of proton bunches.  Pulse pile-up in the detectors 
following injection into the storage ring is one of the
systematic issues requiring careful study in the data analysis.

\subsection{The Proton and Muon Beamlines}

The primary proton beam from the AGS is brought to a water-cooled nickel
production target.  Because of 
mechanical shock considerations, the 
intensity of a bunch is limited to less than 7~Tp.
The beamline, shown in Figure \ref{fg:bmline}, accepts pions
produced at $0^{\circ}$ at the production target. They are collected by
the first two quadrupoles, momentum analyzed,
and brought into the
decay channel by four dipoles.
A pion momentum of 3.115 GeV/$c$, 1.7~\% higher than the magic momentum, 
is selected.
  The beam then enters a straight 80-meter-long
focusing-defocusing quadrupole channel, where those muons from pion 
decays that are emitted approximately parallel to the pion momentum, 
so-called forward decays, are collected and transported downstream.
Muons at the magic momentum of 3.094 MeV/$c$, and having an
 average polarization of 95\%, are separated from the slightly
 higher momentum pions at the second momentum slit.
 However, after this momentum selection  a rather large pion component
 remains in the beam,  whose  composition 
was measured to be 1:1:1,  $e^+:\mu^+:\pi^+$.  The
proton content was calculated to be approximately one-third of the
pion flux\cite{brown1}. The secondary muon beam intensity 
incident on the storage ring was about $2\times 10^6$ per fill of the ring,
which can be compared with $10^8$  particles per fill with ``pion
injection\cite{carey}'' which was used in the 1997 engineering run.

The detectors are exposed to a large flux of background pions, protons,
neutrons, electrons and other particles
associated with the beam injection process, called the injection ``flash''.
The intensity  varies around the ring,
being most intense in the detectors adjacent to
the injection point (referred to below as `upstream' detectors).
The flash consists of prompt and delayed components, with
the prompt component
caused by injected particles passing directly into the detectors.
The delayed component is mainly due to $\gamma$ rays following neutron capture.
The neutrons are produced primarily by the protons and pions
in the beam striking the magnet, calorimeters, etc.
Many neutrons thermalize  in
the calorimeter and surrounding materials over tens of microseconds,
where they can undergo nuclear capture.  These $\gamma$ rays 
cause a DC baseline offset in the PMT signal which
steadily decays
with a lifetime of $\sim 50$-100~$\mu$s.
To reduce the decay time
of the neutron background, the epoxy in the upstream subset
 of detectors is doped
with natural boron carbide powder, thus taking advantage 
of the large neutron capture
cross-section on $^{10}B$.

This injection flash is most severe with the pion injection scheme.
The upstream photomultiplier tubes had to
be gated off for 120~$\mu$s (1.8 muon lifetimes)  following injection, 
to allow the  signals to return to the 
nominal baseline. 
To reduce this ``flash'' and to increase significantly
the number of muons stored
per fill of the storage ring, a  fast muon kicker was developed 
which permitted direct muon injection into the storage ring.

\subsection{The Inflector Magnet}

\begin{figure}[h]
\begin{center}
\subfigure[ ] {\includegraphics[width=0.4\textwidth,angle=0]{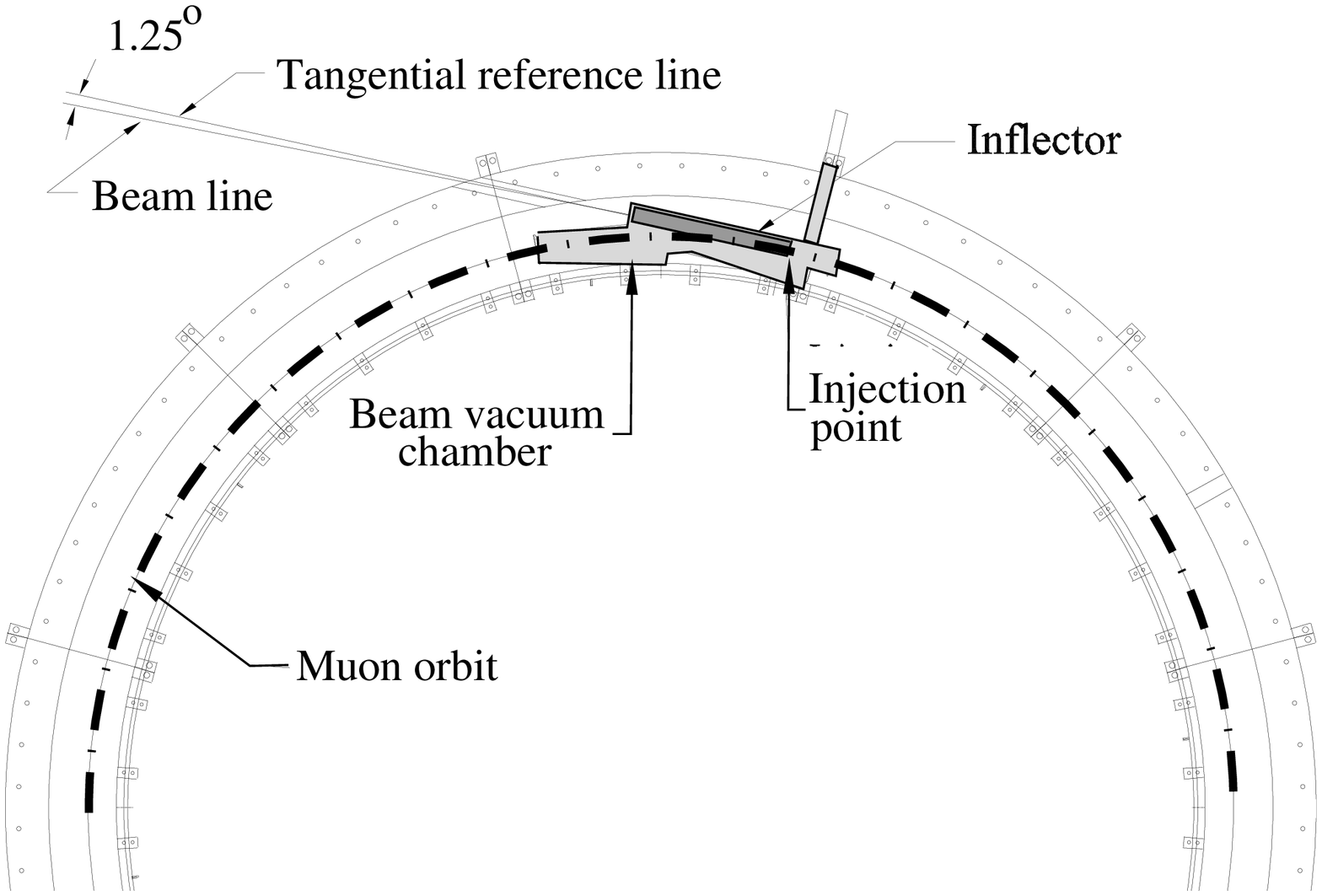}}
\subfigure[ ] {\includegraphics[width=0.58\textwidth,angle=0]{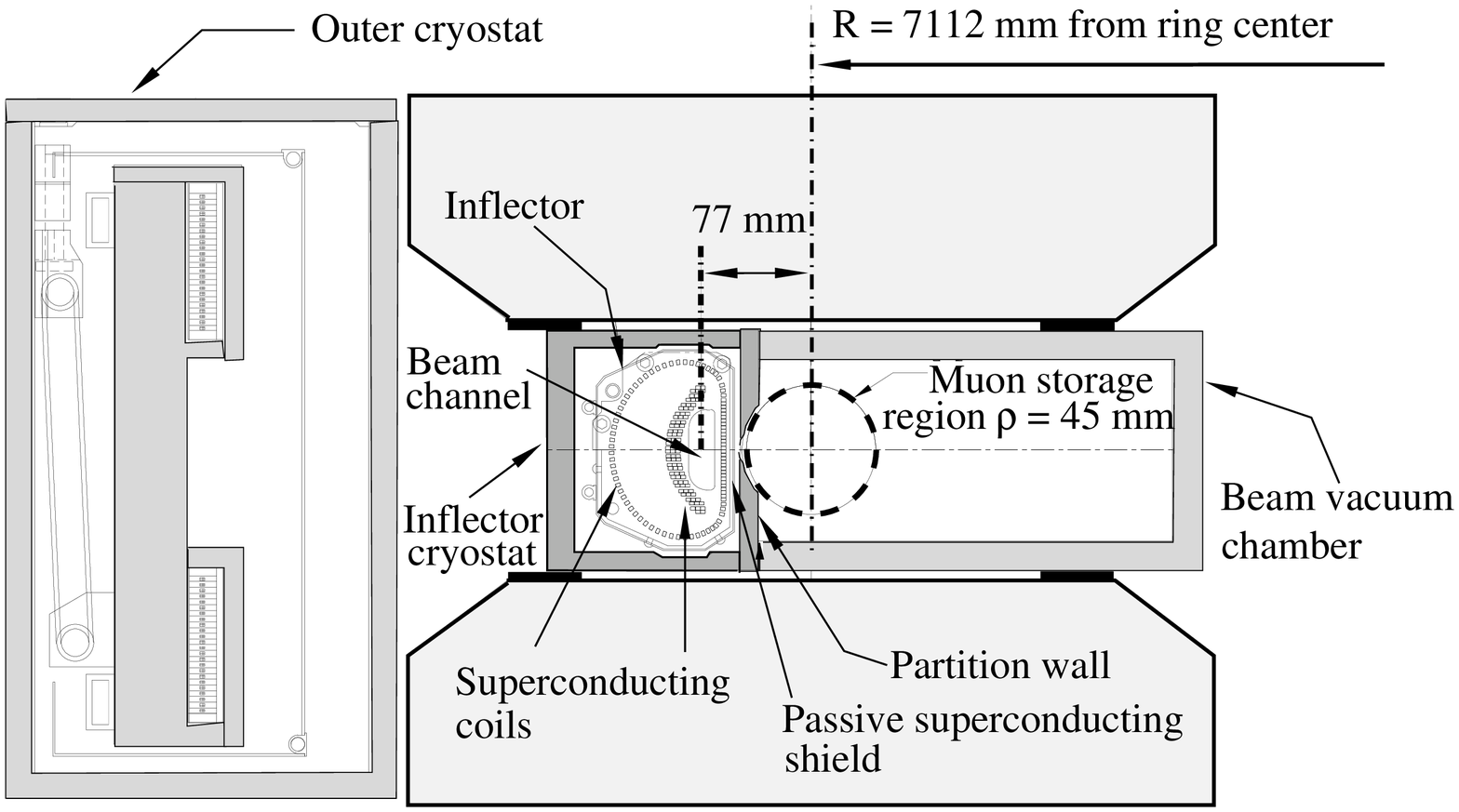}}%
 \caption{ (a) A plan view of the inflector-storage-ring
geometry. The dot-dash line shows the central muon orbit at 7112 mm. 
The beam enters through a hole in the back of the magnet yoke,
then passes into the inflector. The inflector cryostat has a separate vacuum
from
the beam chamber, as can be seen in the cross sectional view.  The 
cryogenic services for the inflector are provided through a radial
penetration through the yoke at the upstream end of the inflector.
(b) A cross-sectional view of the pole pieces, the outer-radius
coil-cryostat arrangement,  and the downstream
end of the superconducting inflector. The muon beam direction
at the inflector exit is into the page. The center of the 
storage ring is to the right. The outer-radius
coils which excite the storage-ring magnetic field are shown, but 
the inner-radius coils are omitted.
  \label{fg:infl}}
\end{center}
\end{figure}

\begin{figure}[h]
\begin{center}
\subfigure[ ] 
{\includegraphics[width=0.45\textwidth,angle=0]{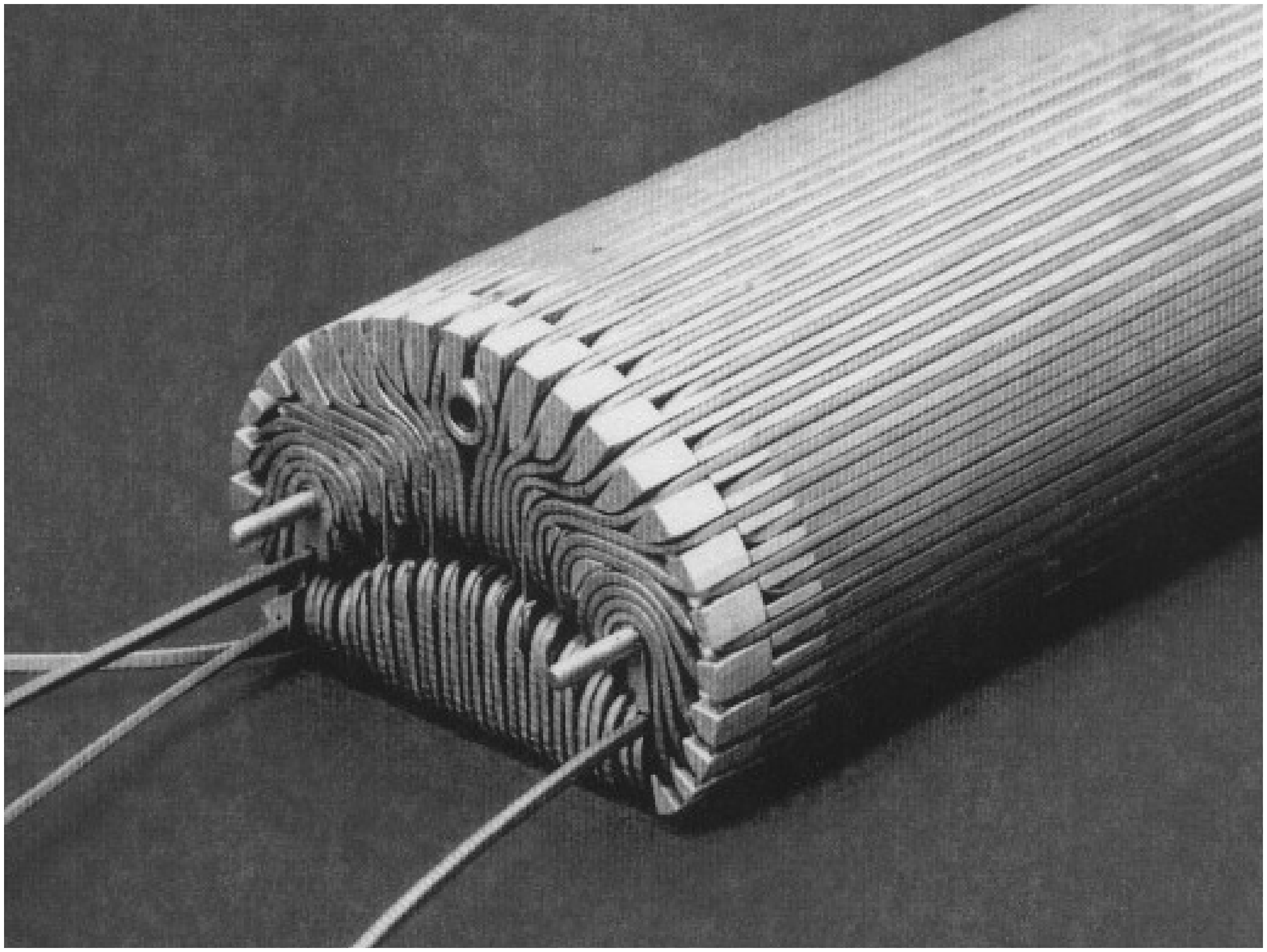}}
\subfigure[ ] 
{\includegraphics[width=0.45\textwidth,angle=0]{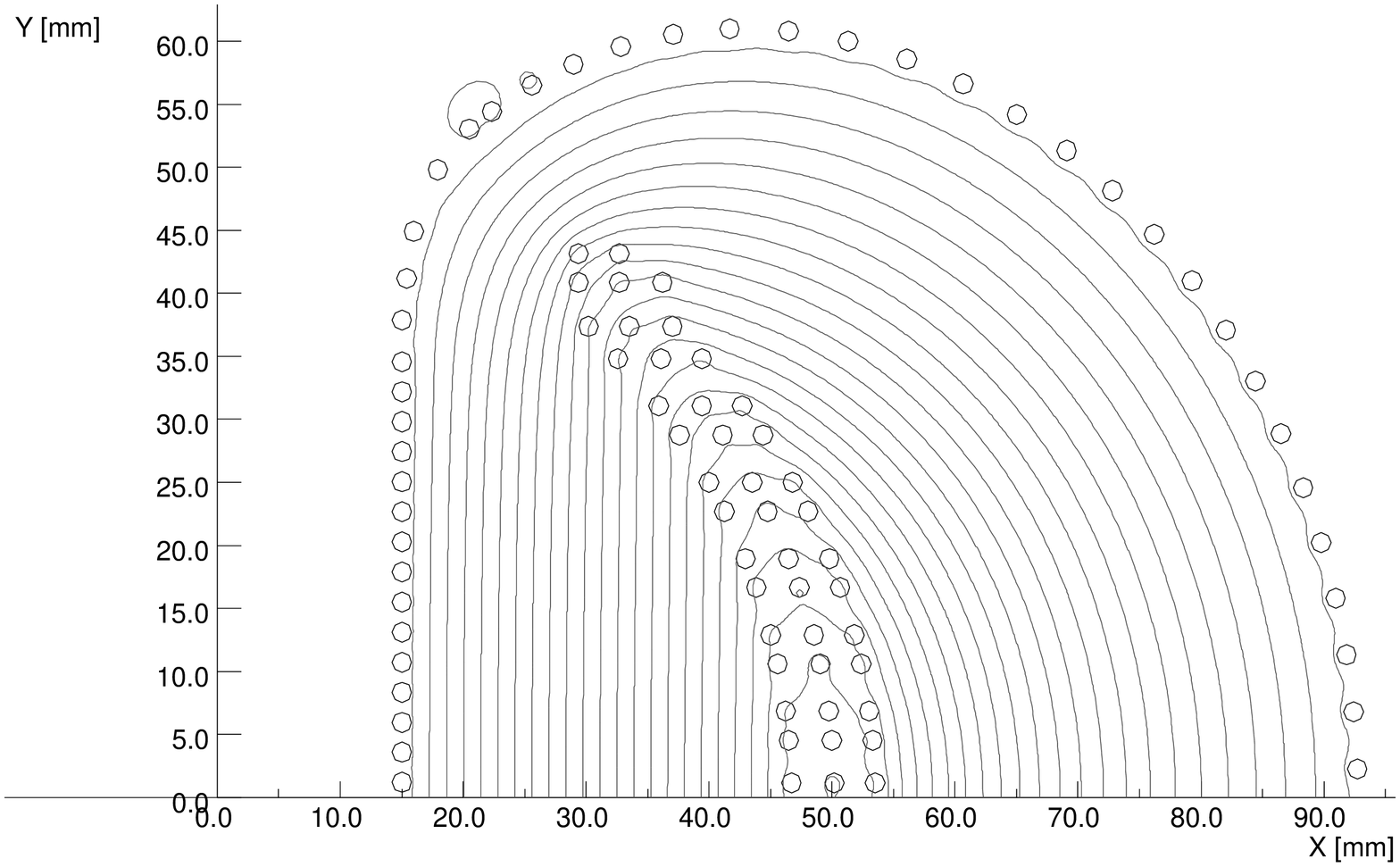}}%
 \caption{ (a) A photo of the prototype inflector showing the crossover
between the two coils.  The beam channel is covered by the lower
crossover.
(b) The magnetic design of the inflector. Note that the magnetic
flux is largely contained inside of the inflector volume.
  \label{fg:inflpik}}
\end{center}
\end{figure}

The geometry of the incoming beam is shown in Figure \ref{fg:infl}.
A unique superconducting inflector magnet\cite{inflector1,inflector2}
was built to cancel the magnetic field permitting the beam to 
arrive undeflected at the edge of the storage 
region (see Figure~\ref{fg:inflpik}(b)).
The  inflector is a truncated, double-cosine theta magnet, 
shown in Figure \ref{fg:infl}(b) at its
downstream end, with the muon velocity going {\it into} the page.  
In the inflector, 
the current flows into the page down the central ``C''-shaped layer of 
superconductor, then out of the page 
through the ``backward-D''-shaped outer conductor layer. 
At the inflector exit, the center of
 the injected beam is 77~mm from the central orbit.
For $\mu^+$ stored in the ring, the main field points up, and 
$q \vec v \times \vec B$ points to the right in Figure \ref{fg:infl}(b),
toward the ring center.

In this design shown in Figure~\ref{fg:infl}(b)
 and Figure~\ref{fg:inflpik}(b), the beam channel aperture
is rather small compared to the flux return area. The
 field is $\sim 3$~T in the return
area (inflector plus central storage-ring field), and 
the flux density is sufficiently high to
lower the critical current in superconductor placed in 
that region. If the beam channel
aperture were to be increased by pushing the coil further
 into the flux return area,
the design would have to be changed, either by employing
 a superconductor with larger
critical current, or by using more conductors in a 
revised geometry, further complicating the fabrication of this
 magnet. The result of the small inflector aperture is a rather poor
phase-space match between the inflector and the storage ring and,
 as a consequence, a
loss of stored muons.

As can be seen from Figure~\ref{fg:inflpik}(a),
the entrance (and exit) to the beam channel is
covered with superconductor, as well as by aluminum windows that 
are not visible in
the photograph. This design was chosen to maximize the mechanical
 stability of the
superconductor in the magnetic field, thus reducing the risk of 
motion which would
quench the magnet. However, multiple scattering in the material
 at both the entrance
and exit windows causes about half the incident muon beam to be lost.

The distribution of conductor on the outer surface of the inflector 
magnet (the ``D-shaped'' arrangement) 
prevents most of the magnetic flux from leaking outside of
the inflector volume, as seen from Figure\ref{fg:inflpik}(b). 
To prevent flux leakage 
from entering the beam storage region, 
the inflector is wrapped with a passive superconducting shield that extends
beyond both inflector ends, with a
2~m seam running longitudinally along the inflector
side away from the storage region.
With the inflector at zero current and the shield warm, 
the main storage ring magnet is energized.  
Next the inflector is cooled down, so that the shield
goes superconducting and pins the precision field inside the inflector region.
When the inflector magnet is powered, the supercurrents in the shield
prevent the leakage field from penetrating into
the storage region behind the
shield.  

A 0.5~m long prototype of the inflector was fabricated and then tested,
first by itself, then in an external magnetic field.  Afterward the 
full-size 1.7~m
inflector was fabricated.  Unfortunately, the full-size inflector
was damaged during initial tests, making it necessary to cut through
 the shield to repair the interconnect between
the two inflector coils. A patch was placed over the cut, and
this repaired inflector was used in the
1997 $\pi$-injection run as well as in the 1998 and  1999 run periods. 
Magnetic flux leakage around the patch reduced the main
storage-ring field locally 
by $\sim 600~\mathrm{ppm}$.  This inhomogeneity
complicated the field measurement due to the degradation in NMR performance,
and contributed
an additional uncertainty of $0.2~\mathrm{ppm}$ to the average magnetic field
seen by the muons. The leakage  field from
a new inflector, installed before the 2000 running period,
 was immeasurably small.

\subsection{The Fast Muon Kicker}

\begin{figure}[h]
\begin{center}
{\includegraphics[width=0.3\textwidth,angle=0]{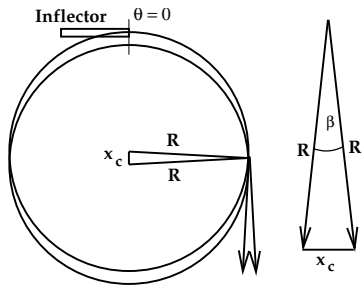}}
\end{center}
  \caption{ A sketch of the beam geometry. $R=7112$~mm is the 
storage ring radius,
$x_c=77$~mm is the distance between the inflector center and the center of the
storage region.  This is also the distance between the centers of the
circular trajectory that a particle entering at the inflector center 
(at $\theta=0$ with $x' = 0$) will follow,
and the circular trajectory a particle at the center of the
storage volume (at $\theta=0$ with $x' = 0$) will follow.  
  \label{fg:kick}}
\end{figure}

Left undisturbed, the injected muon beam would pass once around
the storage ring, strike the inflector and be lost. As shown in
Figure~\ref{fg:kick}, the role of the fast muon kicker is to briefly reduce
a section of the main storage field and move the center of
the muon orbit to the geometric center of the storage ring.
The 77-mm offset at the injection point, between the center of
the entering beam and the central orbit, requires that the beam be kicked
outward by approximately 10 mrad.
The kick should be made at about 90 degrees around the ring, plus a few
degree correction due to the defocusing effect of the electric quadrupoles
between the injection point and the kicker, as shown in 
Figures~\ref{fg:infl}(b) and \ref{fg:kick}.

The requirements on the fast muon kicker are rather stringent.  
While electric, magnetic, and combination electromagnetic kickers were 
considered, the collaboration
settled on a magnetic kicker design\cite{kicker} because it was
thought to be technically easier and more robust than the other options. 
 Because of the very
stringent requirements on the storage ring
magnetic field uniformity, no magnetic materials
could be used.  Thus the kicker field had to be generated 
and shaped solely with 
currents, rather than using ferrite cores.
 Even with the kicker field generated by currents,
there existed the potential
problem of inducing eddy currents which might affect the magnetic field
seen by the stored muons.

\begin{figure}[h]
\begin{center}
{\includegraphics[width=0.4\textwidth,angle=0]{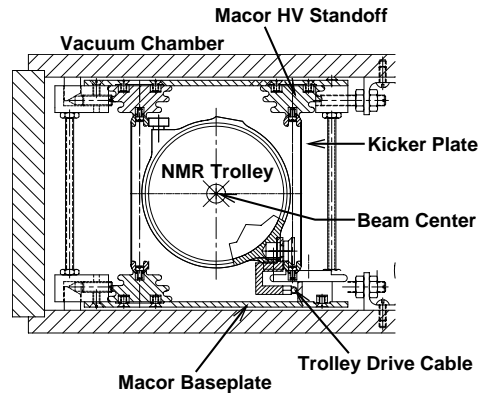}}
\end{center}
  \caption{An elevation view of the kicker plates, showing 
the ceramic
cage  supporting the kicker plates, and 
the NMR trolley riding on the kicker plates. (The trolley is removed
during  data collection).
  \label{fg:kick-plate}}
\end{figure}

The length of the kicker is limited to the  $\sim 5$~m azimuthal space 
between the electrostatic quadrupoles (see Figure~\ref{fg:ring}), so each
of the three sections is 1.76~m long. 
The cross section of the kicker is shown in Figure \ref{fg:kick-plate}.
The  two parallel conductors are connected with cross-overs at 
each end,  forming a single current loop. 
The kicker plates also have to serve
as ``rails'' for the NMR field-mapping trolley (discussed below), 
and the trolley is shown
riding on the kicker rails in Figure \ref{fg:kick-plate}.

The kicker current pulse is formed by an under-damped LCR circuit.  A 
capacitor is charged to 95~kV through a resonant charging circuit.
Just before the beam enters the storage ring, the
capacitor is shorted to ground by firing a deuterium thyratron.
The peak current in an LCR circuit is given by $I_0 = V_0/( \omega_d L)$
making it necessary to keep the system inductance, $L$, low to maximize
the magnetic field for a given voltage $V_0$.  For this reason, the kicker
was divided into three sections, each powered by a separate pulse-forming
network.

\begin{figure}[h!]
\begin{center}
\subfigure[ ]{\includegraphics[width=0.4\textwidth,angle=0]{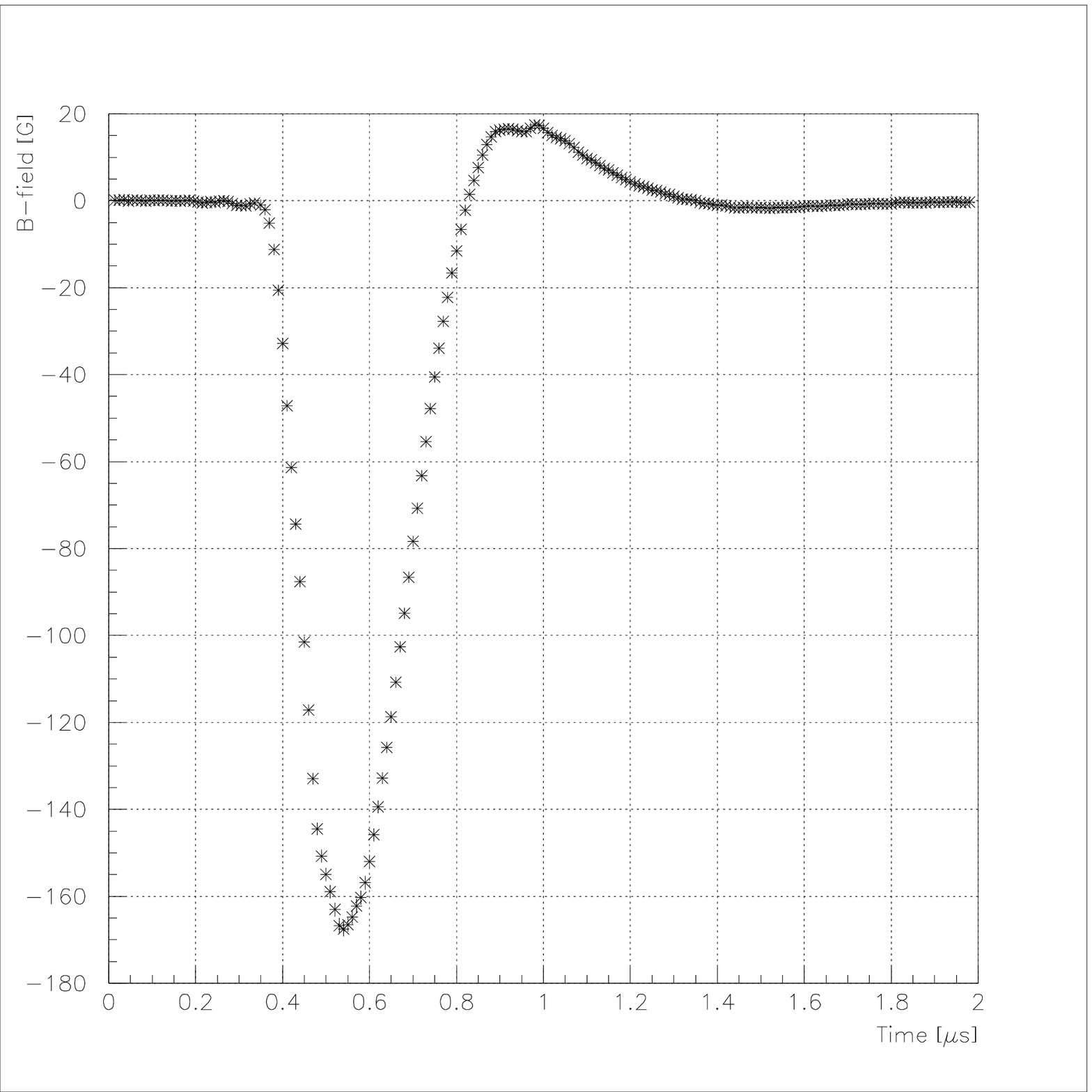}}
\subfigure[ ]{\includegraphics[width=0.4\textwidth,angle=0]{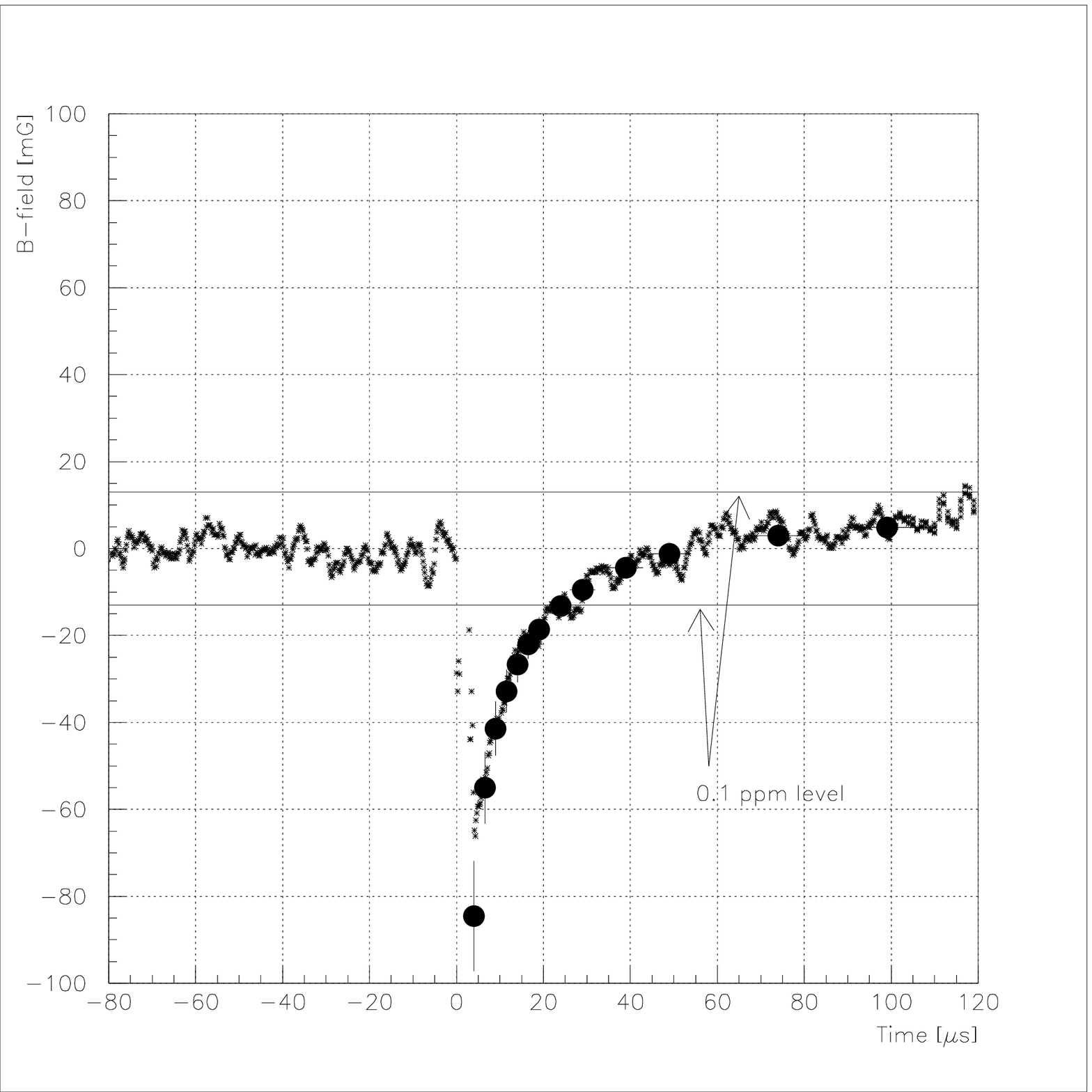}}
\end{center}
  \caption{(a)The main magnetic field of the kicker measured with the 
Faraday effect. (b) The residual magnetic field measured using the 
Faraday effect. The solid points are calculations from 
{\tt OPERA}, and the small $\times$ are the experimental points
measured with the Faraday effect. The solid horizontal lines show 
the $\pm 0.1$~ppm band
for affecting $\int \vec B \cdot d \vec \ell$.
  \label{fg:kickpulse}}
\end{figure}

The electrode design and the time-varying fields were modeled using  
the package {\tt OPERA 2d}\cite{opera}. 
 Calculations for electrodes made of Al, Ti and Cu, were carried out
with the geometry shown in  Figure~\ref{fg:kick-plate}. 
The aluminum electrodes were best for
minimizing eddy currents 20~$\mu$s after injection,
which was fortunate, since mechanical
tests on prototype electrodes made from the three metals showed 
that only aluminum was satisfactory.  This was also true from the
detector point of view, since the low atomic number of aluminum
 minimizes the multiple scattering and showering of the decay electrons
 on their way to the calorimeters.
 The magnetic field
produced by the kicker, along with the residual field, was measured 
using the Faraday effect\cite{kicker}, and these are shown in 
Figure~\ref{fg:kickpulse}. 

The cyclotron period of the ring is 149 ns, substantially less than the 
kicker pulse base-width of $\sim400$~ns, so that the
injected beam is kicked on the first few turns.  Nevertheless, 
approximately $10^4$ muons are stored per fill of the ring, corresponding
to  an
injection efficiency of about 3 to 5\% (ratio of stored to incident muons).
The storage efficiency with muon injection is much greater than that
obtained with pion injection for a given proton flux,
with only 1 percent of the hadronic
flash.

\subsection{The Electrostatic Quadrupoles}

\begin{figure}[h]
  \begin{center}
    \includegraphics*[width=0.5\textwidth]{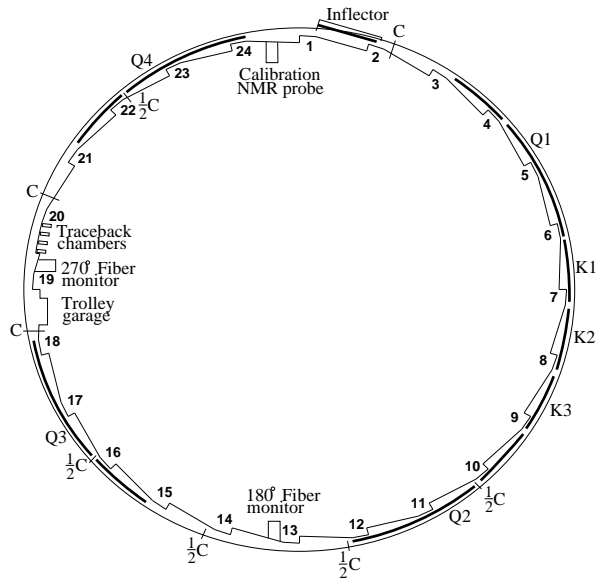}
    \caption{The layout of the storage ring, as seen from above,
 showing the location of the
inflector, the kicker sections (labeled K1-K3), and the quadrupoles 
(labeled Q1-Q4). The beam circulates in a clockwise direction.
Also shown are the collimators, which are labeled
``C'', or ``${1\over2}{\rm C}$'' indicating whether 
the  Cu collimator covers the full aperture, or half the aperture. 
  The collimators are rings with
inner radius: 45~mm, outer radius: 55~mm, thickness: 3~mm.   
The scalloped vacuum chamber consists of 12 sections joined by bellows.
The chambers containing the inflector, the NMR trolley garage, and
the trolley drive mechanism
are special chambers. 
 The other chambers are standard, with
either quadrupole or kicker assemblies installed inside.  
An electron calorimeter is placed behind each of the radial windows,
at the postion indicated by the calorimeter number.
             \label{fg:ring}}
  \end{center}
\end{figure}

\begin{figure}[h!]
\begin{center}
  \includegraphics[width=0.6\textwidth,angle=0]{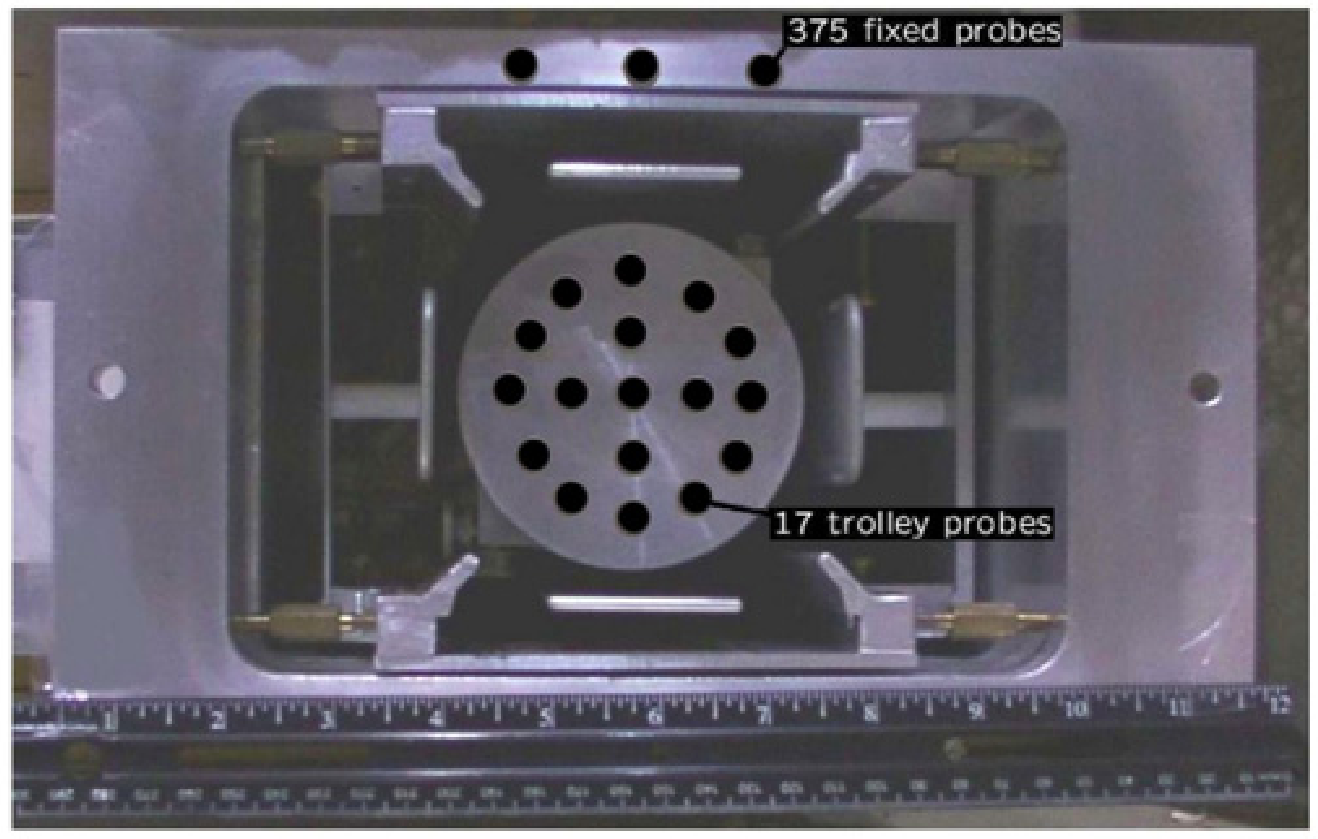}
  \caption{A photograph of an electrostatic quadrupole assembly
inside a vacuum
chamber.  The cage assembly doubles as a rail system for the
 NMR trolley which is resting on the rails. 
The location of the NMR probes inside the 
trolley are shown as black circles (see \S\ref{sct:field}). 
The probes are located just behind
the front face. The inner (outer) circle of probes has a 
diameter of 3.5~cm (7~cm) at the probe centers.  The storage region
has a diameter of 9~cm.  The vertical location of 
three upper fixed probes is also
shown.  The fixed probes are located symmetrically
above and below the vacuum chamber.
  \label{fg:quadpik}}
\end{center}
\end{figure}

The electric quadrupoles, which are arranged around the ring with four-fold
symmetry, provide vertical focusing for the stored muon beam.
The quadrupoles cover 43\% of the ring in azimuth, as shown in
Figure~\ref{fg:ring}.   While the ideal vertical profile for a
quadrupole electrode would be hyperbolic, beam dynamics 
calculations determined that the higher multipoles present with
flat electrodes, 
 which are much easier to fabricate,
 would not cause an unacceptable level of beam losses.
The flat electrodes are  shown
in Figure~\ref{fg:quadpik}.  Only certain multipoles are permitted
by the four-fold symmetry, and a judicious choice of the electrode width 
relative to the separation between opposite plates minimizes
 the lowest of these.
 With this configuration, the 20-pole is the largest,
being 2\% of the quadrupole component and an order of magnitude greater
than the other allowed multipoles\cite{quads}.

In the quadrupole regions, the combined electric and magnetic
fields can lead to electron trapping. The electron orbits  run longitudinally
along the inside of the electrode, and then return on the outside.
Excessive trapping in the relatively modest vacuum of the storage
ring can cause sparking. To minimize trapping, the 
leads were arranged to introduce a dipole field at the end 
of the quadrupole
 thus sweeping away trapped electrons. In addition,
the quadrupoles are pulsed, so that after each fill of the ring 
all trapped particles could escape.  Since some
protons (antiprotons) were stored in each fill of $\mu^+$ ($\mu^-$),
they were also released at the end of each storage time.  This lead
arrangement worked so well in removing trapped electrons that 
for the $\mu^+$ polarity it would have been possible 
to operate the quadrupoles
 in a dc mode.  For the storage of $\mu^-$, this was not true;
the trapped electrons necessitated an order of magnitude better vacuum,
and limited the storage time to less than 700~$\mu$s.

Beam losses during the measurement period, which could distort the
expected time spectrum of decay electrons, had to be minimized.
Beam scraping is used to remove, just after injection, those
muons which would likely be lost later on. To this end, the quadrupoles
are initially
powered asymmetrically, and then brought to their final symmetric voltage
configuration.  
The asymmetric voltages lower the beam and move it sideways in the
storage ring. 
 Particles whose trajectories reach too near the boundaries
of the storage volume (defined by collimators placed at the
ends of the quadrupole sectors) are lost. 
The scraping time was 17~$\mu$s during all data collection runs except
2001, where 7~$\mu$s was used.  The muon loss rates without scraping
were on the order of 0.6\% per lifetime at late times in a fill, 
which dropped to $\sim 0.2$\%  with scraping.

\subsection {The Superconducting Storage Ring}

The storage ring magnet combined with the electrostatic quadrupoles form
a Penning trap that, while very different in scale, has common features
with the electron $g$-value measurements\cite{eg2,fp}. In the electron
experiments, a single electron is stored for long periods of time, and
the spin and cyclotron frequencies are measured. 
Since the muon decays in 2.2~\ms, the study of a single muon is impossible.
Rather, an ensemble of muons is stored at relativistic
energies, and and  their spin precession in the magnetic field is 
measured using the parity-violating decay to analyze the 
spin motion(see Equations~\ref{eq:cyc-E}, \ref{eq:fullfreq}, \ref{eq:fivep}).

The design goal of $\pm 1$~ppm was placed on the field uniformity
when averaged over azimuth in the storage ring.  
A ``superferric'' design,  where the field configuration is largely
determined  by the  shape and magnetic properties of the 
iron, rather than by the current distribution in the superconducting coils,
was chosen.
To reach the ppm level of uniformity it was important to
minimize discontinuities such as holes in the yoke, spaces between 
adjacent pole pieces, and especially the spacing between pole pieces
across the magnet gap containing the beam vacuum chamber.
 Every effort was made to minimize penetrations in 
the yoke, and where they are necessary, such as for the beam entrance 
channel, additional iron is placed around the hole to minimize the
effect of the hole on the magnetic flux circuit.

\begin{figure}[h!]
\begin{center}
\includegraphics[width=0.8\textwidth]{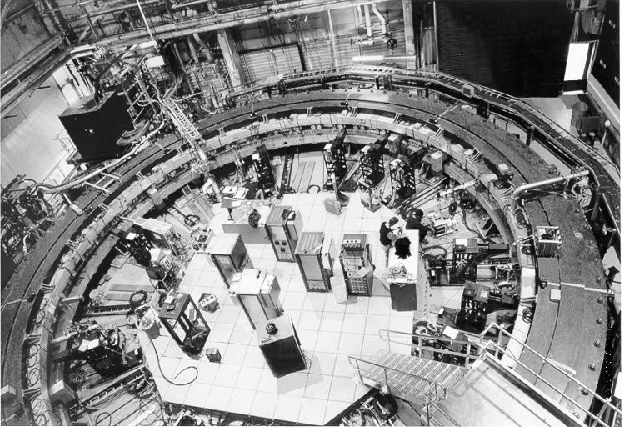}
\end{center}
\caption{The storage-ring magnet.  The cryostats for the
  inner-radius coils are clearly visible. The kickers have not yet been
installed.  The racks in the center are the quadrupole pulsers, and a 
few of the detector stations are installed, especially the quadrant of the ring closest to the person.  The magnet power supply 
is in the upper left, above the plane of the ring. (Courtesy of 
Brookhaven National Laboratory)
 \label{fg:ring-pik}}
\end{figure}

\begin{figure}[h!]
\begin{center}
\subfigure[ ]{\includegraphics[width=0.53\textwidth]{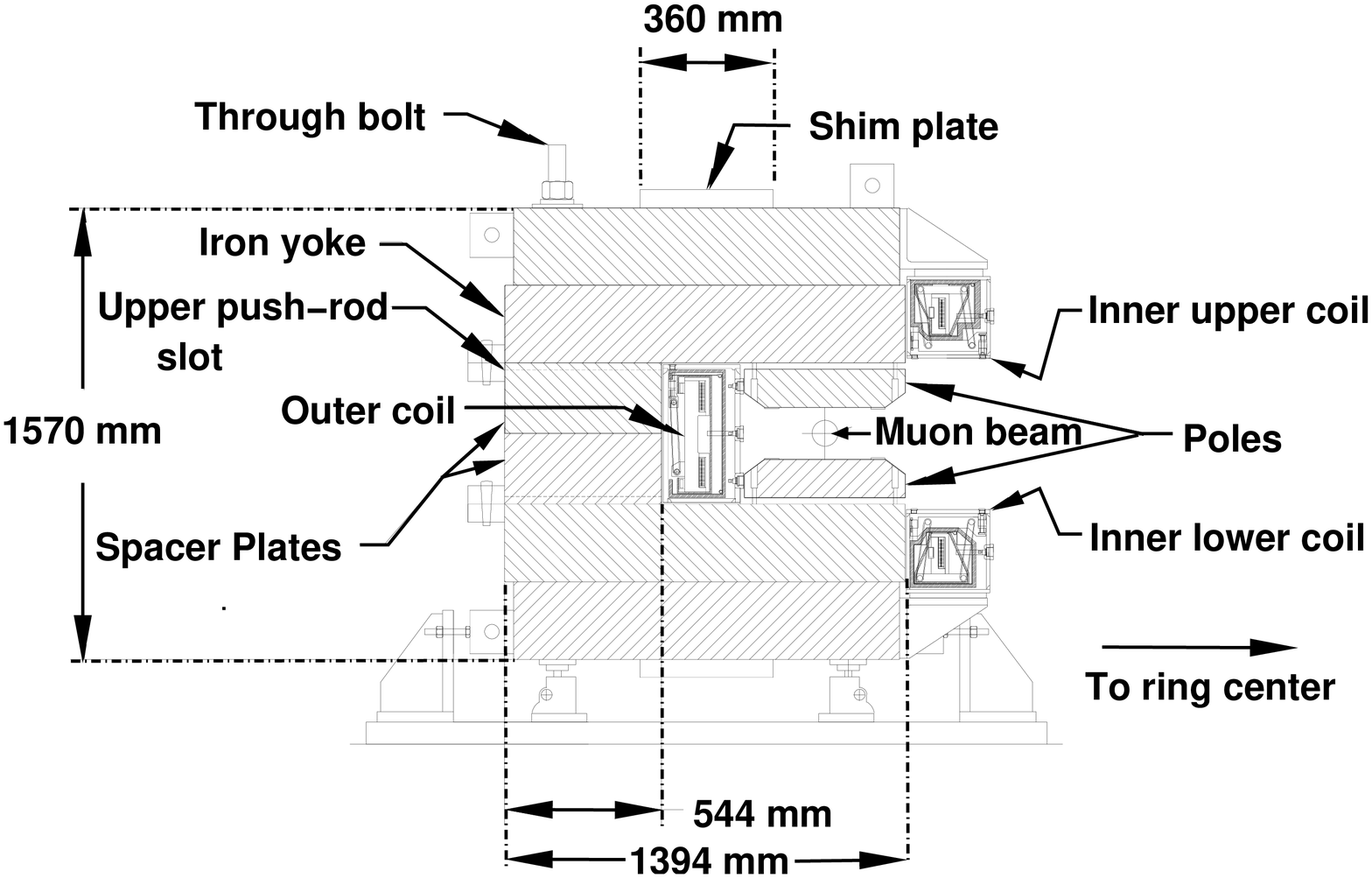}}
\subfigure[ ]{\includegraphics[width=0.46\textwidth]{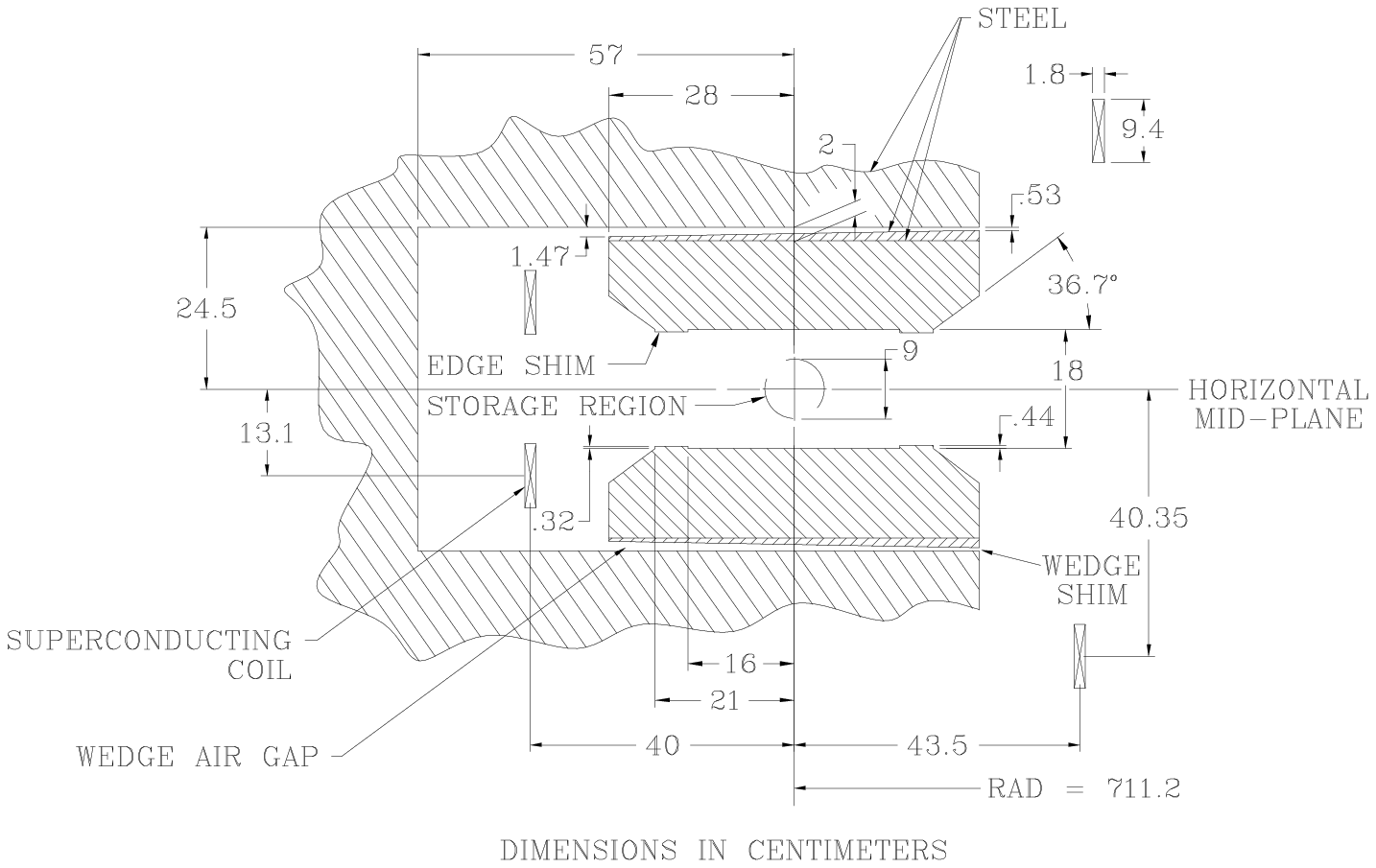}}
\end{center}
\caption{(a) A Cross-sectional view of the storage ring magnet.  The 
beam center is at a radius of 7112 mm.  The pole pieces are separated
from the yoke by an air gap. (b) An expanded view of the pole pieces,
which shows the edge shims and the wedge shim.  
 \label{fg:ringxct}}
\end{figure}

The storage ring, shown in Figure~\ref{fg:ring-pik}, is designed as a
continuous C-magnet\cite{danby} with the yoke made up of twelve sectors
with minimum gaps where  the yoke pieces come together.  
A cross-section of the magnet is shown in
Figure \ref{fg:ringxct}.
The largest gap between adjacent
yoke pieces after assembly is 0.5~mm.  The pole pieces 
are built in 36 pieces, with keystone rather than radial boundaries to 
ensure a close fit.  They
are electrically isolated from each other with 80~$\mu$m 
kapton to prevent eddy currents from running around the ring, 
especially during a quench or energy extraction
from the magnet.  The vertical 
mismatch from one pole piece to the next when going
around the ring in azimuth is held to $\pm 10$~$\mu$m, since the field 
strength depends critically on the pole-piece spacing across the magnet gap.

The field is excited by  14~m-diameter
superconducting coils, which 
in 1996 were the largest-diameter such coils ever fabricated.  
The coil at the outer radius  consists of two
identical coils on a common mandrel, above and below the 
plane of the beam, each with 
24 turns.  Each of the inner-radius coils, which
are housed in separate cryostats,
also consists of 24 turns
(see Figures~\ref{fg:infl}(b) and \ref{fg:ringxct}(a)).
The nominal operating current is 5200 A, which is driven by a power supply.
The choice of using an
extremely stable power supply, further stabilized with  feedback 
from the NMR system, was chosen
over operating in a ``persistent mode'', for two reasons.  The switch
required to change from the powering mode to persistent mode was
technically very complicated, and unlike the usual superconducting magnet
operated in persistent mode, we anticipated the need to 
cycle the magnet power a number of times during a three-month running period.

The pole pieces are fabricated from continuous vacuum-cast
low-carbon magnet steel (0.0004\% carbon),
and the yoke from standard AISI 1006 (0.07\% carbon)
magnet steel\cite{danby}.  At the design stage, 
calculations suggested that the
field could be made quite uniform, and that when averaged over azimuth,
a uniformity of $\pm 1$ ppm could be achieved.  It was anticipated that, at
the initial turn-on of the magnet, the field would have a uniformity of
about 1 part in $10^4$, and that an extensive program of shimming would
be necessary to reach a uniformity of one ppm.

A number of tools for shimming the magnet were therefore built into the design.
The air gap between the yoke and pole pieces dominates the reluctance
of the magnetic circuit outside of the gap that includes the storage region, 
and decouples the field in the storage region from
possible voids, or other defects in the yoke steel.  Iron wedges placed
in the air gap were ground to the wedge angle needed to cancel the
quadrupole field component inherent in a C-magnet.  The dipole can be tuned
locally by moving the wedge radially.
 The edge shims
screwed to the pole pieces were made oversize. Once the
initial field was mapped and the sextupole content was measured,
 the edge shims were
reduced in size individually in two steps of grinding, followed by
 field measurements, to
cancel the sextupole component of the field.

 After mechanical shimming,
these higher multipoles  were found to be 
quite constant in azimuth. They
are shimmed out on average by adjusting currents in 
conductors placed on 
printed circuit boards going around the ring in concentric
circles spaced by 0.25~cm. These  boards are
glued to the top and bottom pole faces between the edge shims and
connected at the pole ends to form a total of 240 concentric
circles of conductor, connected in groups of four, to sixty $\pm 1$~A
power supplies.  These correction coils
 are quite effective in shimming multipoles up through the octupole.
 Multipoles higher than octupole 
are less than 1 ppm at the edge of the storage aperture, and, 
 with our use of a circular storage aperture, are
unimportant in determining the average magnetic field seen by the 
muon beam (see \S~\ref{sct:avB}).

\subsection{Measurement of the Precision Magnetic Field \label{sct:field}}

\begin{figure}[h!]
\centering
\subfigure[~Absolute calibration probe]
{\label{fg:probes:a}\includegraphics*[width=0.5\textwidth]{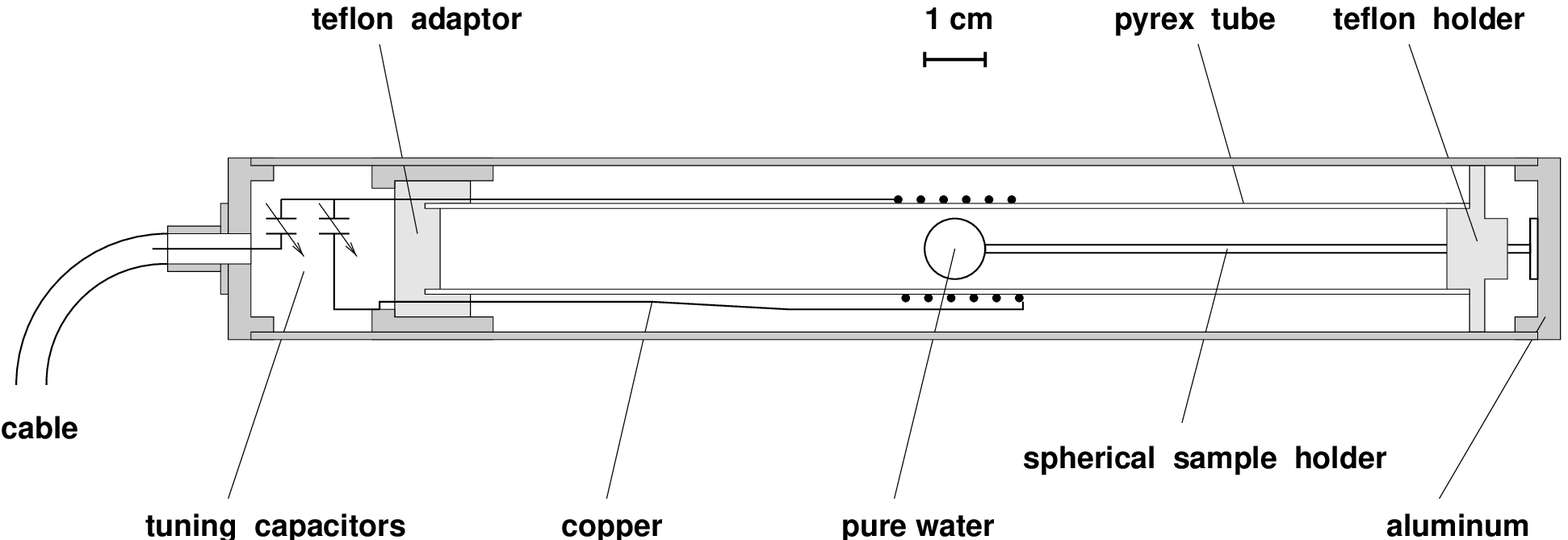}}
\hspace{0.5in}
\subfigure[~Plunging probe]
{\label{fg:probes:b}\includegraphics*[width=0.5\textwidth]{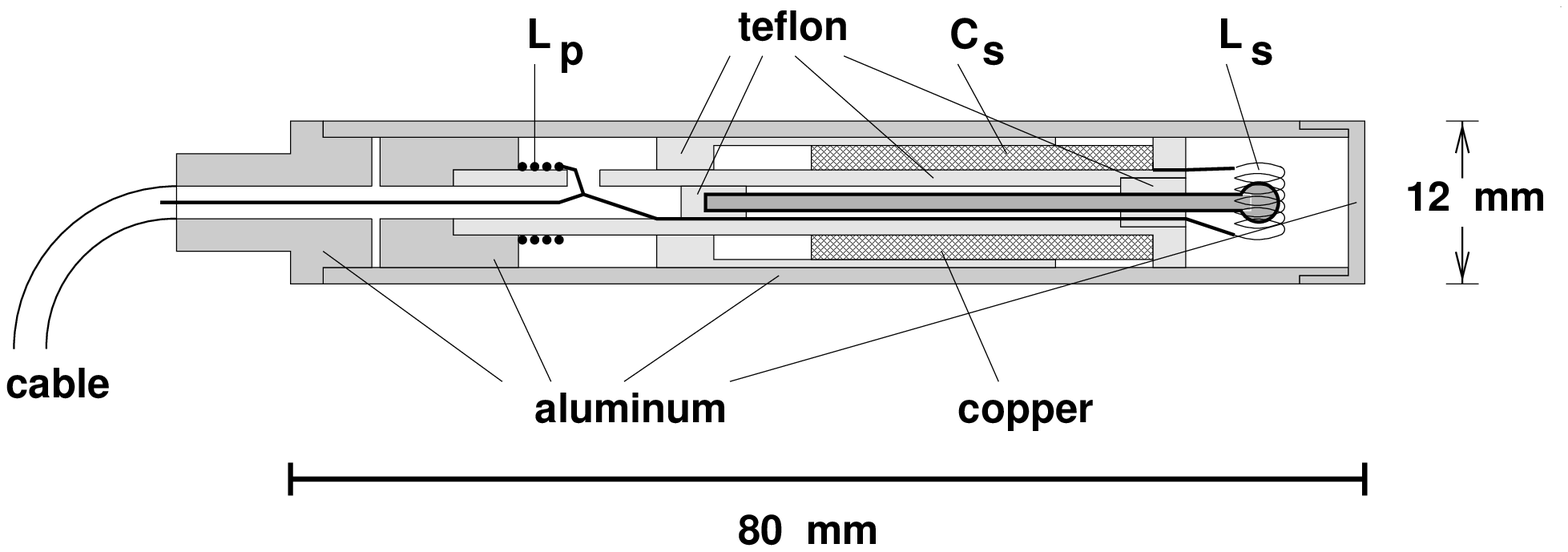}}
\hspace{0.5in}
\subfigure[~Trolley and fixed probe]
{\label{fg:probes:c}\includegraphics*[width=0.5\textwidth]{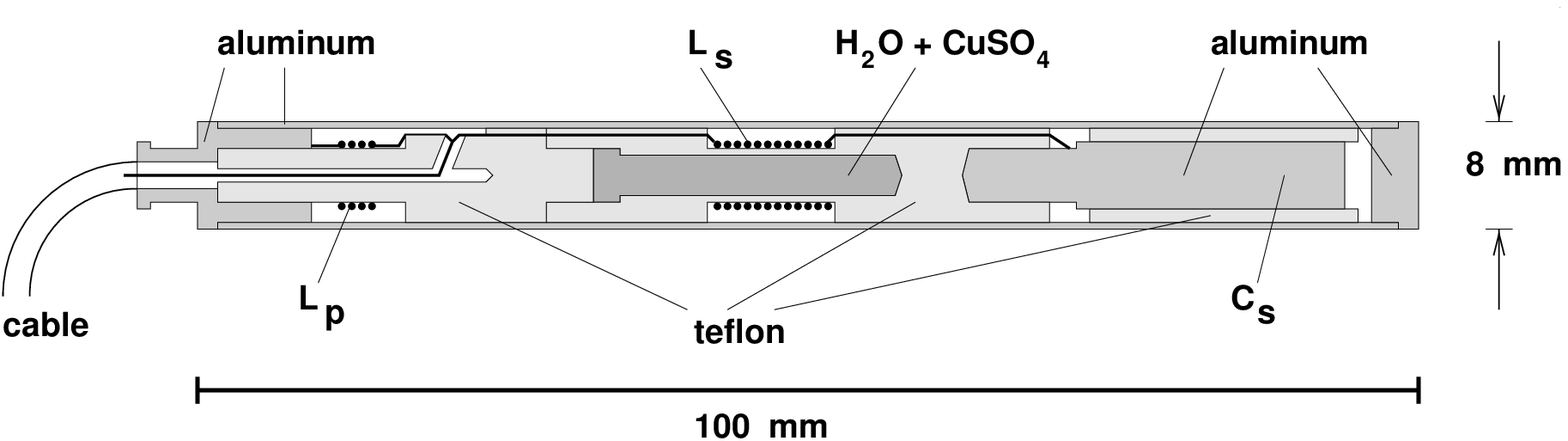}}

\caption{Schematics of different NMR probes. (a) Absolute probe
featuring a spherical sample of water.  This probe was the same
one used in reference~\cite{liu} to determine $\lambda$, the muon-to-proton
magnetic-moment ratio. (b) Plunging probe, which can be inserted
into the vacuum at a specially shimmed region of the storage ring
to transfer the calibration to the trolley probes. (c) The
standard probes used in the trolley and as fixed probes. The resonant
circuit is formed by the two coils with inductances $L_s$ and $L_p$ and a 
capacitance $C_s$ made by the Al-housing and a metal electrode. }
\label{fg:probes}
\end{figure}

The magnetic field is measured and monitored by pulsed
Nuclear Magnetic Resonance techniques\cite{nmr}. 
 The free-induction decay
is  picked up by the coil $L_S$ in Figure~\ref{fg:probes}
after a pulsed excitation rotates the proton spin in the 
sample by $90^{\circ}$ to the magnetic field,
and is mixed with the reference frequency to form the frequency $f_{\rm FID}$.
The reference frequency, $f_{\rm ref} = 61.74$~MHz, 
is obtained from a frequency synthesizer phase locked to
the LORAN C standard\cite{loran}, that is
 chosen with $f_{\rm ref} < f_{\rm NMR}$
such that for all probes  $f_{\rm FID} \simeq 50$~kHz.  The relationship
between the actual field $B_{\rm real}$ and the field corresponding to the
reference frequency is given by 
\be
B_{\rm real} = B_{\rm ref}\left( 1 + {f_{\rm FID} \over f_{\rm ref}}\right).
\label{eq:Bfield-nmr}
\ee

The field measurement process has
three aspects: calibration, monitoring the field during data collection,
 and mapping the field. The 
probes used for these purposes are shown in Figure~\ref{fg:probes}. 
To map the field,
an NMR trolley was built with an array of 17 NMR probes arranged in 
concentric circles, as shown in Figure~\ref{fg:quadpik}.  While
it would be preferable to have information over the full 90-mm aperture,
space limitations inside  the vacuum chamber,
which can be understood by 
examining Figures~\ref{fg:kick-plate} and \ref{fg:quadpik}, 
prevent a larger diameter
trolley.  The trolley is pulled around the storage ring by two cables,
one in each direction circling the ring.  One of these cables is
 a standard thin (Lemo-connector compatible)
co-axial signal cable, and the other 
is non-conducting because of
its proximity to the  kicker high voltage.  Power, data
in and out,
and a reference frequency all are carried on the single Lemo cable.

During data-collection periods, the trolley is parked in a garage
(see Figure~\ref{fg:ring}) in a 
special vacuum chamber.  Every few days, at random times, the field is mapped
using the trolley. 
To map the field, the trolley is moved into the storage
region and pulled through the vacuum chamber, measuring the field at 
over 6000 locations.  During the data collection periods, 
the storage-ring magnet remains  powered continuously for
periods lasting from five to twenty days, 
thus the conditions during mapping are
identical to those during the data collection.  

To cross calibrate the trolley probes, 
bellows are placed at one location in the ring to permit a special
NMR 
plunging probe, or a absolute calibration probe with a 
spherical water sample\cite{fei}, to plunge into the vacuum chamber.
 Measurements of the field
at the same spatial point with the plunging, calibration and trolley probes
provides both relative and absolute calibration of the trolley probes.
During the calibration measurements before and after each running period,
the spherical water probe is used to calibrate the plunging probe, the
center probe of the trolley, and several other trolley probes.
The absolute calibration probe provides the calibration 
to the Larmor frequency of the free proton\cite{phillips}, which
is called $\omega_p$ below.
 
To monitor the field on a continuous basis during data collection, a total of
378 NMR probes are placed in fixed locations
above and below the vacuum chamber around the ring.   Of
these, about half provide data useful in monitoring the field with 
time.   Some  of the others are noisy, or have bad cables or other problems, 
but a significant number of fixed probes are
located in regions near the pole-piece boundaries where the 
magnetic gradients are sufficiently large to reduce the 
free-induction decay time in the
probe, limiting the precision on the frequency measurement. 
The number of probes
at each azimuthal position around the ring alternates between two and
three, at radial positions arranged symmetrically about the 
magic radius of 7112~mm.
Because of this geometry, the fixed probes provide a good monitor
of changes in the dipole and quadrupole components of the field 
around the storage ring.

Initially the trolley and fixed probes contained a water sample. Over the
course of the experiment, the water samples in many of the probes
were replaced with
petroleum jelly. The jelly has several advantages
over water: low
evaporation, favorable relaxation times at room
temperature, a proton NMR signal almost comparable 
to that from water, and a chemical shift (and the accompanying
NMR frequency shift) with a temperature coefficient much smaller
than that of water, and thus negligible for our experiment.

%% file: det-jim-lee-20feb07.tex
\subsection{Detectors and Electronics}

The detector system consists of a variety of particle detectors:
calorimeters, position-sensitive hodoscope detectors, and a set of tracking
chambers. There are also horizontal and vertical arrays of scintillating-fiber
hodoscopes which could be temporarily inserted into the storage region.
A number of custom electronics modules were developed, including
event simulators, multi-hit time-to-digital converters(MTDC), and the 
waveform digitizers (WFD) which are at the heart of the measurement.
We refer to the data collected from one muon injection
pulse in one detector as a `spill,' and we will speak of
`early-to-late' effects: namely, the gain or time
stability requirements in a given detector at early compared to late
decay times in a spill.

\subsubsection{Electromagnetic Calorimeters: Design Considerations
\label{sct:calodesign}}

The electromagnetic calorimeters, together with the custom WFD readout
system,
are the primary source of data for
determining the precession frequency. They provide the energies and 
arrival times of the electrons, and
they also provide signal
information immediately before and after the electron pulses, 
allowing studies of 
baseline changes and pulse pile-up.

There are 24 calorimeters placed evenly around the 45-m circumference
of the storage region, adjacent to the inside radius of the storage vacuum
region as shown in Figures~\ref{fg:ring} and ~\ref{fg:ring-pik}.
Nearly all decay electrons
have momenta ($0<p(lab)<3.1$ GeV/c, see Figure~\ref{fg:differential_na})
below the stored muon momentum ($3.1$ 
GeV/c $\pm 0.2\%$), and they are swept by 
the B-field to the inside of the ring where they
can be intercepted by the calorimeters. The storage-region vacuum chamber is
scalloped so that electrons pass nearly perpendicular to the
vacuum wall before entering the calorimeters, minimizing electron 
pre-showering (see Figure~\ref{fg:ring}).
The calorimeters are positioned
and sized in order to maximize the acceptance of the highest-energy
electrons, which have the largest statistical figure of merit $NA^2$.
The variations of
$N$ and $A$ as a function of electron energy are shown in
Figures~\ref{fg:differential_na}
and \ref{fg:integral_na}.

The electrons with the lowest laboratory energies,
while more numerous than high energy electrons, generally have a lower
figure of merit and
therefore carry relatively little
information on the precession frequency.
 These electrons have relatively small radii of curvature,
and exit the the ring
vacuum chamber closer to the radial direction than electrons at 
higher energies, with most of them 
 missing the detectors entirely.
 Detection of these electrons would require detectors that cover a
much
larger portion of the circumference than is needed for high-energy electrons,
and is not cost effective.

Consequently the detector system is designed to maximize the acceptance
of the high energy decay electrons above approximately 1.8 GeV,
with the acceptance falling rapidly below this energy.
The detector acceptance
reaches a maximum of 87\% at 2.3 GeV, decreases to 70\% at 1.8 GeV, and
continues to decrease
roughly linearly to zero as the energy decreases.
With increasing energy above 2.3 GeV, the
acceptance also decreases because the highest-energy
electrons tend to enter the calorimeters at 
the outer radial edge, increasing
the loss of registered energy due to shower leakage, 
and reducing the acceptance to 80\% at 3 GeV.

In a typical analysis,
the full data sample consists of all electrons above a
threshold energy of about 1.8 GeV, where $NA^2$ is approximately a maximum,
with about 65\% of the electrons
above that energy detected (Figure~\ref{fg:integral_na}).  The
 average asymmetry is about 0.35. The loss of efficiency
is from the low-energy tail in the detector
response characteristic of
electromagnetic showers in calorimeters, and from
lower energy electrons missing the detectors altogether.
The statistical error improves by only 5\% if the data sample contains
all electrons
above 1.8 GeV compared to all above 2.0 GeV.
For threshold energies below 1.7~GeV, the decline in the average
asymmetry more than cancels the additional number of electrons in $NA^2$,
and
the statistical error actually increases.
Some of the independent
analyses fit time spectra of data formed from electrons in
narrow energy bands (about
200 MeV wide). When the results of the separate fits are combined, there was
is a 10\% reduction in the statistical error on $\omega_a$. however, there
is also a slight increase in the systematic error contribution from gain
shifts, because the relative number of events
moved by a gain shift from one energy band to
another increased. One analysis used data weighted by the asymmetry as a
function of energy. It can be shown\cite{redin07}
that this produces the same statistical
improvement as dividing the data into energy bands.

Gain and timing shift limitations are
 much more stringent within a single spill
than from spill to spill.
Shifts at late decay times
compared to early times in a given spill, so-called `early-to-late' shifts,
can lead directly to serious
systematic errors on $\omega_a$.
Shifts of gain or the $t=0$ point from one spill to the next
are generally much less serious; they will usually only change the asymmetry,
average energy, phase, 
etc., {\it but to first order, $\omega_a$ will be unaffected}.

The calorimeters should have pulses with  narrow
time widths to minimize the probability of two pulses overlapping
(pile-up) during the very high electron decay
data rates encountered at early decay times, which can reach 
a MHz in a single detector.
The scintillator is chosen to have minimal long-lived components
to reduce the afterglow from
the intense detector flash associated with beam injection.
Laser calibration studies show that the timing stability
for a typical detector over 
any 200 microsecond time interval  
is better than 15~ps, easily meeting the demands of the measurement of
$a_{\mu}$. For example, a 20 ps timing shift 
would lead to an uncertainty in $a_{\mu}$ of about 0.1 ppm, which is
small compared to the final error.
Modest detector
energy resolution ($\approx 10-15$\% at 2 GeV)
is required in order to select the desired high energy
electrons for analysis. Better energy resolution also
reduces the amount of calibration data needed
to monitor the stability of the detector gains. 

The stability requirement for the electron energy measurement (`gain')
versus time in the spill
is largely determined by the energy dependence of the phase
of the $(g-2)$ oscillation. In a fit of the data to
the 5-parameter function, the
oscillation phase is highly correlated to $\omega_a$. Therefore
a shift in the gain
from early to late decay times, combined with an energy dependence in the
(g-2) phase, can lead to a systematic
error in the determination of $\omega_a$. There are two main
contributing factors to the energy dependence,
which appear with opposite signs:
1) The phase $\phi$ in the 5-parameter function (Equation~\ref{eq:fivep})
depends on the electron drift time.
High-energy electrons must travel further,
on average, from the
point of muon decay to the detector and therefore have longer drift times.
The change in drift time with energy implies a corresponding energy
dependence
in the $(g-2)$ phase.
2) For decay electrons at a given energy,
those with positive (radially outward)
components of momentum at the muon decay
point travel further to reach the detectors than electrons with negative
(radially inward) 
components. They spread out more in the vertical direction
and  may  miss the
detectors entirely. Consequently,
electrons with positive (outward) radial momentum components will
have slightly lower acceptance than those with negative components,
causing the average spin direction to rotate slightly, leading to a shift
in $\phi$.
Recalling the correlation between electron direction and muon spin, the
overall effect is to shift the time when the number oscillation reaches its
maximum, causing a shift in
the precession phase in the 5-parameter function.
The size of the shift depends on the electron energy.
From studies of the data sample and simulations, it is established that
the detector gains need to be stable to better than
$0.2$\% over any $200\ \mu$s
time interval in a spill,
in order to keep the systematic error contribution to $\omega_a$
less than 0.1 ppm from gain shifts. This requirement is met by all of
the calorimeters. The gain
from one spill to the next is not coupled to the the precession frequency and
therefore the requirement on the spill to spill
stability is far less stringent than
the stability requirement within an individual spill.

\subsubsection{Electromagnetic Calorimeters: Construction}

\begin{figure}[h]
\begin{center}
\includegraphics[width=0.5\textwidth,angle=90]{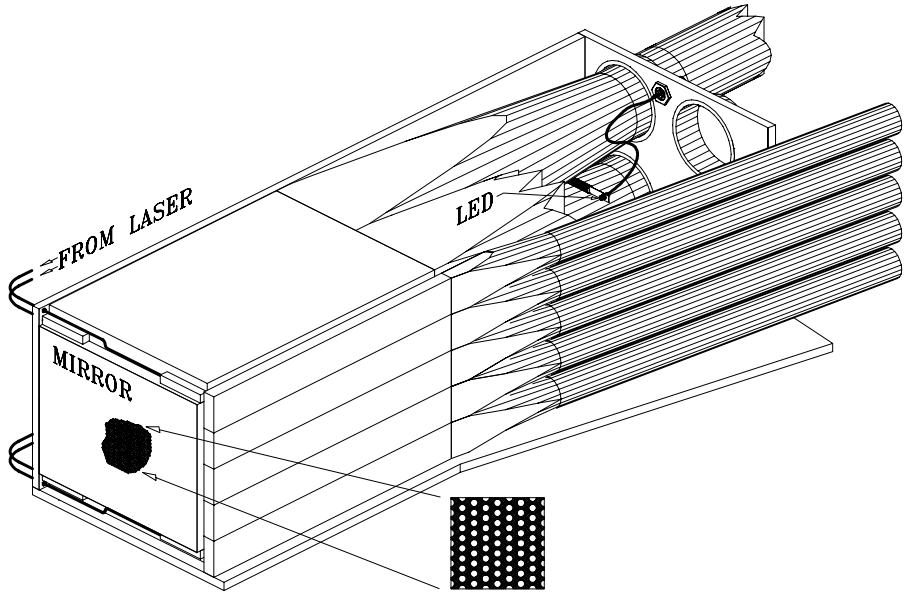}
\caption{Schematic of a detector station. The calorimeter consists of
scintillating fibers embedded in lead, with the fibers being directed
radially toward the center of the storage ring. The
electron
entrance face of the calorimeter is covered with an array of
five horizontally-oriented FSD scintillators. Six of the detector
stations also have
a PSD scintillating $xy$ hodoscope covering the entrance face.\label{fg:cm_na}}
\end{center}
\end{figure}

The calorimeters (see Figure~\ref{fg:cm_na})
consist of 1~mm diameter
scintillating fibers embedded in a
lead-epoxy matrix.\cite{sedyk} The final dimensions
are 22.5~cm (radial) $\times$ 14~cm high $\times$ 15~cm deep, where 
the vertical dimension
is limited by the vertical magnet gap. The  radiation length is
$X_0=1.14$~cm with a Moliere radius of about 2.5~cm.
This gives a calorimeter depth of 13 radiation lengths, which means between
96\% and 93\% of the energies from 1 to 3~GeV are deposited in the calorimeter,
with the remaining energy escaping. 
The fibers are oriented along the radial direction in the ring, so that
incoming electrons are incident approximately perpendicular to the fibers. 
The large-radius ends of the fibers are reflective in order
to improve the uniformity of the
pulse height response as a function of the radial direction.
There is an average 6\% change in the photomultiplier
(PMT) pulse height depending on the radial electron entrance point,
and there is a change of a few percent depending on the vertical position.
The measured versus actual electron energy deviates from a linear relationship
by about 1\% between 0 and 3 GeV, primarily due to shower leakage.

Of the electron shower energy deposited in
the calorimeter, about 5\%  is deposited in the scintillator,
producing on the order of 500 photo-electrons per GeV of incident electron
energy, with the remainder being deposited in the lead or epoxy.
The average overall energy resolution is ${\sigma(E) / E}\approx
10\% / \sqrt{E({\rm GeV})}$, dominated by
the statistics of shower sampling (i.e. variation in the energy that
gets deposited in the scintillator vs. the other calorimeter material),
with some contribution from photo-electron
statistics and shower leakage out the
back or sides of the calorimeter.

Scintillation light
is gathered by four 52.5~cm$^2$, nearly square,
acrylic light-guides which cover
the inner radius side of the calorimeter. Each light-guide necks down
to a circle which mates to a 12-dynode 5~cm-diameter
Hamamatsu R1828
photo-multiplier tube. The photomultiplier
gains in the four segments of each calorimeter are balanced initially using
monochromatic 2 GeV electrons from a BNL test beam.
Some detector calibrations shifted when the detectors were moved to 
their final storage ring locations. In these cases the gains in the
quadrants were balanced using the energy spectra from the muon decay 
electrons themselves.
The electron signals,
which are the sum of the 4 signals from the photomulutipliers
on the calorimeter quadrants, are about 5 ns wide (FWHM), and are sent to
waveform digitizers for readout.

In one spill during the time interval from  a few tens of microseconds
to
640 $\mu s$ after injection,   approximately 20 decay-electrons above 1.8 GeV
are recorded on average in each detector. The instantaneous rate
of decay electrons above 1 GeV changes from about 300 kHz to
almost zero over this period. The gain and timing of photo-multipliers
can depend on the data rate.
The necessary gain and timing stability is achieved with
custom, actively stabilized
photo-multiplier bases\cite{ouyang}.

To prevent paralysis of the photo-multipliers due to the injection flash,
the amplifications in the photo-multipliers are
temporarily reduced by a factor of about
1 million during the beam injection.  
Depending on the intensity of the flash and 
the duration of the background levels encountered at a particular
detector station, 
the amplifications are restored
at times between 2 to 50~$\mu$s after injection.
 The switching of the gain is
accomplished in the Hamamatsu
R1828 photo-multipliers by swapping the bias voltages
 on dynodes 4 and 7. 
With the proper selection of the delay time after injection to let
backgrounds die down,
the gains typically return to
 99.8\% of their steady-state value within several microseconds after
the tube is turned back on. Other gating schemes, such as switching
the photocathode voltage,
were found  to have
either required a much longer time for the gain to recover,
or failed to give the necessary
reduction in the gain when the tube is gated off.

\subsubsection{Special Detectors: FSD, PSD, Fiber Harps,
  Traceback Chambers}

Several specialized detector systems,
the Front Scintillation Detectors (FSD), Position-sensitive Detectors (PSD),
Fiber Harp Scintillators, and Traceback Detectors,
are all designed to provide information on the phase-space parameters
of the stored muon beam and their decay electrons.
Such measurements are compared to simulation
results, and are important, for example,
in the study of coherent betatron motion of the stored beam and detector
acceptances, and in placing a limit on the electric dipole moment of the 
muon (see \S~\ref{sct:beam-dyn} and \S~\ref{sct:EDM}).
A modest knowledge of the beam phase space is necessary
in order to calculate the average magnetic and electric fields
seen by the stored muons.

The FSDs cover the front (electron entrance)
faces
of the calorimeters at about half the detector stations.
They provide information on the vertical positions
of the electrons when they enter the calorimeters. Each consists of
an array of five horizontally-oriented strips of scintillator 23.5~cm long
(radial dimension) $\times$ 2.8~cm high $\times$ 1.0~cm thick. Each strip
is coupled to a 2.8-cm-diameter Hamamatsu R6427 PMT. 
The signal is discriminated and
sent to multi-hit time-to-digital converters (MTDC) for time registration.

The FSDs are an integral part of the muon loss
monitoring system. A lost muon is identified as a particle
which passes through three
successive detector stations, firing the three successive
FSD detectors in coincidence.
The candidate lost muon is also required to leave minimal energy in
each of the
calorimeters, since a muon loses energy by ionization, rather 
than creating an electromagnetic shower.  This technique only 
identifies  muons lost in the inward radial direction.
Those lost in other directions, for example, above or below the storage region,
are not monitored. The scaling between registered and actual muon losses
is estimated based on models of the muon loss mechanisms. 

Five detector stations are equipped with
PSDs. They provide better 
vertical position resolution than the FSDs, and also
 provide radial position information. Each PSD
consists of an $xy$ array of
scintillator sticks which, like the FSDs, cover the electron entrance face
of the calorimeters. There are 20 horizontally- and 32 vertically-oriented
7~mm-wide, 8~mm-thick elements  read out with multi-anode photomultipliers
followed by  discriminators
and MTDCs.

The traceback-chamber system consists of four sets of drift 
chambers which can provide
up to 12 vertical and 12 horizontal position measurements along a track,
with about 300-micron resolution. It is positioned  270
degrees around the ring from the injection point, where
a thin window is installed for the electrons to pass through the vacuum wall
with
minimal scattering. A regular calorimeter is placed behind the array to
serve as a trigger. Using the momentum derived from the
track information, plus a knowledge of the B-field, it is possible to
extrapolate the track back to the point where the trajectory is tangent to the
storage ring. Since the decay electrons are nearly parallel to the
muon momentum, and the muons are traveling with small angles
relative to the tangent, the extrapolation point
is a very good approximation ($\sigma_V \simeq 9$~mm, $\sigma_R \simeq
15$~mm) to the
position of muon decay. Thus, the traceback provides information
on the muon distribution for that portion of the ring just upstream of
the traceback position.

\begin{figure}[ht]
\centering
{\includegraphics[width=0.7\textwidth]{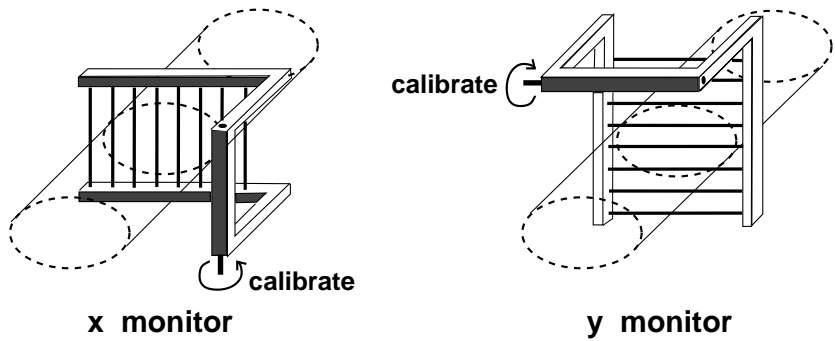}}
  \caption{A sketch of the fiber beam monitors.  The fibers can be 
rotated into the beam for calibration such that each fiber sees the
same portion of the beam.

       \label{fg:fbm}}
\end{figure}

The storage ring is equipped with two sets of scintillating-fiber
beam monitors inside the vacuum chamber.
Their purpose is to measure the beam profile as a function of time, and
they can be inserted at will into the storage region.  Each set
consists of an $x$ and $y$ plane of seven 0.5-mm diameter scintillating
fibers with 13~mm spacing that covers the central part of the beam 
region as shown in
Figure~\ref{fg:fbm}.  These fiber ``harps'' are located at 
$180^\circ$ and $270^\circ$
 around the ring from the injection point as shown in Figure~\ref{fg:ring}.
Each fiber is mated to a clear fiber, which connects to 
a vacuum optical feed-through and then to a PMT which is
read out with a WFD that digitizes continuously during the measurement time.
The  scintillating fibers are
sufficiently thin that the beam can be monitored for many tens of
microseconds after injection, as shown in Figure~\ref{fg:fbm2},
before the stored beam is degraded (effective
lifetime
about 30 microseconds
), and therefore
they are very helpful in providing information on
the early-time position and width distributions of
the stored muon beam.

\subsubsection{Electronics}

For each calorimeter, the arrival times of the signals from the four
photo-multipliers are matched to within
a few hundred picoseconds, and the analog sum is formed.
The resulting signal is fed to a custom
waveform digitizer (WFD) with 400 MHz equivalent sampling rate, which
provides several pulse height
samples from each candidate electron.

WFD data are added to the data stream
only if a trigger is formed, i.e. when  the energy associated with a pulse 
exceeds a pre-assigned threshold, usually taken to be
900 MeV. When a trigger occurs, WFD samples from about 15 ns before
the pulse to about 65 ns after the pulse are recorded. There is the
possibility of two or more electron
pulses being over-threshold in the same 80-ns time window. In that case, the
length of the readout period is extended to include both pulses.
At the earliest
decay times, the detector signals have a large pedestal due to the `flash,' and
some of the upstream detectors are continuously over-threshold at early times.

The energy and time of an electron is obtained by fitting a standard
pulse shape to the WFD pulse using a conventional
$\chi^2$ minimization. The standard pulse shape is established for each
calorimeter. It is based
on an average of the shapes of
a large number of late-time pulses where the
problems associated with overlapping pulses and backgrounds are
greatly reduced.
There are three fitting parameters, time and height of the pulse,
and the constant pedestal, with the fits typically spanning 15 samples centered
on the pulse. The typical time resolution of an
individual electron approached 60 ps.

The period after each pulse is searched for any
additional pulses from other electrons. These accidental
pulses have the advantage that they do not need to be over the
hardware threshold ($\sim 900$~mV), but rather over the
much lower ($\sim 250$ MeV)
minimum pulse height that can  be discriminated from background
by the pulse-fitting algorithm.
A pile-up spectrum is constructed by combining the triggering pulses and the
following accidental pulses as described later in this document.

Zero time for a given fill
is defined by the trigger pulse to the AGS kicker magnet that extracts
the proton bunch and sends it to the pion production target.
The resolution of the zero time
needs only to be much better than the $(g-2)$ precession period in order to
minimize loss of the asymmetry amplitude.

A pulsed UV laser signal is fanned out simultaneously to
all elements of the calorimeter stations
to monitor the gain and time stabilities.
For each calorimeter quadrant,
an optical fiber carries the laser signal to
a small bunch of the detector's scintillating fibers at
the outer-radius edge. The laser pulses
produce an excitation in the scintillators which is a good
approximation to that produced by passing charged particles.
The intensity and timing of the 
laser pulses are monitored with a separate solid-state photo-diode and a PMT,
which are shielded from the beam in order to avoid beam-related
rate changes and background which might cause shifts in the
registered time or pulse height.
The times of the laser pulses are chosen to appropriately map the gain and time
stability during the 6 to 12 injection bunches
per AGS cycle, and over the 10 muon lifetimes per injection.
Dedicated laser runs are made 6 times per day, for about 20 minutes each.
The average timing stability is typically found to be better
than 10~ps in any 200~$\mu$s-interval when averaged over a number of
events, with many stable to 5~ps. This level of timing instability 
contributes less than a 0.05 ppm systematic error on $\omega_a$.

%% file: beam-dyn20feb07.tex
\section{Beam Dynamics \label{sct:beam-dyn}}

The behavior of the beam in the $(g-2)$ storage ring directly affects 
the measurement of $a_\mu$.   Since the detector acceptance
for decay electrons depends on the radial coordinate of the muon 
at the point where it
decays,  coherent radial motion of the stored beam can produce an
amplitude modulation in the observed electron time spectrum.  Resonances
in the storage ring can cause particle losses, thus distorting the observed
time spectrum, and must be avoided when choosing the operating
parameters of the ring.  Care must be taken in setting
the frequency of coherent radial beam motion, the 
 ``coherent betatron oscillation'' (CBO) frequency, which lies close to
the second harmonic of $f_a = \omega_a/(2\pi)$. If $f_{\rm CBO}$
is too close to $2f_a$ the difference frequency $f_{-} = f_{CBO} - f_a$ 
complicates the extraction of $f_a$ from the data, and
can introduce a significant systematic error.

A pure quadrupole electric field provides a 
linear restoring force in the vertical direction, and the combination of
the (defocusing) electric field and the central magnetic field provides a
linear restoring force in the radial direction. 
The $(g-2)$ ring is a weak focusing ring\cite{wied,edwards,cp} with the field
index
\begin{equation}
n = {\kappa R_0 \over \beta B_0},
\label{eq:n}
\end{equation}
where $ \kappa$ is the electric quadrupole  gradient.  For a ring with a
uniform vertical dipole magnetic field and a uniform quadrupole field 
that provides vertical focusing 
covering the full azimuth, the stored particles undergo simple harmonic
motion called betatron oscillations,
in both the radial and vertical dimensions.

The horizontal and vertical motion are given by
\be
x = x_e + A_x \cos (\nu_x {s \over R_0} + \delta_{x})\quad {\rm and} \quad
y = A_y \cos (\nu_y {s\over R_0} + \delta_{y}),
\ee
where $s$ is the arc length along the trajectory, and 
$R_0= 7112$~mm is the radius of the central orbit in the storage ring.
The horizontal and vertical tunes are given by $\nu_x= \sqrt{1-n}$ 
and $\nu_y= \sqrt n$.
Several $n$~-~values
were used in E821 for data acquisition: 
$n = 0.137,\ 0.142$ and 0.122.
The horizontal and 
vertical betatron frequencies are given by
\begin{equation}
 f_x = f_C \sqrt{1-n}\simeq 0.929 f_C \quad {\rm and} \quad
f_y = f_C \sqrt{n} \simeq 0.37 f_C,
\label{eq:betafreq}
\end{equation}
where $f_C$ is the cyclotron frequency and the numerical values 
assume that $n=0.137$.  The corresponding betatron wavelengths are
$\lambda_{\beta_x} = 1.08(2 \pi R_0)$ and $\lambda_{\beta_y} = 2.7(2 \pi R_0)$.
It is important that the betatron wavelengths are not simple multiples of the
circumference, as this minimizes the ability of ring imperfections
and higher multipoles
to drive resonances that would result in particle losses from the ring.

The field index, $n$, also determines the acceptance of the ring.  
The maximum horizontal and vertical angles of the muon momentum
are given by
\be
\theta^x_{\rm max} = { x_{\rm max} \sqrt{1-n} \over R_0},
\quad {\rm and} \quad 
\theta^y_{\rm max} = { y_{\rm max} \sqrt{n} \over R_0},
\label{eq:max-angle}
\ee
where $x_{\rm max},y_{\rm max} =  45$~mm is the radius of the storage aperture.
For a betatron amplitude $A_x$ or $A_y$ less than 45~mm, the 
maximum angle is reduced, as can be seen from the above equations.  

\begin{figure}[htb]
\centering
  \includegraphics[width=0.5\textwidth]{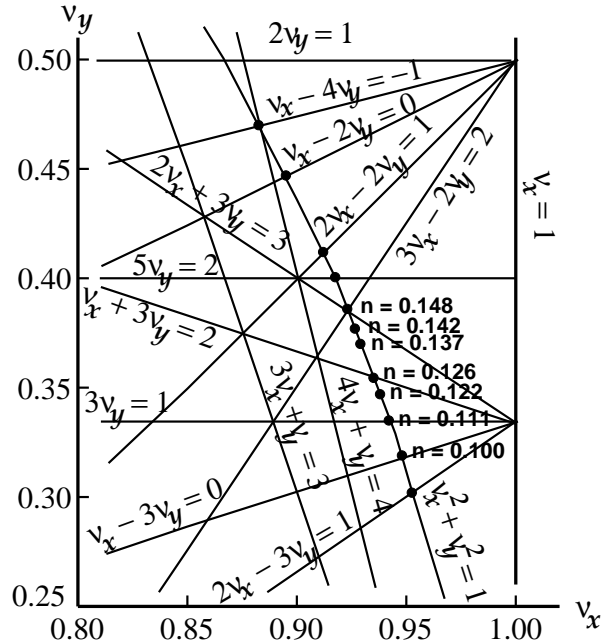}
  \caption{The tune plane, showing the three operating points used 
during our three years of running.
       \label{fg:tunepl}}
\end{figure}

Resonances in the storage ring
will occur if $L \nu_x + M \nu_y = N$, where
$L$, $M$ and $N$ are integers,  which must be avoided in choosing the
operating value of the field index.
 These resonances form straight lines on
the tune plane shown in Figure~\ref{fg:tunepl}, which shows resonance lines up
to fifth order.  The operating point lies on the circle
 $\nu^2_x + \nu^2_y = 1$.

For a ring with discrete quadrupoles, the focusing strength changes as a 
function of azimuth, and the equation of motion looks like
an oscillator whose spring constant changes as a function of
azimuth $s$. The motion is described by
\be
x(s) = x_e + A \sqrt{\beta(s)} \cos (\psi(s) + \delta),
\ee
where $\beta(s)$ is one of the three Courant-Snyder parameters.\cite{edwards}
The layout of the storage ring is shown in Figure~\ref{fg:ring}.  The
four-fold symmetry of the quadrupoles was chosen because it provided
quadrupole-free regions for the kicker, traceback chambers, fiber monitors,
and trolley garage; but the   most important benefit of four-fold
symmetry over the two-fold used at CERN\cite{cern3b} is that
 $\sqrt{\beta_{\rm max}/\beta_{\rm min}} = 1.03$.  The two-fold symmetry
used at CERN\cite{cern3b} gives $\sqrt{\beta_{\rm max}/\beta_{\rm min}} =
  1.15$.  The CERN magnetic field had significant non-uniformities on the
outer portion of the storage region, which when combined with the 15\%
beam ``breathing'' from the quadrupole lattice 
made it much more difficult to determine the average
magnetic field 
weighted by the muon distribution (Equation \ref{eq:omega}).

The detector acceptance depends on the radial position of the muon
when it decays, so that any {\it coherent} radial beam
motion will amplitude modulate the decay $e^{\pm}$ distribution.
The principal frequency will be the
 ``Coherent Betatron Frequency,''
\begin{equation}
f_{\rm CBO} = f_C - f_x = (1 - \sqrt{1-n})f_C \simeq 470\ {\rm kHZ},
\label{eq:cbo}
\end{equation}
which is the frequency at which a single fixed detector sees the beam
coherently moving
  back and forth radially.  This CBO frequency  
is close to the second harmonic of the
$(g-2)$ frequency, $f_a= \omega_a/2\pi\simeq 228$~Hz. 

 An alternative 
way of thinking about the CBO motion
is to view the ring as a spectrometer where the inflector exit is
imaged at each successive betatron wavelength, $\lambda_{\beta_x}$.
  In principle, an inverted
image appears at half a betatron wavelength; but the radial image is spoiled
by the $\pm 0.3$\% momentum dispersion of the ring.  A given
detector will see the beam move radially with the CBO frequency, which is
also the frequency at which the horizontal waist precesses around the ring.
Since there is no dispersion in the vertical dimension, the vertical
waist  (VW) is reformed every half wavelength $\lambda_{\beta_y}/2$. 
A number of frequencies in the ring are tabulated in Table~\ref{tb:freq}
\begin{table}[hbt]
\begin{center}
\caption{Frequencies in the $(g-2)$ storage ring, assuming
that the quadrupole field is uniform in azimuth and that  $n~=~0.137$.}
\begin{tabular}{|l|l|l|l|} \hline
{\it Quantity } & {\it Expression} &{\it Frequency } & {\it Period } 
 \\
 & & &  \\
\hline
$f_a$ &  ${e \over 2 \pi mc} a_{\mu}  B$ & 0.228~MHz & 4.37 $\mu$s  \\
\hline
$f_c$ & ${v \over 2 \pi R_0}$ & 6.7~MHz & 149~ns \\
\hline 
$f_x$ & $\sqrt{1-n}f_c$ & 6.23~MHz & 160~ns \\
$f_y$ & $\sqrt{n}f_c$ & 2.48~MHz & 402~ns \\
\hline
$f_{\rm CBO}$ & $f_c - f_x$ & 0.477~MHz & 2.10  $\mu$s \\
$f_{\rm VW}$  & $f_c - 2f_y$ & 1.74~MHz & 0.574  $\mu$s \\
\hline
\end{tabular}
\label{tb:freq}
\end{center}
\end{table}

\begin{figure}[h]
\begin{center}
\subfigure[ ] {\includegraphics[width=0.5\textwidth,angle=0]{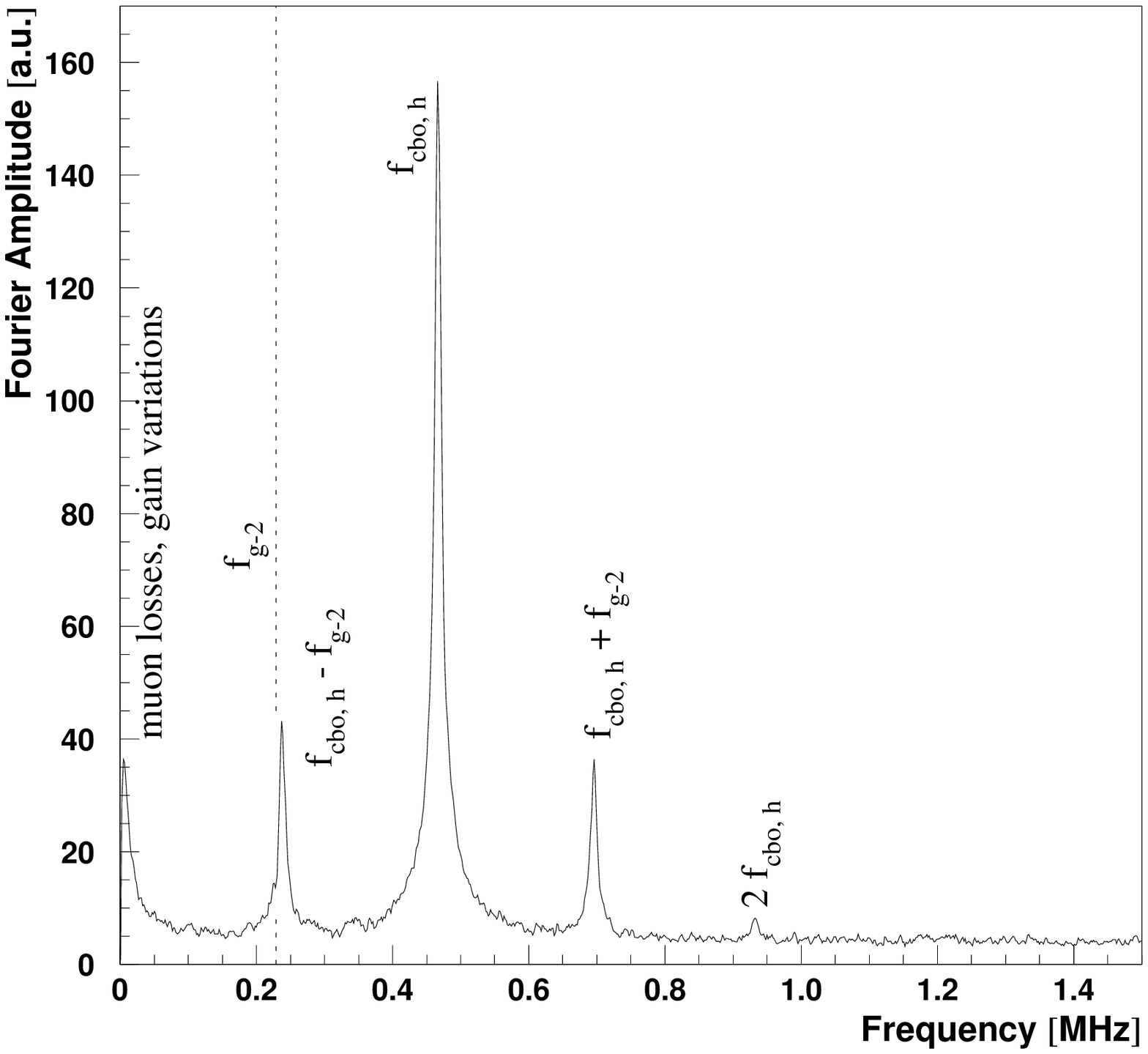}}
\subfigure[ ]
{\includegraphics[width=0.45\textwidth,angle=0]{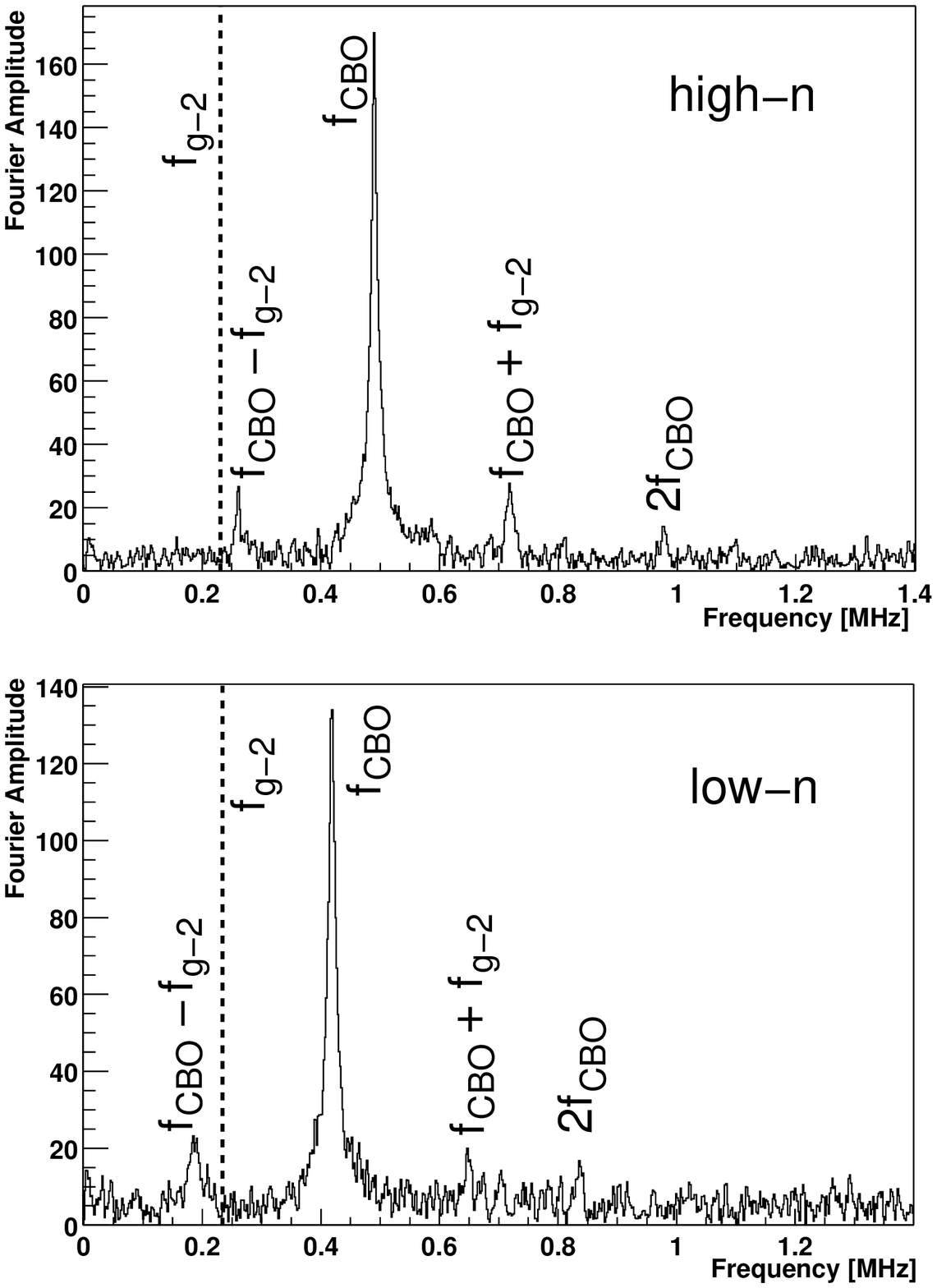}}
\caption{The Fourier transform to the residuals from a fit to the
  five-parameter function, showing clearly the coherent beam frequencies.
(a) is from 2000, when the CBO frequency was close to $2\omega_a$, 
and (b) shows the Fourier transform for the two
n-values used in the 2001 run period.
 \label{fg:fourier}}
\end{center}
\end{figure}

 The CBO frequency and its
sidebands are clearly visible in the Fourier transform to the residuals
from a fit to the five-parameter fitting function Equation~\ref{eq:fivep}, and
are shown in Figure~\ref{fg:fourier}. 
The vertical waist frequency is barely visible. 
In 2000, the
quadrupole voltage was set such that the CBO frequency was uncomfortably
close to the second harmonic of $f_a$, thus placing the difference 
frequency $f_{-} = f_{CBO} - f_a$ next to $f_a$. 
 This nearby sideband
forced us to work very hard to
understand the CBO and how its related phenomena affect the value of 
$\omega_a$ obtained from fits to the data.  In 2001, we carefully set
$f_{CBO}$ at two different values, one well above, the other well 
below $2f_a$, which greatly 
reduced this problem.  

\subsection{Monitoring the Beam Profile \label{sct:mubeam}}

Three tools are available to us to monitor the muon distribution. 
Study of the beam de-bunching after injection yields information on the
distribution of equilibrium radii in the storage ring. 
The FSDs provide information on the vertical centroid of the beam.
 The wire chamber system
and the fiber beam monitors, described above, also provide
valuable information on the properties of the stored beam.

\begin{figure}[ht]
\centering
{\includegraphics[width=0.8\textwidth]{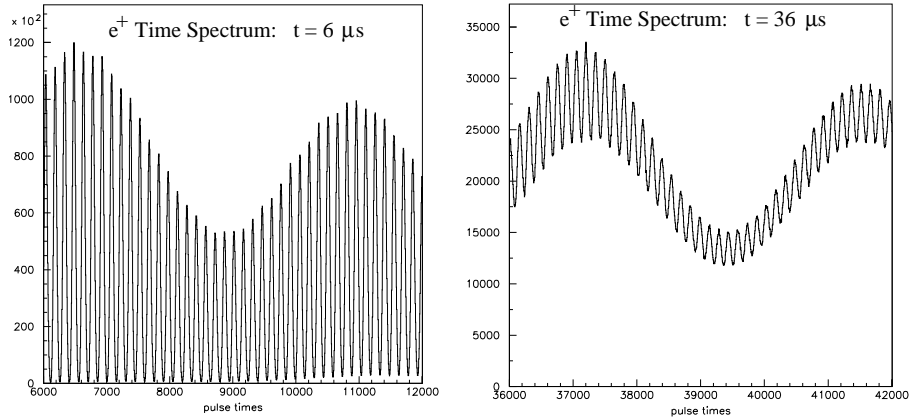}}
  \caption{The time spectrum of a single calorimeter soon after injection.
The spikes are separated by the cyclotron period of 149~ns.
       \label{fg:fastrot}}
\end{figure}

The beam bunch that enters the storage ring has a time spread with
$\sigma \simeq 23 $~ns, while the cyclotron period is 149~ns. 
The momentum distribution of stored muons produces a corresponding
distribution in radii of curvature. The distributions depend on the
phase-space 
acceptance of the ring, the phase space of the beam at the injection point,
and the kick given to the beam at injection.
The narrow horizontal dimension of the beam at the injection point, 
about 18~mm, restricts the stored momentum distribution to about 
$\pm 0.3\%$.  As the muons circle the ring,
  the muons at smaller radius
(lower momentum) eventually pass those at larger radius repeatedly after
multiple transits around the ring,
and the bunch structure largely disappears after 60~\ms.  This de-bunching can
be seen in 
Figure~\ref{fg:fastrot} where the signal from a single detector is shown at two
different times following injection.  The bunched beam is seen very clearly
in the left figure, with the 149~ns cyclotron period being obvious.  The slow
amplitude modulation comes from the $(g-2)$ precession.  By 36~\ms the
beam has largely de-bunched.

\begin{figure}[h!]
\centering
  \includegraphics[width=0.35\textwidth,angle=-90]{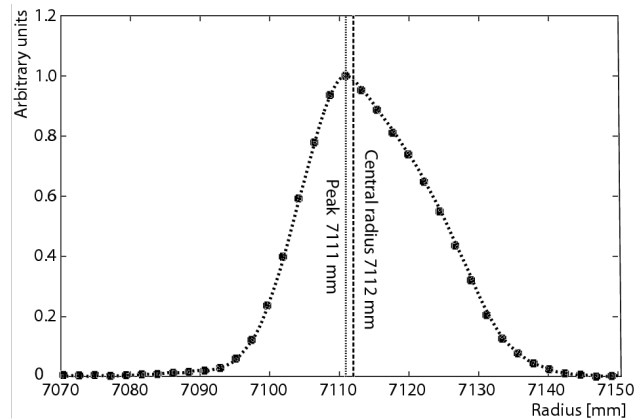}
  \caption{The distribution of equilibrium radii obtained from the 
beam de-bunching.  The solid circles are from a de-bunching model fit to
the data, and the dotted curve is obtained from a modified Fourier analysis.
       \label{fg:rdist}}
\end{figure}

Only muons with orbits centered at
the central radius have the
``magic'' momentum, so knowledge of the momentum distribution, or
equivalently
the distribution of equilibrium radii, is important in determining the
correction to $\omega_a$ caused by the radial electric field used for
vertical focusing.
Two methods of obtaining the distribution of equilibrium radii
from the beam debunching are employed in
E821.  One method uses a model of the time evolution of the bunch structure.
A second, alternative procedure uses modified Fourier
techniques\cite{orlov}.  The results from these
analyses are shown
in Figure~\ref{fg:rdist}. The discrete points were obtained using the model,
and the dotted curve was obtained with the modified
Fourier analysis. The two analyses agree.
The measured
distribution is used both in determining the average magnetic field
seen by the muons and the
radial electric field correction discussed below.

\begin{figure}[ht]
\centering
\subfigure[ ] {\includegraphics[width=0.44\textwidth,angle=0]{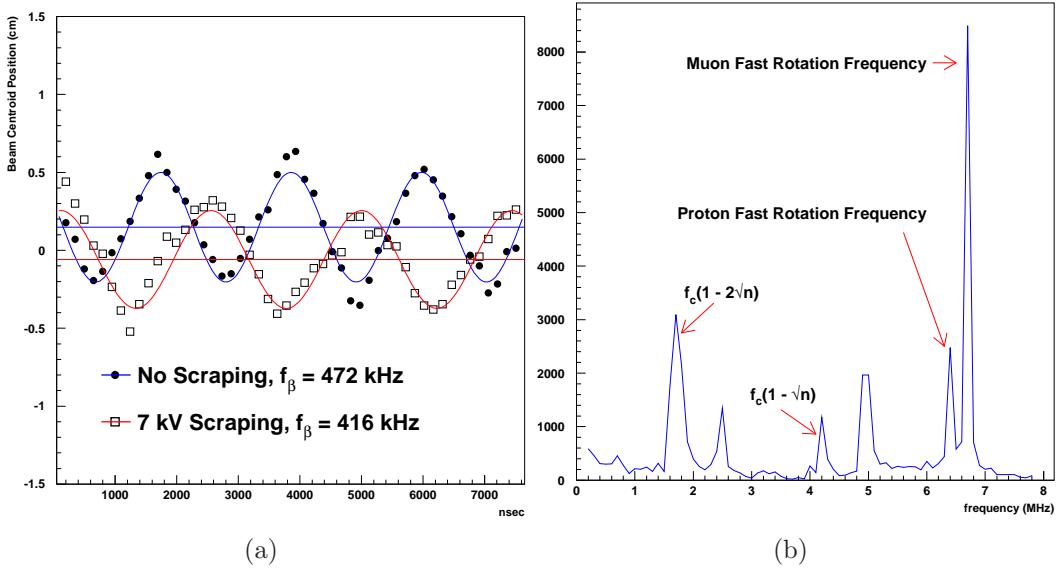}}
\subfigure[ ] {\includegraphics[width=0.44\textwidth,angle=0]{fig/frpeaks.epsi}}%

  \caption{(a) The horizontal
beam centroid motion with beam scraping and without, using
data from the scintillating fiber hodoscopes; note
the tune change between the two. (b) A Fourier transform
of the pulse from a single horizontal fiber, which shows clearly the vertical
waist motion, as well as the vertical tune. The presence of
stored protons is clearly seen in this frequency spectrum.
       \label{fg:fbm2}}
\end{figure}

The scintillating-fiber monitors show clearly the vertical and horizontal
tunes as expected.  In Figure~\ref{fg:fbm2}, the horizontal
beam centroid motion is
shown, with the quadrupoles powered asymmetrically during scraping, and
then
symmetrically after scraping.  A Fourier transform of the latter signal shows
 the expected frequencies, including the cyclotron frequency of protons
stored in the ring.  The traceback system also  sees the
CBO motion.

\subsection{Corrections to $\omega_a$: Pitch and Radial Electric 
Field\label{sct:p-E}}

If the velocity is not transverse to the magnetic field, or if a muon is
not at $ \gamma_{\rm magic}$, the difference frequency is modified
as indicated in Equation~\ref{eq:Ediffreq}. Thus the measured
frequency $\omega_a$ must be corrected for the 
effect of a radial electric field (because of the $\vec \beta \times \vec E$
term),
and for the vertical pitching motion of the muons (which enters through
the $\vec \beta \cdot \vec B$ term). 
These are the \underbar{only} corrections made to the $\omega_a$ 
data.  We sketch the derivation, for E821 below\cite{paley}.
For a general derivation  the reader is 
referred to
References~\cite{fp,farley-pitch}.

First we calculate the effect of the electric field, for the moment
neglecting the $\vec \beta \cdot \vec B$ term. 
If the muon momentum is different from the magic momentum, the precession 
frequency is given by
\be
\omega'_a = \omega_a\left[ 1 - \beta {E_r \over B_y} 
\left( 1 - {1 \over a_{\mu} \beta^2\gamma^2 }\right) \right].
\ee
Using $p = \beta \gamma m = (p_m + \Delta p)$, after some algebra one finds
\be
{\omega_a' - \omega_a \over \omega_a} = 
{\Delta \omega_a \over \omega_a} = -2 {\beta E_r \over B_y }
\left({ \Delta p \over p_m}\right).
\ee
Thus the effect of the radial electric field reduces the observed frequency
from the simple frequency $\omega_a$ given in Equation~\ref{eq:diffreq}.
 Now
\be
{\Delta p \over p_m} = (1-n){\Delta R \over R_0} = (1-n) {x_e\over R_0},
\ee
where $x_e$ is the muon's equilibrium radius
of curvature  relative to the central orbit.
The electric quadrupole field is
\be
E = \kappa x = { n \beta B_y \over R_0} x.
\ee
We obtain
\be 
{\Delta \omega \over \omega} = - 2n(1-n)\beta^2 {x x_e \over R^2_0B_y},
\ee
so clearly
the effect of muons not at the magic momentum is to lower the observed
frequency.
For a quadrupole focusing field plus a uniform magnetic field,
the time average of $x$ is just $x_e$, so 
the electric field correction is given by
\be
C_E = {\Delta \omega \over \omega} 
=  - 2n(1-n)\beta^2 {\langle x^2_e\rangle \over R^2_0B_y},
\ee
where $\langle {x^2_e}\rangle$
 is determined from the fast-rotation analysis (see Figure \ref{fg:fastrot}).
 The uncertainty on  $\langle {x^2_e}\rangle$ is
added in quadrature with the uncertainty in the placement of the 
quadrupoles of $\delta R = \pm 0.5$~mm ($\pm0.01$~ppm), and
with the uncertainty in the mean vertical position of the beam,
$\pm 1$~mm ($\pm 0.02$~ppm).  For the  low-$n$
2001 sub-period, $C_E = 0.47 \pm 0.054$~ppm.

\begin{figure}[ht]
\centering
\includegraphics[width=0.25\textwidth,angle=0]{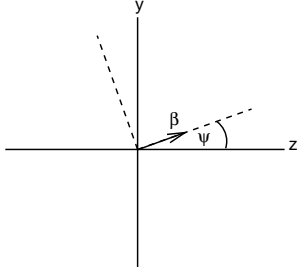}

  \caption{The coordinate system of the pitching muon.  The angle $\psi$
varies harmonically. The vertical direction is $\hat y$ and
 $\hat z$ is the azimuthal (beam) direction.
       \label{fg:pitch-cs}}
\end{figure}

The vertical betatron oscillations of the stored muons lead to 
$\vec \beta \cdot \vec B \neq 0$. 
Since the $\vec \beta \cdot \vec B$ term in Equation~\ref{eq:fullfreq}
 is quadratic
in the components of $\vec \beta$, its contribution to $\omega_a$ will not 
generally average to zero. Thus the
spin precession frequency has a small dependence on the betatron 
motion of the beam. It turns out that the only significant 
correction comes from the vertical betatron oscillation; 
therefore it is called the pitch correction (see Equation~\ref{eq:Ediffreq}).
As the muons undergo vertical
betatron oscillations, the ``pitch'' angle between the momentum and
the horizontal (see Figure~\ref{fg:pitch-cs})
varies harmonically as $\psi = \psi_0 \cos \omega_y t$, where $\omega_y$ is
the vertical betatron frequency $\omega_y = 2 \pi f_y$, given in Equation 
\ref{eq:betafreq}. 
In the approximation that all muons are at the magic $\gamma$,
we set $a_{\mu}-1/(\gamma^2-1)=0$ in Equation~\ref{eq:Ediffreq} and obtain
\be
\vec \omega_a'
=   -{q \over m}
\left[ a_{\mu} \vec B -  a_{\mu}\left( {\gamma \over \gamma + 1}\right)
(\vec \beta \cdot \vec B)\vec \beta 
 \right],
\label{eq:diff-pitch}
\ee
where the prime indicates the modified frequency as it did in the discussion
of the radial electric field given above, and 
$\vec \omega_a = -(q/m)a_{\mu}\vec B $.  We adopt the (rotating) 
coordinate system
shown in Figure~\ref{fg:pitch-cs}, where $\vec \beta$ lies in the $zy$-plane, 
$z$ being the direction of propagation, and $y$ being vertical in the storage
ring. Assuming $\vec B = \hat y B_y$, 
$\vec \beta = \hat z \beta_z +  \hat y \beta_y 
=\hat z \beta \cos \psi  + \hat y \beta \sin \psi $,  we find
\be
\vec \omega_a' = - {q \over m}
 [  a_{\mu}\hat y  B_y -  a_{\mu}\left( {\gamma \over \gamma + 1}\right)
 \beta_y B_y( \hat z \beta_z +\hat y \beta_y )].
\ee
The small-angle approximation $\cos \psi \simeq 1$ and $\sin \psi \simeq 
\psi$ gives the component equations 
\be
\omega_{ay}' = \omega_a\left[ 1 - \left( {\gamma -1 \over \gamma}\right)
\psi^2 \right ] 
\ee
and
\be
\omega_{az}' = -\omega_a \left({\gamma-1 \over \gamma}\right) \psi.
\ee

Rather than use the components given above, we can resolve
 $\omega_a'$ into components along the coordinate system
defined by $\vec \beta$ (see Figure~\ref{fg:pitch-cs}) using the
standard rotation formula.  The transverse component of $\omega'$ is given
by
\be
\omega_{\perp} = \omega_{ay}' \cos \psi - \omega_{az}' \sin \psi.
\ee
Using the small-angle expansion for $\cos \psi \simeq 1 - \psi^2/2$,
we find 
\be
\omega_{\perp} \simeq \omega_a \left[ 1 - {\psi^2 \over 2}\right].
\ee
As can be seen from Table~\ref{tb:freq},
the pitching frequency $\omega_y$ is an order of magnitude larger than the
frequency $\omega_a$, so that
in one \g2 period $\omega_{\parallel}$ oscillates
more than ten times, thus averaging out its effect on $\omega_a'$ so
$\omega_a' \simeq \omega_{\perp}$.  Thus 
\begin{equation}
\omega_a \simeq 
- {q \over m} a_{\mu} B_y \left( 1 - {\psi^2 \over 2} \right)
=   - {q \over m} a_{\mu} B_y 
\left( 1 - {{\psi_0^2 cos^2 \omega_yt } \over 2}\right).
\end{equation}
Taking the time average yields a pitch correction
\be
C_p = - {\langle \psi^2 \rangle \over 2} = - {\langle \psi_0^2\rangle \over 4}
=  - { n\over 4} {{ \langle y^2 \rangle }  \over R_0^2},
\ee
where we have used Equation~\ref{eq:max-angle}
$\langle \psi_0^2 \rangle = { { n \langle y^2 \rangle } / R_0^2}$.
The quantity $ \langle y_0^2 \rangle $  was both 
determined experimentally and from
simulations.  
 For the 2001 period, $C_p = 0.27 \pm 0.036$~ppm,  the
amount the precession frequency is lowered from that given in 
Equation~\ref{eq:omega} because $\vec \beta \cdot \vec B \neq 0$.

We see that both the radial electric field and the
vertical pitching motion {\it lower} the observed
frequency from the simple difference frequency 
$\omega_a = (e/m)a_{\mu}B$, which enters into our determination
of $a_{\mu}$ using Equation~\ref{eq:lambda}. Therefore
 our observed frequency must be
{\it increased} by these corrections to obtain the measured value
of the anomaly.  Note that if $\omega_y \simeq \omega_a$ the 
situation is more complicated, with a resonance behavior that is discussed
in References~\cite{fp,farley-pitch}.

%% file: omega-p-anal20feb07.tex
\section{Analysis of the Data for $\omega_p$ and $\omega_a$}

In the data analysis for E821, great care was taken to insure that the
results were not biased by previous measurements or the theoretical
value expected from the standard model.  This was achieved by a blind
analysis which guaranteed that no single member of the collaboration could
calculate the value of $a_\mu$ before the analysis was complete. 
Two frequencies, $\omega_p$, the Larmor frequency of a free proton which is
proportional to the B field, and
$\omega_a$, the frequency that  muon spin precesses relative to its momentum
are measured.  The analysis was divided into two separate efforts,
$\omega_a$, $\omega_p$, with no collaboration member permitted to work on
the determination of both frequencies.

In the first stage of each year's analysis, each
independent $\omega_a$ (or $\omega_p$) analyzer presented intermediate results
with his own concealed offset on $\omega_a$ (or $\omega_p$). 
Once the independent analyses of  $\omega_a$
appeared to be mutually consistent, an offset common to all independent
$\omega_a$ analyses was adopted, and a similar step was taken by the
independent analyses of $\omega_p$.
The $\omega_a$ offsets were kept strictly concealed, especially from the
$\omega_p$ analyzers.
Similarly, the $\omega_p$ offsets were kept strictly concealed,
especially from the
$\omega_a$ analyzers. The nominal values of $\omega_a$ and $\omega_p$ were
known at best to many ppm error, much larger than the eventual result, and
could not be guessed with any precision.
No one person was allowed to know both offsets, and
it was therefore impossible to calculate the value
of $a_\mu$ until the offsets were publicly revealed, after all analyses were
declared to be complete.

For each of the four yearly data sets, 1998-2001,
 there were 
 between four and five largely independent analyses of $\omega_a$,
and two independent analyses of $\omega_p$.
Typically, on $\omega_a$ there were one or
two physicists conducting independent analyses in two successive years, 
and one on $\omega_p$,  providing continuity between the analysis of
 the separate data sets.  Each of the
fit parameters, and each of the potential sources of systematic error were
studied in great detail. 
For the high statistics data sets, 1999, 2000 and 2001 it was
necessary in the $\omega_a$ analysis
to modify the five-parameter function given in
Equation~\ref{eq:fivep}
to account for a number of small effects.  Often different approaches
were developed to account for a given effect, although there were common 
features between some of the analyses.

All intermediate results for $\omega_a$
were presented in terms of ${\Re}$, which is defined by
\be
\omega_a=2\pi \cdot 0.2291 MHz \cdot 
[1\pm({\Re}\pm \Delta{\Re})\times 10^{-6}]
\ee
 where $\pm \Delta {\Re}$ is the concealed offset.
Similarly, the $\omega_p$ analyzers maintained a constant offset which
was strictly concealed from the rest of the collaboration.

To obtain the value of ${\mathcal R}=\omega_a / \omega_p$ to use
in Equation~\ref{eq:lambda},
 the  pitch and radial
electric-field corrections discussed in \S\ref{sct:p-E} were added to
the measured frequency $\omega_a$ obtained from the least-squares 
fit to the time spectrum.  Once these two corrections were made,
the value of $a_{\mu}$ was obtained from Equation~\ref{eq:lambda} and was
published  with no other changes.

\subsection{The Average Magnetic Field: The $\omega_p$ Analysis}

The magnetic field data consist of three separate sets of measurements:
the calibration data taken before and after each running period, maps of
the magnetic field obtained with the NMR trolley at intervals of 
a few days, and the field measured by each of the fixed NMR probes placed
outside  the vacuum chamber.  It was these latter measurements, taken 
concurrent with the muon spin-precession data and then tied
to the field mapped by the trolley, which were used to determine the
average magnetic field  in the storage ring, and subsequently
the value of $\omega_p$
to be used in Equation \ref{eq:lambda}.

\subsubsection{Calibration of the Trolley Probes}

\begin{figure}
\begin{center}
\subfigure[Calibration position]
{\includegraphics[width=3in]{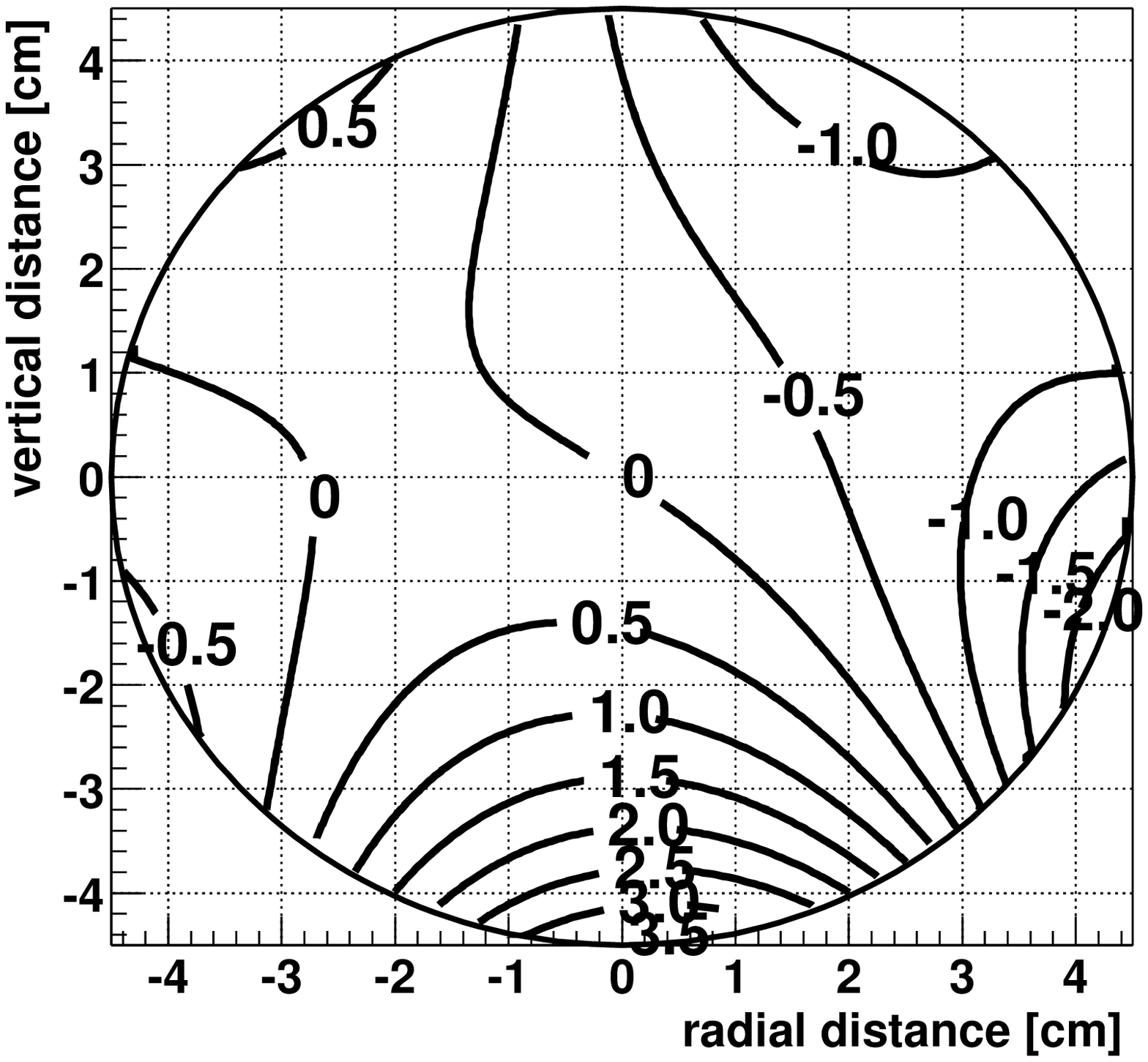}}
\subfigure[Azimuthal average]
{\includegraphics[width=3in]{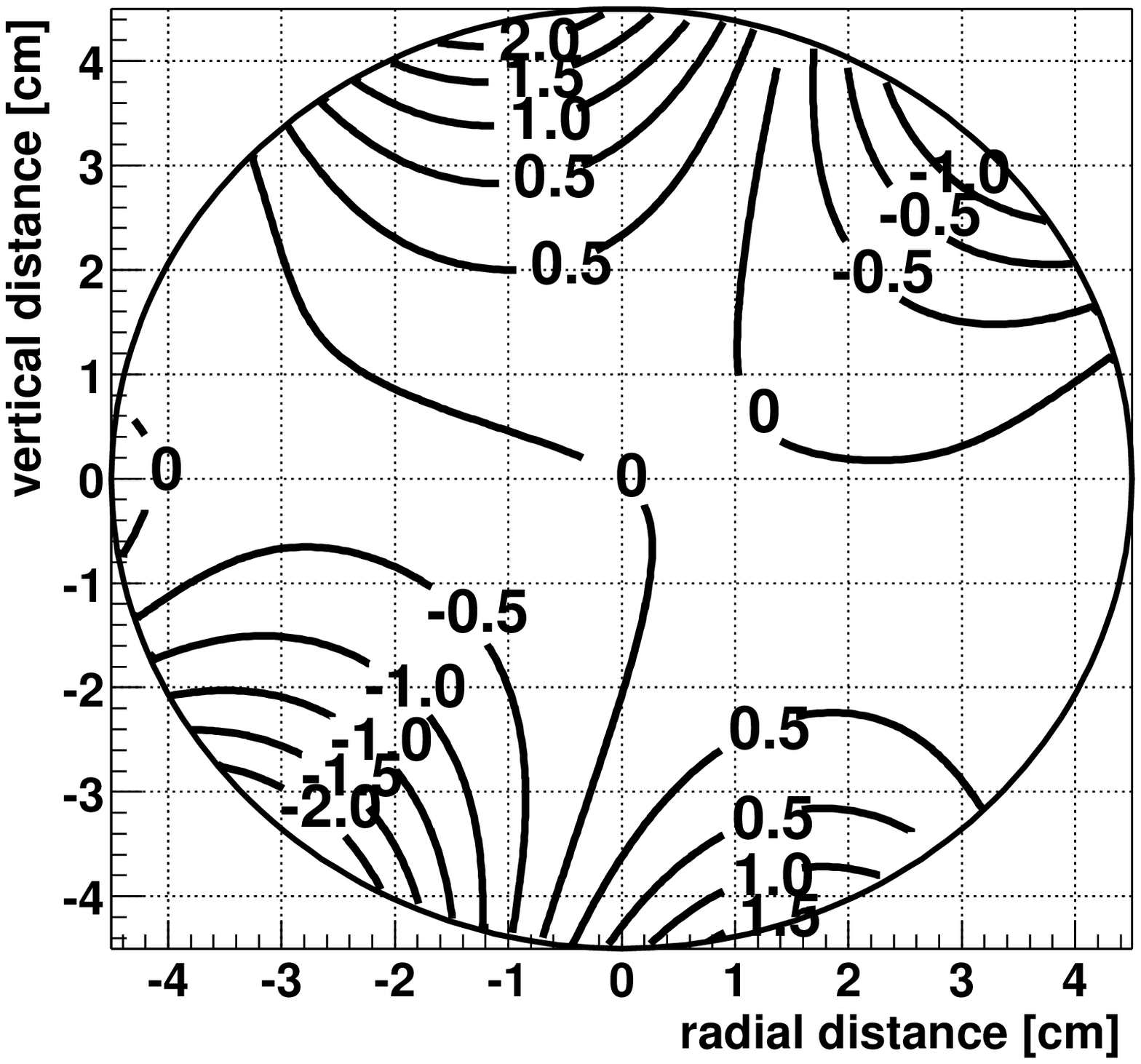}}
\end{center}
\caption{Homogeneity of the field   (a) at the calibration position
and (b) for the 
azimuthal average for one trolley run during the
2000 period.  In both figures, the contours correspond to 0.5~ppm field
differences between adjacent lines.} \label{fg:multipoles}
\end{figure}

 The errors arising from the
cross-calibration of the trolley probes with the plunging probes
 are caused both by
the uncertainty in the relative positioning of the trolley probe
and the plunging probe, and by the local field inhomogeneity. At this
point in azimuth, trolley probes are fixed with respect to the frame
that holds them, and to the rail system on which the trolley
rides. The vertical and radial positions of the trolley probes
with respect to the plunging probe are determined by applying a
sextupole field and comparing the change of field measured by the
two probes. The field shimming at the calibration location minimizes
the error caused by the relative-position uncertainty,  which
in the vertical and radial directions has an
inhomogeneity less than 0.2 ppm/cm, as shown in
Fig.~\ref{fg:multipoles}(b).   The full multipole components at the
calibration position are given in Table~\ref{tb:multipoles},
along with the multipole content of the full magnetic field averaged
over azimuth.   For
the estimated 1~mm-position uncertainty, the uncertainty on the
relative calibration is less than 0.02~ppm.

The absolute calibration utilizes a probe with a spherical 
water sample (see Figure \ref{fg:probes:a})\cite{fei}. The
Larmor frequency of a proton in a spherical water sample is related to
that of the  free proton through\cite{abragam,mohr}
\begin{equation}
f_{\rm L}({\rm sph-H_2O}, T) 
= \left[ 
1 - \sigma({\rm H_2O}, T)
\right] f_{\rm L} ({\rm free}),
\end{equation}
where $\sigma({\rm H_2O}, T)$ is from the diamagnetic shielding of the 
proton in the water molecule, determined from\cite{phillips}
\bea
\sigma( {\rm H_2O}, 34.7^{\circ}C) &=&
1 - {g_p({\rm H_2O}, 34.7^{\circ}C)\over g_J(H)}{ g_J(H) \over g_p(H)}
    {g_p(H) \over g_p({\rm free})} \\
   &=& 25.790(14)\times 10^{-6}. 
\eea
The $g$-factor ratio of the proton in a spherical water sample to
the electron in the hydrogen ground state ($ g_J(H)$) is measured
to 10 parts per billion (ppb)\cite{phillips}.  The ratio of electron 
to proton $g$-factors in hydrogen is known to 9~ppb\cite{winkler}.
The bound-state correction relating the $g$-factor of the proton 
bound in hydrogen to the free proton are calculated in 
References\cite{lamb41,grotch}.  
The temperature dependence of $\sigma$ is corrected for using
$d\sigma({\rm H_2O},T)/dT = 10.36(30) 
\times 10^{-9}/^{\circ}{\rm C}$\cite{british}.
The free proton frequency is determined to an accuracy of
0.05~ppm. 

The fundamental constant $\lambda_+ = \mu_{\mu^+}/\mu_p$ 
(see Equation \ref{eq:lambda}) can be computed from the hyperfine structure
of muonium (the $\mu^+ e^-$ atom)\cite{mohr}, or from the Zeeman splitting in 
muonium\cite{liu}. The latter experiment 
 used the same calibration probe as
was used in our $(g-2)$ experiment; however the magnetic environments of
the two experiments
were different, so that perturbations of the probe 
materials on the surrounding magnetic field 
 differed by a few ppb between the two experiments.

 The errors in the calibration procedure
result both from the uncertainties on the positions of the 
water samples inside the trolley and
the calibration probe, and from magnetic field inhomogeneities.
The precise location of the trolley in 
azimuth, and the location of the probes within the trolley, are not
known better than a few mm.  The uncertainties in the relative calibration
resulting from position uncertainties are 0.03~ppm.  Temperature and 
power-supply voltage dependences contribute 0.05~ppm, 
and the paramagnetism of the ${\rm O}_2$ in the air-filled trolley
causes a 0.037-ppm-shift in the field.

\begin{table}[h]
\begin{center} \caption{Multipoles at the outer edge of the storage
volume (radius = 4.5~cm).  The left-hand set are
for the plunging station where the plunging probe and the calibration
are inserted.  The right-hand set are the multipoles obtained by
averaging over azimuth for a representative trolley run during the
2000 period. 
\label{tb:multipoles}}
\begin{tabular}{lccccc}
 Multipole & \multicolumn{2}{c}{Calibration}
          & ~~~~~~~~~
          & \multicolumn{2}{c}{Azimuthal Averaged} \\
~~$[{\rm ppm}]$     & Normal & Skew & ~~~~~~~~~~~~~& Normal & Skew \\
\hline
Quadrupole~~~~~~~~& -0.71 & -1.04 & ~~~~~~~~~& 0.24 & 0.29 \\
Sextupole & -1.24 & -0.29 & ~~~~~~~~~& -0.53 & -1.06 \\
Octupole  & -0.03 & 1.06  & ~~~~~~~~~& -0.10 & -0.15 \\
Decupole  &  0.27 & 0.40  & ~~~~~~~~~&  0.82 &  0.54 \\
\end{tabular}
\end{center}
\end{table}

\subsubsection{Mapping the Magnetic Field}

\begin{figure}[h!]
  \includegraphics[width=0.75\textwidth,angle=0]{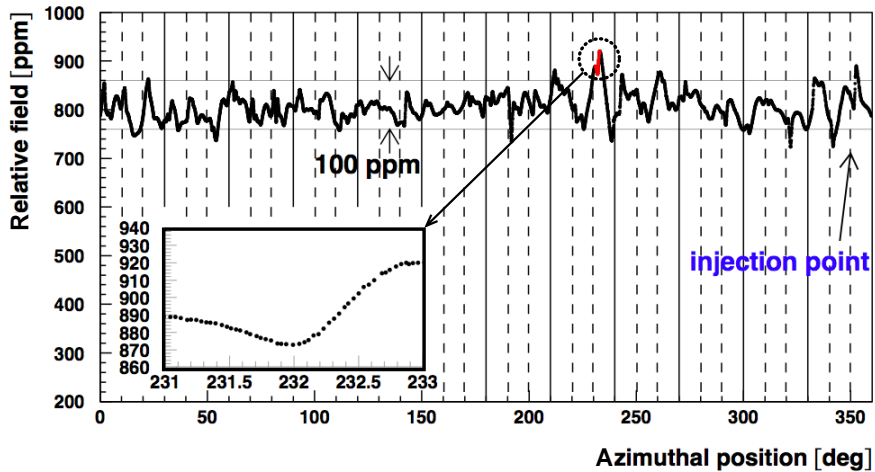}
  \caption{The magnetic field measured at the center of the storage
region vs. azimuthal position.  Note that while the sharp fluctuations 
appear to be noise, when the scale is expanded the variations are quite
smooth and represent true variations in the field.
  \label{fg:field_azimuth}}
\end{figure}

During a trolley run, the value of $B$ is measured by each probe at
approximately 6000 locations in azimuth around the ring.  The magnitude of
the field measured by the central probe is shown as a function of azimuth
in Figure \ref{fg:field_azimuth} for one of the trolley runs.
The insert shows that the fluctuations in this map that appear 
quite sharp are in fact quite smooth, and are not noise.
The field maps from the trolley are used to construct the field
profile averaged over azimuth.  This contour plot for one of the field
maps is shown in Figure~\ref{fg:multipoles}(b).  
Since the storage ring has weak focusing, the average over azimuth
is the important quantity in the analysis.  
Because NMR is only sensitive to the
magnitude of $B$ and not to its direction, the multipole distributions are
must be determined from azimuthal magnetic field averages, where
the field can be written as
\be
B(r, \theta) = \sum_{n=0}^{n=\infty}r^n \left( c_n \cos n \theta +
s_n \sin n \theta    \right),
\label{eq:mult-decomp}
\ee
where in practice the series is limited to 5 terms.

\subsubsection{Tracking the Magnetic Field in Time}

\begin{figure}[h!]
\begin{center}
  \includegraphics[width=0.65\textwidth,angle=0]{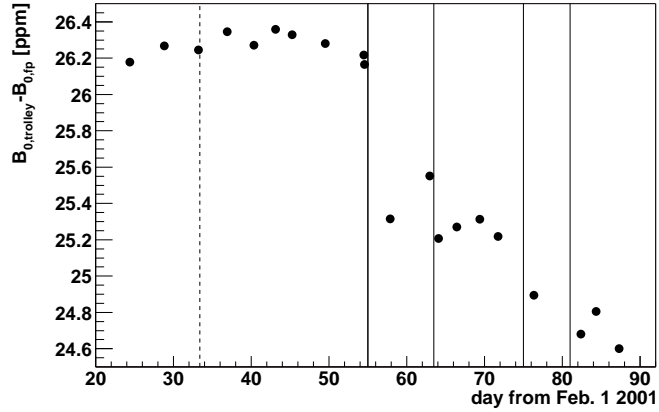}
  \caption{The difference between the average magnetic field measured
by the trolley and that inferred from tracking
the magnetic field with the fixed probes between
trolley maps. The vertical lines show when the magnet was powered
down and then back up. After each powering of the magnet,
the field does not exactly come back to its
previous value, so that only trolley runs taken between magnet powerings can
be compared directly.  
  \label{fg:Btrack}}
\end{center}
\end{figure}

During data-collection periods the field is monitored with the 
fixed probes.   To determine how
well the fixed probes permitted us to monitor the field felt by
the muons, the measured field, and that predicted by
the fixed probes is compared for each trolley run.  The results of
this analysis for the 2001 running period is shown in Figure~\ref{fg:Btrack}.
 The rms distribution of these differences is 0.10~ppm.

\subsubsection{Determination of the Average 
Magnetic Field: $\omega_p$ \label{sct:avB}}

The value of $\omega_p$ entering into the determination of
 $a_{\mu}$ is the field profile weighted by the muon distribution.
The multipoles of the field, Equation~\ref{eq:mult-decomp},
are folded with the muon distribution
\be
M(r, \theta) = \sum [\gamma_m(r) \cos m \theta + \sigma_m(r) \sin m \theta ],
\ee
to produce the average field, 
\be
\langle B \rangle_{\mu-{\rm dist}} = \int M(r,\theta)B(r, \theta) rdr d\theta,
\ee
where the moments in the muon distribution couple moment-by-moment to
the multipoles of $\vec B$.
Computing $\langle B \rangle$ is greatly simplified
if the field is quite uniform (with small higher multipoles),
and the muons are stored in a circular
aperture, thus reducing the higher moments of $M(r, \theta)$.
This worked quite well in E821, and the uncertainty on
$\langle B \rangle$ weighted by the muon distribution was
$\pm  0.03$ ppm.  

The weighted average was determined both by a tracking calculation that
used a field map and  calculated the field seen by each muon,
and also by using the quadrupole component of the field and the beam center
determined from a fast-rotation analysis to determine the average field.
These two agreed extremely well, vindicating the choice of a
circular aperture and the $\pm 1$~ppm specification on the field 
uniformity, that were set in the design stage of the 
experiment.\cite{bennett3}

\subsubsection{Summary of the Magnetic Field Analysis}

The limitations on our knowledge of the magnetic field come from measurement
issues, not statistics, so in E821 
the systematic errors from each of these sources
had to be evaluated and understood. The results and errors 
are summarized in
Table~\ref{tb:FinalFields}.

\begin{table}[h!]
\begin{center} \caption{Systematic errors for the magnetic field for the
different run periods. $^\dagger$Higher multipoles, trolley
temperature and its power-supply voltage response, and eddy currents
from the kicker. \label{tb:FinalFields}}
\begin{tabular}{|l|c|c|c|}
 \hline
Source of errors & 1999 & 2000 & 2001 \\
                 & [ppm]&[ppm]&[ppm]\\
\hline
Absolute calibration of standard probe~~~~~~~~~~ & 0.05 & 0.05 & 0.05\\
Calibration of trolley probes & 0.20 & 0.15 & 0.09\\
Trolley measurements of $B_0$ & 0.10 & 0.10 & 0.05\\
Interpolation with fixed probes & 0.15 & 0.10 & 0.07\\
Uncertainty from muon distribution & 0.12 & 0.03 & 0.03\\
Inflector fringe field uncertainty & 0.20 & -- & -- \\
Other$\dagger$ & 0.15 & 0.10 & 0.10 \\
\hline
Total systematic error on $\omega_p$ & 0.4 & 0.24 & 0.17\\
\hline Muon-averaged field [Hz]: $\omega_p /2\pi$ & ~$61\,791\,256$~ &
~$61\,791\,595$~ & ~$61\,791\,400$~ \\
\hline
\end{tabular}
\end{center}

\end{table}

%% file: omega-a-20feb07.tex
\subsection{The Muon Spin Precession Frequency: The $\omega_a$ Analysis}

To obtain the muon spin precession frequency $\omega_a$
given in Equation~\ref{eq:diffreq},
\be
\vec \omega_a = - a_\mu { q \vec B \over m}, 
\ee
which is observed as an oscillation of the number of detected electrons
with time
\be
N(t,E_{th}) = 
N_{0}(E_{th})e^{-t/\gamma\tau}[1+A(E_{th})\cos(\omega_{a}t+\phi(E_{th}))],
\ee
it is necessary to:
\begin{itemize}
\item Modify the five-parameter function above to include
small effects such as the coherent betatron
oscillations (CBO), pulse pile-up, muon losses, and gain changes, without
adding so many free parameters that the statistical power for determining
$\omega_a$ is compromised.
\item Obtain an acceptable $\chi_R^2$ per degree of freedom
in all fits, i.e. consistent with 1, where
$\sigma(\chi^2_R)=\sqrt{2 / NDF}$.
\item Insure that the  fit parameters are stable independent of
the starting time of the least-square fit. This was found to be a very
reliable means of testing the stability of fit parameters as a function
of the time after injection.
\end{itemize}
In general, fits are made to the data
out to about 640~$\mu$s, about 10 muon lifetimes.

\subsubsection{Distribution of Decay Electrons \label{decay}}

Decay electrons with the highest laboratory energies,
typically $E>1.8$ GeV, are
used in the analysis for $\omega_a$, as discussed in
\S~\ref{sct:mudecay}.
In the (excellent) approximations that $\vec \beta \cdot \vec B=0$ and the
effect of the electric field on the spin is small,
the average spin direction
of the muon ensemble (i.e., the polarization vector)
precesses, relative to the momentum vector, in the plane perpendicular
to the magnetic field, $\vec B=B_y\hat y$, according to
\be
\hat s= (s_{\perp}\sin{(\omega_a t + \phi)} \hat x + s_y\hat y +
s_{\perp}\cos{(\omega_a t+\phi)} \hat z).
\label{eq:shat}
\ee
The unit vectors $\hat x$, $\hat y$, and $\hat z$ are directed along the
radial, vertical and azimuthal directions respectively.
The (constant) components of the spin parallel and perpendicular
to the B-field are  $s_y < < 1$ and $s_{\perp}$, with 
$\sqrt{(s_{\perp}^2+s_y^2)}=1$. 
The polarization vector precesses in the plane perpendicular to the magnetic
field at a rate independent of the ratio ${s_y / s_{\perp}}$.
Since $a_{\mu}=(g-2)/{2}>0$, the spin vector rotates in the same plane,
but slightly faster than the momentum vector. Note that
the present experiment is,
apart from small detector acceptance
effects, insensitive to whether the spin vector
rotates faster or slower
than the momentum vector rotation, and therefore it is insensitive to the
sign of $a_{\mu}$. There are small geometric acceptance
effects in the detectors which demonstrate that our result is consistent
with $a_{\mu}>0$. 
 
The value for $\omega_a$ is determined from the data using
a least-square $\chi^2$ minimization fit to the
time spectrum of electron decays,
$\chi^2=\sum_i {(N_i-N(t_i))^2 / N(t_i)}$, where the $N_i$ are the 
data points, and $N(t_i)$ is the fitting function.
The statistical uncertainty, in the limit where data are taken over
an infinite number of muon lifetimes, is given by Equation~\ref{eq:fracterr}.
The statistical figure of merit is $F_M=A\sqrt{N}$, which
reaches a maximum at about $y=0.8$, or
$E\approx 2.6$ GeV/c (see Figure~\ref{fg:differential_na}).
If all electrons are taken above some minimum
energy threshold, $F_M(E_{th})$ reaches a maximum at about
$y=0.6$, or $E_{thresh}=1.8$ GeV/c (Figure~\ref{fg:integral_na}).

\begin{figure}[h]
\begin{center}
\includegraphics[width=0.8\textwidth,angle=0]{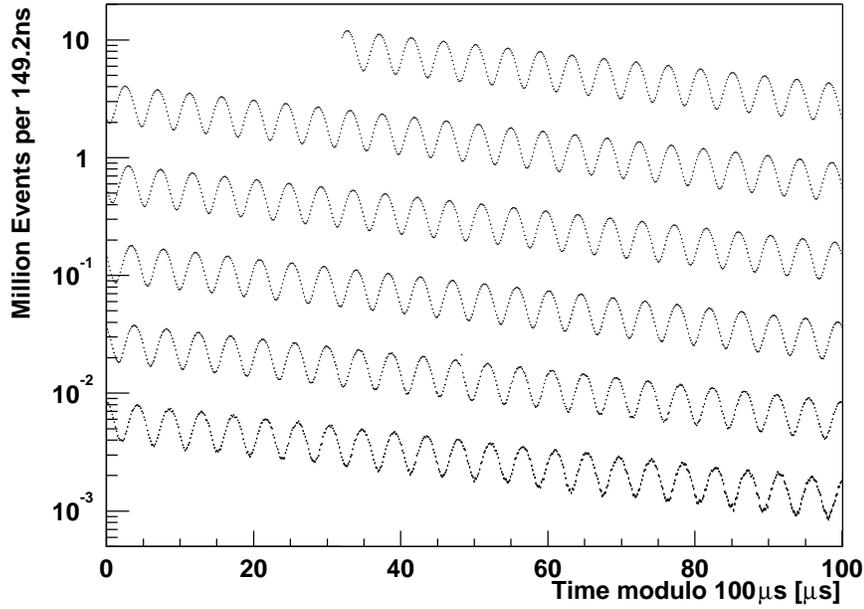}
%fig/wiggle2001_comp.eps}
\caption{Histogram of the total number of electrons above 1.8 GeV versus time
(modulo 100 $\mu$ s) from the 2001 $\mu^-$ data set.  The bin size is the
cyclotron period, $\approx 149.2$~ns, and the total number of electrons
is 3.6 billion.\label{fg:wiggles}}
\end{center}
\end{figure}

The spectra to be fit are in the form of histograms of
the number of electrons detected versus time (see Figure~\ref{fg:wiggles}),
 which in the ideal case
follow the five-parameter distribution function, Equation~\ref{eq:fivep}.
While this is a fairly good approximation for the E821 data sets, 
small modifications, due mainly to detector acceptance effects, must be made 
to the five-parameter function to obtain
 acceptable fits to the data.
The most important of these effects are described in the next section.

The five-parameter function has an important, well-known
invariance property. A sum of
arbitrary time spectra, each obeying the five-parameter
distribution and having the same $\lambda$ and $\omega$, but
different values for $N_0$, $A$, and $\phi$,
also has the five-parameter functional form with the {\it same} values for
$\lambda$ and $\omega$. That is,
\be
\sum_i B_ie^{-\lambda (t_i-t_{i0})}(1+A_i\cos{(\omega (t_i-t_{i0})+\phi_i)})
=
Be^{-\lambda t}(1+A\cos{(\omega t+\phi)}).
\ee

This invariance property 
has significant implications for the way in which data are handled in the
analysis.
The final histogram of electrons versus time is constructed from
a sum over the ensemble of the time spectra produced in individual spills.
It extends in time
from less than a few tens of microseconds after injection
out to $640 \mu s$, a period of about 10 muon lifetimes.
To a very good approximation, the spectrum from each spill 
follows the five-parameter probability distribution.
From the invariance property, the t=0 points and the gains from one
spill to the next do not need to be precisely aligned.

\begin{figure}
\begin{center}
\includegraphics[width=4in]{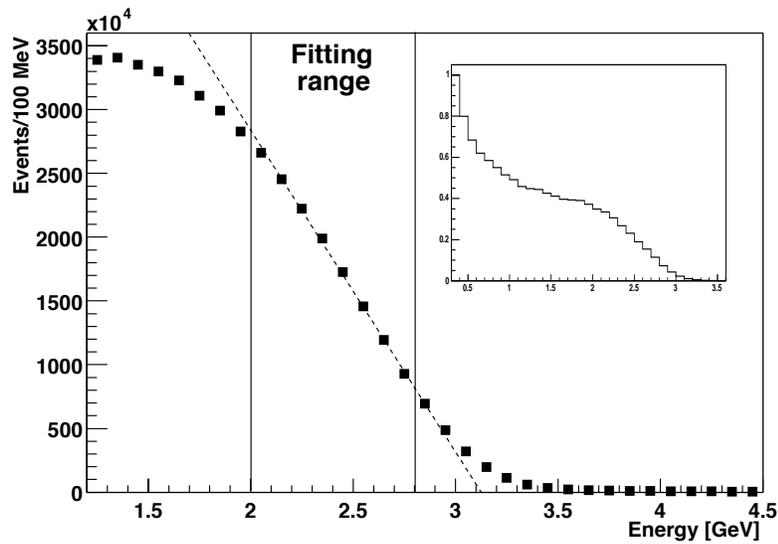}
\end{center}
\caption{Typical calorimeter energy distribution, with an
endpoint fit superimposed. The inset shows the full range of reconstructed
energies, from 0.3 to 3.5~GeV.} \label{fg:eSpectrum}
\end{figure}

Pulse shape and gain stabilities are monitored primarily
using the electron data themselves rather than laser pulses, or some
other external source of pulses. The electron times and energies
are given by fits to standard pulse shapes, which are
are established for each detector by taking an average over many
pulses at late times.
The variations in pulse shapes in all detectors are 
found to be sufficiently small as a function of energy  and
decay time, and contribute
negligibly to the uncertainty in $\omega_a$.
In order to monitor the gains, the energy distributions integrated over one
spin precession period and corrected for pile-up
are collected at various times relative to injection.
The high-energy portion of the energy distribution is well-described
by a straight line between the energy points at heights of
20\% and 80\% of the plateau
in the spectrum (see Figure
~\ref{fg:eSpectrum}).
The position of the x-axis intercept
is taken to be the endpoint energy, 3.1 GeV.
It is found that the energy stability of the detectors on the `quiet' side
of the ring stabilize earlier in the spill than those on
the noisy side of the ring. Some of the gain shift is due to the
PMT gating operation. Since the noisy detectors are gated on later in the spill
than the quiet ones, their gains tend to stabilize later.
The starting times for the detectors are chosen so that
most of their gains are calibrated to better than 0.2\%. On the quiet side of
the ring, data fitting can begin as early as a couple of microseconds
after injection; however it is necessary to delay at least
until the beam-scraping process is completed. For the noisiest
detectors, just downstream of the injection point, the start of fitting
may be delayed to 30 $\mu s$ or more.
The time histograms are then accumulated after applying the gain correction
to the energy of each electron. An uncertainty in the gain
stability on average over a fill
affects $N$, $\lambda$, $\phi$, and $A$ to a small extent.
The result is a systematic error on $\omega_a$ on the order of 0.1 ppm.

The cyclical motion of the bunched beam around the ring (fast-rotation
structure, \S\ref{sct:mubeam}) results in an additional multiplicative
oscillation superimposed on the usual five-parameter spectrum, which
is clearly visible in Figure~\ref{fg:fastrot}.
The amplitude of the modulation, which we refer to
as the `fast-rotation' structure, dies away
with a lifetime $\approx 26  \mu s$ as the muon bunch spreads out around the
ring (see \S~\ref{sct:mubeam}). 
In order to filter the fast-rotation structure from the time spectrum,
two steps are taken. First,
the $t=0$ point for electron data from each spill
is offset by a random time chosen
uniformly over one cyclotron period of 149~ns. Second,
the data are formed into histograms with a bin width
equal to the cyclotron period.
The fast-rotation structure is also
partially washed out when data are summed over all detectors, since the
fast-rotation phase varies uniformly from 0 to $2\pi$ in going from one
detector to the next around the ring.
These measures suppress the fast-rotation
structure from the spectrum by better than a factor of 100.
It is confirmed in extensive simulations
that these fast rotation filters have no significant effect on the
derived value for $\omega_a$. We note that the accidental overlap of
pulses (`pile-up') is enhanced by the fast-rotation structure and is
accounted for in the process of pile-up reconstruction,
as described later in this section.

The five-parameter functional shape is unaffected by the size
of the bin width in the histogram. The number of counts
in each bin of the time histogram
is equal to the integral of the five-parameter function over
the bin width. The integrated five-parameter 
function has the same $\lambda$ and
$\omega$ as the differential five-parameter 
function by the invariance property.
However, we note that unduly large values for the bin width will
reduce the asymmetry and perhaps introduce undesirable
correlations between $\omega_a$ and the other parameters of the fit.
A bin width equal to the cyclotron period
of about 149~ns is chosen in most of the analyses to
help suppress the fast-rotation structure. This is narrow enough
for binning effects to be minimal.

\subsubsection{Modification of the Five-Parameter Fitting Function}

In order to  successfully describe the functional form of the
spectrum of {\it detected} electrons versus time,
the five-parameter function must be modified to include additional
small effects of detector acceptance, muon losses, electron drift time
from the point of muon decay
to the point of detection, pile-up, etc. 
Given the desire to maintain as nearly as possible the
invariance property of the five-parameter function, and
the fact that some of the necessary corrections to the function turn out to
be imprecisely known,
every effort has been made to design the apparatus
so as to minimize the needed modifications.
Fortunately, most of the necessary modifications to the five-parameter function
lead to additional parameters having little correlation with
$\omega_a$ and therefore contribute minimally to the systematic error.
The following discussion will concentrate
on those modifications that contribute
most to the systematic error on $\omega_a$.

Two separate decay electrons arriving at one calorimeter with
a time separation less than
the time resolution of the calorimeters ($\sim 3$ to 5~ns, depending on the
pulse heights) can be mistaken for
 a single pulse (pile-up). The registered energy and
time
in this case will be incorrect, and if not properly accounted for,
it can be a source of a shift in $\omega_a$.
The registered energy and time  of a pileup pulse
is approximately the sum of the energies
and the energy-weighted average time respectively of the two pulses.
The phase of the $(g-2)$ precession depends on energy, and
since the pile-up pulse has an incorrectly registered energy, it has an
incorrect phase. Such a 
phase shift would not be a problem
if pile-up were the same at early and late decay times. 
However, the rate of pile-up is approximately proportional
to the {\it square} of the total decay rate,
and therefore the ratio of its rate to the normal data rate
varies as $\sim e^{-{t/ \gamma \tau}}$.
The early-to-late change in relative rates
leads to an early-to-late shift in the pile-up
 phase $\phi_{\rm pu}(t)$.
When the oscillating term $\cos(\omega_a t+ \phi_{\rm pu}(t)+ \phi)$ 
is fit to $\cos(\omega_a t + \phi)$,
a shift in the value for $\omega_a$ can result. 

To correct for pile-up-induced phase shifts, an average pile-up spectrum is
constructed and then subtracted from the data spectrum.
It is constructed as follows.
When a pulse exceeds a pre-assigned energy threshold, it
triggers the readout of a regular electron pulse, which consists
of about 80~ns of WFD information with
a pulse height reading every 2.5~ns in that time interval. This time segment
is large
enough to include the trigger pulse ($S_T(E,t)$)
plus a significant period beyond it, which usually contains only pedestal.
Occasionally, a second (`shadow', $S_S(E,t)$)
pulse from another electron falls in the pedestal region.
If it falls into a pre-selected time interval after the trigger pulse,
it is used to construct the pileup spectrum by combining the
$S_T$ and $S_S$ pulses into a single pile-up pulse, D.
The width of the pre-selected time interval is chosen
to be equal to the minimum time separation between pulses which
the detectors are capable of resolving (typically around 3~ns, however this
depends slightly on
the relative energies in the two peaks).
The fixed delay from the trigger pulse and the window is subtracted from the
time of $S_S$ and then the pulse $D$ is formed from the sum of
the WFD samples of $S_T$ and the $S_S$.
The
pile-up spectrum is formed from
$P(E,t)=D-S_T-S_S$. By construction, the rate of shadow pulses per unit time
is properly normalized to the trigger rate to a good approximation.
The single
pulses are explicitly subtracted when forming $P(E,t)$
because two single pulses
are lost for every pile-up pulse produced.
To properly account for the impact on the pile-up due to the
rapid variation in data rate produced by
the fast-rotation structure,
the time difference between the trigger pulse and the shadow pulse's
time window
is kept small compared to the fast-rotation period of 149~ns.

\begin{figure}[h]
\begin{center}
\includegraphics[width=0.7\textwidth,angle=0]{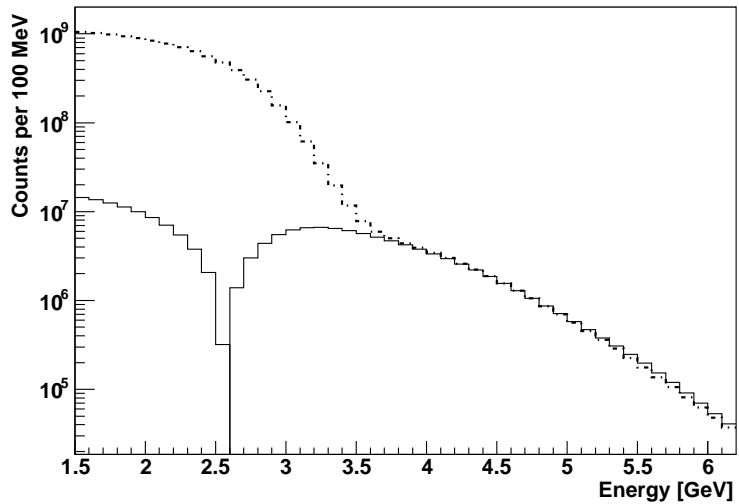}
\end{center}
\caption{Absolute value of the constructed pile-up spectrum 
(solid line) and actual data spectrum
  (dashed line). The two curves agree well at energies well above the
  electron
endpoint at 3.1 GeV, as expected.
The constructed spectrum is negative below 2.5 GeV because
the number of single electrons lost to pile-up exceeds the number of pile-up
pulses.\label{fg:pileup}}
\end{figure}

The resulting pile-up spectrum is then subtracted
from the total spectrum, on average eliminating the pile-up.
As Figure~\ref{fg:pileup} demonstrates, the constructed pile-up spectrum is in
excellent agreement with the data above the endpoint energy of 3.1 GeV, where
only pile-up events can occur. Below about 2.5 GeV, the pile-up spectrum
becomes negative because the number of single pulses, $S_T$ and $S_S$ lost
exceeds the number of pile-up $D$ pulses gained.

The magnitude of the pile-up is reconstructed with an estimated
uncertainty of about $\pm 8\%$. There is also an uncertainty in
the phase of the pile-up.
Further, the pulse-fitting procedure only recognizes shadow pulses
above about 250~MeV.
Based on simulations, the error due to `unseen pile-up'
(pulses below 250~MeV) was
about $0.03 $ ppm in the 2001 data set. The systematic uncertainties
were 0.014 ppm and 0.028 ppm from uncertainties in
the pile-up amplitude and
phase respectively, and a total contribution of 0.08 ppm came
from pile-up effects in the 2001 data set.

The coherent betatron oscillations, or oscillations in the average position
and width of the stored beam, (\S\ref{sct:beam-dyn}) 
 cause  unwanted oscillations
in the muon decay time spectrum. These  effects are generically referred to
as CBO. Some of the important CBO frequencies are given in Table~\ref{tb:freq}.
They necessitate small modifications
to the five-parameter functional form of the spectrum, and,
like pile-up, can cause a shift
in the derived value of $\omega_a$ if they are not properly accounted for
in the analysis.

In particular, the functional form
of the distribution of detector acceptance versus energy
changes as the beam oscillates, affecting the number and average energy
of the detected electrons. The result is that
additional parameters need to be added to
the five-parameter function  to account for CBO oscillations.
The effect of vertical betatron
oscillations is small and dies away much faster than the horizontal
oscillations, so that it can be neglected at the fit start 
times used in most
of the analyses.

The horizontal CBO affects the spectra mainly in two ways:
 it causes oscillations in the number of particles
because of an oscillation in the acceptance of the detectors,
and it causes oscillations in the average detected energy.
For a time spectrum constructed from the decay electrons in a given
energy band, oscillation in the 
parameter $N$ is due primarily to the
oscillation in the the detector acceptance.
Oscillations induced in $A$ and $\phi$, on the other hand,
depend primarily on the oscillation in the average energy.
In either case, each of the parameters $N$, $A$ and $\phi$
acquire small CBO-induced oscillations of the general form
\be
P_i=P_{i0}[1+B_{i}e^{-\lambda_{CBO}}\cos{(\omega_{CBO} t +\theta_{i})}].
\label{eq:CBO1}
\ee

\begin{figure}[htb]
\centering
  \includegraphics[width=0.5\textwidth]{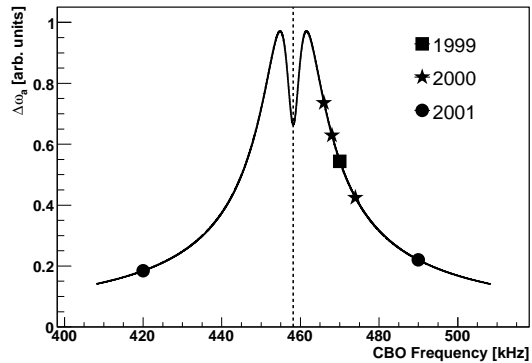}
  \caption{The relative shift in the value obtained for $\omega_a$
as a function of the CBO frequency, when the CBO effects are neglected
in the fitting function.  The vertical line is at $2 f_a$, and the
operating points for the different data collection periods is indicated
on the curve.
       \label{fg:camel1}}
\end{figure}

The presence of the CBO introduces $f_{CBO}$,
its harmonics, and the sum and difference
frequencies associated with beating between $f_{CBO}$ and
$f_a={\omega_a / 2\pi}\approx 229.1 $ kHz.
If the CBO effects are {\it not} included in the fitting function,
 it will pull
the value of $\omega_a$ in the fit by an amount related to how close
$f_{CBO}$ is to the second harmonic of $f_a$, introducing
a serious systematic error (see Table~\ref{tb:freq}).  
This effect is shown qualitatively in Figure~\ref{fg:camel1}.  

The problems posed by the CBO in the fitting procedure
were solved in a variety of ways in the many independent analyses.
All analyses took advantage of the fact that the CBO phase varied fairly
uniformly from 0 to $2\pi$ around the ring in going from one detector to the
next; the CBO oscillations should tend to cancel when data from all detectors
are summed together.
In the 2001 data set,
where the field index $n$ was  chosen to move the difference frequency
 $f_{-}=f_{\rm CBO}- f_a$ well away from  $f_a$,
 one of the analyses relied only on
the factor-of-nine reduction in the CBO amplitude
in the sum of data over all detectors,
and no CBO-induced modification to the five-parameter function was necessary.

The cancellation of CBO from all detectors combined would be
perfect if all the detector acceptances were identical,
or even if opposite pairs of detectors
at $180^\circ$ in the ring were identical. Imperfect cancellation
is due to the reduced performance of some of
the detectors and to slight asymmetries in the storage-ring geometry.
This was especially true of detector number 20, where there were
modifications to the vacuum chamber and
whose position was displaced
 to accommodate the traceback chambers.
The $180^\circ$ symmetry is broken for detectors near the kicker
because the electrons pass through the kicker plates.
Also the fit start times for detectors near the
injection point are inevitably later than for detectors on the other side of
the ring because of the presence of the injection flash.

In addition to relying on the partial CBO cancellation around
the ring, all of the other analysis approaches use
a modified function in which all parameters except $\omega_a$
and $\lambda$ oscillate according to equation ~\ref{eq:CBO1}.
In fits to the time spectra, the CBO parameters
$\omega_{CBO}$ and $\lambda_{CBO}\approx 100 \mu s$
(the frequency and
lifetime of the CBO oscillations, respectively),
are typically held fixed to values determined in separate studies.
They were established in
fits to time spectra formed with independent data from the FSDs and
calorimeters,
in which the
amplitude of CBO modulation is enhanced by aligning the CBO
oscillation phases of the individual detectors and then adding all the
spectra together.

Using these fixed values for $\omega_{CBO}$ and $\lambda_{CBO}$,
fits to the regular spectra are made with
$P_{i0}$, $B_i$ and $\phi_i$ in Equation~\ref{eq:CBO1}
(one set of these three for each of
$N$, $A$, $\phi$) as free parameters.
With the addition of $\omega_a$ and $\lambda$ as parameters, one obtains
a total of 11 free parameters in the global fit to data.
In practice, in some of the analyses, the
smaller CBO parameters are held fixed or eliminated altogether. 

Another correction to the fitting function
is required to account for muons being
lost from the storage volume before they have had a chance to decay.
Such losses lead to a distortion of the spectrum which can
result in an incorrect fitted value for the lifetime and a
poor value for the reduced $\chi^2$ of the fit.
The correlation between the lifetime and the precession frequency is
quite small, so that the loss in number of counts does not really affect
the value of the precession frequency.
The major concern is that the average spin phase of 
muons lost might be different from that of the
muons which remain stored. If this were the case, the average phase of
the stored muons would shift as a function of time,
leading directly to an error in
the fitted value of $\omega_a$.
This uncertainty in the phase forces an inflation in the systematic error 
due to muon losses.
Muon losses from the ring are believed to be
induced by the coupling between the higher moments
of the muon and the
field distributions, driving resonances
that result in an occasional muon striking
a collimator and leaving the ring.
The losses are as
large as 1\% per muon lifetime at early decay times, but
typically settle down to less than 0.1\%
200 $\mu s$ after injection. Such losses require the modification
$N_0\rightarrow N_0(1-A_{\rm loss}  n_{\rm loss} (t))$ in the 
fitting function,
with $A_{\rm loss}$ being an additional
parameter in the fit. The function $n_{\rm loss}$, which represents
the time distribution of lost muons, is obtained from
the muon loss monitors.

An alternative analysis method to determine $\omega_a$
utilizes the so-called `ratio method.'
Each electron event is randomly placed with equal probability
into one of four time histograms, $N_1-N_4$,
each looking like the usual time spectrum, Figure~\ref{fg:wiggles}.
A spectrum based on the ratio of
combinations of the histograms is formed:
\be
r(t_i)={ {N_1(t_i+{1\over 2}\tau_a)+N_2(t_i-{1\over 2}\tau_a)-N_3(t_i)-N_4(t_i)}
\over
{N_1(t_i+{1\over 2}\tau_a)+N_2(t_i-{1\over 2}\tau_a)+N_3(t_i)+N_4(t_i)}}
\label{eq:ratio-1}
\ee
For a pure five-parameter distribution, keeping only the important
large terms, this reduces to
\be
r(t)=A\cos{(\omega_a t+\phi)}+{1\over 16}({\tau_a\over \tau_{\mu}})^2,
\label{eq:ratio-2}
\ee

\noindent
where $\tau_a={2\pi / \omega_a}$ is an estimate ($\sim 10$ ppm is easily
good enough) of the spin precession
period, and the small constant offset produced by the exponential decay is
${1\over 16}({\tau_a\over \tau_{\mu}})^2=0.000287$.
Construction of independent histograms $N_3$ and $N_4$ simplifies the
estimates of the statistical uncertainties in the fitted parameters.

The advantage of fitting $r(t)$, as opposed to the standard
technique of fitting $N(t)$, is that any slowly varying
(e.g. with a period much larger than $\tau_a$)
multiplicative modulation
of the five-parameter function, which includes
the exponential decay of the muon itself,
largely cancels in this ratio. Other examples of slow modulation
are muon losses and small shifts in PMT gains due to the high rates
encountered at early decay times.
There are only 3 parameters, Equation~\ref{eq:ratio-2}, compared
to 5, Equation~\ref{eq:fivep},  in the regular spectrum. Note that
faster-varying effects, such as the CBO, will
not cancel in the ratio and must be handled in ways similar to the standard
analyses. One of the ratio analyses of the 2001 data set
used a fitting function formed from the ratio of functions $h_i$,
consisting of the five-parameter function modified to include
parameters to correct for acceptance effects such as the CBO:

\be
r_{fit}= {
{ h_1(t+{1\over 2}\tau_a)+h_2(t-{1\over 2}\tau_a)-h_3(t)-h_4(t) }
\over
{  h_1(t+{1\over 2}\tau_a)+h_2(t-{1\over 2}\tau_a)+h_3(t)+h_4(t) }}
\label{eq:mod-ratio}
\ee

The systematic errors for three yearly data sets, 1999 and 2000 for
$\mu^+$ and 2001 for $\mu^-$, are given in Table~\ref{tb:syster}.

\begin{table}[hbt]
\caption{Systematic errors for $\omega_a$ in the 1999, 2000 and 2001 data
periods.  In 2001,  systematic errors for
the AGS background, timing shifts, $E$-field
and vertical oscillations, beam de-bunching/randomization, binning
and fitting procedure together equaled 0.11 ppm and this is
indicated by $\ddag$ in the table.}
\begin{center}
\begin{tabular}{|l|l|l|l|}\hline

$\sigma_{\rm syst}$ $\omega_a$  &~~1999~~~  &  ~~2000~~~ &~~2001~~~ \\
 & (ppm) & (ppm) & (ppm) \\
\hline
Pile-up & 0.13 & 0.13  & 0.08    \\
AGS Background~~~ & 0.10 & 0.01  &  $\ddag$   \\
Lost Muons & 0.10 & 0.10 & 0.09  \\
Timing Shifts & 0.10 & 0.02  & $\ddag$  \\
E-field and Pitch & 0.08 & 0.03  & $\ddag$  \\
Fitting/Binning & 0.07 & 0.06  & $\ddag$ \\
CBO & 0.05 & 0.21 & 0.07  \\
Gain Changes & 0.02 & 0.13 & 0.12 \\ \hline
Total for $\omega_a$ & 0.3 & 0.31  &  0.21  \\
\hline
\end{tabular}
\end{center}

\label{tb:syster}
\end{table}

%% file: EDMforreview26feb07a.tex
\section{The Permanent Electric Dipole Moment of the Muon
\label{sct:EDM}}

If the muon were to possesses a permanent electric dipole moment (see
Equation~\ref{eq:dirac-dpm}), an extra term would be added 
to the spin
precession equation (Equation~\ref{eq:omega}):
\be
\omega_{EDM}=
-{q\over m}{\eta\over 2}\left[\vec \beta \times \vec B + {1\over c}
(\vec E-{\gamma\over \gamma +1}(\vec\beta\cdot \vec E)\vec\beta)
\right],
\label{eq:EDMw}
\ee
where the dimensionless constant
$\eta$ is proportional to the muon electric dipole moment,
$\vec d =\eta {q \over 2mc}\vec s$. The dominant 
$\vec \beta \times \vec B$ term
 is directed radially in the storage ring, transverse to $\vec \omega_a$. The
total precession vector, $\vec \omega=\vec \omega_a+\vec\omega_{EDM}$, is
tipped from the vertical direction by the angle
${\delta}=\tan^{-1}{\eta\beta\over 2a}$. Equation~\ref{eq:shat},
with the simplification that initially $s_y=0$, $s_{\perp}=1$ and $\phi=0$,
is modified to
\be
\hat s= (\cos{\delta}\sin{\omega_a t} \ \hat x +
\sin{\delta}\sin{\omega_a t}\ \hat y +
\cos{\omega_a t} \ \hat z).
\label{eq:shata}
\ee
The tipping of the precession plane produces an 
oscillation in the average vertical component of the spin 
which, because of the correlation between the spin and 
electron momentum directions, in turn causes oscillation 
in the average vertical component of the electron momentum.
The average vertical position of the electrons, at the 
 entrance face of the calorimeters, oscillates at angular
frequency $\omega$, 90 degrees out of phase
with the number oscillation depicted in Figure~\ref{fg:wiggles},
with an amplitude proportional to the EDM.
Additionally, the magnitude of of the precession frequency
is increased by the EDM,
\be
\omega=\sqrt{\omega_a^2+ \left({q\eta\beta B\over 2m}\right)^2}\ .
\label{eq:omEDM}
\ee

A measure (or limit) of the muon EDM, independent of the value
of $a_{\mu}$, will be produced from
an on-going analysis of the vertical electron motion using the
FSD, PSD and traceback data from E821.
Furthermore,
a difference between the experimental and standard model values of $a_{\mu}$
could be generated by an electric dipole moment.

No intrinsic electric dipole moment (EDM) has ever been experimentally
detected in the muon
or in any other elementary particle or atom.
The existence of a permanent (EDM)
in an elementary particle would 
violate both  $T$ (time reversal) and $P$ (parity) symmetries.
This is in contrast to the magnetic dipole moment (MDM),
which is allowed by both of these symmetries.
A non-vanishing EDM (or MDM) is consistent with $C$, 
charge conjugation symmetry, and with the
combined symmetry, $CPT$, provided that the magnitudes of
the EDM (or MDM) for a particle and its
anti-particle be equal in magnitude
but opposite in sign. No violation of $CPT$ symmetry
has ever been observed,
and it is strictly invariant in the Standard Model.
Assuming
$CPT$ invariance, the $T$ and $P$ violations of a non-vanishing
EDM imply a violation
of $CP$ and $CT$, respectively.
While $P$ violation in the weak interactions is maximal and has been
observed in  many reactions, 
$CP$ violation has
only been observed in decays of neutral $K$ and $B$ mesons,
with very small amplitudes. The violation of $T$ invariance has only
been observed in the decays of neutral $K$ mesons\cite{CPLEAR}.
An EDM large enough to be detected would
require new physics beyond the standard model.

It is widely believed that CP violation is one of
the ingredients needed to explain the baryon asymmetry of the universe,
however the type and size of CP violation which has been experimentally
observed to date is much too small
to account for it.
Searches for
EDMs are the subject of many past, current and future
experiments since
its detection
would herald the presence of new sources of CP violation
beyond the
Standard Model. Indeed many proposed extensions to the Standard Model
predict or at least allow the possibility of EDMs
and the failure to observe them has historically served as a
powerful constraint on these models.

The current experimental limit on the muon EDM, a by-product of
the third CERN $(g-2)$ experiment, is
$d_{\mu}<1\times 10^{-18}$ e-cm (90\% C.L.)\cite{BaileyEDM}.
This is
based on the non-observation of an oscillation
in the number of electrons above compared to below the calorimeter mid-planes
at the precession frequency, $f_a$, 90 degrees
out of phase with the number oscillation.

A new muon EDM limit will be set using E821 data,
by searching for an oscillation
in the average vertical positions of electrons from the FSD and PSD data.
The new data have the advantage over the CERN data
of containing information on the
shape of the vertical distribution of electrons, not just the number above or
below the detector mid-plane, and the sizes of the data samples are
significantly larger.
Analyses of these data are in progress.

A major systematic error in the EDM measurements by CERN, and by E821 using
the PSD and FSD information,
arises from any misalignment between the
vertical center of the stored muons and the vertical
positions of the detectors. As the spin precesses, the
vertical distribution of electrons entering the calorimeters changes.
In the case of the $\mu^-$ decays,
high energy (in the MRF) electrons
are more likely to be emitted opposite rather than
along the muon spin direction. When the spin has an
inward (outward) radial component in the storage ring, the average electron
will have a positive (negative) radial component of laboratory
momentum, and will have to
travel a greater (lesser) distance to get to the detectors.
When the spin points inward, more electrons are emitted outward,
and on average the electrons travel further
and spread out more in the vertical direction before getting to the detectors.
The change in vertical distribution
combined with a misalignment between the beam and the detectors,
inevitably leads to a vertical oscillation
which appears as a false EDM signal. This is the major limitation to this
approach for measuring the EDM, and substantial improvements in the
EDM limit in
the future will require a dedicated experiment to circumvent this 
problem\cite{farleyedm}.  

In E821, the traceback
chambers are able to measure oscillation of the vertical {\it angles} of the
electron trajectories as
well as average position,
providing a limit on the muon EDM
which is complementary to that obtained
from the FSD and PSD data. The angle information from the
traceback data
is less susceptible to
the systematic error from beam misalignment than up-down data, however
the statistical sample is smaller than the FSD or PSD data.
It is expected that the final EDM result from the
E821 the FSD, PSD, and traceback data will improve
upon the present limit on the muon EDM by a factor of four to five, to the
level of $\approx {\rm few} \ \times\ 10^{-19}$ e-cm.

In the process of deducing the experimental value of $a_{\mu}$
(see equation~\ref{eq:E821wa}) from the
measured values of $\omega_a$ and $\omega_p$, it is assumed that
the muon EDM is negligibly small and can be neglected. It is found under
this assumption and discussed in a later section of this manuscript,
that the experimental value of the anomaly differs from the
standard model prediction (equation ~\ref{eq:sm06}),

\be
\Delta a_{\mu}=295(88)\times 10^{-11}
\label{eq:amushift}
\ee

The extreme view could be taken that this
apparent difference, $\Delta a_{\mu}$, is not due to a shift
in the magnetic anomaly,
but rather is due entirely to a shift of the precession frequency,
$\omega_a$, caused by a non-vanishing EDM. From
Equation~\ref{eq:omEDM}
the EDM would have
to have the value  
$d_{\mu}=2.4(0.4)\times 10^{-19}$ e-cm.
This is a a factor of $\approx 10^8$ larger than the current limit on the
electron EDM. Such a large muon EDM is not predicted by even the most
speculative models outside of the standard model, but cannot be excluded 
by the previous CERN experimental limit on the EDM, and also
may not be definitively excluded
by the eventual E821 experimental result from the PSD, FSD and traceback data.
Of course, if the change in the precession frequency
were due to an EDM rather than a shift in the anomaly,
then this would be also be a very interesting
indication of new physics~\cite{feng}.

%% file: E821summary20feb07.tex
\section{Results and Summary of E821}

The values obtained in E821 for $a_\mu$ are given in 
Table~\ref{tb:measurements}.  However, 
the experiment measures ${\mathcal R}= \tilde \omega_a / \omega_p$,
{\it not}  $a_\mu$ directly, where the tilde over $\omega_a$
means that the pitch and radial electric field corrections have 
been included (see \S\ref{sct:p-E}).  The fundamental constant 
$\lambda_+ = \mu_{\mu^+} / \mu_p$ (see Equation~\ref{eq:lambda}) connects
the two quantities.
The values obtained for $\mathcal R$ are given in 
Table~\ref{tb:frequency-results}.  
 
\begin{table} [h]
\caption{The frequencies $\omega_a$  and $\omega_p$ obtained from the
three major data-collection periods. The radial electric field and 
pitch corrections applied to the $\omega_a$ values are given in the second
column.  Total uncertainties for each
quantity are shown. The right-hand column gives the values of 
$\mathcal R$, where the tilde indicates the muon spin precession frequency
corrected for the radial electric field and the pitching motion.
The error on the average includes
correlations between the systematic uncertainties of the three measurement
periods.}
\label{tb:frequency-results}
\begin{center}
\begin{tabular}{lllll}
 Period~~~~~~~~~ & $\omega_a/(2\pi)$ [Hz]~~~~~~ & $E$/pitch [ppm]~~~& $\omega_p /(2\pi)$ [Hz]~~~~~~ & $\mathcal{R} =
\tilde \omega_a/\omega_p$ \\
 \hline
  1999 ($\mu^+$) & $229\,072.8(3)$ & +0.81(8) & $61\,791\,256(25)$ & $0.003\,707\,204\,1(5\,1)$ \\
  2000 ($\mu^+$) & $229\,074.11(16)$ & +0.76(3) & $61\,791\,595(15)$ & $0.003\,707\,205\,0(2\,5)$ \\
  2001 ($\mu^-$) & $229\,073.59(16)$ & +0.77(6) & $61\,791\,400(11)$ & $0.003\,707\,208\,3(2\,6)$
  \\ \hline
  Average & -- & -- & -- & 0.003\,707\,206\,3(2\,0) \\
   \end{tabular}
\end{center}
\end{table}

The results are
\be
{\mathcal R_{\mu^-}}  = 0.003 \, 707 \, 208 \, 3 (2\,6) 
\ee
 and 
\be
{\mathcal R_{\mu^+}}  = 0.003\, 707\, 204\, 8 (2\,5).
\ee
The {\sl CPT} theorem predicts that the magnitudes of 
$\mathcal R_{\mu^+}$ and  $\mathcal R_{\mu^-} $ should be equal.
The difference is 
\be
\Delta {\mathcal R} = {\mathcal R_{\mu^-}} - {\mathcal R_{\mu^+}}
 = (3.5 \pm 3.4)\times 10^{-9}.
\ee
Note that it is the quantity $\mathcal R$ that must be compared for a 
{\sl CPT} test, rather 
than $a_\mu$, since the quantity $\lambda_+$ which 
connects them is derived from measurements of the
hyperfine structure of the $\mu^+e^-$ atom (see Equation~\ref{eq:lambda}).

\begin{figure}[h!]
\begin{center}
\includegraphics*[angle=-90,width=.75\textwidth]{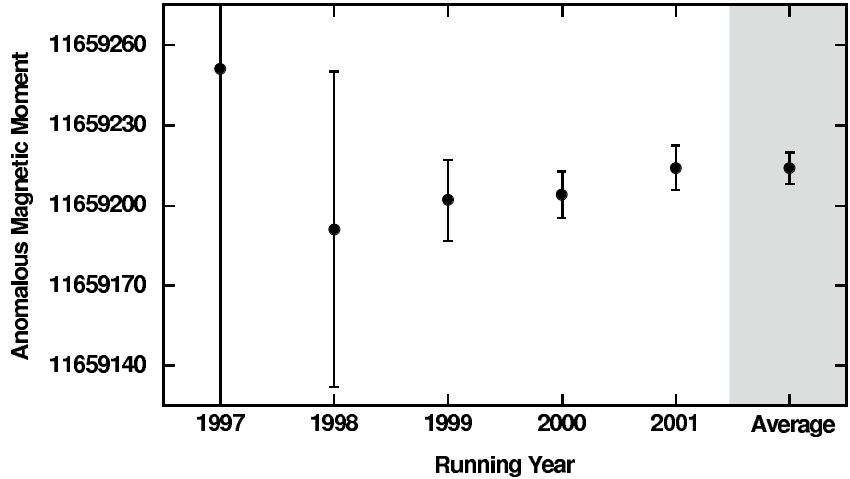}
\caption{Results for the E821 individual measurements of $a_\mu$ by
running year, together with the final average.\label{fg:summaryplot}}
\end{center}
\end{figure}

Using the latest value
$\lambda_+ = \mu_{\mu^+} / \mu_p = 3.183\,345\,39(10)$~\cite{liu}
gives the values for $a_\mu$ shown in Table~\ref{tb:measurements}. 
Assuming  {\sl CPT} invariance, we combine all  measurements 
of $a_\mu^+$ and $a_\mu^-$ to obtain the `world average'\cite{bennett3}
\begin{equation}
  a_\mu(\mathrm{Expt}) = 11\,659\,208.0(6.3) \times
  10^{-10}~~\mbox{(0.54\,ppm)}.
\end{equation}
The values of $a_\mu$ obtained in E821 are shown in
Figure~\ref{fg:summaryplot}. 
The final combined value of  $a_\mu$ represents an improvement in precision 
of a factor of 13.5 over the CERN experiments.  The final error of
0.54~ppm consists of a 0.46~ppm statistical component and a 0.28~systematic
component.  

%% file: theory27feb07.tex
%%%%%%%%%%%%%%%%%%%%%%%%%%%%%%%%%%%
%%%%%%%%%%%%%%%%%%%%%%%%%%%%%%%%%%%%
%%%%%%%%%%%%%%%%%%%%%%%%%%%%%%%%%%%%       
\section{The Theory of the Muon g-2}\label{theory}
%%%%%%%%%%%%%%%%%%%%%%%%%%%%%%%%%%%%
%%%%%%%%%%%%%%%%%%%%%%%%%%%%%%%%%%%%
\noi
As discussed in the introduction, 
the $g$-factor of the muon is the quantity which relates its spin
$\vec{s}$ to its magnetic moment $\vec{\mu}$ in appropriate units:
\be
\vec{\mu}=g_{\mu}\frac{q}{2m_{\mu}}\vec{s}\,,\qquad\annd\qquad
\underbrace{g_{\mu}=2}_{\mbox{\rm \small
Dirac}}(1+a_{\mu})\,.
\ee
In the Dirac theory of a charged spin-$1/2$ particle, $g=2$. Quantum
Electrodynamics (QED) predicts deviations from the Dirac exact value,
because in the presence of an external magnetic field the muon
 can emit and re-absorb virtual photons. The correction
$a_{\mu}$ to the Dirac prediction is called the anomalous magnetic moment.
As we have seen in the previous section, it is a quantity directly accessible to experiment.

In this section, we
shall present a review of the various contributions to
$a_{\mu}$ in the Standard Model, with special emphasis on the
evaluations of the hadronic contributions.

%%%%%%%%%%%%%%%%%%%%%%%%%%%%%%%%%%%%%%%%%%%%%%%%%%
\subsection{The QED Contributions}\label{qed}
%%%%%%%%%%%%%%%%%%%%%%%%%%%%%%%%%%%%%%%%%%%%%%%%%%

\noi
In QED of photons and leptons alone, the Feynman diagrams which contribute to $a_{\mu}$ at a given
order in the perturbation theory expansion (powers of
$\frac{\alpha}{\pi}$), can be divided into four classes:
\goodbreak

\subsubsection{Diagrams with Virtual Photons and Muon Loops.}\hfill

\begin{figure}[h]

\begin{center}
\includegraphics[width=0.3\textwidth]{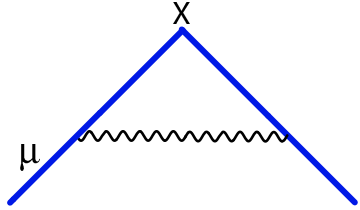}

\vspace*{0.5cm}
\caption { Lowest-order QED contribution. The 
solid blue line represents the 
muon in an external magnetic 
field ({\bf X} in the figure). The wavy black line
 represents the virtual photon.\label{figure:sch}}
\end{center}

\end{figure}

\begin{figure}[h]

\begin{center}
\includegraphics[width=0.9\textwidth]{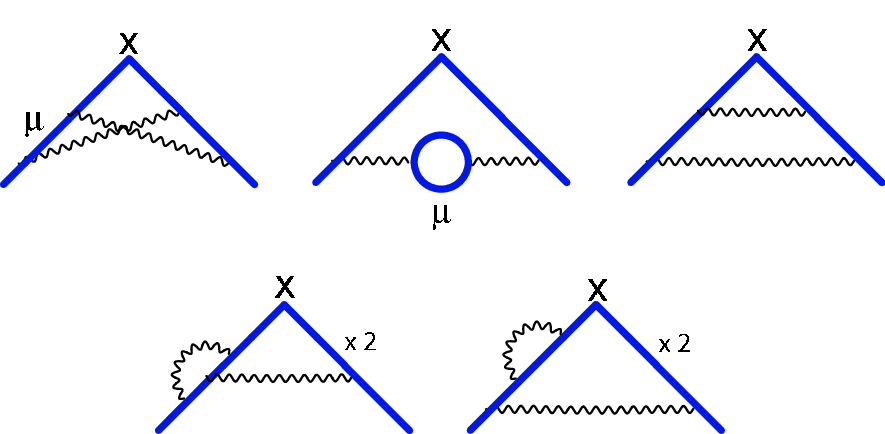}

\vspace*{0.5cm}
\caption{ Two-loop QED Feynman diagrams with 
the same lepton flavor.}\label{figure:peter}
\end{center}

\end{figure}

\noi
Examples are the lowest-order contribution in Figure~\ref{figure:sch} and the
two-loop contributions in Figure~\ref{figure:peter}. In full generality, this class of
diagrams consists of those with virtual photons only (wavy black lines), and those with virtual photons and internal fermion loops (solid blue loops) restricted to be of the same flavor as the
external line (solid blue line) in an external magnetic 
field ({\bf X} in the diagrams). Since
$a_{\mu}$ is a dimensionless quantity, and there is only one type of fermion
mass in these graphs--with no other mass scales-- these contributions
 are purely
numerical, and they are the same for the three charged leptons:
$l=e\,,\mu\,,\tau$. Indeed,
$a_{l}^{(2)}$ from Figure~\ref{figure:sch} is the celebrated 
Schwinger result\cite{Sch48}, mentioned previously,

\be\lbl{sch}
a_{l}^{(2)}=\frac{1}{2}\frac{\alpha}{\pi}\,;
\ee

%%%%%%%%%%%%%%%%%
%%%%%%%%%%%%

\noi
while $a_{l}^{(4)}$ from the seven diagrams in Figure~\ref{figure:peter} (the
factor of $2$ in a diagram corresponds to the contribution from the 
{\it mirror} diagram) gives the
result\cite{Pet57,Som57}

{\setl 
\bea
a_{l}^{(4)}& 
= & {\left\{ 
\frac{197}{144}+\frac{\pi^2}{12}-\frac{\pi^2}{2}\ln
2+\frac{3}{4}\zeta(3)\right\}}\left(\frac{\alpha}{\pi}\right)^2\lbl{2loop} \\
& = & -0.328\ 478\ 965\dots\left(\frac{\alpha}{\pi}\right)^2\,. \lbl{peterman}
\eea}

\noi
%%%%%%%%%%%%%%%%%%%%%%%%%%%%%%%%%%%%%%
%%%%%%%%%%%%%%%%%%%%%%%%%%%%%%%%%%%%%%

\begin{figure}[h]

\begin{center}
\includegraphics[width=0.9\textwidth]{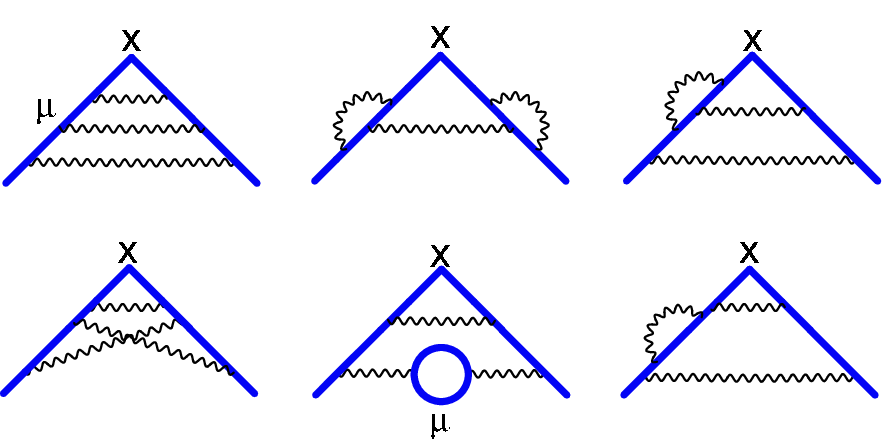}

\vspace*{0.5cm}
\caption{  A few Feynman diagrams of the three-loop 
type. In this class the flavor of the internal fermion loops
 is the same as the external fermion.}\label{figure:lare}
\end{center}

\end{figure}

At the three-loop level there are $72$ Feynman 
diagrams of this type. Only a few representative 
examples are shown in Figure~\ref{figure:lare}. 
%%%%%%%%%%%%%%%%%%%%%%%%%%%%%%%%%%%%%%
%%%%%%%%%%%%%%%%%%%%%%%%%%%%%%%%%%%%%%
Quite remarkably, their  total contribution is also known
analytically~(see Reference~\cite{LR96} and references therein). They bring in transcendental numbers like
$\zeta(3)=1.202~0569\cdots$, the Riemann zeta-function of argument $3$ which already appears at the two-loop level in Equation~\rf{2loop}, as well as transcendentals  of higher
complexity:~\footnote{There is in fact 
an interesting relationship between the appearance of these 
transcendentals in perturbative quantum field theory and 
mathematical structures like knot theory and non-commutative
 geometry, which is under active
 study~(see, e.g., References~\cite{BK97,CK98} and references therein).}

{\setl 
\bea
a_{l}^{(6)} & 
= & \left\{ 
\frac{28259}{5184}+\frac{17101}{810}\pi^2-\frac{298}{9}\pi^2 \ln 2+\frac{139}{18}\zeta(3) -\frac{239}{2160}\pi^4 \right. \nn \\
&  &  +\frac{100}{3}\left[{\rm Li}_4(1/2)+\frac{1}{24}\left(\ln^2 2-\pi^2\right)\ln^2 2 \right] \nn \\ & & \left. +\frac{83}{72}\pi^2\zeta(3) -\frac{215}{24}\zeta(5)\right\}\left(\frac{\alpha}{\pi}\right)^3\nn \\
& = & 1.181\ 241\ 456\dots\left(\frac{\alpha}{\pi}\right)^3\,,\lbl{3l} 
\eea}

\noi
where ${\rm Li}_4(1/2)=0.517~479\cdots$, is a particular case of the 
polylogarithm-function (see, e.g. References~\cite{Le58,KMR69}) 
which often appears in loop calculations in perturbation theory, in the form of the integral representation:

{\setl
\bea
{\rm Li}_k (x) & = & \frac{(-1)^{k-1}}{(k-2)!}\int_0^1 \frac{dt}{t}\ln^{k-2}{t} \ln(1-xt)\,, \quad k\ge 2\\ 
 & = & \sum_{n=1}^{\infty} \left(\frac{1}{n}\right)^{k}x^n\,,\quad \vert x\vert\le 1\,,    
\eea}

\noi
while
\be
\zeta(k)={\rm Li}_k (1) = \sum_{n=1}^{\infty} \frac{1}{n^k}\,;\quad k\ge 2\,.
\ee

At the four-loop
level, there are 891 Feynman diagrams of this type. Some of them are already
known analytically, but in general one has to resort to numerical methods for
a complete evaluation. This impressive calculation, which requires many
technical skills (see the chapter {\it Theory of the anomalous magnetic
  moment of the electron--Numerical Approach} in Reference~\cite{KQED90} for an
overall review),
 is under constant updating due to advances 
in computing technology. The most recent published value for the whole
four-loop contribution, from fermions with the same flavor, gives the
 result~(see Reference~\cite{KN06} for details and references therein)
\be\lbl{4l}
a_{l}^{(8)} = -1.728~3(35) \left(\frac{\alpha}{\pi}\right)^4 ,
\ee
where the error is due to the present numerical uncertainties.

Notice the alternating sign of the results from the 
contributions of one loop to four loops, a simple feature which is not yet
{\it a priori} understood . Also, the fact that the sizes of the
$\left(\frac{\alpha}{\pi}\right)^n$ 
coefficients for $n=1,2,3,4$  remain rather small is an interesting feature, 
 allowing one to expect that the order of magnitude of the five-loop
 contribution, from a total of 12672 Feynman diagrams~\cite{KN06,AHKN06}, is
 likely to be of  $\cO\left (\alpha/\pi\right)^5\simeq 7\times 10^{-14}$.
This magnitude is
 well beyond the accuracy required to compare with the present 
experimental results on the muon anomaly, but eventually needed 
for a more precise determination of the fine-structure
 constant $\alpha$ from the electron anomaly~\cite{gabalpha}.

\subsubsection{\it Vacuum Polarization Diagrams from Electron Loops}\lbl{vprga}\hfill

\begin{figure}[h]

\begin{center}
\includegraphics[width=0.8\textwidth]{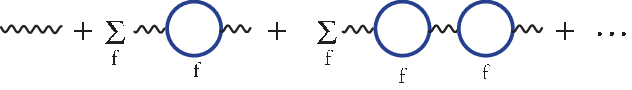}

\vspace*{0.5cm}
\caption{ Diagrammatic representation of the full 
photon propagator in Equation~\rf{fullpp}.}\label{figure:full}
\end{center}

\end{figure}

\noi
Vacuum polarization contributions result from the replacement 
\begin{equation}\lbl{repl}
        -i\frac{g_{\alpha\beta}}{q^2}\Ra
                i\left(g_{\alpha\beta}-\frac{q_\alpha q_\beta}{q^2}\right)\frac{\Pi^{({f})}(q^2)}{q^2}\,,
\end{equation}
whereby a free-photon propagator (here in the Feynman gauge) is dressed with the renormalized proper photon self-energy induced by a loop with fermion $f$. Formally, with $J_{\rm em}^{(f)~\alpha}(x)$ the electromagnetic current generated by the $f$-fermion field  at a space-time point $x$, the proper photon self-energy is defined by the correlation function
\begin{equation}
        i\int d^4 x\  e^{iq\cdot x}\  \langle 0\vert T\{J_{\rm em}^{(f)~\alpha}(x) J_{\rm em}^{(f)~\beta}(0)\} \vert 0\rangle = -\left(g^{\alpha\beta} q^2 -q^{\alpha}q^{\beta} \right)\Pi^{(f)}(q^2)\,.
\end{equation}
In full generality, the renormalized photon propagator involves a summation
 over all  fermion loops as indicated in Figure~\ref{figure:full} and 
 is given by the expression
\be\lbl{fullpp}
D_{\alpha\beta}(q)=-i\left(g_{\alpha\beta}-\frac{q_{\alpha}q_{\beta}}{q^2}\right)\frac{1}{q^2}\sum_f \frac{1}{1+\Pi^{(f)}(q^2)}-ia\frac{q_{\alpha}q_{\beta}}{q^4}\,,
\ee
where $a$ is a parameter reflecting the gauge 
freedom in the free-field propagator ($a=1$ in the Feynman gauge).

%%%%%%%%%%%%%%%%%%%%%%%%%%%%%%%%%%%%%%
%%%%%%%%%%%%%%%%%%%%%%%%%%%%%%%%%%%%%%

\noi  
Since the photon self-energy is transverse in the $q$-momenta, the replacement in Equation~\rf{repl} is unaffected by the possible gauge dependence of the free-photon propagator. 

\begin{figure}[h]

\begin{center}
\includegraphics[width=0.7\textwidth]{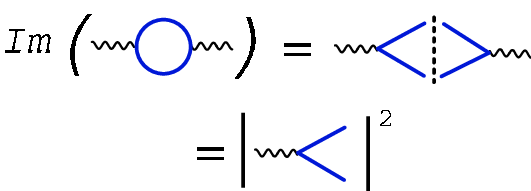}

\vspace*{0.5cm}
\caption{  Diagrammatic relation between the spectral 
function and the cross-section in Equation~\rf{unitar} .}\label{figure:absor}
\end{center}

\end{figure}

On the other hand, the on-shell renormalized photon self-energy obeys a dispersion relation with a subtraction at $q^2=0$ (associated to the on-shell renormalization); therefore
\begin{equation}\lbl{disprel}
\frac{\Pi^{({f})}(q^2)}{q^2}=\int_0^\infty \frac{dt}{t}\frac{1}{t-q^2 }\frac{1}{\pi}
        \Imm\Pi^{({f})}(t)\,, 
\end{equation}where $\frac{1}{\pi}
        \Imm\Pi^{({f})}(t)$ denotes the $f$-spectral function, related to
        the one-photon 
$e^+ e^-$ annihilation cross-section into $f^+f^-$ (see the illustration in Figure~\ref{figure:absor} ) as follows: 
\begin{equation}\lbl{unitar}
        \sigma(t)_{e^+ e^- \ra f^+ f^-}=\frac{4\pi^2\alpha}{t} \frac{1}{\pi}
        \Imm\Pi^{({f})}(t)\,.
\end{equation}
More specifically, at the one-loop level in perturbation theory
\begin{equation}\lbl{sp}
\frac{1}{\pi}   \Imm\Pi^{({f})}(t)=\frac{\alpha}{\pi}\frac{1}{3}\sqrt{1-\frac{4m_{f}^2}{t}}\left(1+2\frac{m_{f}^2}{t} \right)  \theta(t-4m_{f}^2)\,.
\end{equation}

\begin{figure}[h]

\begin{center}
\includegraphics[width=0.3\textwidth]{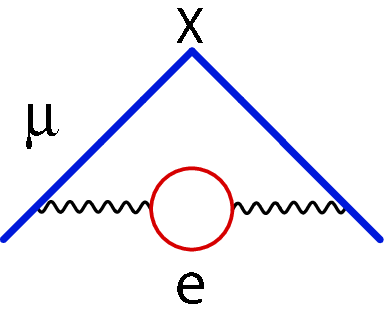}

\vspace*{0.5cm}
\caption{  Vacuum Polarization contribution from a Small Internal
Mass}\label{figure:muon_el}
\end{center}

\end{figure}

The simplest example of this class of Feynman diagrams is the one in
 Figure~\ref{figure:muon_el}, which is the only contribution of this type at
 the two-loop
 level, with the result~\cite{SW57,Pet57}
%%%%%%%%%%%%%%%%%%%%%%%%%%%%%%%%%%%%%%
%%%%%%%%%%%%%%%%%%%%%%%%%%%%%%%%%%%%%%
\be\lbl{vpe}
a_{\mu}^{(4)}(m_{\mu}/m_e )=\Big[\underbrace{\left({\frac{2}{3}}
\right)}_{\beta_{1}}
\left({\frac{1}{2}}\right)
\ln\frac{m_{\mu}}{m_{e}}-\frac{25}{36}
+\cO\left(
\frac{m_{e}}{m_{\mu}}\right)
\Big]\left(\frac{\alpha}{\pi} \right)^2\,.
\ee

%%%%%%%%%%%%%%%%%%%%%%%%%%%%%%%%%%%%%%
%%%%%%%%%%%%%%%%%%%%%%%%%%%%%%%%%%%%%%
\noi
Vacuum polarization contributions from fermions with a mass smaller than 
that of the external line ($m_e \ll m_{\mu}$)
 are enhanced by QED short-distance logarithms of the
ratio of the two masses (the muon mass to the electron mass in this case), and are therefore very
important, since $\ln\left(\frac{m_{\mu}}{m_e}\right)\simeq 5.3$. As shown 
in Reference~\cite{LdeR74}, these contributions are governed by a
Callan-Symanzik-type equation
\be\label{cs}
\left(m_{e}\frac{\partial}{\partial
m_{e}}+\beta(\alpha)\alpha\frac{\partial}{\partial\alpha}
\right)\tilde{a}_{\mu}\left(\frac{m_{\mu}}{m_{e}},\alpha\right)=0\,,
\ee
where $\beta(\alpha)$ is the QED-function associated with charge
renormalization, and $\tilde{a}_{\mu}(\frac{m_{\mu}}{m_{e}},\alpha)$
denotes the asymptotic contribution to $a_{\mu}$ from powers of logarithms of $\frac{m_\mu}{m_e}$ and
constant terms only. This renormalization group equation is at the origin of
the simplicity of the result in Equation~\rf{vpe} for the leading term: the factor $2/3$ in front
of $\ln\frac{m_{\mu}}{m_{e}}$ comes from the first term in the expansion of the QED $\beta$-function in powers of $\frac{\alpha}{\pi}$~\cite{deRR74}:
\begin{equation}
\beta(\alpha)=\frac{2}{3}\left(\frac{\alpha}{\pi}\right)+\frac{1}{2}\left(\frac{\alpha}{\pi}\right)^2-\frac{121}{144}\left(\frac{\alpha}{\pi}\right)^3 +\cO\left[\left(\frac{\alpha}{\pi}\right)^4 \right]\,,
\end{equation}
while the factor $1/2$ in Equation~\rf{vpe} is the lowest-order coefficient of
$\alpha/\pi$ in Equation~\rf{sch}, which fixes the boundary
condition (at $m_{\mu}=m_e$) to solve the differential equation in Equation~(\ref{cs}) at the first
non-trivial order in perturbation theory, i.e., 
$\cO(\frac{\alpha}{\pi})^2$. 
Knowing the QED $\beta$-function at
three loops, and $a_l$ (from
the universal class of diagrams discussed above) also at three loops,
allows one to sum analytically the leading, next-to-leading, and
next-to-next-to-leading powers of
$\ln m_{\mu}/m_e$ to all orders in perturbation theory~\cite{LdeR74}. Of course, these
logarithms can be re-absorbed in a QED running fine-structure coupling 
at the scale of the muon mass
$\alpha_{\rm QED}(m_{\mu})$. Historically, the first
experimental evidence for the running of a coupling constant in quantum
field theory comes precisely from the anomalous magnetic moment of the muon
in QED, well before QCD and well before the measurement of
$\alpha(M_{Z})$ at LEP\cite{ewwg} (a fact which unfortunately has been largely 
forgotten).

The exact analytic representation of the vacuum polarization diagram in Figure~\ref{figure:muon_el} in terms of Feynman parameters is rather simple:
\begin{equation}\lbl{parvp}
\hspace*{-2cm}
a_{\mu}^{(4)}(m_{\mu}/m_e) = \left(\frac{\alpha}{\pi}\right)^2 \int_0^1 \frac{dx}{x}(1-x)(2-x)\int_0^1 dy y(1-y) \frac{1}{1+\frac{m_e^2}{m_{\mu}^2}\frac{1-x}{x^2}\frac{1}{y(1-y)}}\,. 
\end{equation}
This simplicity (rational integrand) is a characteristic feature of the
Feynman parametric integrals, 
which makes them rather useful for numerical integration. It has also been
recently shown~\cite{FGdeR05} that the Feynman parameterization when
combined with a Mellin-Barnes integral representation is very well
 suited to obtain asymptotic expansions in ratios of mass parameters in a systematic way.

The integral in Equation~\rf{parvp} was first computed in ref.~\cite{El66}. A compact form of the analytic result is given in ref.~\cite{Pa04}, which we reproduce below to illustrate the typical structure of an exact analytic expression ($\rho=m_e/m_{\mu}$):

\bea\lbl{exactvp}
\hspace*{-2.5cm}\lefteqn{
a_{\mu}^{(4)}(m_{\mu}/m_e) =\left\{-\frac{25}{36}-\frac{1}{3}\ln \rho +\rho^2 (4+3 \ln \rho)+
\rho^4 \left[ \frac{\pi^2}{3}-2\ln \rho \ln \left(\frac{1}{\rho}-\rho\right)-{\rm Li}_{2}(\rho^2)\right] \right. } \nn \\ & & \left. \! +\frac{\rho}{2}(1-5\rho^2)
\left[\frac{\pi^2}{2}\! -\ln \rho \ln\left(\frac{1-\rho}{1+\rho} \right)- {\rm Li}_{2}(\rho)+{\rm Li}_{2}(-\rho)\right] \right\} \left(\frac{\alpha}{\pi} \right)^2\!\!\,.      
\eea

\noi
Numerically, using the latest CODATA~\cite{MT02} and PDG~\cite{Eietal04} recommended value for the mass ratio $m_{\mu}/m_e=206.768~2838(54)$, one finds
\be\lbl{vper}
a_{\mu}^{(4)}(m_{\mu}/m_e) =1.094~258~3111~(84)\left(\frac{\alpha}{\pi} \right)^2\,,
\ee
where the error in the last two significant figures is the one induced by the present error in the mass ratio.

\begin{figure}[h]

\begin{center}
\includegraphics[width=0.2\textwidth]{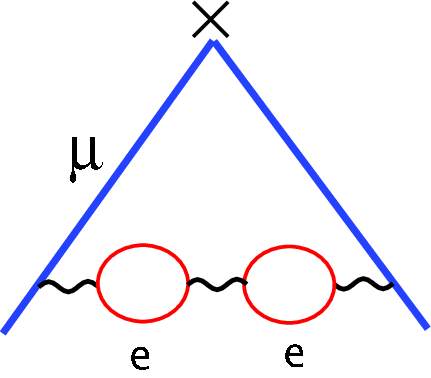}

\vspace*{0.5cm}
\caption{ Three Loop contribution from Vacuum Polarization corrections due to two Electron Loops}
\label{figure:mu_el_el}
\end{center}
\end{figure}

At the three-loop level, there is the diagram
 in Figure~\ref{figure:mu_el_el} generated by 
the insertion in the Schwinger diagram in 
Figure~\ref{figure:sch} of the vacuum polarization due 
to two electron loops~\cite{Kino67}, 
%%%%%%%%%%%%%%%%%%%%%%%%%%%%%%%%%%%%%%
%%%%%%%%%%%%%%%%%%%%%%%%%%%%%%%%%%%%%%

\begin{figure}[h]

\begin{center}
\includegraphics[width=0.8\textwidth]{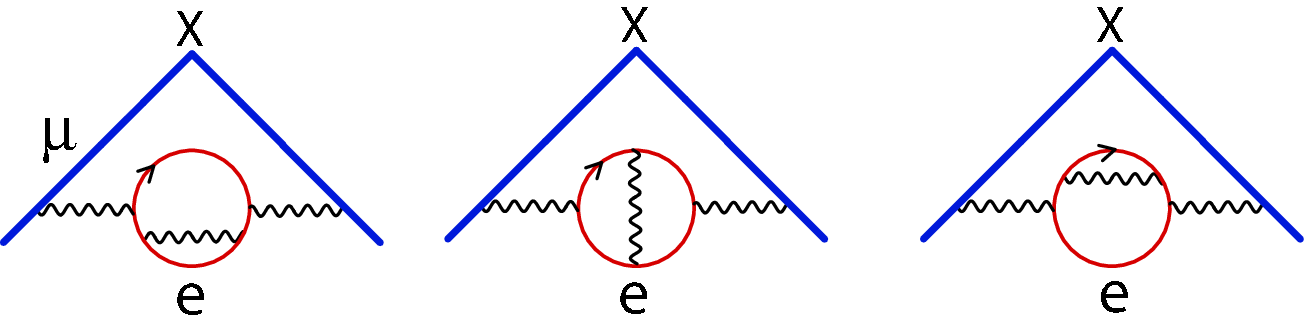}

\vspace*{0.5cm}
\caption{ Three-loop contribution  from the proper fourth-order vacuum polarization corrections due to an electron loop.}\label{figure:mu_el_alpha}
\end{center}
\end{figure}

\noi
as well as the three diagrams in Figure~\ref{figure:mu_el_alpha} 
generated by the proper fourth-order vacuum polarization
 with one electron loop~\cite{LdeR68}.
%%%%%%%%%%%%%%%%%%%%%%%%%%%%%%%%%%%%%%
%%%%%%%%%%%%%%%%%%%%%%%%%%%%%%%%%%%%%%

\begin{figure}[h]

\begin{center}
\includegraphics[width=0.6\textwidth]{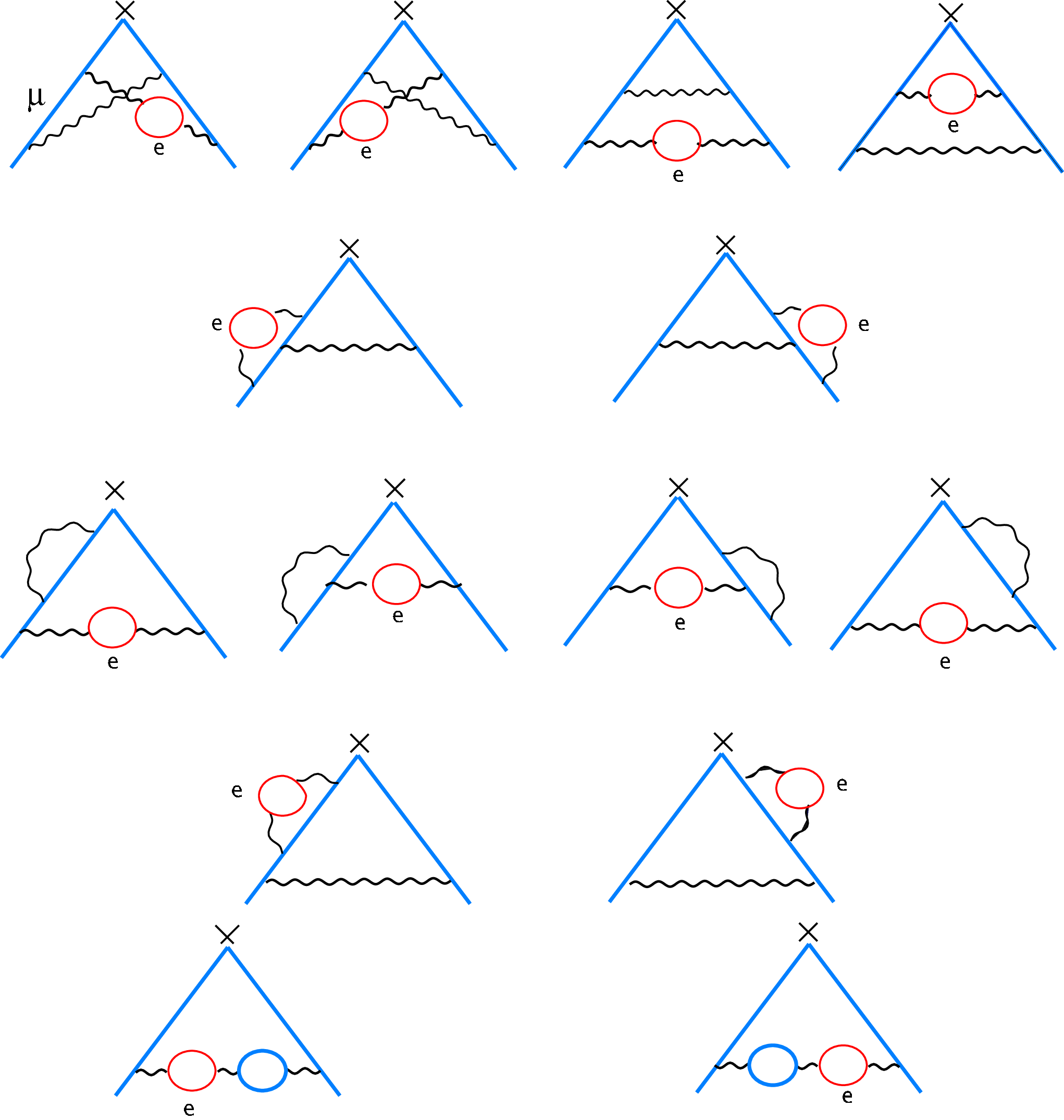}

\vspace*{0.5cm}
\caption { Three-loop contributions from vacuum polarization corrections due to an electron loop; the diagrams in each line define a subclass of gauge invariant contributions.}\label{figure:mu_gamma_el}
\end{center}
\end{figure}

At the same order, one has to add 
the 14 Feynman diagrams generated by the insertion of the lowest-order 
vacuum polarization correction due to an electron loop in the fourth-order 
vertex Feynman diagrams in Figure~\ref{figure:peter}. This generates the
diagrams of Figure~\ref{figure:mu_gamma_el}, which we have collected in 
five subclasses of gauge-invariant diagrams. 
The history and references of the earlier evaluation of these contributions can be found in
the review article in ref.~\cite{LPdeR72}. The exact analytic expression from
these vacuum polarization graphs, for arbitrary values of the masses, was
completed in 1993 by Laporta and Remiddi~\cite{La93, LaRe93}. The expression 
is so lengthy that it is not even reproduced in the original 
papers, where instead, the asymptotic expansion in the 
mass ratio $m_{\mu}/m_e$ is given up to very high order.
 The first few terms in that expansion are
\bea\lbl{assap}
\hspace*{-2.5cm}\lefteqn{
a_{\mu}^{(6)}(m_{\mu}/m_e)_{\rm vp} = \left(\frac{\alpha}{\pi}\right)^3 \left\{   \frac{2}{9}\ln^2 \left(\frac{m_{\mu}}{m_e}\right) +
\left[\frac{31}{27}+\frac{\pi^2}{9}-\frac{2}{3}\pi^2 \ln{2}+\zeta(3)\right]\ln\left(\frac{m_{\mu}}{m_e}\right)\right.}\nn \\
& & \hspace*{-2.5cm} +\frac{1075}{216}-\frac{25}{18}\pi^2 +\frac{5}{3}\pi^2 \ln{2} -3\zeta(3)+\frac{11}{216}\pi^2 -\frac{1}{9}\ln^4 2 -\frac{2}{9}\pi^2\ln^2 2 -\frac{8}{3}{\rm Li}_4 (1/2) \nn \\
& &  \left. +\frac{m_e}{m_{\mu}}\left[\frac{3199}{1080}\pi^2-\frac{16}{9}\pi^2 \ln 2 -\frac{13}{18}\pi^3 \right]+ \cO\left[\left(\frac{m_e}{m_{\mu}}\right)^2 \ln^2 \left(\frac{m_{\mu}}{m_e}\right) \right] \right\}\,.
\eea

\noi
The leading, next-to-leading, and next-to-next-to 
leading terms coincide with the earlier 
renormalization group calculation of ref.~\cite{LdeR74}. 
Numerically, using the exact analytic expression one finds~\cite{Pa04}
\begin{equation}\lbl{vpel4}
       a_{\mu}^{(6)}(m_{\mu}/m_e)_{\rm vp}=1.920~455~130~(33) \left(\frac{\alpha}{\pi}\right)^3 \,,
\end{equation}
while using the asymptotic expression in Equation~\rf{assap} one gets
\begin{equation}
       a_{\mu}^{(6)}(m_{\mu}/m_e)_{\rm vp}\simeq 1.920 ... \left(\frac{\alpha}{\pi}\right)^3 \,,
\end{equation}
which gives an indication of the size of the  terms neglected in Equation~\rf{assap}.

%%%%%%%%%%%%%%%%%%%%%%%%%%%%%%%%%%%%%%
%%%%%%%%%%%%%%%%%%%%%%%%%%%%%%%%%%%%%%

The contributions at the four-loop level with 
the insertion of vacuum-polarization diagrams from electron loops have been carefully analyzed by Kinoshita and Nio in a series of papers, the most recent being Reference~\cite{KN05} where earlier references can also be found. They have classified all the possible Feynman diagrams in three groups of gauge-invariant contributions:
\begin{enumerate}
        \item Group I consists of all the possible electron-loop-type 
vacuum-polarization insertions  in the photon propagator of the Schwinger
graph in Figure~\ref{figure:sch}. All together, 
this group consists of 49 diagrams. Typical examples of
 Feynman graphs of this group are shown in Figure~\ref{figure:mu_el_el_el}.
        
%%%%%%%%%%%%%%%%%%%%%%%%%%%%%%%%%%%%%%
\begin{figure}[h]

\begin{center}
\includegraphics[width=0.5\textwidth]{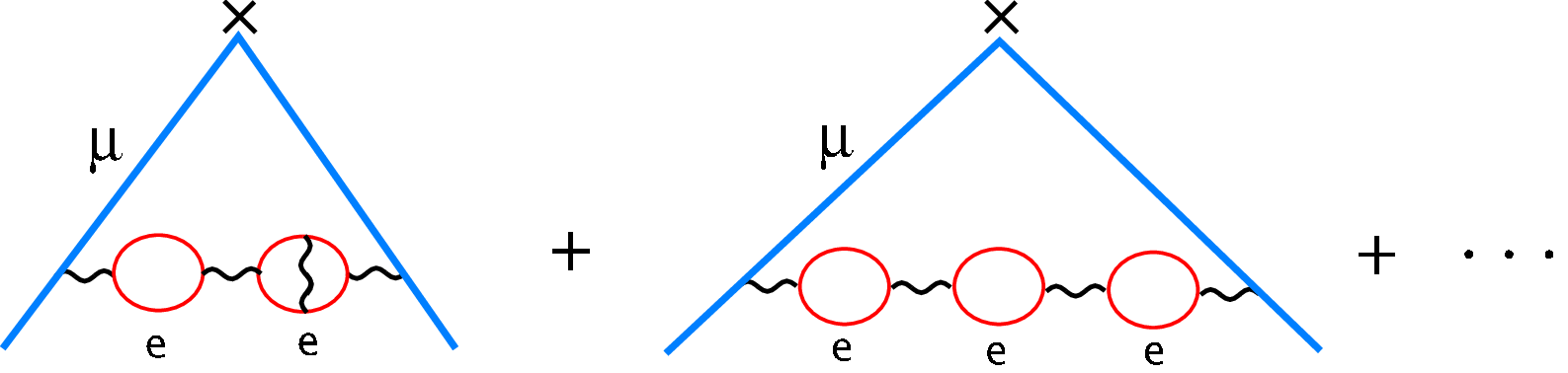}

\vspace*{0.5cm}
\caption{ Examples of four-loop contributions of group I.}
\label{figure:mu_el_el_el}
\end{center}
\end{figure}
%%%%%%%%%%%%%%%%%%%%%%%%%%%%%%%%%%%%%%
        \item Group II consists of the 90  diagrams obtained from all the possible  electron-loop-type vacuum polarization insertions in the proper two-loop vertex diagrams in Figure~\ref{figure:peter}. Examples of Feynman graphs of this group are shown in Figure~\ref{figure:muVP4gamma}.
%%%%%%%%%%%%%%%%%%%%%%%%%%%%%%%%%%%%%%
\begin{figure}[h]

\begin{center}
\includegraphics[width=0.6\textwidth]{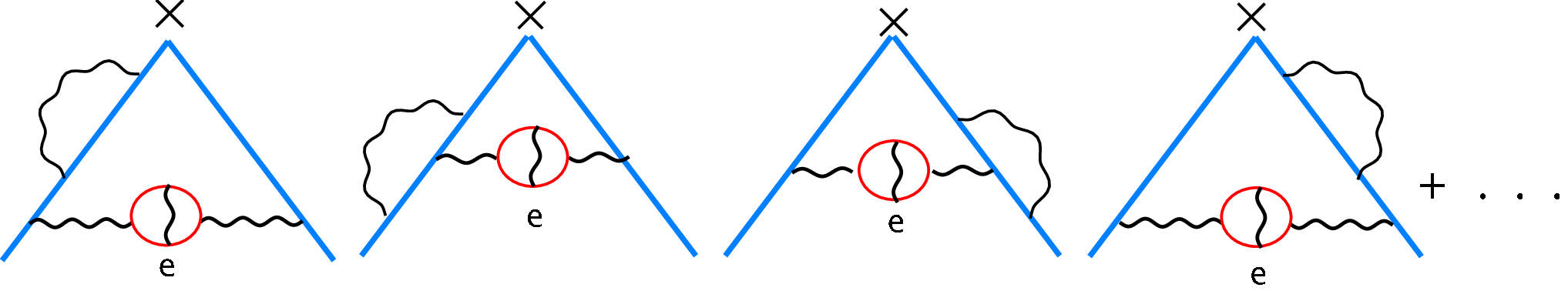}

\vspace*{0.5cm}
\caption { Examples of four-loop contributions of group II.}\label{figure:muVP4gamma}
\end{center}
\end{figure}
%%%%%%%%%%%%%%%%%%%%%%%%%%%%%%%%%%%%%%

        \item Group III consists of the 150 diagrams obtained by 
the lowest-order electron-loop vacuum polarization insertion 
in all possible photon propagators of the proper sixth-order
 vertex diagrams. Examples of Feynman graphs of this 
group are shown in Figure~\ref{figure:mu6VP2.eps}.
        
%%%%%%%%%%%%%%%%%%%%%%%%%%%%%%%%%%%%%%
\begin{figure}[h]

\begin{center}
\includegraphics[width=0.6\textwidth]{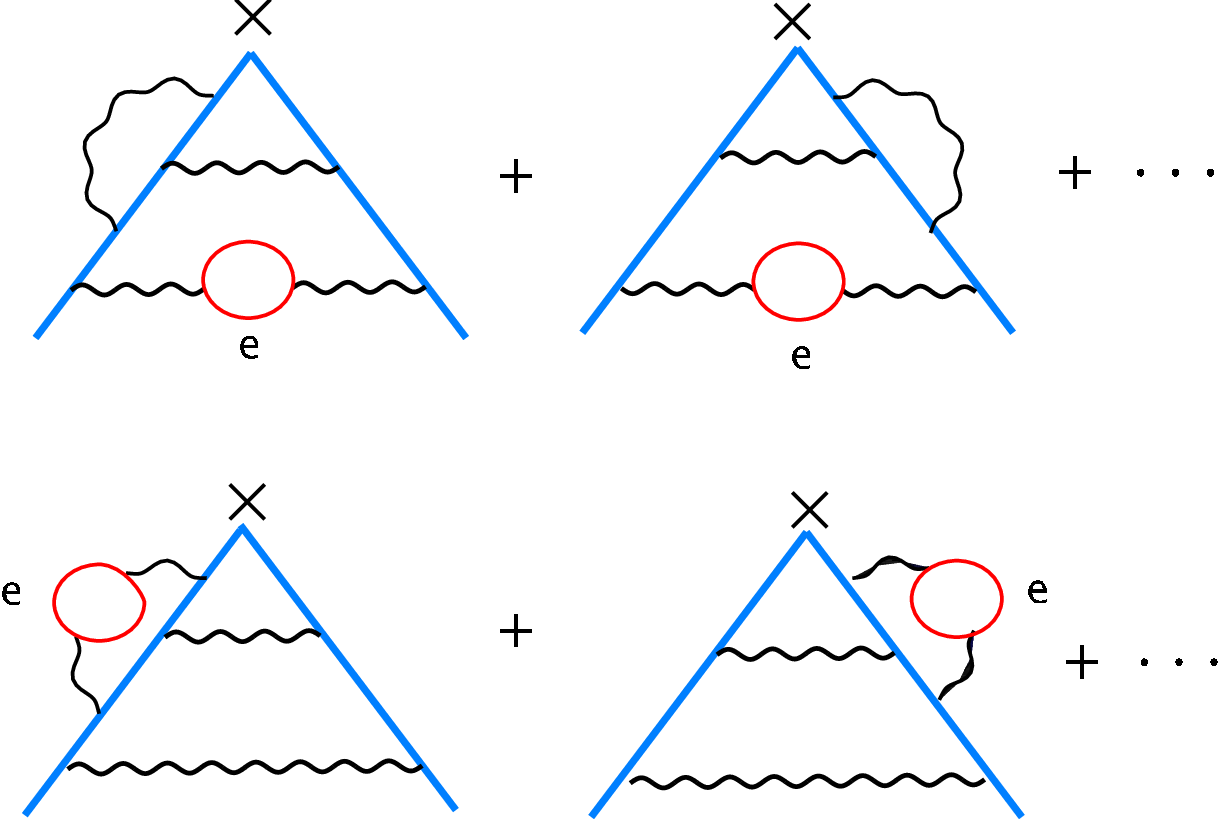}

\vspace*{0.5cm}
\caption { Examples of four-loop contributions of group III.}\label{figure:mu6VP2.eps}
\end{center}
\end{figure}
%%%%%%%%%%%%%%%%%%%%%%%%%%%%%%%%%%%%%%  
\end{enumerate}
All integrals of these three groups have been evaluated numerically, 
although some of them have also been evaluated using their asymptotic expansion in $m_{\mu}/m_e$~\cite{La93,BB95}. The numerical results obtained in Reference~\cite{KN05} are

{\setl
\bea
a_{\mu}^{(8)}(m_{\mu}/m_e)_{\rm vpI} & = & 16.720~359~(20) \lbl{vpI} \left(\frac{\alpha}{\pi}\right)^4\,, \\
a_{\mu}^{(8)}(m_{\mu}/m_e)_{\rm vpII} & = & -16.674~591~(68) \lbl{vpII} \left(\frac{\alpha}{\pi}\right)^4\,, \\  
a_{\mu}^{(8)}(m_{\mu}/m_e)_{\rm vpIII} & = & 10.793~43~(414) \lbl{vpIII} \left(\frac{\alpha}{\pi}\right)^4\,.
\eea}

\noi
The strong cancellation between the vpI and the vpII contributions, as well
as their individual size, can be qualitatively understood  using the 
Callan-Symanzik framework which describes their respective asymptotic contributions, as discussed in Reference~\cite{LdeR74}.

\subsubsection{\it Vacuum polarization diagrams from tau loops.}\hfill
 
\begin{figure}[h]

\begin{center}
\includegraphics[width=0.3\textwidth]{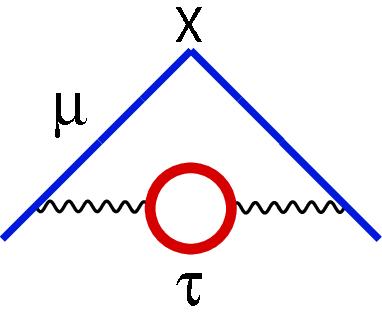}

\vspace*{0.5cm}
\caption{  Vacuum Polarization contribution from a Large Internal
Mass}\label{figure:mu_tau}
\end{center}

\end{figure}

\noi
The simplest example of this class of diagrams is the Feynman graph in Figure~\ref{figure:mu_tau}, 
%%%%%%%%%%%%%%%%%%%%%%%%%%%%%%%%%%%%%%
%%%%%%%%%%%%%%%%%%%%%%%%%%%%%%%%%%%%%%
which gives a
contribution\cite{LdeR68}
\be\lbl{tauvp}
a_{\mu}^{(4)}(m_{\mu}/m_{\tau})=
{\left[\frac{1}{45}\left(\frac{m_{\mu}}{m_{\tau}}\right)^2+\cO\left(
\frac{m_{\mu}^4}{m_{\tau}^4}\log\frac{m_{\tau}}{m_{\mu}}\right)
\right]}\left(\frac{\alpha}{\pi} \right)^2\,.
\ee
In full generality, internal heavy masses in the vacuum polarization
loops (heavy with respect to the external lepton line) decouple; i.e., they give a contribution which vanishes in the limit of an infinitely heavy mass.  This is the reason why the muon anomaly is more sensitive to new Physics than the electron anomaly, roughly by a factor $\left(m_{\mu}/m_e\right)^2$. This is also the reason why the QED theory for the electron anomalous magnetic moment is more precise, since the errors induced by the masses of the heavy leptons appear at a much smaller level. 

From a structural point of view, the Feynman diagram in
Figure~\ref{figure:mu_tau} has much in common with the {\it hadronic vacuum
  polarization} 
contribution which we shall discuss later and therefore it deserves some attention. In particular, one
would like to understand  how the simple leading behavior in
Equation~\rf{tauvp} arises. For this purpose it is convenient to reconsider
 the equations \rf{repl} to \rf{sp} for the case where the reference  fermion is a tauon, i.e., $f=\tau$. 
The contribution to the anomalous magnetic moment of the muon from a 
$\tau$-loop vacuum polarization insertion can then be viewed as the convolution of the $\tau$-spectral function $\frac{1}{\pi}
        \Imm\Pi^{({\tau})}(t)$ with the contribution to the muon anomaly, induced by a {\it fictitious massive photon}  with a free-propagator:
\begin{equation}
        -i\left(g_{\alpha\beta}-\frac{q_\alpha q_\beta}{q^2}\right)\frac{1}{q^2 -t}\,.
\end{equation}
The overall contribution to the muon anomaly from Figure~\ref{figure:mu_tau} can then be written as 
\begin{equation}\lbl{lotauvp}
        a_{\mu}^{(4)}(m_{\mu}/m_{\tau})= \frac{\alpha}{\pi}\int_{4m_{\tau}^2}^{\infty}\frac{dt}{t}\frac{1}{\pi}
        \Imm\Pi^{({\tau})}(t) \int_0^1 dx\frac{x^2 (1-x)}{x^2 + \frac{t}{m_{\mu}^2}(1-x)}\,,
\end{equation}
\noi
explicitly showing the fact that the integrand is positive and monotonically
 decreasing in the integration region $4m_{\tau}^2\le t\le \infty$. (This is
 why the result in Equation~\rf{tauvp} is positive.) Notice also that, for
 large values of $t/m_{\mu}^2$, the integral over the Feynman 
parameter $x$ behaves as $\frac{1}{3}\frac{m_{\mu}^2}{t}+\cO\left[
\left(m_{\mu}^2/t\right)^2 \log (t/m_{\mu}^2)\right]$\,.
Since the threshold in the $\tau$-spectral function is much larger than the
muon mass ($4m_{\tau}^2\gg m_{\mu}^2$), we can approximate the function $
a_{\mu}[{\rm Figure}~\ref{figure:mu_tau}]$ by its leading asymptotic
behavior, 
which results in 
\begin{equation}\lbl{tauvpap}
         a_{\mu}^{(4)}(m_{\mu}/m_{\tau})\simeq \left(\frac{\alpha}{\pi}\right)^2 \frac{1}{3}\int_{4m_{\tau}^2}^{\infty}\frac{dt}{t}\frac{m_{\mu}^2}{t}\frac{1}{\pi}
        \Imm\Pi^{({\tau})}(t)=
        \left(\frac{\alpha}{\pi}\right)^2 \frac{1}{3}\frac{1}{15}\frac{m_{\mu}^2}{m_{\tau}^2}\,,
\end{equation}
where the remaining integral over the $\tau$-spectral function 
is just the slope of the $\tau$-vacuum polarization at the origin, which can
be easily calculated inserting the explicit spectral function in
Equation~\rf{sp}. This is how the leading behavior in Equation~\rf{tauvp}
arises. Numerically, using the latest CODATA~\cite{MT02}- and
PDG-recommended~\cite{Eietal04} 
value for the mass ratio $m_{\mu}/m_{\tau}=5.945~92(97)\times 10^{-2}$, 
one finds  
\begin{equation}
                 a_{\mu}^{(4)}(m_{\mu}/m_{\tau})\simeq 0.000~078~564~(26) \left(\frac{\alpha}{\pi}\right)^2\,,
\end{equation}
using the simple asymptotic result in Equation~\rf{tauvpap};
 while using the exact expression in Equation~\rf{exactvp} 
(with $\rho=m_\tau /m_\mu$), one gets
\begin{equation}\lbl{vptau}
                a_{\mu}^{(4)}(m_{\mu}/m_{\tau})= 0.000~078~064~(25) \left(\frac{\alpha}{\pi}\right)^2\,,
\end{equation}
showing that the leading asymptotic behavior 
already reproduces correctly the first two non-zero
digits of the exact $\left(\frac{\alpha}{\pi}\right)^2$ coefficient.

\begin{figure}[h]

\begin{center}
\includegraphics[width=0.7\textwidth]{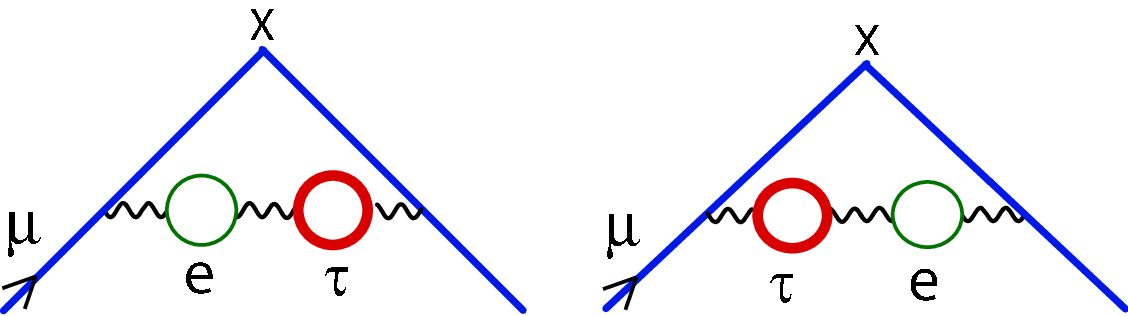}

\vspace*{0.5cm}
\caption {  Mixed vacuum polarization contribution at the three-loop level.}\label{figure:mu_el_tau}
\end{center}

\end{figure}

At the three-loop level, the $\tau$-vacuum polarization diagrams are those of Figs.~\ref{figure:mu_el_el},\ref{figure:mu_el_alpha},\ref{figure:mu_gamma_el} with the electron  replaced by the tau and hence: $m_e\rightleftharpoons m_{\tau}$. The numerical contribution, from the exact analytic calculation of Laporta and Remiddi~\cite{La93, LaRe93}, gives the result
\begin{equation}\lbl{vptau4}
a_{\mu}^{(6)}(m_{\mu}/m_{\tau})_{\rm vp}=-0.001~782~33~(48)\left(\frac{\alpha}{\pi}\right)^3\,.
\end{equation}
To this one has to add the mixed vacuum polarization diagrams in Figure~\ref{figure:mu_el_tau}, a priori enhanced because of the electron loop. 
%%%%%%%%%%%%%%%%%%%%%%%%%%%%%%%%%%%%%%
%%%%%%%%%%%%%%%%%%%%%%%%%%%%%%%%%%%%%%

\noi
A recent analytic evaluation of the asymptotic contributions up to terms of 
$\cO\left[\left(\frac{m_{\mu}^2}{m_{\tau}^2}\right)^5 \ln\frac{m_{\tau}^2}{m_{\mu}^2}\ln\frac{m_{\tau}^2\ m_{\mu}^2}{m_e^4} \right]$ and $\cO\left(
\frac{m_e^2}{m_{\tau}^2} \frac{m_{\mu}^2}{m_{\tau}^2}\right)$ 
has been made in Reference~\cite{FGdeR05} using a new technique,
 which appears to be very powerful, to obtain asymptotic expansions of Feynman graphs. The result agrees with the asymptotic terms calculated in Reference~\cite{CzSk99}, using the more complex {\it method of regions} developed by Smirnov~\cite{Smir02}.
 Numerically, this results in a contribution
\begin{equation}\lbl{vpeltau4}
        a_{\mu}^{(6)}(m_{\mu}/m_e \,, m_{\mu}/m_{\tau})_{\rm vp}  =0.000~527~66~(17)\left(\frac{\alpha}{\pi}\right)^3\,.
\end{equation}

At the four-loop level, the vacuum polarization contributions from
 $\tau$-loops 
that are still potentially important are those enhanced by 
electron vacuum polarization loops. They have been calculated 
numerically by Kinoshita and Nio~\cite{KN05,Ki05,kno06b} with the result
\begin{equation}\lbl{vpetau6}
        a_{\mu}^{(8)}(m_{\mu}/m_e \,, m_{\mu}/m_{\tau})_{\rm vp}  =-0.046~188~(37)\left(\frac{\alpha}{\pi}\right)^4\,.
\end{equation}
\subsubsection{\it Light-by-Light Scattering Diagrams from Lepton Loops.}\hfill

\begin{figure}[h]

\begin{center}
\includegraphics[width=0.6\textwidth]{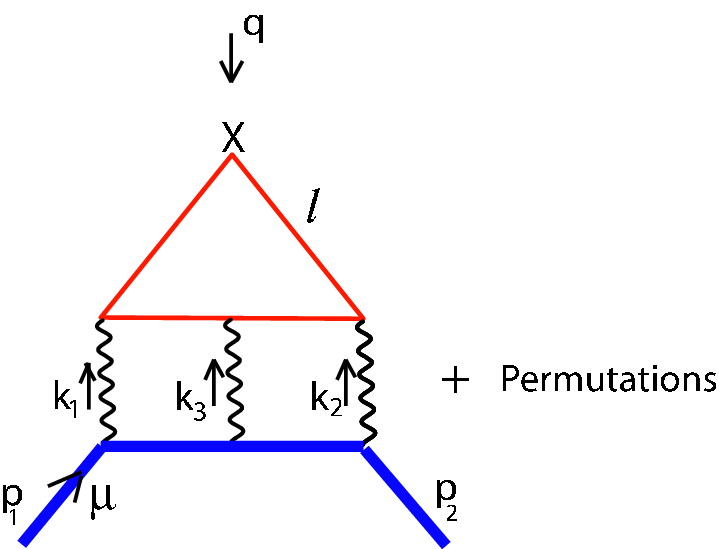}

\vspace*{0.5cm}
\caption{  Light-by-light scattering contribution from an
Internal Lepton $l$.}\label{figure:llbyl}
\end{center}

\end{figure}

\noi
It is well known that the light-by-light diagrams in QED, once the full set of gauge-invariant combinations is
considered. do not require new renormalization counter-terms. Because of that, historically, it came as a
 big surprise to find out that
the set of diagrams in Figure~\ref{figure:llbyl},
%%%%%%%%%%%%%%%%%%%%%%%%%%%%%%%%%%%%%%
%%%%%%%%%%%%%%%%%%%%%%%%%%%%%%%%%%%%%%
\noi
when the lepton $l$ in the loop is an electron, produced a contribution
proportional to
$\log(m_{\mu}/m_{e})$; and, in fact, with a large
coefficient\cite{Kinetal69}, which was first evaluated numerically.
Much progress has been made since then, and this contribution
 is now known analytically for arbitrary
values of the lepton masses\cite{LR93}. 

From a theoretical point of view, there are some general basic features
of the leading light-by-light contributions to the muon anomaly,
which will also be useful later for the discussion of the {\it hadronic
light-by-light 
contribution}, and which  we wish to recall here. 
Basically, one is confronted with a vertex function (see Figure~\ref{figure:llbyl} for reference to the routing of momenta~\footnote{We use the following conventions: $\{\gamma_{\mu},\gamma_{\nu}\}=2g_{\mu\nu}$, with $g_{\mu\nu}$ the Minkowski metric tensor of signature $(+,-,-,-)$; $\sigma_{\mu\nu}=(i/2)[\gamma_\mu,\gamma_\nu]$, $\gamma_5 =i\gamma^0 \gamma^1 \gamma^2 \gamma^3$, and with the totally antisymmetric tensor  $\epsilon_{\mu\nu\rho\sigma}$ chosen so that $\epsilon_{0123}=+1$.}):

{\setl
\bea
\Gamma_{\mu}^{(l)}(p_2 , p_1) & = & i e^6 \int\frac{d^4 k_1}{(2\pi^4)}\int\frac{d^4 k_2}{(2\pi^4)}\frac{\Pi_{\mu\nu\rho\sigma}^{(l)}(q,k_1, k_3, k_2)}{k_1^2 k_2^2 k_3^2}\times \nn \\ & & \gamma^{\nu}(\pslsout+\ksls_{2}-m_{\mu})^{-1}\gamma^{\rho}(\pslsin -\ksls_{1}-m_{\mu})^{-1}\gamma^{\sigma}\,,
\eea}

\noi
where $\Pi_{\mu\nu\rho\sigma}^{(l)}(q,k_1, k_3, k_2)$ denotes the off-shell
photon-photon scattering amplitude ($q+k_1 +k_2 +k_3 =0$) induced 
by the lepton $l$-loop, and $q=p_2-p_1$. An ingenious use of gauge-invariance,
first proposed by the authors of Reference~\cite{Kinetal69}, makes the extraction
of the anomalous magnetic moment much simpler, at the same time explicitly
removing the spurious logarithmic ultraviolet divergence inherent to
 each individual photon-photon scattering amplitude.
 Current conservation (or external gauge-invariance)
 gives rise to the Ward identity
\begin{equation}
        q^{\mu}\Pi_{\mu\nu\rho\sigma}^{(l)}(q,k_1, k_3, k_2)=0\,,
\end{equation}
from which one obtains by differentiation
\begin{equation}
        \Pi_{\mu\nu\rho\sigma}^{(l)}(q,k_1, k_3, k_2)=-q^{\lambda}\frac{\partial}{\partial q^{\mu}}\Pi_{\lambda\nu\rho\sigma}^{(l)}(q,k_1, k_3, k_2)\,.
\end{equation}
Therefore, one can write
\begin{equation}
\Gamma_{\mu}^{(l)}(p_2 , p_1) =q^{\lambda}\Gamma_{\mu\lambda}^{(l)}(p_2 , p_1)\,,   
\end{equation}
with
{\setl
\bea
\Gamma_{\mu\lambda}^{(l)}(p_2 , p_1) & = & -i e^6 \int\frac{d^4 k_1}{(2\pi^4)}\int\frac{d^4 k_2}{(2\pi^4)}\frac{\frac{\partial}{\partial q^{\mu}}\Pi_{\lambda\nu\rho\sigma}^{(l)}(q,k_1, k_3, k_2)}{k_1^2 k_2^2 k_3^2}\times \nn \\ & & \gamma^{\nu}(\pslsout+\ksls_{2}-m_{\mu})^{-1}\gamma^{\rho}(\pslsin -\ksls_{1}-m_{\mu})^{-1}\gamma^{\sigma}\,.
\eea}

\noi
The muon anomaly is then extracted from this expression, via the (formally) simple projection
\begin{equation}
     a_{\mu}^{(6)}(m_{\mu}/m_{l})_{\rm lxl}=\frac{1}{48}\frac{1}{m_{\mu}}\tr\left\{
        (\psls + m_{\mu})[\gamma^{\rho},\gamma^{\sigma}](\psls +m_{\mu}) \Gamma_{\rho\sigma}^{(l)}(p,p)\right\}\,.
\end{equation}

The first few terms in the asymptotic expansion of $ a_{\mu}^{(6)}(m_{\mu}/m_{l})_{\rm lxl}$ when the lepton in the loop is an electron are

{\setl 
\bea
a_{\mu}^{(6)}(m_{\mu}/m_e)_{\rm lxl} & = & \left\{\frac{2}{3}\pi^2 \ln\frac{m_{\mu}}{m_e}+\frac{2}{3}-\frac{10}{3}\pi^2-3\zeta(3) +\frac{59}{270}\pi^4 \right.\\
 &+& \frac{m_e}{m_{\mu}}\left[\frac{4}{3}\pi^2 \ln\frac{m_{\mu}}{m_e}+\frac{424}{9}\pi^2-\frac{196}{3}\pi^2\ln 2 \right]\\      & + & \left. \cO\left[\left(\frac{m_e}{m_{\mu}} \right)^2 \ln^3\left(\frac{m_e}{m_{\mu}} \right)  \right]\right\}\left(\frac{\alpha}{\pi}\right)^3 \,,
\eea}

\noi
explicitly showing the large coefficient of the $\ln\frac{m_{\mu}}{m_e}$ term $\left(\frac{2}{3}\pi^2 = 6.58 \right)$, first noticed numerically by the authors of Reference~\cite{Kinetal69}.
Numerically, using the exact expression of Laporta and Remiddi~\cite{LR93}, one gets
\begin{equation}\lbl{llel}
a_{\mu}^{(6)}(m_{\mu}/m_e)_{\rm lxl} =20.947~924~89~(16)\left(\frac{\alpha}{\pi}\right)^3 \,.      
\end{equation}

The contribution when the lepton in the loop 
in Fig~\ref{figure:llbyl} is heavy decouples, as it does for the 
case of the $\tau$. The first few terms 
in the asymptotic expansion are

{\setl 
\bea
a_{\mu}^{(6)}(m_{\mu}/m_{\tau})_{\rm lxl}& = & \left\{\left(\frac{m_{\mu}}{m_{\tau}}\right)^2 \left[-\frac{19}{16}+\frac{3}{2}\zeta(3) -\frac{161}{810}\left(\frac{m_{\mu}}{m_{\tau}}\right)^2 \ln^2 \frac{m_e}{m_{\mu}}\right]\right. \lbl{mutau6}\\
 & & \left.+\cO\left[\left(\frac{m_{\mu}}{m_{\tau}}\right)^4 \ln \frac{m_{\mu}}{m_{\tau}}\right]\right\} \left(\frac{\alpha}{\pi}\right)^3 \,.
\eea}

\noi
Numerically, from the exact expression~\cite{LR93}, one obtains a contribution
\begin{equation}
a_{\mu}^{(6)}(m_{\mu}/m_{\tau})_{\rm lxl}=0.002~142~83~(69)\left(\frac{\alpha}{\pi}\right)^3\,.
\end{equation}

At the four-loop level there are 180 muon vertex diagrams containing a
 light-by-light-scattering electron-loop 
sub-diagram  with second-order radiative corrections. 
They have been calculated numerically~\cite{KN05} using
 two different techniques for the more complicated 
subclasses, with the total result
\be\lbl{lle8}
a_{\mu}^{(8)}(m_{\mu}/m_e)_{\rm lxl}=121.843~1~(59)\left(\frac{\alpha}{\pi}\right)^4 \,.
\ee

%%%%%%%%%%%%%%%%%%%%%%%%%%%%%%%%%%%%%%%%%%%%%%
\begin{table*}[h]
\begin{center}

\caption[]{ QED Contributions to the Muon Anomalous Magnetic Moment. The two
numerical values in the same entry correspond to the CODATA
value~\cite{MT02}, and to the more recent~\cite{gabalpha} value of $\alpha$.
The preliminary $\left(\frac{\alpha}{\pi}\right)^5$ estimate is not listed
(see text) but it is used in the final comparison of theory with experiment.
}
\label{table:qed}

\begin{tabular}{|c|r|c|} \hline \hline {\sc\small Contribution} &
{\sc\small  Result in Powers of $\frac{\alpha}{\pi}$} & {\sc\small Numerical
Value in $10^{-11}$ Units}
\\ \hline \hline   & & \\
$a_{\mu}^{(2)}$ Eq.~\rf{sch} & $0.5\left(\frac{\alpha}{\pi}\right)\ $ &
$116~140~973.27~(0.39)$\\ & & $116~140~972.76~(0.08)$\\
\hline
$a_{\mu}^{(4)}$ Eq.~\rf{peterman}  &
$-0.328~478~965~(00)\left(\frac{\alpha}{\pi}\right)^2$ &\\
$a_{\mu}^{(4)}(m_{\mu}/m_e)$ Eq.~\rf{vper} &
$1.094~258~311~(08)\left(\frac{\alpha}{\pi}\right)^2$ & \\
$a_{\mu}^{(4)}(m_{\mu}/m_{\tau})$ Eq.~\rf{vptau} &
$0.000~078~064~(26)\left(\frac{\alpha}{\pi}\right)^2$ &\\
& & \\
$a_{\mu}^{(4)}(\rm total)$ & $0.765~857~410
~(27)\left(\frac{\alpha}{\pi}\right)^2$ & 413~217.62~(0.015)\\ & &\\
\hline
$a_{\mu}^{(6)}$ Eq.~\rf{3l}  &
$1.181~241~46~(00)\left(\frac{\alpha}{\pi}\right)^3$ & \\
$a_{\mu}^{(6)}(m_{\mu}/m_e)_{\rm vp}$ Eq.~\rf{vpel4} &
$1.920~455~13~(03)\left(\frac{\alpha}{\pi}\right)^3$ &  \\
$a_{\mu}^{(6)}(m_{\mu}/m_{\tau})_{\rm vp}$ Eq.~\rf{vptau4} &
$-0.001~782~33~(48)\left(\frac{\alpha}{\pi}\right)^3$ &  \\
$a_{\mu}^{(6)}(m_{\mu}/m_e\,, m_{\mu}/m_{\tau})_{\rm vp}$ Eq.~\rf{vpeltau4}
&
$0.000~527~66~(17)\left(\frac{\alpha}{\pi}\right)^3$ &  \\
$a_{\mu}^{(6)}(m_{\mu}/m_e)_{\rm lxl}$ Eq.~\rf{llel} &
$20.947~924~89~(16)\left(\frac{\alpha}{\pi}\right)^3$ &  \\
$a_{\mu}^{(6)}(m_{\mu}/m_{\tau})_{\rm lxl}$ Eq.~\rf{llel} &
$0.002~142~83~(69)\left(\frac{\alpha}{\pi}\right)^3$ &  \\ & & \\
$a_{\mu}^{(6)}(\rm total)$ &
$24.050~509~64~(87)\left(\frac{\alpha}{\pi}\right)^3$ & 30~141.90é
(0.001)°\\ & & \\
\hline
$a_{\mu}^{(8)}$ Eq.~\rf{4l} &
$-1.728~3~(35)\left(\frac{\alpha}{\pi}\right)^4$ & \\
$a_{\mu}^{(8)}(m_{\mu}/m_e)_{\rm vp}$ Eqs.~\rf{vpI},\rf{vpII},\rf{vpIII} &
$10.839~2~(41)\left(\frac{\alpha}{\pi}\right)^4$& \\
$a_{\mu}^{(8)}(m_{\mu}/m_e \,, m_{\mu}/m_{\tau})_{\rm vp} Eq.~\rf{vpetau6}$
& $-0.046~2~(00)\left(\frac{\alpha}{\pi}\right)^4$ &   \\
$a_{\mu}^{(8)}(m_{\mu}/m_e)_{\rm lxl}$ Eq.~\rf{lle8} &
$121.843~1~(59)\left(\frac{\alpha}{\pi}\right)^4$  & \\
$a_{\mu}^{(8)}(m_{\mu}/m_e\,, m_{\mu}/m_{\tau})_{\rm lxl}$ Eq.~\rf{lletau} &
$0.083~8~(01)\left(\frac{\alpha}{\pi}\right)^4$ &  \\  & & \\
$a_{\mu}^{(8)}(\rm total)$  &
$130.991~6~(80)\left(\frac{\alpha}{\pi}\right)^4$ &
381.33~(0.023) \\ &  & \\
\hline & & \\
$a_{\mu}^{(2+4+6+8)}(\rm QED)$ &  & $116~584~714.12~(0.39)(0.04)$\\
& & $116~584~713.61~(0.08)(0.04)$\\
\hline\hline

\end{tabular}

 \end{center}
\end{table*}
%%%%%%%%%%%%%%%%%%%%%%%%%%%%%%%%%%%%%%%%%%%%%%%
\noi
At the same four-loop level there are still diagrams with mixed
 electron-loops and tau-loops, with one of them of the light-by-light
 type. Although suppressed by a $\left(m_{\mu}/m_{\tau} \right)^2$ factor,
 they are still enhanced by a $\ln\frac{m_{\mu}}{m_e}$ factor. They have also been calculated  numerically by Kinoshita and Nio~\cite{KN05}, with the result
\begin{equation}\lbl{lletau} 
a_{\mu}^{(8)}(m_{\mu}/m_e\,,m_{\mu}/m_{\tau} )_{\rm    lxl}=0.083~782~(75)\left(\frac{\alpha}{\pi}\right)^4 \,. 
\end{equation}

\vspace*{1cm}
All together, the purely QED contributions to the muon anomalous magnetic
moment, including electron- and tau-loops of the vacuum polarization type
 and/or of the light-by-light scattering type,
are known to four loops, to an accuracy which is certainly good enough for the present
comparison between theory and experiment. 
In Table~\ref{table:qed} we have collected all the results of the QED
contributions that we have discussed. The numerical results 
have been obtained using the latest recommended
 CODATA value for the fine-structure constant~\cite{MT02},
\begin{equation}
        \alpha^{-1}=137.035~999~11(46)\ [3.3~{\rm ppb}]\,,
\end{equation}
which, as discussed in the introduction, is dominated by the electron
 anomalous magnetic moment measurements. In fact, the most recent
 determination of $\alpha$ from the comparison between QED
 and a new measurement of the electron $g$-factor~\cite{gabeg2} 
gives the result~\cite{gabalpha}
\begin{equation}
        \alpha^{-1}=137.035~999~710(96)\ [0.70~{\rm ppb}]\,,
\end{equation}
which is quite a remarkable improvement.
 The numerical results using this more precise value of $\alpha$
 as an input are also given in the Table~\ref{table:qed}. Use of this
new value only
 affects the determination of the lowest-order term $a_{\mu}^{(2)}$
 (the second-line value) and hence the total at the bottom in 
Table~\ref{table:qed}. 

  Notice that the errors in $a_{\mu}^{(4)}(\rm total)$, $a_{\mu}^{(6)}(\rm
  total)$ and $a_{\mu}^{(8)}(\rm total)$ are very small and dominated by the
  error in the $\tau$-mass, and in the 
$\mu$-mass to a lesser degree. The first error in the total
 sum $a_{\mu}^{(2+4+6+8)}(\rm QED)$, which is still the dominant error, 
 is the one induced by the error in the determination of  the 
fine-structure constant; the second error is the one induced by the lepton
 mass ratios.

The most recent estimate of the five-loop contribution by 
Kinoshita~\cite{KN06,Ki05} gives a result
\begin{equation}\lbl{fiveloops}
        \hspace*{-1cm} a_{\mu}^{(10)} 
(m_{\mu}/m_e\,,m_{\mu}/m_{\tau} )_{\rm  vp~and~  lxl}
\simeq 663~(20)~\left(\frac{\alpha}{\pi}\right)^5 = 4.48~(0.14)\times 10^{-11}\,.
\end{equation}
Because  this estimate is not yet at the level of rigor of the 
lowest-order determinations, we have not included it in
Table~\ref{table:qed}; however, we shall take it into account when making a
 final comparison between theory and experiment. 

The
question that naturally arises is whether or not the discrepancy between
the experimental result  on the one
hand, and the QED contribution from leptons alone which we 
have discussed, can be understood in
terms of the extra hadronic and electroweak contributions predicted by the
Standard Model. This will be the subject of the following subsections.

\subsection{Contributions from Hadronic Vacuum Polarization}\label{HVPlowest}

\begin{figure}[h]

\begin{center}
\includegraphics[width=0.3\textwidth]{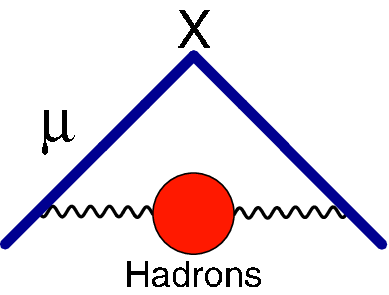}

\vspace*{0.5cm}
\caption{\it\small  Hadronic Vacuum Polarization Contribution}\label{figure:hvp}
\end{center}

\end{figure}

\begin{figure}[h]

\begin{center}
\includegraphics[width=0.6\textwidth]{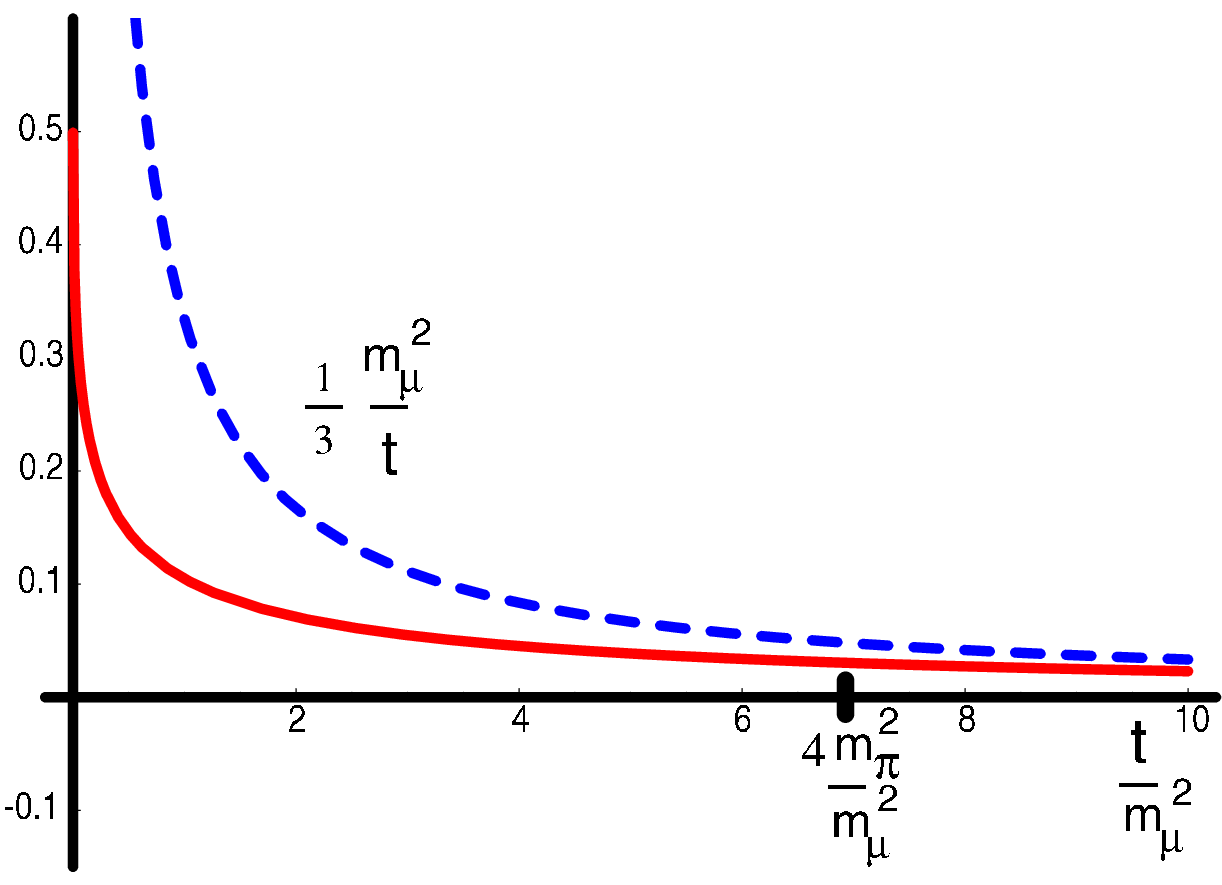}

\vspace*{0.5cm}
{\label{fg:Figure11} \caption{  Behavior of the
 function $K^{(2)}(t/m_{\mu}^2)$ in Eq.~\rf{kfun2}. 
The dotted blue line represents its leading asymptotic
 behavior in Eq.~\rf{aslo}.}}
\end{center}

\end{figure}

All calculations of the lowest-order hadronic vacuum polarization
contribution to the muon anomaly (see Figure~\ref{figure:hvp}) are based on 
 the spectral 
representation~\cite{BM61} 
\begin{equation}\lbl{lohvp}
        a_{\mu}^{(4)}({\rm H})_{\rm vp}= \frac{\alpha}{\pi}\int_0^{\infty}\frac{dt}{t}\frac{1}{\pi}
        \Imm\Pi^{({\rm H})}(t) \int_0^1 dx\frac{x^2 (1-x)}{x^2 + \frac{t}{m_{\mu}^2}(1-x)}\,,
\end{equation}
with the hadronic spectral function $\frac{1}{\pi}
        \Imm\Pi^{({\rm H})}(t)$  related to the \underline{\it one-photon} $e^+ e^-$ annihilation cross-section into hadrons as follows: 
\begin{equation}\lbl{onephoton}
        \sigma(t)_{\{e^+ e^- \ra (\gamma)\ra {\rm hadrons}\}}=\frac{4\pi^2\alpha}{t} \frac{1}{\pi}
        \Imm\Pi^{({\rm H})}(t)\,.
\end{equation}
%%%%%%%%%%%%%%%%%%%%%%%%%%%%%%%%%%%%%%
%%%%%%%%%%%%%%%%%%%%%%%%%%%%%%%%%%%%%%
As already explained in the previous subsection when discussing the contribution from vacuum polarization due to a heavy lepton, Equation~\rf{lohvp} results from the replacement
\begin{equation}
        -i\frac{g_{\alpha\beta}}{q^2}\Ra
                i\left(g_{\alpha\beta}-\frac{q_\alpha q_\beta}{q^2}\right)\frac{\Pi^{({\rm H})}(q^2)}{q^2}
\end{equation}
in the free-photon propagator of the lowest-order QED diagram in
Figure~\ref{figure:sch} by the one corrected by the proper hadronic photon
self-energy in Figure~\ref{figure:hvp}. Since the photon self-energy is
transverse in the $q$-momenta, the replacement is unaffected by the gauge
dependence in the free-photon propagator. The on-shell renormalized photon
self-energy obeys a dispersion 
relation, with a subtraction at $q^2=0$ associated with 
the on-shell charge renormalization; therefore
\begin{equation}\lbl{disprel2}
\frac{\Pi^{({\rm H})}(q^2)}{q^2}=\int_0^\infty \frac{dt}{t}\frac{1}{t-q^2 }\frac{1}{\pi}
        \Imm\Pi^{({\rm H})}(t)\,. 
\end{equation}
The fact that only one subtraction is needed in this dispersion relation
follows from the QCD short-distance behavior 
of $\Pi^{({\rm H})}(q^2)$ in the deep  Euclidean region ($-q^2\ra\infty$). 
The r.h.s. of Equation~\rf{lohvp} can thus be viewed as the convolution (the
integral over $t$) of the hadronic spectral function  with the contribution
to the muon anomaly, induced by a {\it fictitious massive photon} ($m^2=t$)
with
 a free-propagator:
\begin{equation}
        -i\left(g_{\alpha\beta}-\frac{q_\alpha q_\beta}{q^2}\right)\frac{1}{q^2 -t}\,.
\end{equation}
This {\it massive photon} contribution to the muon anomaly~\cite{BdeR68}, which we shall denote by $\left(\frac{\alpha}{\pi} \right)K^{(2)}(t/m_{\mu}^2)$, results in a simple Feynman parametric integral (the integral over $x$ in Equation~\rf{lohvp}):
\begin{equation}\lbl{kfun2}
K^{(2)}\left(\frac{t}{m_{\mu}^2}\right)=\int_0^1 dx\frac{x^2 (1-x)}{x^2 + \frac{t}{m_{\mu}^2}(1-x)}\,.
\end{equation}
The integrand on the r.h.s. of Equation~\rf{kfun2} is positive and
monotonically decreasing in the integration region $0\le t\le \infty$. This is
why the 
lowest-order 
hadronic vacuum polarization contribution to the muon anomaly is positive.
 This integral was first evaluated analytically in Reference~\cite{BdeR68}. A convenient representation for numerical (and analytical) evaluations, which has often been used in the literature,  is the one given in Reference~\cite{LdeR69}:

For $t\ge 4m_{\mu}^2$, and with $\beta_\mu =\sqrt{1-\frac{4m_{\mu}^2}{t}}$ and  $x=\frac{1-\beta_{\mu}}{1+\beta_{\mu}}$,
\begin{equation}
\hspace*{-2.3cm}
K^{(2)}\left(\frac{t}{m_{\mu}^2}\right)=\frac{1}{2}x^2 (2-x^2) +\frac{(1+x)^2 (1+x^2)}{x^2}[\ln(1+x)-x+\frac{1}{2}x^2]+\frac{1+x}{1-x}x^2 \ln x\,.      
\end{equation}
 Let us also remark that, for large values of $t/m_{\mu}^2$, the function $K^{(2)}(t/m_{\mu}^2)$ decreases as $m_{\mu}^2/t$:
\begin{equation}\lbl{aslo}
\lim_{t\ra\infty}K^{(2)}(t/m_{\mu}^2)=\frac{1}{3}\frac{m_{\mu}^2}{t}+\cO\left[
\left(m_{\mu}^2/t\right)^2 \log (t/m_{\mu}^2)\right]\,,
\end{equation}
and this is why the bulk of the contribution to $a_{\mu}[{\rm Fig.}~10]$
 comes from the low-energy region-- in particular from the prominent
 lowest $\rho$-resonance. The shape of the function $K^{(2)}(t/m_{\mu}^2)$,
 as well as its leading asymptotic behavior in Equation~\rf{aslo}, are
 plotted in \ref{fg:Figure11}.
%%%%%%%%%%%%%%%%%%%%%%%%%%%%%%%%%%%%%%
%%%%%%%%%%%%%%%%%%%%%%%%%%%%%%%%%%%%%
\noi
One can see from this figure that, already at the $\rho$-mass value
$t=M_{\rho}^2$, the leading asymptotic behavior is a rather good
approximation to the exact function. In fact, as first pointed out in
Reference~\cite{BdeR69}, the leading asymptotic behavior of the function
$K^{(2)}(t/m_{\mu}^2)$
 provides an {\it upper bound} to the size of $a_{\mu}[{\rm Fig.}~10]$:
\begin{equation}
\hspace*{-1cm}
                a_{\mu}^{(4)}({\rm H})_{\rm vp}\le \frac{\alpha}{\pi}\frac{1}{3}m_{\mu}^2\int_{0}^{\infty}dt\frac{1}{t^2}\frac{1}{\pi}
        \Imm\Pi^{({\rm
                H})}(t)
=\frac{\alpha}{\pi}\frac{1}{3}m_{\mu}^2\left(\frac{\partial\Pi^{({\rm
                H})}(q^2)}{\partial q^2}\right)_{q^2=0}\, .
\end{equation}
In other words,  to lowest-order in powers of the fine-structure constant, the
hadronic vacuum-polarization contribution to the muon anomaly has an
 upper bound which is governed by the slope at the origin of the
 hadronic photon self-energy, a quantity which is one of the 
low-energy constants (a constant of $\cO(p^6)$) of the effective
 chiral Lagrangian of QCD.

From a theoretical point of view, it is more convenient to
 convert Equation~\rf{lohvp} into an integral over the hadronic 
photon self-energy in the Euclidean region,
 i.e., an integral over $Q^2=-q^2\ge 0$ instead of an integral over the
 Minkowski region $t\ge 4m_{\pi}^2$. This is easily achieved
 combining Equation~\rf{lohvp} and the dispersion relation
 in Equation~\rf{disprel}, with the result~\cite{LdeR69,LPdeR72}:
\begin{equation}
a_{\mu}^{(4)}({\rm H})_{\rm vp}=\frac{\alpha}{\pi}\int_0^1 dx (1-x)\left[-\Pi^{({\rm H})}\left(-\frac{x^2}{1-x}m_{\mu}^2 \right) \right]\,.      
\end{equation}
Another useful representation, which follows from this one
 by partial integration, is the one in terms of the Adler-function
\begin{equation}
        \cA^{(H)}(Q^2)=\int_0^\infty dt \frac{Q^2}{(t+Q^2)^2}\frac{1}{\pi}\Imm \Pi^{({\rm H})}(t)\,,
\end{equation}
with the result
\begin{equation}\lbl{adleran}
a_{\mu}^{(4)}({\rm H})_{\rm vp}=\frac{\alpha}{\pi}\frac{1}{2}\int_0^1\frac{dx}{x} (1-x)(2-x)
        \cA^{(H)}\left(\frac{x^2}{1-x}m_{\mu}^2\right)        \,.
\end{equation}
These representations are better adapted to theoretical analysis and, eventually, to lattice QCD numerical evaluations. In fact, some exploratory lattice QCD work in this direction has already started~\cite{latticeVP}.

It is possible to make a quick analytic estimate of
 $a_{\mu}^{(4)}({\rm H})_{\rm vp}$ using a Large-$N_c$ QCD framework
 (see, e.g. References~\cite{deR02,Pe02} for review articles 
and Reference~\cite{FGdeR04}). The spectral function in the {\it minimal hadronic
 approximation} (MHA) to Large-$N_c$ QCD consists of a lowest narrow state
 (the $\rho$) plus a perturbative QCD continuum starting at a threshold
 $s_0$, 
yielding a simple parameterization for the Adler function (for three flavors):
\begin{equation}\lbl{MHAadler}
\hspace*{-2cm}  \cA^{(H)}(Q^2)=\frac{2}{3}e^2
        \left\{2f_{\rho}^2 M_{\rho}^2\frac{Q^2}{(Q^2+M_{\rho}^2)^2} + \frac{N_c}{16\pi^2}\frac{4}{3}\left(1+ \frac{3}{8}\frac{N_c \als (s_0)}{\pi}+ \cdots \right)\frac{Q^2}{Q^2+s_0}\right\}\,,
\end{equation}
where $f_{\rho}$ denotes the $\rho$ to vacuum coupling, related to the electronic width of the $\rho$ as follows:
\begin{equation}
        \Gamma_{\rho\ra e^+ e^-}=\frac{4\pi\alpha^2}{3}f_{\rho}^2 M_{\rho}\,,
\end{equation}
and the dots stand for higher-order terms in powers of $\als$. The short-distance scale $s_0$ denotes the threshold of the perturbative QCD (pQCD) continuum. The actual value of $s_0$ is fixed from the requirement that, in the chiral limit, there is no $1/Q^2$-term in the short-distance behavior of the Adler function, a property which follows from the  operator product expansion (OPE). This leads to the constraint
\begin{equation}\lbl{constr}
        2f_{\rho}^2 M_{\rho}^2=\frac{N_c}{16\pi^2}\frac{4}{3}s_0 \left(1+ \frac{3}{8}\frac{N_c \als (s_0)}{\pi}+ \cdots \right)\,.
\end{equation}
Inserting the large-$N_c$ ansatz given in \rf{MHAadler}, with $Q^2=\frac{x^2}{1-x}m_{\mu}^2$ and the constraint \rf{constr} in Equation~\rf{adleran}, results in the simple estimate
\begin{equation}\lbl{resMHA}
                a_{\mu}^{(4)}({\rm H})_{\rm vp}\sim (57.2\pm 15.0)\times 10^{-9}\,,
\end{equation}
 where we have set $f_{\rho}^2=2\frac{F_0^2}{M_{\rho}^2}$, as predicted by
 the MHA to Large-$N_c$ QCD, with $F_0$ the pion to vacuum coupling in the
 chiral limit with: $F_0\simeq 87~\MeV$ and $M_{\rho}\simeq
 750~\MeV$. The error in Equation~\rf{resMHA} is a generous estimate of the
 systematic error of the approximations involved. One can, in principle,
 obtain a much more  refined theoretical evaluation, but not with the high
 precision which is nowadays required. Regrettably, the present lattice QCD
 estimates~\cite{latticeVP} are not more precise
 than this simple large-$N_c$ QCD estimate.

 The phenomenological evaluation of the hadronic vacuum polarization to the
 muon anomaly, using $\sigma(e^+ e^- \ra {\rm hadrons})$ data, goes back to
 earlier papers in References~\cite{GdeR69,cern3a,CNPdeR76} with the results
 quoted in the first part of Table~\ref{table:hvp}. (The question-mark
 error in Reference~\cite{GdeR69} is due to
 ignorance at that time of the high-energy contributions, since this
 determination was prior to QCD.) The advent of more and more precise data on
 $e^+ -e^-$ annihilations, as well as the new $g_{\mu}-2$ results from the
 Brookhaven E821 experiment, have motivated more refined 
re-evaluations of these
 hadronic contributions, which also incorporate QCD
 asymptotic behavior properties.  In the second
 part of Table~\ref{table:hvp} we collect recent detailed determinations
 including, in particular, the re-analysis of earlier measurements by the
 CMD-2 detector at the VEPP-2M collider in 
Novosibirsk~\cite{Akhetal}. It is also possible to measure
 $\frac{1}{\pi}\Imm\Pi^{({\rm H})}(t)$ in the region $0\leq t\leq M_{\tau}^2$
 using hadronic $\tau$-decays~\cite{ADH98}. Two representative results, which partly
 use $\tau$-data, and which also take into account corrections to the
 conserved-vector-current-symmetry (CVC) limit where the spectral functions
 from $\tau$-decays and $e^+ e^-$ are directly related, are quoted in the
 third part of Table~\ref{table:hvp}.  
As a test of the CVC hypothesis,
the $e^+e^-$ data can be used to predict the $\tau$ branching 
fractions which, unfortunately, leads to branching ratios in disagreement
with the data from LEP and CLEO~\cite{DEHZ03}.  As additional 
$e^+e^-$ and $\tau$ data have become available, this discrepancy has 
increased~\cite{Davier06}. 
 
 A preliminary update of the hadronic vacuum polarization,
  contributions reported by M.~Davier~\cite{Davier06} 
at the Tau06 conference incorporates the new 
 CMD-2 results~\cite{Akhmetshin_06a,Akhmetshin_06b}, 
new SND results~\cite{Achasov_06}, as 
well as  BaBar data for some exclusive channels other 
than $\pi^+ \pi^-$. Hagiwara, et al.,~\cite{Hagiwara_06}
have also included these new data, as well as the KLOE data, in a recent reanalysis and obtain a
similar result.
  We quote these recent evaluations at the bottom part of 
Table~\ref{table:hvp}.  The overall results quoted in this Table illustrate the 
progress which has been made in this field, especially in the last few years.
 Note that the new data allow for a significant
 reduction of the experimental error as well as the 
error due to radiative corrections.

\begin{table*}[h]
\caption[]{  Compilation of Results on $a_{\mu}^{(4)}({\rm H})_{\rm vp}$}
\label{table:hvp}

\begin{tabular}{|c|c|} \hline \hline {\bf { Authors}} &
{\bf { Contribution to $a_{\mu}\times 10^{11}$}} 
\\ \hline \hline
{ Gourdin, de Rafael}~\cite{GdeR69} (1969) & $6500\pm 500 \pm {\rm (?)}$ \\
Bailey {\it et al.}~\cite{cern3a} (1975) & $7300\pm 1000 $\\
{ Calmet-Narison-Perrottet-de Rafael}~\cite{CNPdeR76} (1976) & $7020\pm 800$ \\
$\cdots$ & $\cdots$ \\
\hline \hline
{ Davier {\it et al.,} [$e^+ e^-$-data]}~\cite{DEHZ03} (2003) & $6963 \pm 62_{\mbox{\rm
\tiny exp}}
\pm 36_{\mbox{\rm
\tiny rad}}$
\\
{ Jegerlehner}~\cite{Jeger04} (2004) & { $6948 \pm 86$} \\ 
{ de Troc\'oniz, Yndur\'ain }~\cite{deTY04} (2004) &
{$6944\pm 48_{\mbox{\rm
\tiny exp}}
\pm 10_{\mbox{\rm
\tiny rad}}$}\\
{ Hagiwara {\it et al.,} }~\cite{HMNT04} (2004) & $6924 \pm 59_{\mbox{\rm
\tiny exp}}
\pm 24_{\mbox{\rm
\tiny rad}}$ \\
\hline\hline
{ Davier {\it et al.,} [$\tau$-data]}~\cite{DEHZ03} (2003) & $7110 \pm
50_{\mbox{\rm
\tiny exp}}
\pm 8_{\mbox{\rm
\tiny rad}}\pm 28_{\mbox{\rm
\tiny $SU(2)$}}$ \\
{ de Troc\'oniz, Yndur\'ain }~\cite{deTY04} (2004) &
{$7027\pm 47_{\mbox{\rm
\tiny exp}}
\pm 10_{\mbox{\rm
\tiny rad}}$}\\
\hline\hline
{Davier, [$e^+e^-$-data]}~\cite{Davier06} (2006) & $6908 \pm 30_{\mbox{\rm
\tiny exp}} \pm19_{\mbox{\rm
\tiny rad}} \pm 7_{\mbox{\rm
\tiny QCD}}$ \\
{ Hagiwara  {\it et al.,} [$e^+e^-$-data]}~\cite{Hagiwara_06}
 (2006) & $6894 \pm
42_{\mbox{\rm
\tiny exp}}
\pm 18_{\mbox{\rm
\tiny rad}} $ \\
 \hline\hline
\end{tabular}
\end{table*}

 Several remarks, however,  concerning the results in Table~\ref{table:hvp} are in order:

\begin{itemize}

        \item  There still remains a problematic discrepancy
 between the most recent $e^+ e^-$ determinations and those 
using $\tau$-hadronic data. This prevents one from using a
 straightforward average of these different determinations. 
        
        \item  Various sources of isospin breaking effects, when transforming
        $\tau$-hadronic spectral functions to the required $e^+ e^-$ 
spectral function, have been identified. A straightforward correction
 is the one due to the electroweak short-distance contribution to the
 effective four-fermion coupling
 $\tau\ra \nu_{\tau}d\bar{u}$~\cite{S82,MS88,BNP92}.
 Effects due to the mass differences between the
 $\pi^+$ and the $\pi^0$, as well as mass and width
 differences between the charged and the neutral $\rho$,
 have also been considered~\cite{ADH98,GJ03}.
        
        \item   A more problematic issue is the one 
of radiative corrections for a specific channel. 
Calculations concerning the mode $\nu_{\tau}\pi^- \pi^0$ 
have been reported in References~\cite{CEN01,CEN02}.
        
        \item
       The most precise $e^+ e^-\ra \pi^+ \pi^-$ data
 come from the CMD-2
       detector, which are now in agreement with data from SND, but not with
       the recent data from KLOE at Frascati. In fact, the two most
       recent compilations\cite{Davier06,Hagiwara_06} 
       differ on the use of
       KLOE's results, which agree with CMD on the integral but
       not the shape of the pion form factor. 
       The CMD-2 results have been corrected for leptonic and hadronic
 vacuum polarization effects, so that the measured final state
 corresponds to $\pi^+ \pi^-$, including pion-radiated photons and 
virtual final-state QED effects. 
                
        \item
        By contrast, hadronic $\tau$-data are available from the four LEP
detectors, CLEO, OPAL, L3, and ALEPH, and show good consistency
 when measuring the $\tau^-\ra \nu_{\tau}\pi^- \pi^0$ branching ratio. 
        
        \item
         One can hope that the forthcoming results from the KLOE 
detector~\cite{KLOE} and BaBar~\cite{BABAR}, using the radiative return
         technique,  will eventually provide the necessary consistency check
         of the $e^+ e^-$ input. The BELLE $\tau$-data, with one hundred
times the statistics of the LEP experiments,
should also provide a test of the LEP and CLEO results.
          
        \item
        Finally, let us comment on the fact that the contribution from the
        high-energy tail to $a_{\mu}^{(4)}({\rm H})_{\rm vp}$ in
        Equation~\rf{lohvp} is well under control because of the asymptotic
        freedom property of QCD. The perturbative QCD prediction is known to
        next-to-next-to-leading order with 
second-order quark mass corrections included. (See Reference~\cite{CKS96} and references therein for earlier calculations.)    
\end{itemize}
      
\subsubsection{Higher-Order Hadronic Vacuum Polarization Contributions}\hfill

\begin{figure}[h]

\begin{center}
\includegraphics[width=0.6\textwidth]{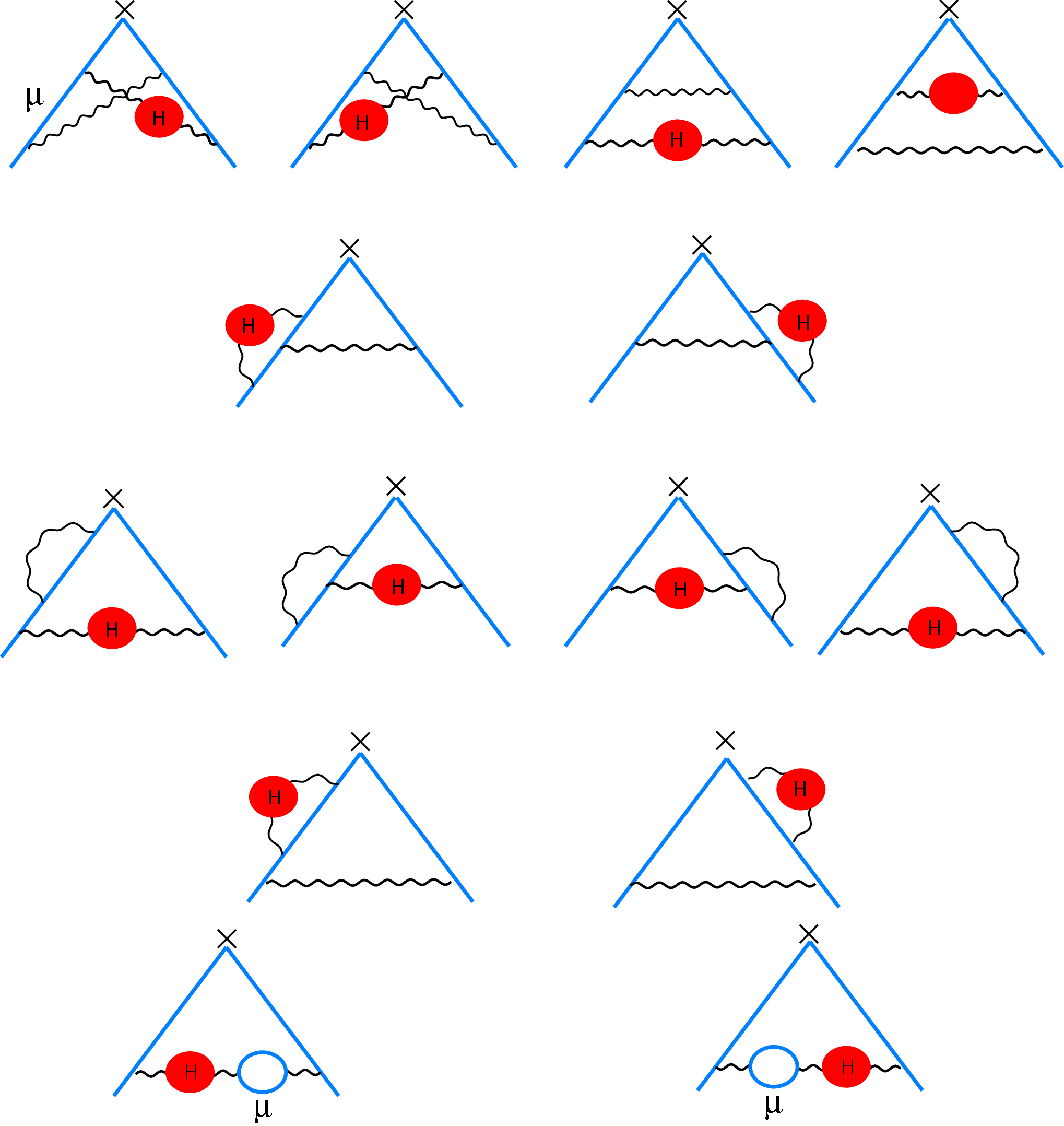}

\vspace*{0.5cm}
\caption{  Hadronic Vacuum Polarization Corrections to the 
two-loop graphs in Figure~\ref{figure:peter}.}\label{figure:hvp4}
\end{center}
\end{figure}

\begin{figure}[h]

\begin{center}
\includegraphics[width=0.6\textwidth]{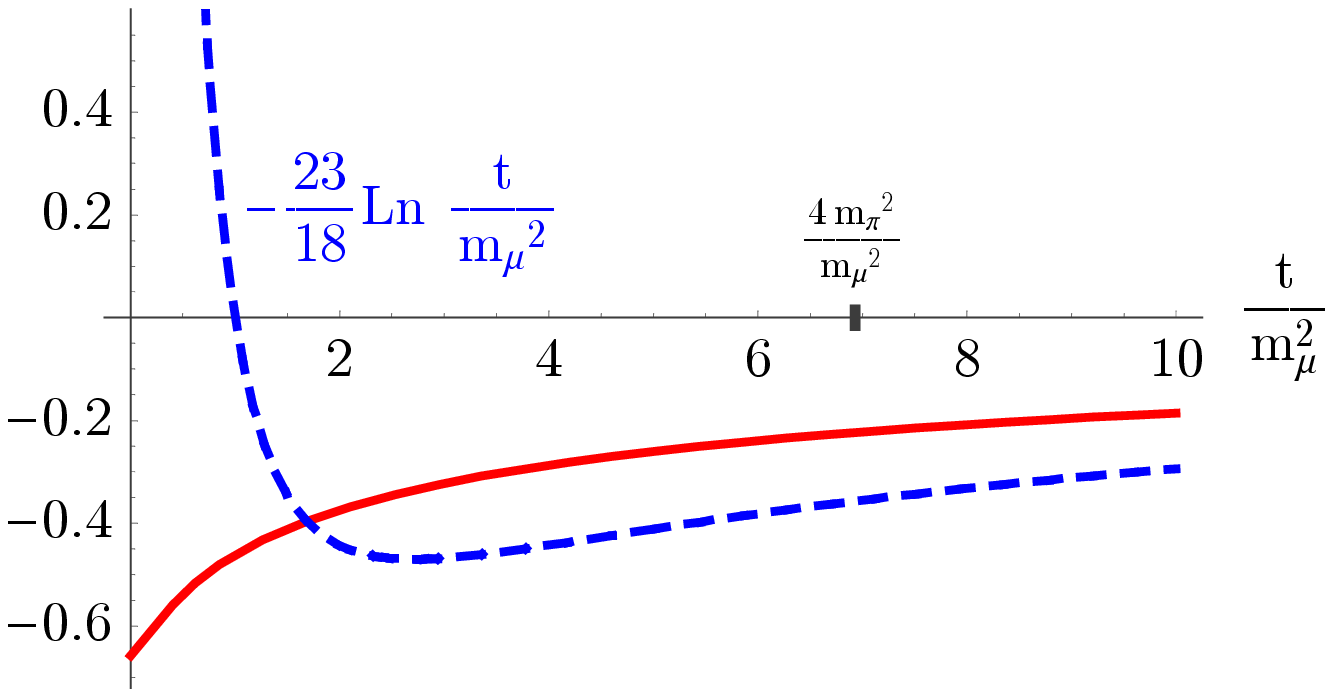}

\vspace*{0.5cm}
\caption{  Behavior of the function $K^{(4)}(t/m_{\mu}^2)$ 
in Eq.~\rf{hohvp}. The dotted blue line represents its leading asymptotic behavior in Eq.~\rf{asho}.}\label{figure:K4}
\end{center}

\end{figure}

\noi
There are three classes of hadronic vacuum polarization contributions of an
overall $\cO\left[\left(\frac{\alpha}{\pi}\right)^{3}\right]$ which were
first considered in Reference~\cite{CNPdeR76}. A first class is the set of Feynman
diagrams in Figure~\ref{figure:hvp4}. They are obtained by inserting one
hadronic vacuum polarization correction into 
each of the virtual photons of the purely QED fourth-order
 diagrams in Figure~\ref{figure:peter}. 
%%%%%%%%%%%%%%%%%%%%%%%%%%%%%%%%%%%%%%
%%%%%%%%%%%%%%%%%%%%%%%%%%%%%%%%%%%%%%
\noi
The diagrams in each line in Figure~\ref{figure:hvp4} are well-defined
 gauge-invariant subsets. One can easily see that their overall contribution,
 by analogy to our discussion of the lowest-order hadronic
 vacuum polarization, will be governed by an integral
\begin{equation}\lbl{hohvp}
        a_{\mu}^{(6)}(\rm H)_{\rm vp}= \left(\frac{\alpha}{\pi}\right)^2\int_0^{\infty}\frac{dt}{t}\frac{1}{\pi}
        \Imm\Pi^{({\rm H})}(t)K^{(4)}\left(\frac{t}{m_{\mu}^2}\right)\,,
\end{equation}
where $\left(\frac{\alpha}{\pi}\right)^2 K^{(4)}\left(\frac{t}{m_{\mu}^2}\right)$ is now the contribution to the muon anomaly from a {\it fictitious massive photon} in the presence of $\cO\left(\frac{\alpha}{\pi}\right)$ QED corrections. 
The exact form of the function $K^{(4)}\left(\frac{t}{m_{\mu}^2}\right)$,
which is a rather complicated expression, was obtained by Barbieri and
Remiddi~\cite{BR75} and has not been fully checked since then. We reproduce
in Figure~\ref{figure:K4} the shape of this function and compare it with its leading
asymptotic behavior, which agrees with the one that has been extracted~\cite{FGdeR03} from
an independent effective field-theory calculation of anomalous dimensions of
composite operators 
in Reference~\cite{CMV03},
\begin{equation}\lbl{asho}
\lim_{t\ra\infty}K^{(4)}\left(\frac{t}{m_{\mu}^2}\right)=-\frac{23}{18}\frac{m_{\mu}^2}{t}\ln\left(\frac{t}{m_{\mu}^2} \right)+\cO\left(\frac{m_{\mu}^2}{t} \right)\,.  
\end{equation}

%%%%%%%%%%%%%%%%%%%%%%%%%%%%%%%%%%%%%%
%%%%%%%%%%%%%%%%%%%%%%%%%%%%%%%%%%%%%%

\noi
The fact that the function  $K^{(4)}\left(\frac{t}{m_{\mu}^2}\right)$ is negative for $4\mu^2\le t\le \infty$ implies that, contrary to the lowest-order case,  the contribution to $a_{\mu}^{(6)}(\rm H)_{\rm vp}$ must be {\it negative}.

\begin{figure}[h]

\begin{center}
\includegraphics[width=0.3\textwidth]{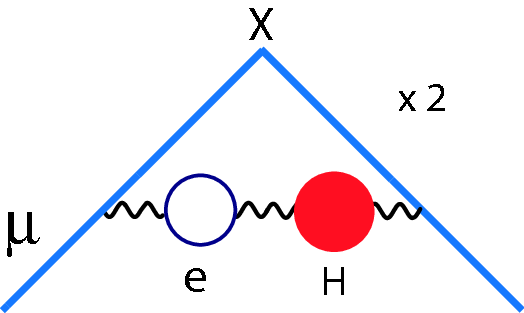}

\vspace*{0.5cm}
\caption { Second Class of Higher-Order Hadronic Vacuum Polarization Contributions.}\label{figure:ehvp}
\end{center}

\end{figure}

A second class of potentially important contributions are the mixed vacuum
polarization insertions in Figure~\ref{figure:ehvp} due to an electron loop
and a hadronic loop. [The similar contribution where the electron loop is
replaced by a $\tau$ loop decouples as $m_{\mu}^2/m_{\tau}^2$ and
 is negligible at the present required level of accuracy.] 
%%%%%%%%%%%%%%%%%%%%%%%%%%%%%%%%%%%%%%
%%%%%%%%%%%%%%%%%%%%%%%%%%%%%%%%%%%%%%
\noi
This contribution, can be expressed as a double convolution~\cite{CNPdeR76}:
\be
\lbl{mixedvp}\hspace*{-2cm}
        a_{\mu}^{(6)}\left(\frac{m_{\mu}}{m_e}\,,{\rm H}\right)_{\rm vp}=\frac{\alpha}{\pi}\int_{4
M_{\pi}^2}^{\infty}\frac{dt}{t}\!{
\frac{1}{\pi}\Imm\Pi(t)}
\underbrace{{\int_{0}^{1}\frac{x^2 (1-x)}{x^2
+{\frac{t}{m_{\mu^2}}}(1-x)}}dx}_{\sim\ \frac{1}{3}
\ \frac{m_{\mu}^2}{t}} 
\ \underbrace{(-2)\Pi\left(\frac{x^2}{1-x}
\frac{m_{\mu}^2}{m_{e}^2}
\right)}_{\sim (-2)\frac{\alpha}{\pi}\left(-\frac{1}{3}\right)\log
\frac{m_{\mu}^2}{m_{e}^2}}\,;
\ee
where, in underbraces, we have indicated the leading asymptotic behaviors in
$m_{\mu}^2 /t$ and $m_{e}^2 /m_{\mu}^2$-- indicating
 that this contribution must be positive. The question remains: which of the
 two contributions from Figure~\ref{figure:hvp4} (which, we concluded,
 is {\it negative}) and from Figure~\ref{figure:ehvp}  (which is {\it
   positive}) dominates? A simple answer to this
 can be found, without doing the full calculation~\cite{FGphd05},
 by combining the asymptotic behaviors in Eqs.~\rf{asho} and \rf{mixedvp}:   

{\setl
\bea
\hspace*{-2cm}
a_{\mu}^{(6)}\left({\rm H~and}~\frac{m_{\mu}}{m_e}\,,{\rm H}\right)_{\rm vp} & \simeq & 
\left(\frac{\alpha}{\pi}\right)^2\int_{4
M_{\pi}^2}^{\infty}\frac{dt}{t}\frac{m_{\mu}^2}{t} {
\frac{1}{\pi}\Imm\Pi(t)}\left[-\frac{23}{18}\ln\frac{t}{m_{\mu}^2}+\frac{2}{9}\ln\frac{m_{\mu}^2}{m_e^2}\right]\\
 & \sim & \left(\frac{\alpha}{\pi}\right)^3 \frac{m_{\mu}^2}{M_{\rho}^2}\ \frac{2}{3}\frac{16\pi^2 F_{0}^2}{M_{\rho}^2}\left[-\frac{23}{18}\ln\frac{M_{\rho}^2}{m_{\mu}^2}+\frac{2}{9}\ln\frac{m_{\mu}^2}{m_e^2}\right]\,,
\eea}

\noi
where in the second line we have use the QCD Large-$N_c$ dominance of the $\rho$ contribution. The overall result is indeed negative, because of the larger weight in the hadronic term. This cancellation, however,  also shows the need of a more refined estimate. Indeed, the most recent evaluation, using the same $e^+e^-$ data as in the lowest-order calculation, gives~\cite{HMNT03}
\begin{equation}\lbl{eq:hagi}
a_{\mu}^{(6)}\left({\rm H~and}~\frac{m_{\mu}}{m_e}\,,{\rm H}\right)_{\rm vp}=\left(-97.9\pm 0.9_{\mbox{\rm
\tiny exp}}
\pm 0.3_{\mbox{\rm
\tiny rad}}\right)\times 10^{-11}\,.    
\end{equation}
This determination goes down to $a_{\mu}^{(6)}({\rm H~and}~m_{\mu}/m_e\,,{\rm
 H})_{\rm vp}=-101\pm 1 \times 10^{-11}$ if the $\tau$ data mentioned
 previously are used.

\begin{figure}[h]

\begin{center}
\includegraphics[width=0.6\textwidth]{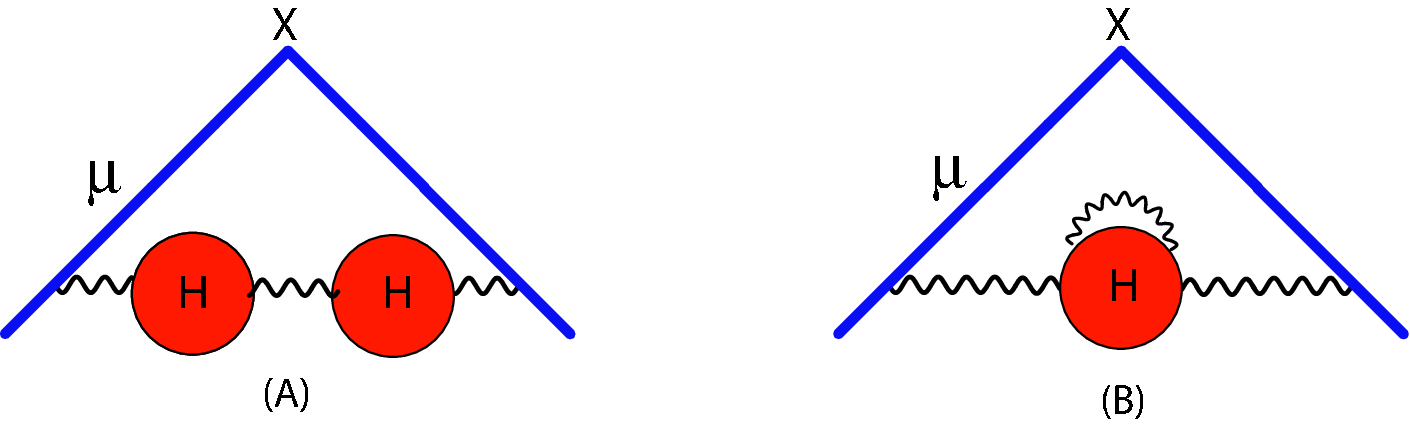}

\vspace*{0.5cm}
\caption { Further contributions from higher-order hadronic vacuum polarization.}\label{figure:hvphvp}
\end{center}

\end{figure}

 As first mentioned in Reference~\cite{CNPdeR76}, at
 $\cO\left[\left(\frac{\alpha}{\pi}\right)^{3}\right]$,  there are still two
 extra contributions from hadronic vacuum polarization which have to be
 considered. They are shown in Figure~\ref{figure:hvphvp}. The evaluation of
 the diagram in Figure~\ref{figure:hvphvp}(A)  can, in principle, be achieved
 by lumping 
together in the lowest-order hadronic vacuum polarization,
 the effect of replacing the {\it one-photon} exchange cross-section
 in Equation~\rf{onephoton} by the {\it physical} cross-section which
 incorporates the full hadronic vacuum polarization (i.e., not just the
 proper hadronic self-energy). The contribution from the diagram in
 Figure~\ref{figure:hvphvp}(B) can also be incorporated in the lowest-order
 evaluation, if the one-photon-exchange hadronic cross-section also takes
 into account final states with one-photon emission {\it as well as} the one
 photon radiative corrections to the hadronic vertex. Unfortunately, extracting the
 latter from the full data, involves a certain amount of model dependence,
 with uncertainties which are difficult to ascertain from
 the discussions in the literature. In principle, these uncertainties should be reflected in the {\it radiative correction} error quoted in Eq.~\rf{eq:hagi}.
 %%%%%%%%%%%%%%%%%%%%%%%%%%%%%%%%%%%%%%
   
\subsection{Contributions from 
Hadronic Light-by-Light Scattering}\label{HLLS}

\begin{figure}[h]

\begin{center}
\includegraphics[width=0.6\textwidth]{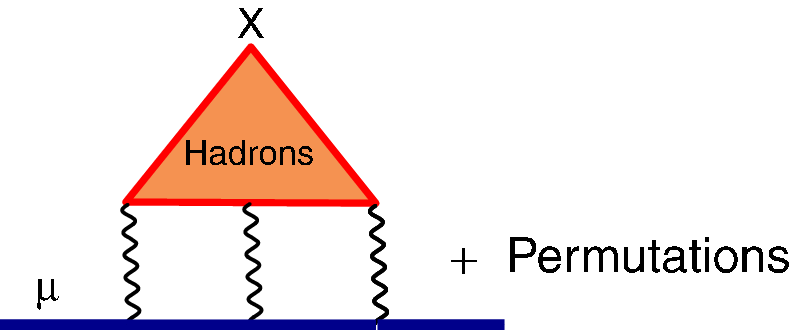}

\vspace*{0.5cm}
\caption{  Hadronic Light-by-Light Contributions}\label{figure:hll}
\end{center}
\end{figure}

\noi
These are the contributions illustrated by the 
diagrams in Figure~\ref{figure:hll}.
All the estimates of these contributions made so far are model dependent.
There has been progress, however, in identifying the dominant regions of
virtual momenta, and in  using models which incorporate some of the
required features of the underlying QCD theory. The combined frameworks
of QCD in the $1/N_c$-expansion and of chiral perturbation
theory\cite{deR94} have been very useful in providing a
guideline to these estimates.

%%%%%%%%%%%%%%%%%%%%%%%%%%%%%%%%%%%%%%
%%%%%%%%%%%%%%%%%%%%%%%%%%%%%%%%%%%%%%

\begin{figure}[h]

\begin{center}
\includegraphics[width=0.8\textwidth]{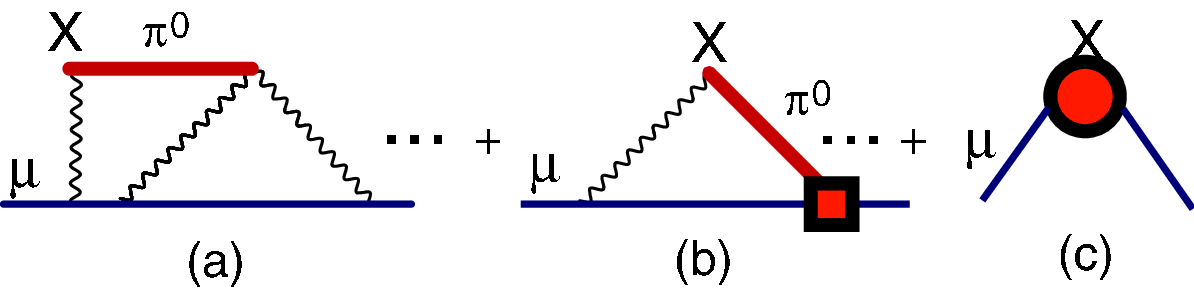}

\vspace*{0.5cm}
\caption {  One Goldstone Reducible Diagrams in Chiral
Perturbation Theory}\label{figure:hllch}
\end{center}

\end{figure}

So far, the only rigorous result in this domain has come from the
observation\cite{KNPdeR02} that, in large-$N_c$ QCD and to leading-order
in the chiral expansion, the dominant contribution to the muon $g-2$ from
hadronic light-by-light scattering comes from the contribution of the
diagrams which are one-particle (Goldstone-like) reducible.
The diagrams in Figure~\ref{figure:hllch}(a) produce a
$\ln^2\left(\mu/m\right)$-term modulated by a coefficient which has been calculated analytically\cite{KNPdeR02}, with the result:
\be\lbl{piefft}
a_{\mu}[{\rm Fig.}~\ref{figure:hllch}]=\left(\frac{\alpha}{\pi}
\right)^3\left\{{\frac{N_c^2}{48\pi^2}
\frac{m_{\mu}^2}{F_{\pi}^2}}\ln^2\left(\frac{{\mu}}{m} \right) +
\cO\left[\ln\left(\frac{{\mu}}{m}\right)+\kappa(\mu)
\right]\right\}\,.
\ee
Here, $F_{\pi}$ denotes the pion coupling constant in the chiral limit
($F_{\pi}\sim 90~\MeV$);  the
$\mu$-scale in the logarithm is an arbitrary ultraviolet (UV)-scale, and the
$m$-scale is an infrared (IR) mass, either $m_{\mu}$ or $m_{\pi}$ if only the leading term in Equation~\rf{piefft} is known.

%%%%%%%%%%%%%%%%%%%%%%%%%%%%%%%%%%%%%%
%%%%%%%%%%%%%%%%%%%%%%%%%%%%%%%%%%%%%%%
\noi
The
arbitrariness on the UV and IR scales in Equation~\rf{piefft} would be removed, if one
knew the terms linear in $\log\mu$ from Figure~\ref{figure:hllch}(b), as well as the constant
$\kappa(\mu)$ from the local counter-terms generated by Figure~\ref{figure:hllch}(c).
Unfortunately, neither the determination of  the coefficient of the
$\log\mu$-term, nor the constant $\kappa(\mu)$-term, can
 be made in a completely
 model-independent
way\footnote{Reference~\cite{KNPdeR02} provides a discussion of this point
using a renormalization group approach. Essentially the same arguments
have been subsequently emphasized in Reference~\cite{RW02}.}. 

The nice feature about 
Equation~\rf{piefft} is that, as discussed in Reference~\cite{KNPdeR02}, it plays a fundamental role in fixing the overall sign
of the hadronic light-by-light scattering contribution to the muon
$g-2$. Indeed, in the various hadronic model calculations of this contribution,
there appear hadronic scales (like the $\rho$-mass), which act
as a UV-regulator and  play the role of $\mu$ in Equation~\rf{piefft}.
Therefore, letting the hadronic leading mass--scale become large, and provided that
the model incorporates correctly the basic chiral properties of
the underlying QCD theory, must reproduce the characteristic
universal
$\ln^2(\mu)$ behavior  with the \underline{same}
coefficient as in Equation~\rf{piefft}. This test, when applied to the most recent earlier
calculations~\cite{HK98,BPP96} (prior to the Knecht-Nyffeler calculation
in Reference~\cite{KNb01}), {\it failed} to reproduce the sign of the coefficient
of the
$\ln^2(\mu)$-term in Equation~\rf{piefft}. The authors of References~\cite{HK98,BPP96} have later
found errors in their calculations which, when corrected, reproduce the
effective field theory test. Their results now agree with the
Knecht-Nyffeler calculation~\cite{KNb01} on which we next report.

\begin{description}

\item[]{\it The Knecht-Nyffeler Calculation~\cite{KNb01}}

In full generality, the pion pole contribution to the muon anomaly has
an hadronic structure, as represented by the shaded blobs in Figure~\ref{figure:hllpi}.
The authors of Reference~\cite{KNb01} have shown that, for a large class of
off-shell $\pi^{0}\gamma\gamma$ form factors (which includes the
large-$N_c$ QCD class), this contribution has an
integral representation over two Euclidean invariants $Q_1^2$ and
$Q_2^2$ associated with the two loops in Figure~\ref{figure:hllpi}:
\be
a_{\mu}[{\rm Fig.}~\ref{figure:hllpi}]=\int_{0}^\infty \!\!dQ_{1}^2\int_{0}^{\infty}\!\!
dQ_{2}^2\ \ \cW(Q_{1}^2,Q_{2}^2)
\cH(Q_{1}^2,Q_{2}^2)\,,
\ee
where $\cW(Q_{1}^2,Q_{2}^2)$ is a skeleton kernel which they
calculate explicitly, and   
$\cH(Q_{1}^2,Q_{2}^2)$ is a convolution of
two generic 
$\cF_{\pi^{0}\gamma^{*}\gamma^{*}}(q_{1}^2,q_{2}^2)$
form factors. In Large-$N_c$ QCD,
\be\lbl{pigg}
\cF_{\pi^{0}\gamma^{*}\gamma^{*}}
(q_{1}^2,q_{2}^2)\big\vert_{N_c\ra\infty}=\sum_{ij}
\frac{c_{ij}(q_{1}^2,q_{2}^2)}{(q_{1}^2-M_{i}^2)(q_{2}^2-M_{j}^2)}\,,
\ee
with the sum extended to an infinite set of narrow states. 
%%%%%%%%%%%%%%%%%%%%%%%%%%%%%%%%%%%%%%
\begin{figure}[h]

\begin{center}
\includegraphics[width=0.6\textwidth]{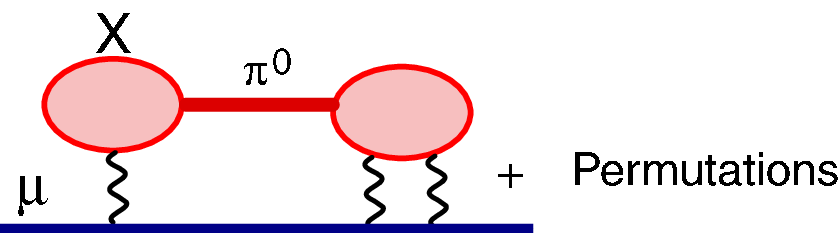}

\vspace*{0.5cm}
\caption {  Hadronic light-by-light from a $\pi^{0}$ intermediate 
state.}\label{figure:hllpi}
\end{center}

\end{figure}
%%%%%%%%%%%%%%%%%%%%%%%%%%%%%%%%%%%%%%%
In practice, the calculation in\cite{KNb01} has been done by restricting
the sum in Equation~\rf{pigg} to one- and two-vector states, and fixing the
polynomial
$c_{ij}(q_{1}^2,q_{2}^2)$ from short-distance and
long-distance QCD properties. This way, they obtained the result
\be
{a_{\mu}^{(6)}(\pi^0 )_{\rm lxl}=(5.8 \pm 1.0)\times 10^{-10}}\,,
\ee
where the error also includes an estimate of the hadronic approximation.
Further inclusion of the $\eta$ and $\eta\prime$ states results in a
final estimate
\be
{a_{\mu}^{(6)}(\pi^0 +\eta+\eta' )_{\rm lxl}=(8.3 \pm 1.2)\times 10^{-10}}\,, 
\ee
in agreement with the earlier calculations by Bijnens {\it at al}~\cite{BPP96} and Kinoshita {\it et al }~\cite{HK98} (after correcting for the overall sign).

\item[]{\it Comment on the Constituent Quark Model (CQM)}

%%%%%%%%%%%%%%%%%%%%%%%%%%%%%%%%%%%%%%
\begin{figure}[h]

\begin{center}
\includegraphics[width=0.6\textwidth]{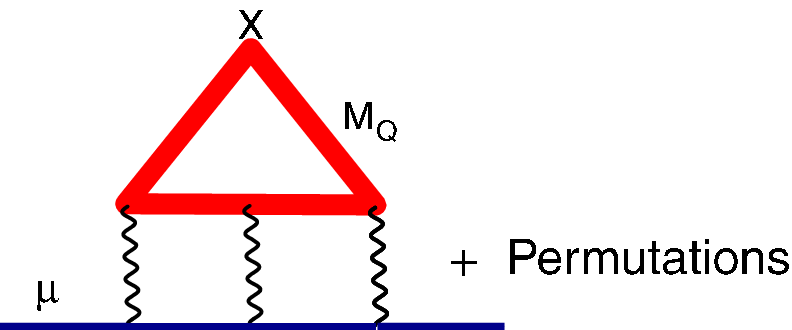}

\vspace*{0.5cm}
\caption {  Hadronic light-by-light in the constituent quark
mode}\label{figure:hllcq}
\end{center}

\end{figure}
%%%%%%%%%%%%%%%%%%%%%%%%%%%%%%%%%%%%%%%

We would like to comment on an argument which is often used
in favor of the {\it Constituent Quark Model} as a {\it ``simple''} way to
 fix the sign and size of the hadronic light-by-light scattering contribution to
the muon $g-2$. Since the argument has even been advocated in
some publications, we feel obliged to refute it here, with the 
hope that it will stop further confusion.
The constituent quark model contribution from the diagram in Figure~\ref{figure:hllcq},
can be easily extracted from the work of Laporta and Remiddi~\cite{LR93} for a heavy lepton contribution (see Equation~\rf{mutau6} above), with the result
\be
\hspace*{-2cm}
a_{\mu}^{(6)}({\rm CQM})_{\rm lxl}\!=\!\left(\frac{\alpha}{\pi}
\right)^3\!N_{c}{\frac{2}{9}}
\left\{{\underbrace{\left[\frac{3}{2}\zeta(3)
-\frac{19}{16}
\right]}_{0.616} }\left(\frac{m_{\mu}}{{M_{Q}}}
\right)^2 \!\!+ \cO\!\!\left[
\left(\frac{m_{\mu}}{{M_{Q}}}
\right)^4 \!\log^2 \left(\frac{{M_{Q}}}{m_{\mu}}
\right) \right]\right\}\,.
\ee
Seen from a low-energy-effective-field-theory point of view, the
constituent quark mass
$M_Q$ in the CQM should provide the UV-regulating scale. However, the
model {\it is not} a good effective theory of QCD and, therefore, it fails
to reproduce the characteristic QCD
$\ln^2 M_Q$ behavior when $M_Q$ is allowed to become arbitrarily large;
the CQM contribution decouples in the large $M_Q$-limit.
 Incidentally, the contribution from free u and d quarks with a
 small mass $m\ll \mu$ goes as   
$\sim\left(\frac{\alpha}{\pi}
\right)^3\!N_{c}{\frac{7}{27}}\times \frac{2}{3}\pi^2 \ln\frac{m_{\mu}}{m}$, 
which is also incompatible with the QCD result in Equation~\rf
{piefft}. 
Therefore, 
arguments based on the fact that the CQM (and/or pQCD) 
gives a {\it positive} contribution are certainly 
{\it ``simple,''} but also {\it incorrect}.
Notice however, that, contrary to the naive CQM, the constituent
chiral quark model of Georgi and Manohar\cite{GM84} (see also
Reference~\cite{EdeRT90}) does indeed reproduce the correct $\ln^2 M_Q$
behavior in the
$M_Q\ra\infty$ limit. This is because, in these models, the  Goldstone
particles couple with the constituent quarks in a way which respects
chiral symmetry, and the pion pole diagram appears then explicitly. The
same happens in the extended version of the Nambu--Jona-Lasinio model~\cite{BBdeR93}. These models, however, suffer
from other diseases~\cite{PPdeR98}, and therefore
they are not fully reliable to compute
the hadronic light-by-light scattering contribution. 
\end{description}

Hopefully, hadronic models of the
 light-by-light scattering contribution which respect the QCD constraint
 in Equation~\rf{piefft}
will be progressively improved, so as to incorporate further and further QCD
features; in particular, short-distance constraints, following the
lines discussed in References~\cite{deR02,Pe02,deR02a}. An interesting 
 contribution in this direction has  been reported 
in Reference~\cite{MV05}. Unfortunately, as recently discussed in Reference~\cite{BP07} their numerical evaluation
 is incomplete and model dependent with largely underestimated errors. We believe that, at present, one can only
claim to know the hadronic light-by-light scattering contribution with
a {\it cautious} error, which takes into account the
uncertainties from the integration regions which remain model
 dependent. While awaiting further improvement, the educated
 value one can quote
at present, based on the combined work of
References~\cite{KNb01} and\cite{HK98,BPP96} (appropriately corrected) 
as well as Reference~\cite{MV05}, is
\be
a_{\mu}^{(6)}({\rm H})_{\rm lxl}=(11\pm
4)
\times 10^{-10}\,.
\ee  
 
\subsection{Electroweak Contributions}\label{EW}

\begin{figure}[h]

\begin{center}
\includegraphics[width=0.8\textwidth]{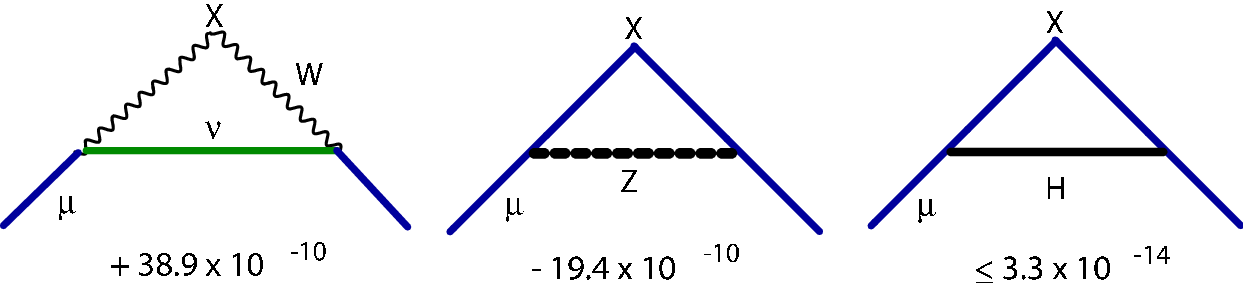}

\vspace*{0.5cm}
\caption { Weak interactions at the one-loop level}\label{figure:electroweak}
\end{center}

\end{figure}

\noi
The leading contribution to the anomalous magnetic moment of the muon from the
electroweak Lagrangian of the Standard Model, 
originates at the one-loop level. The relevant Feynman diagrams, which for simplicity we draw in the unitary gauge, are shown in Figure~\ref{figure:electroweak}, where we also indicate
the size of their respective contributions.
%%%%%%%%%%%%%%%%%%%%%%%%%%%%%%%%%%%%%%
%%%%%%%%%%%%%%%%%%%%%%%%%%%%%%%%%%%%%%%
\noi
 The analytic evaluation of the overall contribution 
gives the result~\cite{EW's}
 
{\setl
\bea\lbl{EW1}
a_{\mu}^{\mbox{\rm\tiny
 W}} & = &
\frac{G_{\mbox\tiny F}}{\sqrt{2}}\frac{m_{\mu}^2}{8\pi^2}
\left[{\frac{5}{3}\!+\!
\frac{1}{3}(1\!-\!4\sin^{2}\theta_{W})^2 \!+\!\cO\left({
\frac{m_{\mu}^2}{M_{Z}^2}\log\frac{M_{Z}^2}{m_{\mu}^2}}\right)}\right.
\nn
\\
& & \left.
\hspace*{2.5cm}+{\frac{m_{\mu}^2}{M_{H}^2}}
\int_{0}^{1}dx 
\frac{2x^2
(2-x)}{1-x+{\frac{m_{\mu}^2}{M_{H}^2}}x^2}
\right]=19.48\times 10^{-10}\,, 
\eea}

\noi
where the weak mixing angle is defined by $\sin^2 \theta_{W}
=1-M^2_W/M_Z^2=0.223$ (for a Higgs mass $M_H=150~\GeV$),
and where $G_F=1.16639(1)\times 10^{-5}$ is the Fermi constant.
Notice that the contribution from the Higgs boson, shown
in parametric form in the second line, is of $\cO\left(\frac{G_{\mbox\tiny F}}{\sqrt{2}}\frac{m_{\mu}^2}{4\pi^2}\frac{m_{\mu}^2}{M_H^2}\ln\frac{M_H^2}{m_{\mu}^2}\right)$, which is already very
small for the present lower bound of $M_H$. Closed analytic expressions of
the electroweak one-loop contribution can be 
found in Reference~\cite{Stu99} and references therein.

Let us recall that the present {\it world average} experimental error
in the determination of the muon anomaly is\cite{bennett3} 
$
\Delta a_{\mu}\big\vert_{\mbox{\rm\tiny Exp.}}\!\!\!=\pm 6.3\times
10^{-10}\,,
$
and, hoping for a continuation of the BNL experiment, it is expected to
be further reduced. The {\it a priori} possibility that the two-loop
electroweak corrections may bring in enhancement factors due to large
logarithmic factors, like $\ln(M_Z^2 /m_{\mu}^2)\simeq 13.5$, has motivated a
thorough theoretical effort for their evaluation, which has been quite a
remarkable 
achievement, and which we next discuss. 

It is convenient to separate the two-loop electroweak contributions into
two sets of Feynman graphs: those containing closed fermion loops,
which we denote by $a_{\mu}^{EW(2)}({\mbox{\rm\footnotesize ferm}})$,
and the others which we denote by 
$a_{\mu}^{EW(2)}({\mbox{\rm\footnotesize bos}})$. With this notation,
the electroweak contribution to the muon anomalous magnetic moment becomes
\be
a_{\mu}^{EW}=a_{\mu}^{W}+a_{\mu}^{EW(2)}({\mbox{\rm\footnotesize
bos}})+a_{\mu}^{EW(2)}({\mbox{\rm\footnotesize ferm}})\,,
\ee
with $a_{\mu}^{W}$ the one-loop contribution in Equation~\rf{EW1}.

\begin{description}
\item[]{\it Bosonic Contributions to $a_{\mu}^{EW(2)}({\mbox{\rm\footnotesize
bos}})$.}

The leading logarithmic terms of the two-loop electroweak bosonic
corrections have been extracted using asymptotic expansion
techniques. In fact, these contributions have
now been evaluated analytically, in a systematic expansion in powers of
$\sin^2\theta_{W}$, up to 
$\cO[(\sin^2\theta_{W})^3]\,,$ where $\log\frac{M_{W}^2}{m_{\mu}^2}$
terms, $\log\frac{M_{H}^2}{M_{W}^2}$ terms, $\frac{M_{W}^2}{M_{H}^2}
\log\frac{M_{H}^2}{M_{W}^2}$ terms, $\frac{M_{W}^2}{M_{H}^2}$ terms,
and constant terms are kept~\cite{CKM96}. Using 
$\sin^2\theta_{W}=0.224$ and $M_{H}=250\,\GeV\,,$ the authors of
Reference~\cite{CKM96} find

{\setl
\bea\lbl{bos}
a_{\mu}^{EW(2)}({\mbox{\rm\footnotesize
bos}}) & = & \frac{\GF}{\sqrt{2}}\,\frac{m_{\mu}^2}{8\pi^2}\,
\frac{\alpha}{\pi}\times 
\left[-5.96\log\frac{M_{W}^2}{m_{\mu}^2}+0.19\right]\nn  \\ & = &
\frac{\GF}{\sqrt{2}}\,\frac{m_{\mu}^2}{8\pi^2}\,
\left(\frac{\alpha}{\pi}\right)\times (-79.3)\,.
\eea}

\item[]{\it Fermionic Contributions to $a_{\mu}^{EW(2)}({\mbox{\rm\footnotesize ferm}})$.}

The discussion of the two-loop electroweak fermionic corrections is
more delicate. Because of the $U(1)$ anomaly
cancellation between lepton loops and quark loops in the electroweak theory, one cannot separate
hadronic from leptonic effects any longer  in  diagrams like the ones shown
in Figure~\ref{figure:WA}, where a triangle with two vector currents and an 
axial-vector current appears (the so-called VVA-triangle). 
Individually, the lepton-loop and quark-loop contributions are each 
gauge-dependent;  depending on the gauge choice,
they can even lead to UV-divergent contributions.
 Only the sum of contributions within each family of leptons 
and quarks is free from these ambiguities. As first discussed in
References\cite{PPdeR95,CKM95}, it is this anomaly cancellation which
 eliminates some
of the large logarithms that were incorrectly kept in a previous
calculation in Reference~\cite{KKSS92}. It is therefore appropriate to separate
the two-loop electroweak fermionic corrections into two classes. One is
the class arising from Feynman diagrams like the
ones in Figure~\ref{figure:WA}, where a subgraph with a VVA-triangle of 
leptons and quarks appears, including the graphs  where the
$Z$ lines are replaced by
$\Phi^{0}$ lines, if the calculation is done in the $\xi_{Z}$-gauge.  We
denote this class by
$a_{\mu}^{EW(2)}(l,q)\,.
$ The other class is defined by the rest of the diagrams, where quark
loops and lepton loops can be treated separately, which we call
$a_{\mu}^{EW(2)}({\mbox{\rm\footnotesize ferm-rest}})$
i.e.,
$$
a_{\mu}^{EW(2)}({\mbox{\rm \footnotesize
ferm}})=a_{\mu}^{EW(2)}(l,q)+
a_{\mu}^{EW(2)}({\mbox{\rm\footnotesize ferm-rest}})\,.
$$

%%%%%%%%%%%%%%%%%%%%%%%%%%%%%%%%%%%%%%
\begin{figure}[h]

\begin{center}
\includegraphics[width=0.6\textwidth]{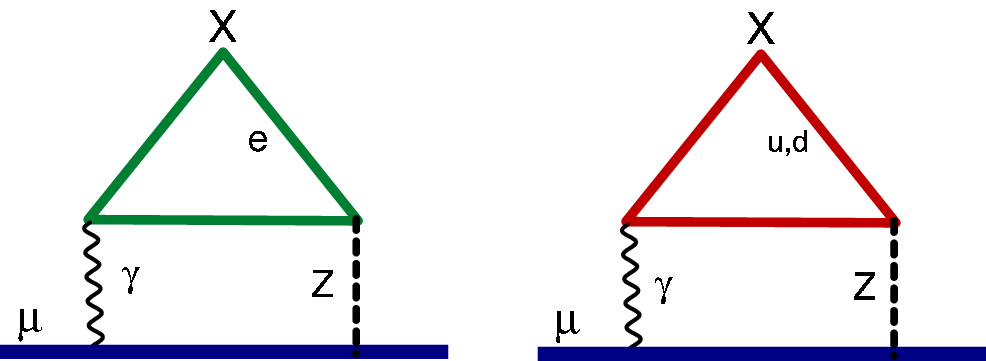}

\vspace*{0.5cm}
\caption { Two-loop electroweak diagrams generated by the
$\gamma\gamma Z$-Triangle. There are similar diagrams corresponding to the $\mu\,; c,s$ and $\tau\,; t,b$ generations. }\label{figure:WA}
\end{center}

\end{figure}
%%%%%%%%%%%%%%%%%%%%%%%%%%%%%%%%%%%%%%%

The contribution from $a_{\mu}^{EW(2)}({\mbox{\rm\footnotesize
ferm-rest}})$ brings in $m_{t}^2/M_{W}^2$ factors. It has been
estimated, to a very good approximation, in Reference~\cite{CKM95} with the
result,
\be\lbl{ferrest}
a_{\mu}^{EW(2)}({\mbox{\rm\footnotesize
ferm-rest}})=\frac{\GF}{\sqrt{2}}\,\frac{m_{\mu}^2}{8\pi^2}\,
\frac{\alpha}{\pi}\times\left(-21\,\pm\,4 \right)\,,
\ee
where the error here is the one induced by diagrams with Higgs 
propagators with an allowed Higgs mass in the
range $114~\GeV < M_H < 250~\GeV$.

Concerning the contributions to
$a_{\mu}^{EW(2)}(l,q)$, it is convenient to treat
the three generations separately. The
contribution from the third generation can be calculated in a
straightforward way, because all the fermion masses in the triangle loop are large with respect to the muon mass,  with the result~\cite{PPdeR95,CKM95}

{\setl
\bea\lbl{3rdg}
\hspace*{-1cm}
a_{\mu}^{EW(2)}(\tau,t,b) & = & \frac{\GF}{\sqrt{2}}\,\frac{m_{\mu}^2}
{8\pi^2}\,
\frac{\alpha}{\pi}  \times
 \left[-3\log\frac{M_{Z}^2}{m_{\tau}^2}-\log\frac{M_{Z}^2}{m_{b}^2}-
\frac{8}{3}\log\frac{m_{t}^2}{M_{Z}^2}+\frac{8}{3}\right. \nn \\
 &+ & \left. 
\cO\left(\frac{M_{Z}^2}{m_{t}^2}\log\frac{m_{t}^2}{M_{Z}^2}
\right)
\right] =\frac{\GF}{\sqrt{2}}\,\frac{m_{\mu}^2}
{8\pi^2}\,
\frac{\alpha}{\pi}\times (-30.6)\,.
\eea}

\noi
The leading terms can be easily obtained using effective field theory techniques~\cite{PPdeR95}; but
terms of $\cO\left(\frac{M_{Z}^2}{m_{t}^2}
\log\frac{m_{t}^2}{M_{Z}^2}
\right)$ and  $\cO\left(\frac{M_{Z}^2}{m_{t}^2}\right)$ have also been
calculated in Reference~\cite{CKM95}. There are in principle QCD
perturbative corrections to this estimate which have not been
calculated, but the result in Equation~\rf{3rdg} is good enough for the
accuracy required at present.

As emphasized in Reference~\cite{PPdeR95}, an appropriate QCD calculation
when the quark in the loop of Figure~1 is a {\it light quark} should
take into account the dominant effects of spontaneous chiral-symmetry
breaking. Since this involves the $u$ and $d$ quarks, as well as the
second-generation $s$ quark, it is convenient to lump together the
contributions  from the first and second generations. An
 evaluation of these contributions, which
incorporates the QCD long-distance chiral realization~\cite{PPdeR95,KPPdeR02} as well as perturbative~\cite{CMV03} and non-perturbative~\cite{KPPdeR02} 
short-distance constraints, gives the result
\be\lbl{12gs}
a_{\mu}^{EW(2)}(e,\mu,u,d,s,c)=\frac{\GF}{\sqrt{2}}\,\frac{m_{\mu}^2}
{8\pi^2}\,
\frac{\alpha}{\pi} \times (-24.6\pm 1.8)\,.
\ee

\noi
From the theoretical point of view, this calculation has revealed
surprising  properties concerning the {\it non-anomalous} component of the
VVA-triangle~\cite{Vain03},
 with a new set of {\it non-renormalization theorems} in perturbation theory~\cite{Vain03,KPPdeR04} (see also References~\cite{JT05,PT06,Dorok05}). The physical meaning of these perturbation theory results when considered in the full QCD vacuum is, however,  still unclear~\cite{KPPdeR04}.

\end{description}
   
Putting together the numerical results in Eqs.\rf{bos}, \rf{ferrest},
\rf{3rdg} with the more recent result in Equation~\rf{12gs}, one finally obtains the
value

{\setl
\bea
a_{\mu}^{EW} & = & \frac{\GF}{\sqrt{2}}\,\frac{m_{\mu}^2}
{8\pi^2}\left\{\frac{5}{3}\!+\!\frac{1}{3}
\left(1\!\!-\!\! 4\sin^2\theta_{W}\right)^2\!\!-\!\left(
\frac{\alpha}{\pi}\right) [155.5(4)(1.8)]\right\} \\ & = & 15.4(0.2)(0.1)
\times 10^{-10}\,, 
\eea}

\noi
where the first error is essentially due to the Higgs mass
 uncertainty, while the second comes from higher-order
 hadronic effects in the VVA loop evaluation. The overall
 result shows indeed that the two-loop correction represents
 a sizable reduction
of the one-loop result by an amount of $21\%\,.$ This result has 
prompted the evaluation of the electroweak leading-log three-loop
 effects of $\cO\left[\frac{\GF}{\sqrt{2}}\,\frac{m_{\mu}^2}
{8\pi^2}\left(\frac{\alpha}{\pi}\right)^2 \ln\frac{M_Z}{m_{\mu}}\right]$,
 using renormalization group arguments similar to the ones discussed in 
\S\rf{vprga}. The results~\cite{DG98,CMV03} show that those effects
 are negligible [$\sim\cO(10^{-12})$] for the accuracy needed at present.

%%%%%%%%%%%%%%%%%%%%%%%%%%%%%%%%%%%%%%%%%%%%%%%%%%
\subsection{Summary of the Standard Model Contributions to the Muon Anomaly}
%%%%%%%%%%%%%%%%%%%%%%%%%%%%%%%%%%%%%%%%%%%%%%%%%%

The present status in the evaluation of the anomalous
magnetic moment of the muon in the Standard Model can be summarized as
follows:

\begin{itemize}
\item {\it Leptonic QED Contributions}
\be
\hspace*{-1.5cm}
a_{\mu}({\mbox{\rm\small
QED}})=({116~584~718.09\pm 0.14_{5\rm loops}\pm 0.08_{\alpha}
\pm 0.04_{\rm masses})\times 10^{-11}}\,.
\ee
This is the number one obtains by
adding to the final result at the bottom of Table~\ref{table:qed}
(using recent~\cite{gabalpha} value of $\alpha$) to
the estimate of the five-loop contribution in Equation~\rf{fiveloops}.

\item {\it Hadronic Contributions}
\begin{itemize}
\item{\it Hadronic Vacuum Polarization with EM-Data}

From the two most recent determinations in Table~\ref{table:hvp}, 
which take into account the new data on $e^+ e^-$ annihilation 
into hadrons, we get
\be\lbl{preeidel}
a_{\mu}[\rm HVP(06)]=(6901 \pm 42_{\mbox{\rm
\tiny exp}}
\pm 19_{\mbox{\rm \tiny rad}} \pm 7_{\mbox{\rm \tiny QCD}})\times 10^{-11}\,,
\ee
where we have averaged the central values, but kept the largest uncertainties
in the two determinations. Notice that this result is consistent
 with the average of the central values
 of the earlier  determinations from $e^+ e^-$ data
 in the second part of
 Table~\ref{table:hvp}, which when keeping the largest error from experiment  
and from radiative corrections give:
$a_{\mu}[\rm HVP]=(6945 \pm 62_{\mbox{\rm
\tiny exp}} \pm 36_{\mbox{\rm
\tiny rad}})\times 10^{-11}$.

\item{\it Higher-Order Hadronic Vacuum Polarization with EM-Data}

\begin{equation}
a_{\mu}[\rm HVP~h.o.]=\left(-97.9\pm 0.9_{\mbox{\rm
\tiny exp}}
\pm 0.3_{\mbox{\rm
\tiny rad}}\right)\times 10^{-11}\,.    
\end{equation}

\item {\it Hadronic Light-by-Light Scattering}
\be
a_{\mu}[\rm HLLS]=(110\pm
40)
\times 10^{-11}\,. \label{eq:hlls}
\ee
\end{itemize}
\item {\it Electroweak Contributions }
\be
a_{\mu}[EW]=(154\pm
2\pm1)\times 10^{-11} 
\ee
\end{itemize}

\noi
The sum of these contributions, using the HVP06 result in Equation~\rf{preeidel} and adding experimental and theoretical
errors in quadrature, gives then a total
\be\label{eq:sm06}
a_{\mu}^{\mbox{\rm\tiny SM(06)}}=11~659~1785~(61)\times 10^{-11}\,.
\ee
These determinations are to be compared to the experimental world average in Equation~(\ref{eq:E821wa})
\be
a_{\mu}^{\mbox{\rm\tiny
exp}}~=11~659~2080~(63)\times
10^{-11}\,. 
\ee
Therefore, we conclude that, with the input for the Standard-Model contributions discussed
above, one finds at present a  $3.4~\sigma$ discrepancy. 
 
%%%%%%%%%%%%

%% file: future-exp20feb07.tex
\section{Future Experimental Possibilities}

The discrepancy between the experiment and the theory 
gives strong motivation to improve the experimental number.
An upgraded experiment, E969\cite{E969}, with
goals of $\sigma_{\rm syst} = 0.14$~ppm
and $\sigma_{\rm stat} = 0.20$~ppm received scientific approval from
from Brookhaven Laboratory.   Despite the
3.4~$\sigma$ difference between theory and experiment,
the new experiment remains unfunded. Such an experiment would stimulate
additional theoretical work which would further decrease the theoretical
error. The ultimate experimental sensitivity of a future experiment depends
both on the available beam intensity, and on the ability to further 
reduce the systematic errors. 
 
The systematic and statistical errors by year for E821, 
along with  the goals for E969,  are given in Table~\ref{tb:error-all}.
There is a clear improvement in the systematic errors with time, and the
improvement required for E969 is modest and should be readily achievable
with planned upgrades in the detector system, electronics, and the
field mapping system. In order to achieve the required
number of muons for E969,
a plan to improve the beamline in order to increase the
stored muon flux by a factor of five has been proposed.

\begin{table}[h]
\begin{center} \caption{Systematic and statistical errors
in ppm for each of the
e821 running periods. }
\label{tb:error-all}
\begin{tabular}{|l|c|c|c|c|c|}
\hline
Systematic uncertainty  &  1998 &  1999 &  2000 &  2001 &  E969 Goal\\
\hline
Magnetic Field $\omega_p$ (ppm)  & 0.5 &      0.4  &  0.24 &  0.17 &  0.1 \\
\hline
Anomalous Precession $\omega_a$ (ppm) &  0.8 &   0.3 &   0.31 &  0.21 &  0.1 \\
\hline
Statistical Uncertainty (ppm) & 4.9 &   1.3 &   0.62 &  0.66 &  0.20 \\
\hline
Total Uncertainty (ppm)   &     5.0 &   1.3 &   0.73 &  0.72 &  0.25 \\
\hline
\end{tabular}
\end{center}
\end{table}

The E821 experimental approach, which combines the ``magic $\gamma$,'' 
electrostatic focusing, 
and a fast muon kicker has the potential to push the experimental error
below the 0.1 ppm level, perhaps as far as about 0.05~ppm
before too many technical barriers are reached.  To improve the
statistical error on $\omega_a$
the muon flux would have to be increased substantially, and
presently such a source does not exist.
As the muon flux is increased,
systematic errors from pulse pile-up and detector instabilities generated 
by high rates become more severe.  To improve the $\omega_p$ measurement,
additional work would have to be done on the absolute calibration, which 
at present depends on the measurement of Phillips, et al.,\cite{phillips}
to relate the magnetic moment of a proton in a spherical water sample to
that of the free proton.  An alternate calibration based on a
$^3He$ probe has been suggested\cite{KJpc}.  
To go beyond 0.1 ppm, it would also be necessary to further
refine the measurement of $\lambda_+$ with a new muonium hyperfine
splitting measurement\cite{KJpc}. 

Progressing beyond 0.05~ppm might require a new technique.
For example, increasing
$\gamma \tau$ by going to a
higher muon momentum and increasing $f_a$ by increasing
the magnetic field would reduce the statistical error
without requiring more muons (see Equation~\ref{eq:fracterr}). However,
a higher momentum would require abandoning the advantages of the
magic gamma. Increasing the magnetic field would  require the
elimination of the iron yoke and pole tips in favor of a
magnetic field driven
entirely by superconducting coils.
A proposal to
use sector focusing dipoles has been made\cite{future-ex}, but the
need to know  $\int \vec B \cdot d \vec \ell$
for the muons  to a precision of 10~ppb with magnetic gradients  present
 is challenging, to say the least.

%% file: summary20fev07.tex
\section{Summary and Conclusions}

Measurement of the muon's anomalous magnetic moment has a history 
stretching back to the late 1950s.  In the paper from Nevis
Laboratory that presented evidence
for parity violation in muon decay~\cite{garwin1}, it was reported 
that the data were consistent with $g=2$.  In a subsequent paper
from Nevis~\cite{garwin2}, it was
demonstrated that the muon anomaly was consistent with the
QED calculation of Schwinger of  $\sim \alpha/2 \pi$,
and thus provided the first evidence, confirmed by all existing 
subsequent data, that the muon behaves like a heavy 
electron.

The result from E821 at the Brookhaven National Laboratory
is sensitive to extremely refined 
effects predicted by the Standard Model, such as the hadronic
 light--by--light scattering and the electroweak 
contributions. If these two effects were removed from the 
theoretical prediction obtained in Equation~\ref{eq:sm06},
the discrepancy with the experimental
result would be $7.1~\sigma$.  If only the electroweak contribution is
removed, the discrepancy is 5.1~$\sigma$, clearly showing the sensitivity of the E821 experiment to the electroweak contribution. 
The largest uncertainties in the theoretical prediction of
$a_{\mu}$
come,  from the systematic errors in the experimental data
from $e^+ + e^-$ annihilation to hadrons
 (in spite of the fact that the accuracy in the present
 determination  $(< 1\%)$ is quite remarkable);
and from the theoretical error in the hadronic light-by-light contribution.  

There is progress to be expected as
additional $e^+  e^-$ data become available. 
The good news is that the $\pi\pi$ data from two 
independent experiments at the Novosibirsk collider, CMD-2 and SND, are
in agreement.  There seems to be some disagreement between
the shape of the pion form factor obtained by the KLOE experiment
at Frascati and that from the Novosibirsk experiments. 
Further experimental input in the near future is
 expected  on $e^+e^- \rightarrow \pi^+ \pi^-$ 
from a new KLOE analysis
 with a tagged photon in the detector, and from BaBar using initial-state
 radiation.  At present the data from Novosibirsk dominate the
determination of $a_\mu[{\rm HVP}]$ in the $\rho$-meson region,
making an independent confirmation of the $\pi\ \pi$ cross section in this
energy region clearly desirable.

On the other hand, the puzzle centered on hadronic $\tau$-decay data 
versus $e^+  e^-$ data remains in spite of efforts to quantify
the CVC corrections. Problems exist in comparisons between
the two that are quite
independent of the contribution to $a_\mu$.
For example the failure to predict the measured 
$\tau$ branching fractions from 
the $e^+e^-$ data and the conserved vector current hypotheses,  at present 
invalidate attempts to use these $\tau$ data to obtain 
the hadronic contribution\cite{Davier06}. 
Work on these issues in $\tau$ physics is continuing.
We conclude that until the issues with $\tau$-decay data are resolved,
the $e^+e^-$ data are preferred to evaluate 
the lowest-order hadronic contribution to $a_\mu$. 

One should stress the impressive achievement reached
 in the calculation of the QED contribution, where the dominant error at present comes
 from the five-loop contribution.  
Thanks to the dedicated efforts of T.~Kinoshita and collaborators, this
uncertainty is likely to go down significantly
 in the near future.  Great progress has also been made in understanding 
the electroweak contribution, which is now fully calculated at the two-loop 
level with an accuracy of $1.5\%$, an uncertainty dominated by the unknown
Higgs mass. The main effort
 should now be concentrated on a more
 accurate determination of the hadronic light--by--light contribution.
 A goal of $15\%$ accuracy in its determination seems possible
 with a dedicated effort from theorists. 

An improvement in the experimental value of the
muon anomaly by at least a 
factor of two appears to be quite feasible, and is the goal of the
proposed E969 experiment at Brookhaven National Laboratory. 
Given the current 3.4-standard-deviation difference 
between theory and experiment, and the prospects for the uncertainty
in the theoretical value to decrease over the next few years, 
such a measurement appears timely and highly desirable.

Historically the muon anomaly has provided an important 
 constraint on speculative
theories of physics beyond the Standard
 Model~\cite{Czar01,Martin03,Stockinger_07}.  
As particle physics moves into the Large Hadron Collider era 
in the next few years at CERN, the muon anomaly will provide an
independent constraint on the interpretation of the discoveries
made at the LHC.

\section{Acknowledgments}

The Brookhaven based
experimental work reported on in this paper was carried out over a
twenty-year period by
the E821 collaboration, and we gratefully acknowledge the many contributions
that each member~\cite{bennett3,carey,brown1,brown2,bennett1,bennett2} made
to its success. In this review of E821
we have tried to  report accurately the work that led to the final precision
of 0.54~ppm on the muon anomaly, and we (JPM and BLR) accept full
responsibility for any shortcomings of this exposition.  Earlier reviews
on aspects of muon $(g-2)$ have been written by several of our 
colleagues\cite{orev}.
We thank our colleagues on E821 and E969 for many
helpful conversations on various aspects of the g-2 experiments.
We are especially thankful to R.M. Carey and K. Ellis
for their significant editorial suggestions. We thank
F.J.M. Farley, D.W. Hertzog and K.R. Lynch
for their comments on this manuscript.  BLR acknowledges useful
comments by Andrzej Czarnecki on non-standard model contributions.
We acknowledge K.J. Jungmann and  R. Prigl for their
helpful comments on the sections about the magnetic field.
The preparation of this manuscript was supported in part by the U.S.
National Science Foundation.
E821 was supported in part by the U.S. Department of
Energy, the U.S. National Science Foundation, the German
Bundesminister fu¨r Bildung und Forschung, the Russian
Ministry of Science, the US-Japan Agreement in High
Energy Physics, the NATO Office of Scientific Research.
The work of EdeR has been supported in part by TMR, EC-Contract No.
HPRN-CT-2002-0031 (EURIDICE) and by the European Community's Marie Curie
Research Training Network program under contract No. MRTN-CT-2006-035482
(FLAVIAnet).